\documentclass[coverpage,hyper,s5paper,11pt]{su_thesis}		
\usepackage{pdfpages}
\usepackage{ifthen}
\usepackage{amssymb}
\usepackage[fleqn]{amsmath}
\usepackage{graphicx}
\usepackage[bf, sf, FIGTOPCAP, normal, tight]{subfigure}
\usepackage{changepage}
\usepackage{calc}
\usepackage{cancel}
\usepackage{feynmf}
\usepackage{multicol}
\usepackage{multirow}
\usepackage{bigstrut}
\usepackage{rotating}
\usepackage{booktabs}
\usepackage{hhline}
\usepackage[applemac]{inputenc}
\def\jhep{JHEP}%
\def\jcap{JCAP}%

\newcounter{papercount}

\newlength{\paperboxheight}
\newlength{\paperboxwidth}
\newlength{\verticalpaperboxoffset}

\newcommand{\paperboxbackgroundcolor}{black}
\newcommand{\paperboxtextcolor}{Goldenrod}
\newcommand{\papersummary}[8]{

  \begin{description}
    \item[\bfseries\sffamily #8] {#1}. \emph{{#2}}, {#3}~\textbf{{#4}}, {#5} ({#6}) {#7}.
  \end{description}

}

\newcommand{\paper}[9]{

  \ifodd \value{page}
  \else
    \newpage
    \null
    \newpage
  \fi
  
  \addtocounter{papercount}{1}
  \vspace*{\verticalpaperboxoffset}

  \begin{adjustwidth*}{}{-6.2em}
    \begin{flushright}
      \huge\sffamily\bfseries \color{\paperboxtextcolor} \colorbox{\paperboxbackgroundcolor}
        {\parbox[c][\paperboxheight]{\paperboxwidth}{\hspace{1cm}Paper \Roman{papercount}}}
    \end{flushright}
  \end{adjustwidth*}

  \vfill
  
  \begin{flushleft}
    {#1}\\
    \emph{{#2}}\\
    {#3}~\textbf{{#4}}, {#5} ({#6}) {#7}.
  \end{flushleft}

  \phantomsection
  \addcontentsline{toc}{chapter}{Paper \Roman{papercount}: #2}
  \label{#9}

  \newpage
  \null
  \newpage

  \addtolength{\verticalpaperboxoffset}{1.2\paperboxheight} 
  
}

\definecolor{purple}{RGB}{85,26,139}
\hypersetup{linkcolor=purple}
\newcommand{\beq}{\begin{equation}}
\newcommand{\eeq}{\end{equation}}
\newcommand{\bea}{\begin{eqnarray}}
\newcommand{\eea}{\end{eqnarray}} 
\newcommand{\ba}{\begin{array}}
\newcommand{\ea}{\end{array}}

\renewcommand{\hat}{\widehat}

\title{Supersymmetry vis-\`{a}-vis Observation}
\subtitle{Dark Matter Constraints, Global Fits and Statistical Issues}
\author{Yashar Akrami}
\date{June 2011}
\shortdate{2011}
\type{Doctoral Thesis in Theoretical Physics}

\division{Oskar Klein Centre for Cosmoparticle Physics\\\vskip 0.5em
  and\\\vskip 0.5em
  Cosmology, Particle Astrophysics and String Theory\\\vskip 0.5em}
\department{Department of Physics}
\address{SE-106 91 Stockholm}
\city{Stockholm}
\country{Sweden}
\cplogo{\includegraphics[height=4cm, trim = 0 0 250 200, clip=true]{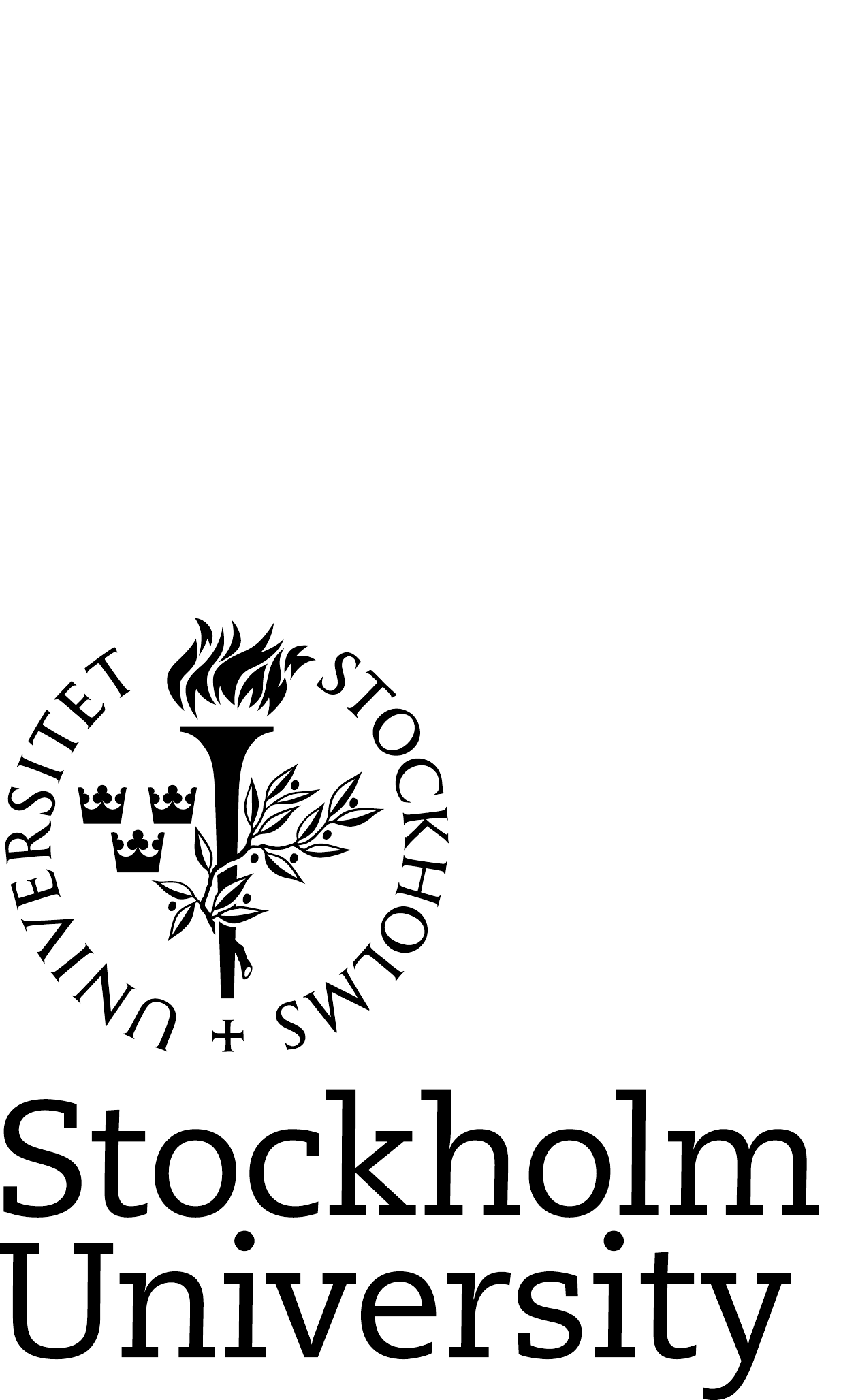}}

\publisher{\vspace{2mm}\\Printed by Universitetsservice US-AB, Stockholm, Sweden, 2011.\vspace{5mm}
\vspace{5mm}\\Typeset in pdf\LaTeX}
\copyrightline{pp.\ i--xx, 1--142 \copyright\ Yashar Akrami, 2011\vspace{2mm}}
\trita{xx}
\issn{aa}
\isrn{bb}
\isbn{978-91-7447-312-4 (pp.\ i--xx, 1--142)}
\comment{\emph{Cover image:}  Artist's impression of `supersymmetry vis-\`{a}-vis observation'. The image is composed of: (1) A real photograph from the dome of H\={a}fezieh (the tomb of the persian poet H\={a}fez) in Shiraz, Iran. The roof is decorated by enamelled mosaic tiles. Credit: The author's father. (2) The cosmic microwave background (CMB) temperature fluctuations as observed by the Wilkinson Microwave Anisotropy Probe (WMAP). Credit: NASA/WMAP Science Team.}
\innerlogo{\includegraphics[height=4cm, trim = 0 0 200 500, clip=true]{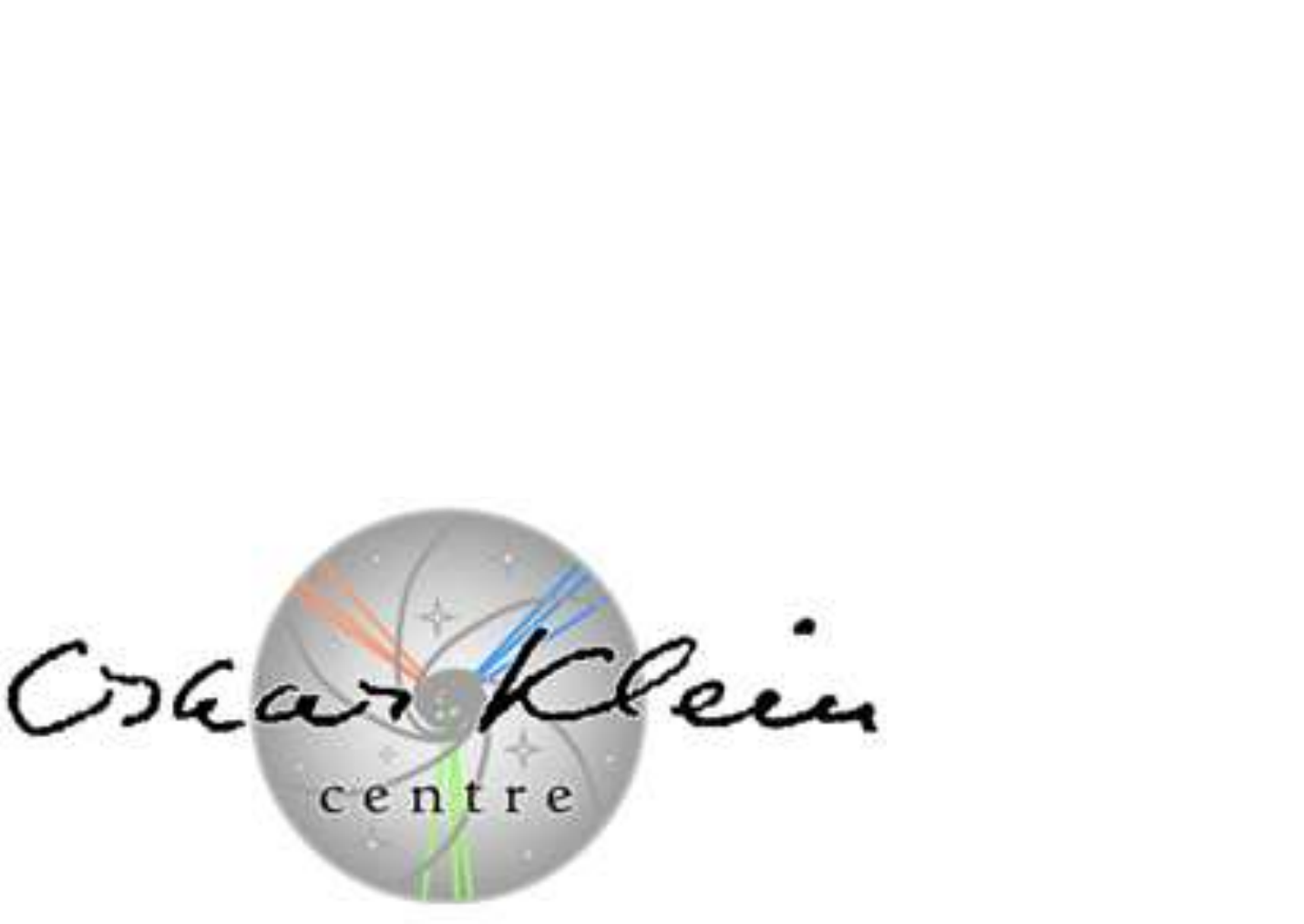}}
\extrainnerlogo{\includegraphics[height=4cm, trim = 0 0 250 400, clip=true]{sulogo}}

\newcommand{\Scott}{\textbf{Paper I}}
\newcommand{\AkramiGA}{\textbf{Paper II}}
\newcommand{\AkramiDD}{\textbf{Paper III}}
\newcommand{\AkramiCOV}{\textbf{Paper IV}}

\def\signofmetric{1}

\if0\signofmetric
\def\BDpos{}

\fi

\if1\signofmetric
\def\BDpos{-}

\fi

\if2\signofmetric
\def\BDpos{\oplus}

\fi

\if3\signofmetric
\def\BDpos{\ominus}

\fi

\def\beq{\begin{eqnarray}}
\def\eeq{\end{eqnarray}}
\def\bea{\begin{eqnarray*}}
\def\eea{\end{eqnarray*}}

\def\sbar{\overline}
\def\stilde{\widetilde}

\def\lagr{{\cal L}}

\def\msbar{\overline{\rm MS}}
\def\conj{{{\rm c.c.}}}

\def\MPlanck{M_{\rm P}}
\def\MGUT{M_{\rm GUT}}
\def\GGUT{g_{\rm GUT}}

\def\sigmabar{\overline\sigma}

\def\half{{1\over 2}}
\def\FX{F}

\def\centeron#1#2{{\setbox0=\hbox{#1}\setbox1=\hbox{#2}\ifdim
\wd1>\wd0\kern.5\wd1\kern-.5\wd0\fi
\copy0\kern-.5\wd0\kern-.5\wd1\copy1\ifdim\wd0>\wd1
\kern.5\wd0\kern-.5\wd1\fi}}
\def\ltap{\;\centeron{\raise.35ex\hbox{$<$}}{\lower.65ex\hbox{$\sim$}}\;}
\def\gtap{\;\centeron{\raise.35ex\hbox{$>$}}{\lower.65ex\hbox{$\sim$}}\;}

\def\slashchar#1{\setbox0=\hbox{$#1$}           
   \dimen0=\wd0                                 
   \setbox1=\hbox{/} \dimen1=\wd1               
   \ifdim\dimen0>\dimen1                        
      \rlap{\hbox to \dimen0{\hfil/\hfil}}      
      #1                                        
   \else                                        
      \rlap{\hbox to \dimen1{\hfil$#1$\hfil}}   
      /                                         
   \fi}                                        %

\def\fig#1{Fig.~\ref{#1}}

\def\eq#1{Eq.~\ref{#1}}
\def\eqs#1{Eqs.~\ref{#1}}

\newcommand{\newc}{\newcommand}

\newcommand{\met}{\mbox{\ensuremath{\slash\kern-.7emE_{T}}}}
\newcommand{\pet}{\mbox{\ensuremath{\slash\kern-.7emP_{T}}}}

\newc{\be}{\begin{equation}}
\newc{\ee}{\end{equation}}
\newc\ps{\mbox{ ps}}
\newc{\mev}{\mbox{ MeV}}
\newc{\gev}{\mbox{ GeV}}
\newc{\tev}{\mbox{ TeV}}
\newc{\GeV}{\gev}
\newc{\MeV}{\mev}
\newc{\TeV}{\tev}
\newc{\cl}{\text{CL}}
\newc\BR{BR}
\newc{\alphaemmz}{\alpha_{\text{em}}(m_Z)^{\overline{MS}}}
\newc{\alphas}{\alpha_s(m_Z)^{\overline{MS}}}
\newc\zetah{\zeta_h}
\newc\eg{{\rm {e.g.~}}}
\newc\etal{{\rm {et al.}}}
\newc\ie{{\rm i.e.~}}
\newc\etc{{\rm {etc}}}
\newc{\mhalf}{m_{1/2}}
\newc{\mzero}{m_0}
\newc{\tanb}{\tan\beta}
\newc{\azero}{A_0}
\newc{\sgn}{{\rm sgn}}
\newc{\deltaamususy}{\delta a_{\mu}^{\text{SUSY}}}
\newc{\bsg}{\bsgamma}
\newc\gmtwo{(g-2)_{\mu}}
\newc\deltaamu{\Delta a_{\mu}}
\newc{\abundchi}{\Omega_{\chi} h^2}
\newc{\mtop}{m_t}
\newc{\mtpole}{m_t}
\newc{\hl}{h}
\newc{\mhl}{m_{\hl}}  
\newc{\mpl}{M_{\text{Pl}}}
\newc{\msusy}{M_{\rm SUSY}}
\newc{\ms}{M_{\text{S}}}
\newc{\VEV}[1]{\langle #1 \rangle}
\newc{\sineff}{\sin^2 \theta_{\rm{eff}}}
\newc\MN{{\sf {MultiNest}}}

\newc\bsgamma{b\rightarrow s \gamma }
\newc\brbsgamma{\BR(\overline{B}\rightarrow X_s\gamma)}
\newc\bsmumu{\overline{B}_s\to\mu^+\mu^-}
\newc\brbsmumu{\BR(\overline{B}_s\to\mu^+\mu^-)}
\newc\bdmmumu{\overline{B}_d\to\mu^+\mu^-}
\newc\bbbarmix{\overline{B}_s\mbox{--}B_s}
\newc\delmbs{\Delta M_{B_s}}
\newc\brbtaunu{\BR(\overline{B}_u\to \tau \nu)}
\newc{\mbmbmsbar}{m_b(m_b)^{\msbar} }

\newc\AIPCP[3] {{\em AIP Conf. Proc.} {\bf #1} (#2) #3}
\newc\AJ[3] {{\em Astrophys. J.} {\bf #1} (#2) #3}
\newc\AMS[3] {{\em Ann. Math. Statist.} {\bf #1} (#2) #3}                
\newc\AP[3] {{\em Ann. Phys.} {\bf #1} (#2) #3}
\newc\APJ[3] {{\em Astropart. J.} {\bf #1} (#2) #3}
\newc\APP[3] {{\em Astropart. Phys.} {\bf #1} (#2) #3}
\newc\APS[3] {{\em Astrophys. J. Suppl.} {\bf #1} (#2) #3}
\newc\ARNPS[3] {{\em Ann. Rev. Nucl. Part. Sci.} {\bf C#1} (#2) #3}
\newc\BA[3] {{\em Bayesian Anal.} {\bf C#1} (#2) #3}              
\newc\CPC[3] {{\em Comput. Phys. Commun.} {\bf C#1} (#2) #3}
\newc\CP[3] {{\em Contemp. Phys.} {\bf #1} (#2) #3}                     
\newc\EPJ[3] {{\em Euro. Phys. Journ.} {\bf C#1} (#2) #3}
\newc\JCAP[3] {{\em JCAP} {\bf #1} (#2) #3}
\newc\JHEP[3] {{\em JHEP} {\bf #1} (#2) #3}
\newc\JPG[3] {{\em J. Phys.} {\bf G #1} (#2) #3}
\newc\IJMP[3] {{\em Int. J. Mod. Phys.} {\bf A #1} (#2) #3}
\newc\MNRAS[3] {{\em Mon. Not. Roy. Astron. Soc.} {\bf #1} (#2) #3}
\newc\MPL[3] {{\em Mod. Phys. Lett.} {\bf A #1} (#2) #3}
\newc\NAR[3] {{\em New Astron. Rev.} {\bf #1} (#2) #3}                  
\newc\NCA[3] {{\em Nuovo Cimento} {\bf #1} (#2) #3}
\newc\NIM[3] {{\em Nucl. Instrum. Methods} {\bf #1} (#2) #3}
\newc\NIMA[3] {{\em Nucl. Instrum. Methods} {\bf A #1} (#2) #3}
\newc\NAT[3] {{\em Nature} {\bf #1} (#2) #3}
\newc\NPB[3] {{\em Nucl. Phys.} {\bf B #1} (#2) #3}
\newc\NPA[3] {{\em Nucl. Phys.} {\bf A #1} (#2) #3}
\newc\NPPS[3] {{\em Nucl. Phys. Proc. Suppl.} {\bf #1} (#2) #3}                
\newc\PLB[3] {{\em Phys. Lett.} {\bf B #1} (#2) #3}
\newc\PR[3] {{\em Phys. Rep.} {\bf #1} (#2) #3}
\newc\PRL[3] {{\em Phys. Rev. Lett.} {\bf #1} (#2) #3}
\newc\PRD[3] {{\em Phys. Rev.} {\bf D #1} (#2) #3}
\newc\PRC[3] {{\em Phys. Rev.} {\bf C #1} (#2) #3}
\newc\PTP[3] {{\em Prog. Theor. Phys.} {\bf #1} (#2) #3}
\newc\RMP[3] {{\em Rev. Mod. Phys.} {\bf #1} (#2) #3 }
\newc\RPP[3] {{\em Rept. Prog. Phys.} {\bf #1} (#2) #3 }
\newc\SC[3] {{\em Science} {\bf #1} (#2) #3 }
\newc\ZPC[3] {{\em Z. Phys.} {\bf C #1} (#2) #3}
\newc\Err[3] {{\em Erratum-ibid.} {\bf #1} (#2) #3 }

\newcommand{\hu}{H_u}
\newcommand{\hd}{H_d}
\newcommand{\thu}{\widetilde{H}_u}
\newcommand{\thd}{\widetilde{H}_d}
\newcommand{\hc}{\mbox{ h.c.}}

\begin{document}

\begin{poem}

\vspace*{30mm}

\begin{figure}[h]
        \raggedleft
\includegraphics[width=1\textwidth, trim = 0 0 0 0, clip=true]{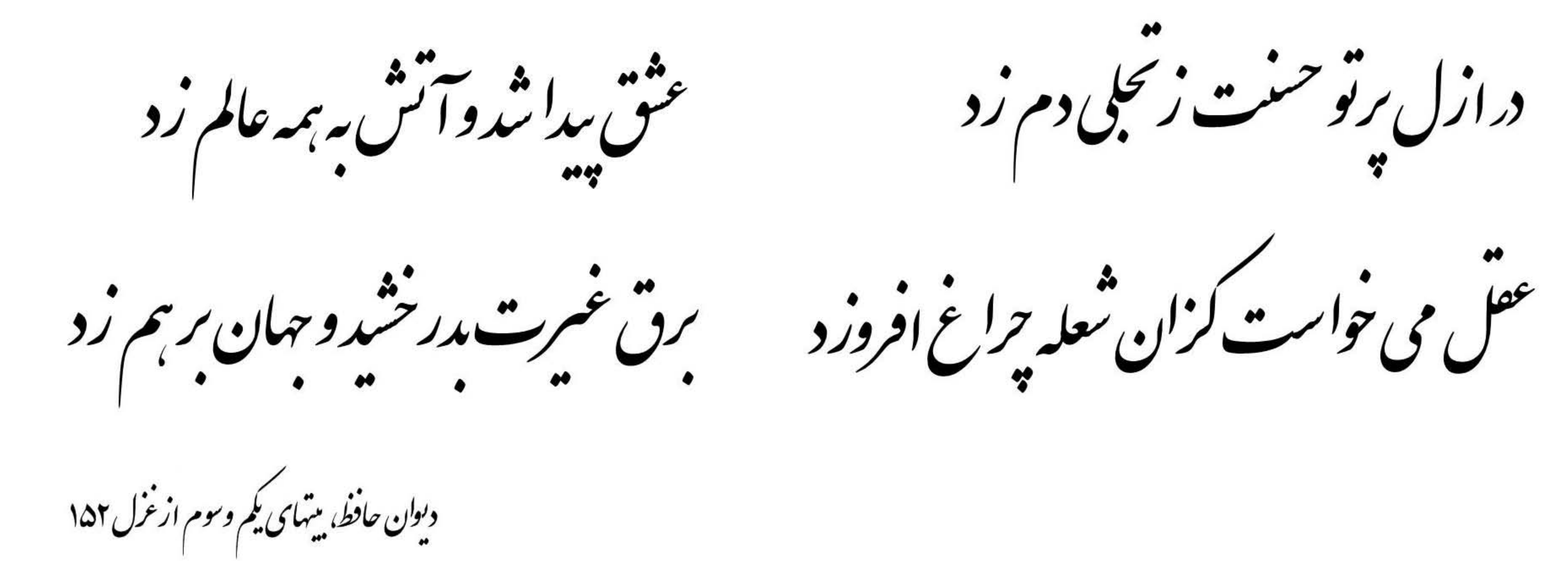}
        \label{fig:rotation}
\end{figure}

\vspace*{20mm}

\noindent

In eternity without beginning, the splendor-ray of Thy beauty boasted\\

\vspace*{-1mm}

Revealed became love; and, upon of the world, fire dashed.\\

\vspace*{2mm}

From that torch, reason wanted to kindle its lamp\\

\vspace*{-1mm}

Jealousy's lightning flashed; and in chaos, the world dashed.\\

\vspace*{5mm}
\raggedleft
The Persian Poet, \textbf{H\={a}fez} (1325/26-1389/90)

\end{poem}

\begin{abstract}

Weak-scale supersymmetry is one of the most favoured theories beyond the Standard Model of particle physics that elegantly solves various theoretical and observational problems in both particle physics and cosmology. In this thesis, I describe the theoretical foundations of supersymmetry, issues that it can address and concrete supersymmetric models that are widely used in phenomenological studies. I discuss how the predictions of supersymmetric models may be compared with observational data from both colliders and cosmology. I show why constraints on supersymmetric parameters by direct and indirect searches of particle dark matter are of particular interest in this respect. Gamma-ray observations of astrophysical sources, in particular dwarf spheroidal galaxies, by the \emph{Fermi} satellite, and recording nuclear recoil events and energies by future ton-scale direct detection experiments are shown to provide powerful tools in searches for supersymmetric dark matter and estimating supersymmetric parameters. I discuss some major statistical issues in supersymmetric global fits to experimental data. In particular, I further demonstrate that existing advanced scanning techniques may fail in correctly mapping the statistical properties of the parameter spaces even for the simplest supersymmetric models. Complementary scanning methods based on Genetic Algorithms are proposed.
\\\noindent \strut 
\end{abstract}

\begin{svensksammanfattning}

Supersymmetri \"{a}r en av de mest v\"{a}lstuderade teorierna f\"{o}r fysik bortom standardmodellen f\"{o}r partikelfysik. Den l\"{o}ser p{\aa} ett elegant s\"{a}tt flera teoretiska och observationella problem inom b{\aa}de partikelfysik och kosmologi. I denna avhandling kommer jag att beskriva de teoretiska fundamenten f\"{o}r supersymmetri, de problem den kan l\"{o}sa och konkreta supersymmetriska modeller som anv\"{a}nds i fenomenologiska studier. Jag kommer att diskutera hur f\"{o}ruts\"{a}gelser fr{\aa}n supersymmetriska modeller kan j\"{a}mf\"{o}ras med observationella data fr{\aa}n b{\aa}de partikelkolliderare och kosmologi. Jag visar ocks{\aa} varf\"{o}r resultat fr{\aa}n direkta och indirekta s\"{o}kanden efter m\"{o}rk materia \"{a}r s\"{a}rskilt intressanta. Observationer av gammastr{\aa}lning fr{\aa}n astrofysikaliska k\"{a}llor, i synnerhet dv\"{a}rggalaxer med Fermi-satelliten, samt kollisioner med atomk\"{a}rnor i kommande storskaliga direktdetektionsexperiment \"{a}r kraftfulla verktyg i letandet efter supersymmetrisk m\"{o}rk materia och f\"{o}r att best\"{a}mma de supersymmetriska parametrarna. Jag diskuterar n{\aa}gra statistiska fr{\aa}gest\"{a}llningar n\"{a}r man g\"{o}r globala anpassningar till experimentella data och visar att nuvarande avancerade tekniker f\"{o}r att skanna parameterrymden ibland misslyckas med att korrekt kartl\"{a}gga de statistiska egenskaperna, \"{a}ven f\"{o}r de enklaste supersymmetriska modellerna. Alternativa skanningsmetoder baserade p{\aa} genetiska algoritmer f\"{o}resl{\aa}s.

\end{svensksammanfattning}

\chapter*{List of Accompanying Papers}
\pagestyle{plain}
\addcontentsline{toc}{chapter}{List of Accompanying Papers}
\papersummary{Pat Scott, Jan Conrad, Joakim Edsj\"o, Lars Bergstr\"om, Christian Farnier \& \textbf{Yashar Akrami}}{Direct constraints on minimal supersymmetry from Fermi-LAT observations of the dwarf galaxy Segue 1}{\jcap}{01}{031}{2010}{\href{http://arxiv.org/abs/0909.3300}{arXiv:0909.3300}}{\protect{\hyperref[papI]{Paper I}}}
\papersummary{\textbf{Yashar Akrami}, Pat Scott, Joakim Edsj\"o, Jan Conrad \& Lars Bergstr\"om}{A profile likelihood analysis of the constrained MSSM with genetic algorithms}{\jhep}{04}{057}{2010}{\href{http://arxiv.org/abs/0910.3950}{arXiv:0910.3950}}{\protect{\hyperref[papII]{Paper II}}}
\papersummary{\textbf{Yashar Akrami}, Christopher Savage, Pat Scott, Jan Conrad \& Joakim Edsj\"o}{How well will ton-scale dark matter direct detection experiments constrain minimal supersymmetry?}{\jcap}{04}{012}{2011}{\href{http://arxiv.org/abs/arXiv:1011.4318}{arXiv:1011.4318}}{\protect{\hyperref[papIII]{Paper III}}}
\papersummary{\textbf{Yashar Akrami}, Christopher Savage, Pat Scott, Jan Conrad \& Joakim Edsj\"o}{Statistical coverage for supersymmetric parameter estimation: a case study with direct detection of dark matter}{\jcap}{07}{002}{2011}{\href{http://arxiv.org/abs/arXiv:1011.4297}{arXiv:1011.4297}}{\protect{\hyperref[papIV]{Paper IV}}}

\begin{acknowledgments}

First and foremost, I would like to thank my supervisor Joakim Edsj\"o for his excellent guidance, encouragement and enthusiastic supervision especially during the completion of this thesis. Thanks also to my secondary supervisor Lars Bergstr\"om for his valuable advices, generous support and providing the opportunity of pursuing my academic interests and goals. Thanks to both of them also for understanding my situation as a foreigner here in Sweden and for their pivotal helps in resolving intricate life-related problems. Many thanks also to Jan Conrad whose various helps and guidance have been crucial for the successful completion of this work. I am also grateful to him for invaluable non-physics advices that will certainly have indisputable influence on my future career. Jan, I do not forget the nice discussions we had during the visit to CERN.

Many thanks to all other professors and senior researchers at Fysikum, Department of Astronomy and KTH for sharing their invaluable knowledge and expertise with me. Thank you Marcus Berg, Claes-Ingvar Bj\"ornsson, Claes Fransson, Ariel Goobar, Fawad Hassan, Garrelt Mellema, Edvard M\"ortsell, Kjell Rosquist, Felix Ryde, Bo Sundborg, Christian Walck and G\"oran \"Ostlin. Special thanks to Marcus Berg for bringing to our group a new and highly enthusiastic ambiance to learn and discuss interesting aspects of high energy physics and cosmology. My warmest thanks to Fawad Hassan for being an excellent teacher and a good friend, and for his great willingness and patience in answering my endless questions. I am grateful to Ulf Danielsson and Stefan Hofmann for broadening my knowledge in theoretical physics with exciting discussions and ideas that made me think about `other' possibilities. I also thank Hector Rubinstein for all the nice conversations I had with him. Although he is no longer with us, he will always be in my mind.

My thanks also to the CoPS, HEAC and guest students Karl Andersson, Michael Blomqvist, Jonas Enander, Michael Gustafsson, Marianne Johansen, Joel Johansson, Jakob J\"onsson, Natallia Karpenko, Maja Llena Garde, Erik Lundstr\"om, David Marsh, Jakob Nordin, Narit Pidokrajt, Anders Pinzke, Sara Rydbeck, Angnis Schmidt-May, Pat Scott, Sofia Sivertsson, Alexander Sellerholm, Stefan Sj\"ors, Mikael von Strauss, Tomi Ylinen, Stephan Zimmer and Linda \"Ostman, the CoPS and OKC postdocs Rahman Amanullah, Torsten Bringmann, Alessandro Cuoco, Tomas Dahlen, Hugh Dickinson, Malcolm Fairbairn, Gabriele Garavini, Christine Meurer, Serena Nobili, Kerstin Paech, Antje Putze, Are Raklev, Joachim Ripken, Rachel Rosen, Martin Sahlen, Chris Savage, Vallery Stanishev and Gabrijela Zaharijas, and all other current or former students and postdocs that I may have forgotten to enumerate here. I have definitely benefited from all the conversations and discussions I have had with them and enjoyed every second I have spent with them. Special thanks to Pat and Chris for good times in the office and for all I have learned from collaborating with them. 

I would also like to thank Ove Appelblad, Stefan Csillag, Kjell Fransson, Mona Holgerstrand, Marieanne Holmberg, Elisabet Oppenheimer and all other people in administration for their valuable helps over the last few years.

Thanks also to the Swedish Research Council (VR) for making it possible for me and all my colleagues at the Oskar Klein Centre for Cosmoparticle Physics to work in such a work-class and highly prestigious institution.

Thanks to my parents and sister for all their continous encouragement and unconditional support. `Baba' \& `Maman' thank you for all troubles you endured stoically over the years. What I learned from you was all eagerness for truth, integrity and wisdom. Thanks to you Athena for being such a kind and supportive sister.

And last but not least, thanks to you Mahshid for all the confidence, independence and strength you have shown in me, for all your support and encouragement and for all great moments we shared over the last four and a half years of my life.

\end{acknowledgments}

\begin{preface}

This thesis deals with the phenomenology of weak-scale supersymmetry and strategies for comparing predictions of supersymmetric models with different types of observational data, in particular the ones related to the identification of dark matter particles. Currently, various experiments, either terrestrial, such as colliders and instruments for direct detection of dark matter, or celestial, such as cosmological space telescopes and dark matter indirect detection experiments, are providing an incredibly large amount of precise data that can be used as valuable sources of information about the fundamental laws and building blocks of Nature. Analysing these data in statistically consistent and numerically feasible ways is now one of the crucial tasks of cosmologists and particle physics phenomenologists. There are several issues and subtleties that should be addressed in this respect, and dealing with those form the bulk of the present work.

The papers included in this thesis can be divided into two general categories: Some (\Scott~and \AkramiDD) mostly aim to illustrate how real data can be used in constraining supersymmetric and/or other fundamental theories, and others (\AkramiGA~and \AkramiCOV) are more about whether existing statistical and numerical tools and algorithms are powerful enough for correctly comparing theoretical predictions with observations.

\subsubsection*{Thesis plan}

This thesis is organised as follows. It is divided into three major parts: Part~\ref{intro} is an introduction to the theoretical and statistical backgrounds relevant to my work, Part~\ref{summaryoutlook} summarises the main results we have obtained in our investigations and Part~\ref{papers} presents the included papers. Part~\ref{intro} is itself divided into 7 chapters: Chapter 1 is a short and non-technical introduction to the field and the main motivations for investigating models of physics beyond the Standard Model of particle physics in particular supersymmetry, Chapters 2 and 3 discuss the motivations for considering supersymmetry as a possible underlying theory of Nature in more detail and in demand for explaining both the dark matter problem in cosmology and theoretical issues with the Standard Model, Chapter 4 introduces supersymmetry and its theoretical foundations in a top-down approach and in a rather technical language,  Chapter 5 details the most interesting supersymmetric models that are being used in current phenomenological studies, Chapter 6 provides a review of different observational sources of information that can constrain supersymmetric models and parameters, and Chapter 7 describes statistical frameworks and techniques for analysing supersymmetry.

Almost all the included papers are written in rather comprehensive, self-contained and self-explanatory manners. Therefore, in order to avoid any unnecessary repetitions, I have written the introductory chapters such that they provide in a rather consistent and coherent way a more general and detailed description of the field to which the papers contribute. This also provides some additional background material that may not have been discussed in detail in the papers. The reader is therefore strongly recommended to consult the papers for more advanced and technical discussions.

\subsubsection*{Contribution to papers}

\Scott~focuses on potential experimental constraints one may place upon supersymmetric models from indirect searches of dark matter (this has been done for the particular case of the Constrained Minimal Supersymmetric Standard Model (CMSSM) as the model, and gamma-ray observations of the dwarf galaxy Segue 1 as the data).  We have assumed that the lightest neutralino is the dark matter particle that annihilates into gamma rays observable by our detectors.  The instrument for observations is the Large Area Telescope (LAT) aboard the \emph{Fermi} satellite.  Conventional state-of-the-art Bayesian techniques are employed for the exploration of the CMSSM parameter space and the model is constrained using the LAT data alone and also together with other experimental data in a global fit setup. In preparing and writing the paper, I was mostly involved in general discussions and edition of the manuscript.  I also helped Pat Scott in setting up \textsf{SuperBayeS} for the numerical calculations.

\AkramiGA~deals with the issue of efficiently scanning highly complex and poorly-understood parameter spaces of supersymmetric models.  It attempts to introduce a new scanning algorithm based on Genetic Algorithms (GAs) that is optimised for frequentist profile likelihood analyses of such models.  In addition to comparing its performance with that of the conventional (Bayesian) methods and illustrating how our results can affect the entire statistical inference, some physical consequences of the results (in terms of the implications for the Large Hadron Collider (LHC) and dark matter searches) are also presented and discussed.  The analyses are done for a global fit of the CMSSM to the existing cosmological and collider data. I have been the main author of the paper.  The use of Genetic Algorithms for exploration of supersymmetric parameter spaces was to a great extent my own initiative.  I modified \textsf{SupeBayeS} and added GA routines to it.  I did the numerical calculations, analysed the results and produced the tables and figures.  I wrote most of the text.  

\AkramiDD~aims to predict how far one can go in constraining supersymmetric models with future dark matter direct detection experiments. The methodology and the main strategy of the paper are very similar to the analysis of \Scott: The studied supersymmetric model is the CMSSM and nested sampling is used as the scanning technique. Both profile likelihoods and marginal posteriors are presented. I have been the main author of the paper, performed the numerical scans, analysed the results and produced tables and plots. Christopher Savage also significantly contributed to the work by providing the background material for direct detection theory and experiments, as well as preparing the likelihood functions for the experiments that I used in the analysis.

\AkramiCOV~studies a rather technical issue in the statistical investigations of supersymmetric models, namely the coverage problem. The analysis of this paper was computationally very demanding and required a substantial amount of computational power; this made the project a rather lengthy and challanging one. I have been the main author for this paper as well. I wrote most of the text and produced the results and all plots and tables. The numerical likelihood function for the analysis was provided by Christopher Savage, but I performed all the scans and interpreted the results.

\hspace{0cm}\\
\raggedleft
\noindent Yashar Akrami\\
Stockholm, April 2011

\end{preface}

\tableofcontents

\clearpage

\mainmatter

\part{Introduction}
\label{intro}

\setlength{\unitlength}{1mm}
\begin{fmffile}{feyn}

\chapter{Why dark matter and why go beyond the Standard Model?}
\label{sec:whyDMwhyBSM}

The visible Universe that we know and love is made up of planets, stars, galaxies and clusters of galaxies. We know that these objects exist mostly because they emit light or other types of electromagnetic radiation which we detect either by eye or by various telescopes. In addition, the celestial objects substantiate their existence through their gravitational effects which impact the motions of other objects in their vicinity. For most nearby astrophysical objects the two sources of information fairly agree and are therefore used as complementary ways in studying interesting properties of their sources. A problem emerges however when we look at scales of the order of galaxies or larger, where the gravitational effects imply the presence of massive bodies that are not detected electromagnetically. These objects that exhibit all the gravitational properties of normal matter but do not emit electromagnetic radiation (and are therefore invisible) are referred to as `dark matter' (DM).

Almost every attempt at explaining the nature of DM with the known types of matter has so far failed. This is mainly because DM seems to be required in order to consistently explain very different astrophysical phenomena that have been observed by completely different methods. This inevitably leads us to the assumption that DM is composed of new types of matter that are beyond our current understanding of the elementary particles and their interactions.

Our present knowledge of the fundamental building blocks of the Universe is summarised in the so-called Standard Model (SM) of particle physics (for an introduction, see \eg ref.~\cite{SM:Burgess}). The SM provides a mathematically consistent (though rather sophisticated) framework for describing different phenomena in a relatively large range of energy scales. At low energies the model describes the everyday life processes in terms of normal atoms, molecules and chemical interactions between them, and at high energies it has been capable of explaining various processes observed in nuclear reactors, particle colliders and high-energy astrophysical processes with remarkably high precision. The SM is a quantum-mechanical description of particles (or fields) and is based on a particular theoretical framework called quantum field theory.

The SM is now extensively tested at colliders and is in excellent agreement with the current data. However, as we stated earlier, the SM does not contain any type of matter with properties similar to the ones we need for DM. This simply implies that if DM exists, the SM has to be appropriately modified or extended so as to include DM particles with required properties. The need for DM is therefore one of the strongest motivations for going `beyond' the SM.

Apart from the lack of any DM candidates in the SM, there are additional reasons in support of the existence of new physics beyond this framework. These reasons are mainly motivated by some theoretically irritating characteristics of the model that cannot be explained otherwise. Perhaps the most notorious one is that the SM does not contain gravity. Currently four different type of force have been known in Nature: the gravitational force between massive objects, the electromagnetic force between charged particles, the strong force that put together neutrons and protons inside atomic nuclei, and the weak force which is responsible for radioactive processes. While three of these forces, \ie electromagnetic, strong and weak are well described quantum mechanically by the SM, the gravitational interactions do not fit consistently into the model. The reason is that when one attempts to quantise gravity with the known mathematical methods of quantum field theory, the resulting theory contains some infinities that cannot be removed in an acceptable manner. This is done for the other interactions through the so-called `renormalisation' procedure, a method that breaks down for gravitational interactions. We are therefore forced to treat gravity as a classical field which is best described by Einstein's theory of general relativity. This distinction between gravity and the other forces does not lead to serious problems provided that we do not want to describe gravitational processes at high energies where the quantum effects become important. There are however interesting high-energy cases where one needs to have a quantum-mechanical description of gravity so as to be able to study the physical systems. Two important examples are (1) extreme objects such as black holes and (2) the physics of the very early Universe. It is therefore commonly accepted that the SM must be modified at least at those high energies where gravity needs to be quantised. 

In addition, the SM possesses a very special mathematical structure that is based on particular types of fields and symmetries. This structure, although being crucial for the model to successfully describe different phenomena in particle physics, does not find any explanation within the theoretical principles of the model. The model also contains some free parameters, such as masses and couplings whose values have been determined experimentally. Some of these parameters take on values that require extensive fine-tuning. All these aesthetically vexatious issues and a few more give us strong hints that the SM is not the fundamental description of Nature and has to be appropriately extended.

Fortunately, several interesting extensions for the SM exist, the best of which are those that address all or most of the aforementioned issues simultaneously. One of these proposals is weak-scale supersymmetry. It is a very powerful framework in which the SM is conjectured to be modified by some new physics that kicks in at energies just above the electroweak scale, \ie the energy scale at which the electromagnetic and weak forces are assumed to be unified into one single electroweak force. This new physics assumes that all particles of the SM are accompanied by some partner particles that are more massive than the original ones. The existence of these so-called superpartners provides elegant solutions to many of the problems listed above, and paves the way for the resolution of many others in some broader theoretical framework. An important example is the inclusion of new matter fields with properties similar to what we need for a viable DM candidate.

Supersymmetric models, like any other theories in physics, need to be tested experimentally. Indeed, there have been many theoretically fascinating ideas in the history of physics that were abandoned only because they have not been consistent with particular experimental data. Fortunately, there are various sources of information from both man-made experiments, such as particle colliders, and astrophysical/cosmological observations that can be used for testing the supersymmetric models. Ideally, all these different types of data should be combined appropriately so as to give the most reliable answers to our questions about the validity of particular models and frameworks. This is however not a trivial task, because there are usually various sources of complication and uncertainty that enter the game and, if not addressed properly, can make any interpretations completely unreliable. This is exactly where the main objectives of the present thesis stand. We would like to examine how a class of interesting supersymmetric models can be compared with observations in the presence of different experimental (and theoretical) uncertainties and statistical/numerical complications.

First, in the following two chapters we give a more thorough (and more technical) description of the problems with the SM, including the need for DM. In each case, we describe in rather general terms how the problem finds appropriate solutions in supersymmetry. The detailed resolutions of some of the problems will be discussed later when supersymmetry is defined and concrete supersymmetric models are presented in chapters~\ref{sec:SUSYfoundations} and~\ref{sec:SUSYreal}, respectively. In chapter~\ref{sec:SUSYobs} we review important observational constraints we have employed in our analyses and describe different uncertainties in each case. Chapter~\ref{sec:SUSYstat} will be devoted to a discussion of the main statistical and numerical issues that we have dealt with in our endeavour. In the last chapter, \ie chapter~\ref{sec:summary}, we will briefly review our major results and present an outlook for future work.

\end{fmffile}

\setlength{\unitlength}{1mm}
\begin{fmffile}{feyn}

\chapter{The cosmological dark matter problem}
\label{sec:cosmology}
\section{The standard cosmological model}

The standard model of cosmology (for an introduction, see \eg refs.~\cite{cosmology:Weinberg,cosmology:Mukhanov}) is a mathematical framework for studying the largest-scale structures of the Universe and their dynamics. In other words, cosmologists attempt to answer various fundamental questions about the origin and evolution of the cosmos using the fundamental laws of physics. The model is based on Einstein's theory of general relativity as the currently best description of gravity at the classical level, as well as two important assumptions about the distribution of matter and energy in the Universe that are usually called together cosmological principles: the homogeneity and isotropy on large scales. The cosmological principles immediately imply that the correct metric for the Universe has to be of a particular form that is known as Friedmann-Lema\^\i tre-Robertson-Walker (FLRW) metric and has the following form:
\beq
ds^2=dt^2-a(t)^2\left[\frac{dr^2}{1-kr^2}+r^2(d\theta^2+sin^2\theta d \phi^2)\right].
\eeq
Here $r$, $\theta$ and $\phi$ denote the spherical coordinates and $t$ is time. $a$ as a function of time, is called the scale factor of the Universe and is an unknown function that can be determined by solving the Einstein field equations
\beq
G_{\mu\nu}=8\pi G T_{\mu\nu} .
\label{einstein}
\eeq
Here $G_{\mu\nu}$ is the Einstein tensor which contains all geometric properties of spacetime and $T_{\mu\nu}$ is the stress-energy-momentum tensor (or simply stress-energy tensor) that includes the information about various sources of matter and energy on that spacetime. $G$ is Newton's gravitational constant. The time evolution of $a$ therefore depends upon the assumptions we make for the matter and energy content of the Universe. $k$ is called the curvature parameter and depending on its value, the Universe may be closed, open or flat (corresponding to $k=+1$, $k=-1$ and $k=0$, respectively).

The assumption for the stress-energy tensor $T_{\mu\nu}$ on the right-hand side of \eq{einstein} is that the matter and energy of the Universe can be well described by a perfect fluid that is characterised by two quantities $\rho$ (its energy density) and $p$ (its pressure). By inserting the stress-energy tensor for such a fluid, $T_{\mu\nu}=diag(\rho,p,p,p)$, into \eq{einstein} we end up with the following simple equations:
\begin{eqnarray}
\left(\frac{\dot{a}}{a}\right)^2+\frac{k}{a^2}=\frac{8\pi G}{3}\rho, \nonumber \\
\frac{\ddot{a}}{a}=-\frac{4\pi G}{3}(\rho+3p).
\label{friedmann}
\end{eqnarray}

By solving these so-called Friedmann equations, one can obtain the dynamics of the Universe in terms of the time evolution of the scale factor $a$. The quantity $\dot{a}/a$ on the left-hand side of the first equation that gives the expansion rate is called Hubble parameter $H(t)$. In order to solve \eqs{friedmann}, it is essential to also know how $\rho$ and $p$ are related, \ie what the equation of state (EoS) is for the perfect fluid. For normal non-relativistic matter, the energy density $\rho$ is much larger than the pressure $p$ and one can therefore reasonably assume that the EoS is simply $p_m=0$. For relativistic matter (or radiation) on the other hand $\rho_{r}=3p_{r}$, and for the vacuum energy $\rho_{V}=-p_V$ (vacuum energy can be effectively written in terms of a cosmological constant $\Lambda$ in which case $\rho_{\Lambda}=\Lambda/8\pi G$).

In cosmology it is useful to write the various energy density contributions to the total density (at present time) in terms of the so-called density parameters $\Omega_m$, $\Omega_{r}$, and $\Omega_\Lambda$ for matter, radiation and vacuum, respectively. The same is usually done for the curvature term in the first Friedmann equation by defining $\Omega_k$ in an analogous way. These density parameters are defined as the ratio of a density $\rho$ at present time ($\rho^0$) to a specific quantity called the critical density $\rho_{c}$. $\rho_{c}$ (defined as $\rho_{c}\equiv 3H_0^2/8\pi G$, where $H_0$ is the present value of the Hubble parameter) is the density for which the Universe has an exact flat curvature:
\beq
\Omega_m\equiv\frac{\rho_m^0}{\rho_c},~~~\Omega_r\equiv\frac{\rho_r^0}{\rho_c},~~~\Omega_k\equiv\frac{k}{H_0^2},~~~\Omega_\Lambda\equiv\frac{\Lambda}{3H_0^2}.
\eeq 
The first Friedmann equation in \eqs{friedmann} can be written in the following simple form in terms of the density parameters:
\beq
H^2=H_0^2\left\{\Omega_r(1+z)^4+\Omega_m(1+z)^3-\Omega_k(1+z)^2+\Omega_\Lambda\right\},
\eeq
where $z\equiv a_0/a-1$ is the redshift with $a_0$ being the present value of the scale factor usually taken to be $1$. There are various ways to measure the Hubble parameter $H$ as a function of time from which one can determine the values for different density parameters and therefore the energy budget of the Universe.

\begin{figure}
        \centering
\includegraphics[scale=0.5, trim = 0 0 0 0, clip=true]{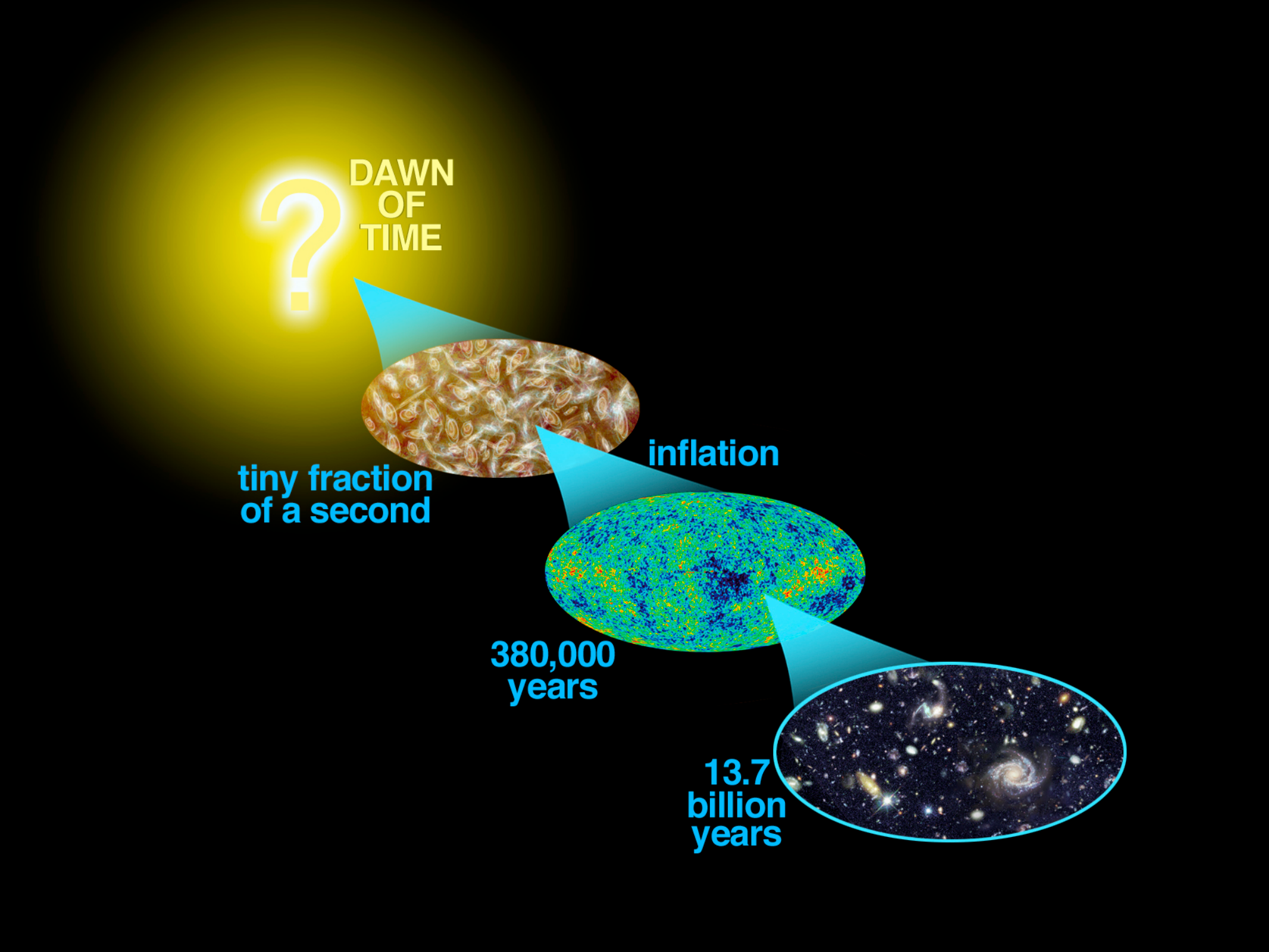}
 \caption{Major change points in the history of the Universe. Credit: NASA/WMAP Science Team.}
        \label{fig:timeline}
\end{figure}

Thanks to different high-precision cosmological observations, we have now been able to not only confirm the relative validity of our standard cosmological model, but also determine the values of different parameters that enter the mathematical formulation of the model to a high degree of accuracy. We now know that (see \eg \fig{fig:timeline}) the Universe started from an extremely hot and dense state about $13.7$ billion years ago (a state that we call the Big Bang) and then expanded, cooled down and became structured by galaxies, stars and other astrophysical objects. We also know that the curvature of the Universe is, to a good approximation, flat and also that it has recently entered an accelerated expansion phase. Although we still need a quantum theory of gravity to understand what exactly happened in the very early moments of the cosmic evolution, we have been able to infer some properties of the Universe at those times. For example there are various reasons to believe that shortly after its birth the Universe has seen a short inflationary phase during which its size has grown exponentially: (1) The Universe is (at least approximately) flat. (2) The observed cosmic microwave background radiation (\ie the relic radiation from the recombination epoch at which photons that were originally in thermal equilibrium with matter could escape the equilibrium and freely travel in the Universe) is to a great degree isotropic. (3) The Universe is not perfectly homogeneous and structures exist. All these features can be gracefully explained by inflation. The underlying mechanism for inflation is yet to be understood, but the evidence for its occurrence is so strong that it has now become one of the main paradigms of modern cosmology.

\section{The need for dark components}

\begin{figure}
        \centering
\includegraphics[scale=0.5, trim = 0 0 0 0, clip=true]{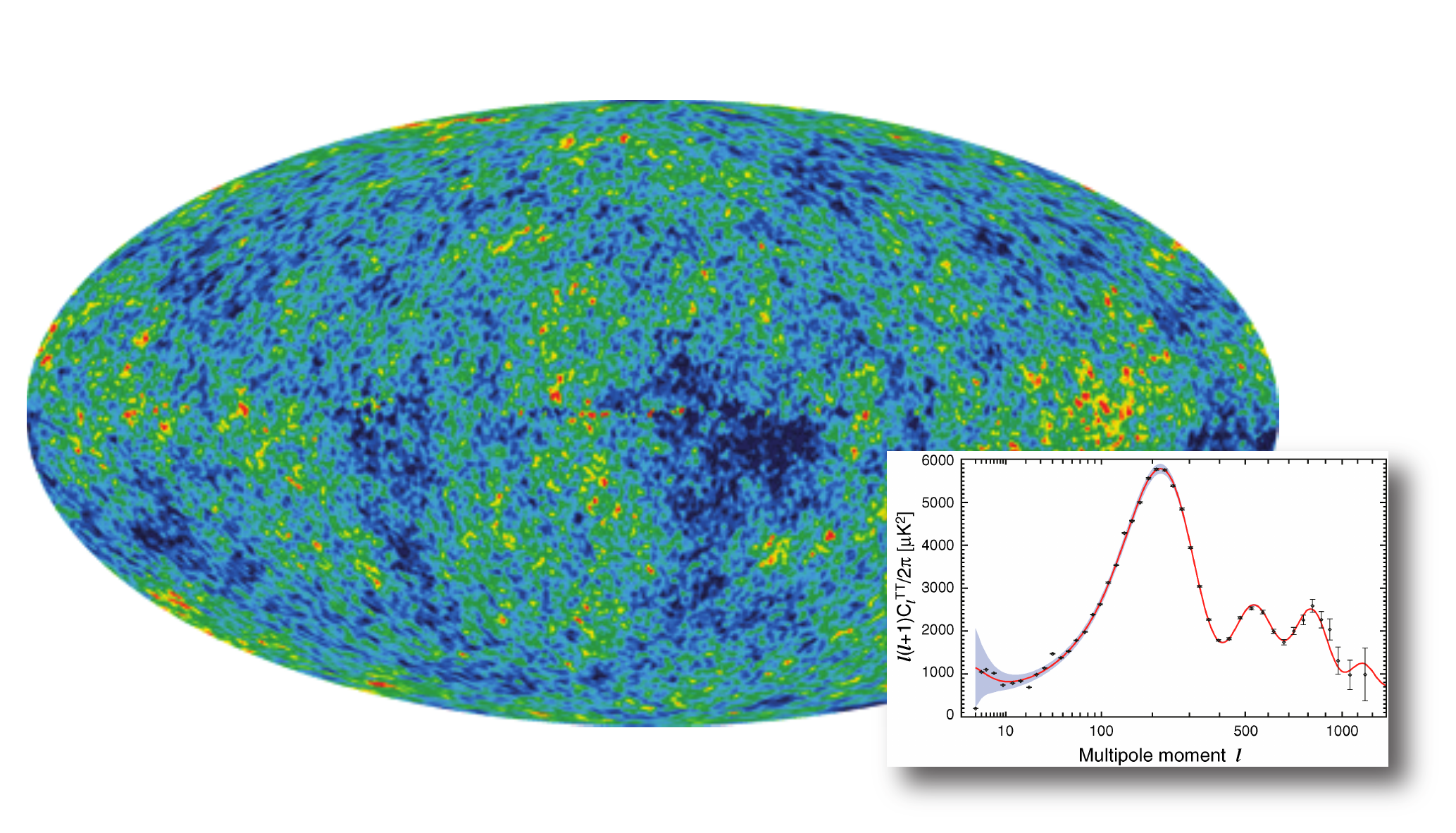}
 \caption{Temperature fluctuations on the cosmic microwave background (CMB) observed by the Wilkinson Microwave Anisotropy Probe (WMAP) satellite (background image), and the angular power spectrum of the fluctuations (inset). Credit: NASA/WMAP Science Team.}
        \label{fig:cmb}
\end{figure}

Perhaps the best confirmation of our cosmological picture to date has been from observations of the cosmic microwave background (CMB). It is extremely difficult (if not impossible) to explain the black-body spectrum of the CMB with alternative cosmological models. The measurements performed by the NASA satellite Wilkinson Microwave Anisotropy Probe (WMAP) have played a central role in this direction~\cite{Komatsu:2010fb}. Not only have such measurements confirmed the fact that the Big Bang theory is a successful description of the Universe, they have also determined the actual values of the density parameters we introduced in the previous section. By fitting the model to the so-called angular power spectrum of the CMB for the tiny temperature fluctuations observed on the 7-year WMAP sky map (see \eg \fig{fig:cmb}), it is now known that, for example, $\Omega_m=0.27\pm 0.03$ and $\Omega_\Lambda=0.73\pm 0.03$.

The first surprising observation is that the vacuum energy (or the cosmological constant) is non-zero and even constitutes about $74\%$ of the total energy budget of the Universe. A similar number was for the first time reported in 1998 by two different measurements of the so-called luminosity distance (a quantity that is defined in terms of the relationship between the absolute magnitude and apparent magnitude of an astronomical object and can be calculated theoretically for a cosmological model in terms of the Hubble parameter for an object with a specific redshift) using Type Ia supernovae (SNe)~\cite{Riess:1998cb,Perlmutter:1998np}. The first explanation for this energy component that implies a recent transition of the Universe to an accelerated expansion epoch was that it is just a cosmological constant. From a particle physics point of view, however, the vacuum energy density of the SM contributes to the cosmological constant and hence affects the expansion history of the Universe. But the value estimated in this way is much larger than the observed one and this poses a serious problem that cannot be explained within the SM~\cite{Weinberg1989}. It was then proposed that perhaps some new physics has made such contributions from the vacuum energy small (or zero) and what we observe cosmologically is not the cosmological constant but rather a new energy source (with an EoS parameter $w\equiv p/\rho$ that is not identically equal to $-1$) that can be detected only gravitationally (hence the name dark energy). There are numerous suggestions for the nature of the dark energy, most of which come from particle physics theories beyond the SM (for a review, see \eg ref.~\cite{Frieman:2008sn}).

\begin{figure}
        \centering
\includegraphics[height=8cm, trim = 0 0 0 0, clip=true]{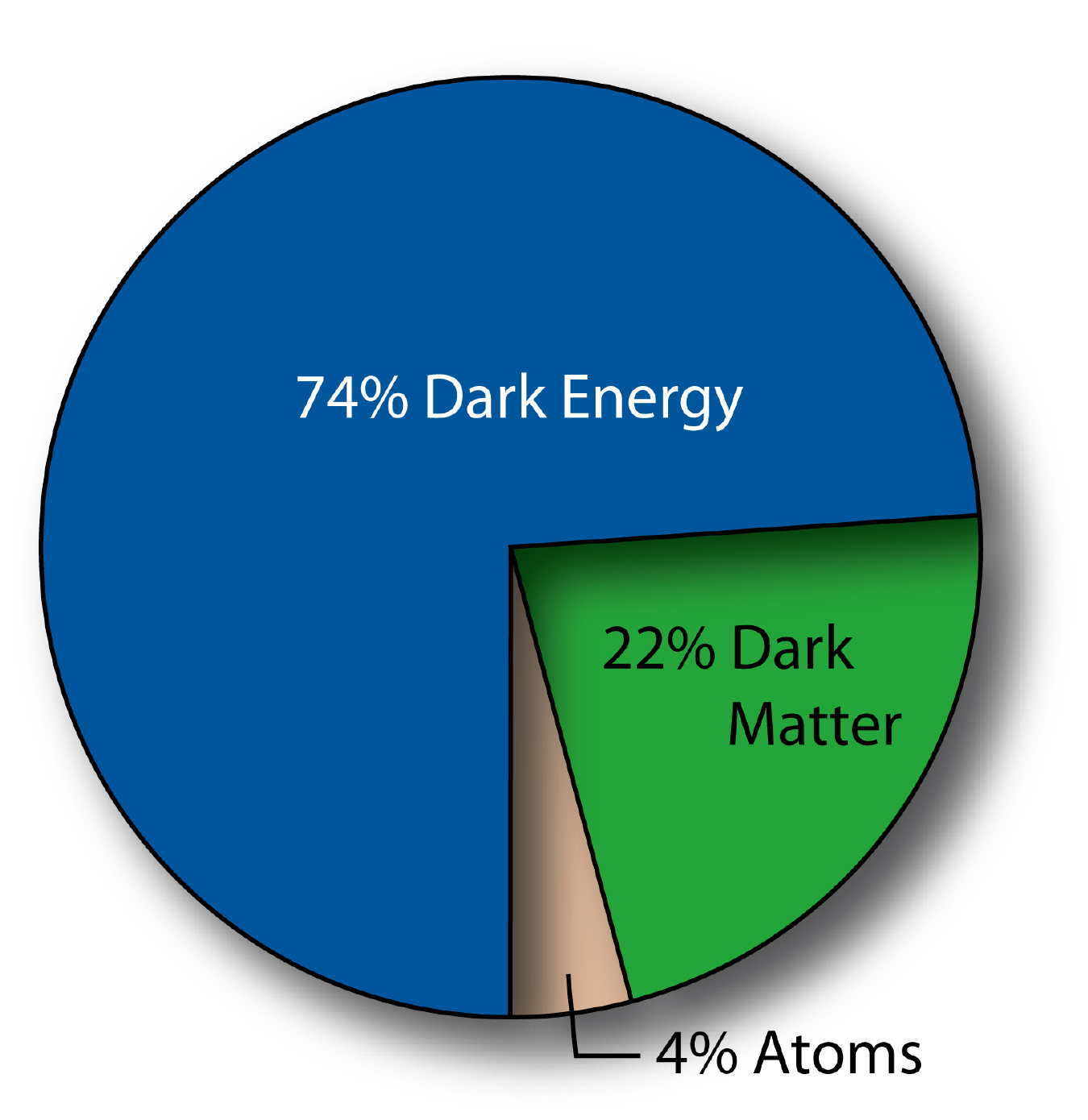}
 \caption{A pie chart of the content of the Universe today. Credit: NASA/WMAP Science Team.}
        \label{fig:darkcomp}
\end{figure}

Although the WMAP results imply that normal matter (with the EoS of $p_m=0$) forms about $26\%$ of the total energy density, the surprise comes from the value it has measured for the energy density of baryonic matter in the Universe. This is the matter that is composed mainly of baryons and includes all types of atoms we know. The baryons' energy density can be measured because the CMB angular power spectrum is sensitive directly to the amount of baryonic matter: While the location of the first peak (see \fig{fig:cmb}) gives us information about the total amount of matter, \ie $\Omega_m$, the second peak tells us about the total amount of baryonic matter $\Omega_b$. Estimations then determine $\Omega_b$ to be $0.045\pm 0.003$. Comparing the values for $\Omega_m$ and $\Omega_b$ indicates that the usual baryonic matter constitutes only about $4\%$ of the energy content of the Universe and about $22\%$ is non-baryonic (see \fig{fig:darkcomp}). All baryons interact with photons and can be detected also through non-gravitational effects whereas the non-baryonic component has been detected only gravitationally and is therefore named dark matter. In order to agree with observations of large-scale structure of the Universe, this non-baryonic dark matter must be dominantly cold (\ie almost non-relativistic). This cold dark matter (CDM) together with the assumption that dark energy is nothing but the cosmological constant $\Lambda$, a hypothesis that is in excellent agreement with all existing observations, contrives the foundations of our current standard model of cosmology that is accordingly called $\Lambda$CDM.

\begin{figure}[tbp]
\centering
\begin{minipage}[c]{0.5\textwidth}
  \includegraphics[width=\textwidth]{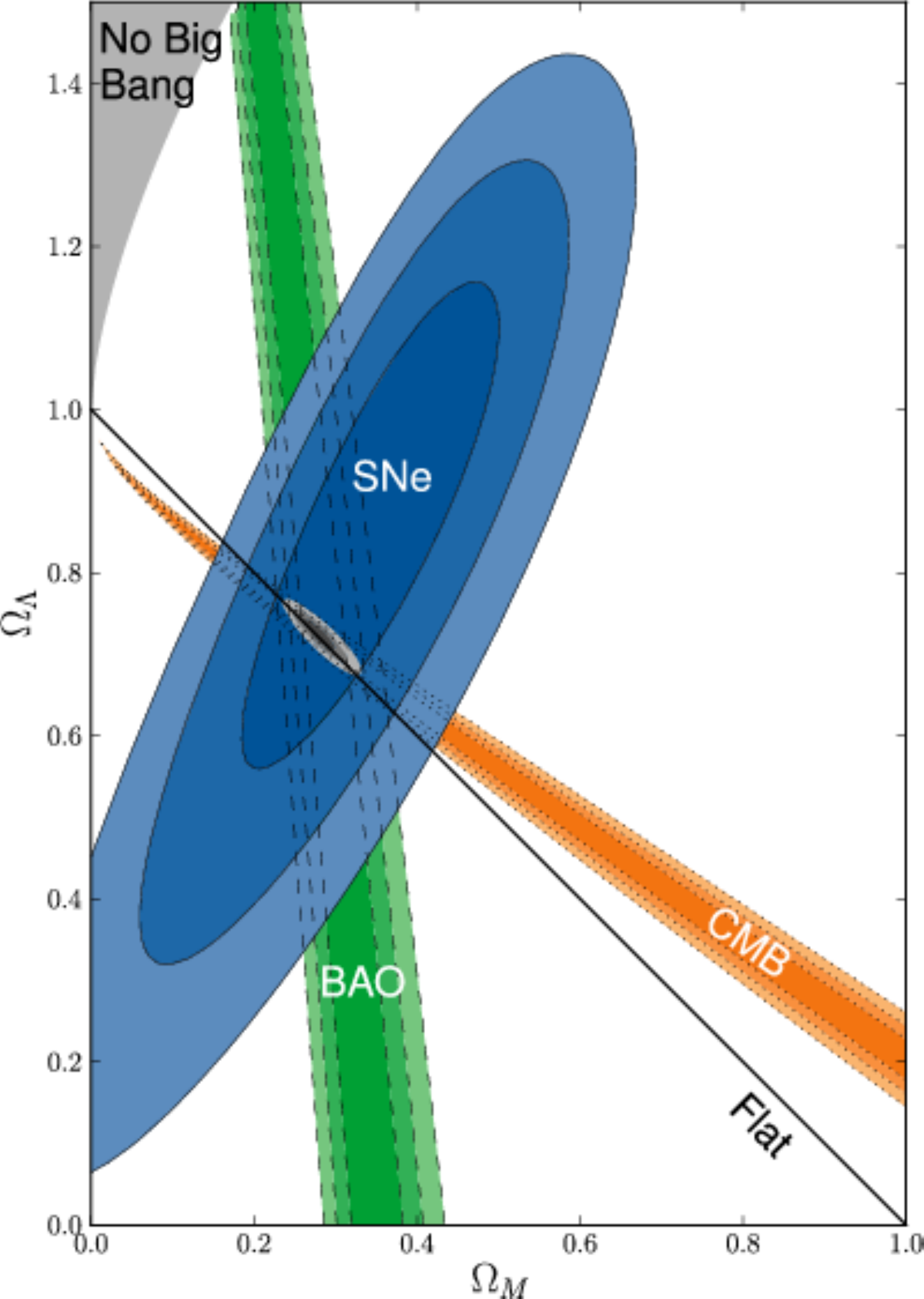}
\end{minipage}%
\begin{minipage}[c]{0.5\textwidth}
  \includegraphics[width=\textwidth]{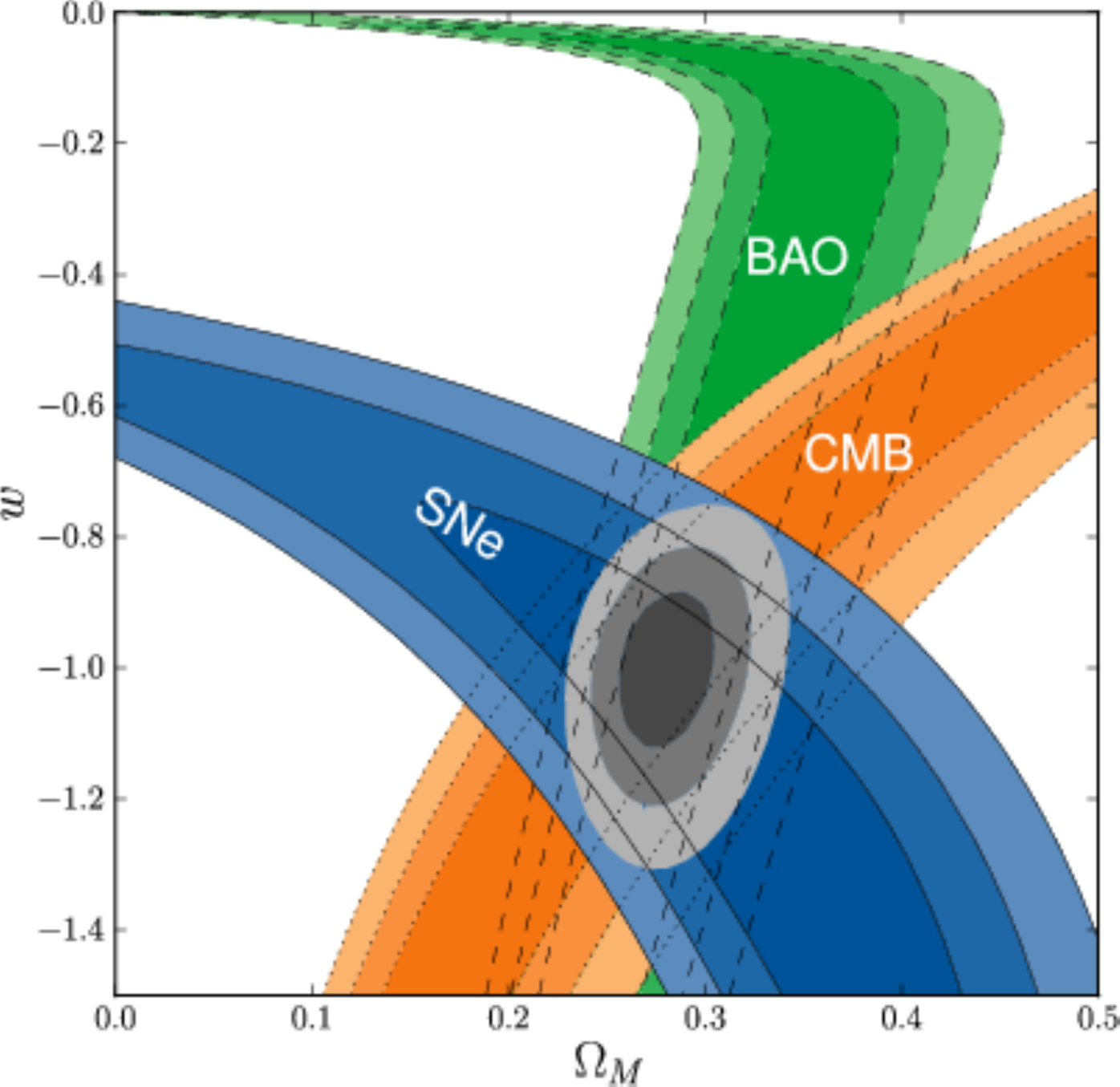}
\end{minipage}
\caption{The concordance cosmological model: $68.3\%$, $95.4\%$, and $99.7\%$ confidence regions in the $\Omega_m$-$\Omega_\Lambda$ (left) and $\Omega_m$-$w$ (right) planes determined by observations of Type Ia supernovae (SNe), baryon acoustic oscillations (BAO) and cosmic microwave background (CMB). Adapted from ref.~\cite{Amanullah:2010vv}.}
\label{fig:concordance}
\end{figure}

The left panel of \fig{fig:concordance} shows the currently best constraints on the energy densities of matter and dark energy from three important types of cosmological observations, \ie the CMB, Type Ia SNe and baryon acoustic oscillations (BAO)~\cite{Amanullah:2010vv}. The latter refers to an overdensity of baryonic matter at certain length scales due to acoustic waves that propagated in the early Universe. BAO can be predicted from the $\Lambda$CDM model and compared with what we have observed from the distribution of galaxies on large scales. The right panel of \fig{fig:concordance} depicts constraints from the same set of data but in terms of $\Omega_m$ versus the equation of state parameter $w$ for dark energy ($w=-1$ is for dark energy being the cosmological constant). By looking at both plots, it is quite interesting to see that the constraints from all these three sources of information are in perfect agreement with each other and also consistent with our theoretical model. This model is also in harmony with many other observations (such as constraints from Big Bang Nucleosynthesis (BBN) on the baryon density~\cite{Iocco:2008va}, gravitational lensing~\cite{Massey:2007wb} and X-ray data from galaxy clusters~\cite{Allen:2002eu}), and is accordingly called the concordance model of cosmology.

\begin{figure}
        \centering
\includegraphics[width=0.5\textwidth, trim = 0 140 0 0, clip=true]{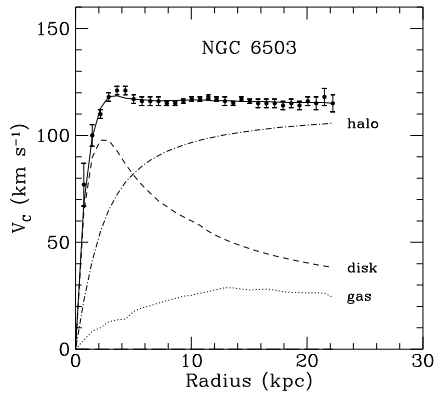}
 \caption{An example of the rotation curves of galaxies (for NGC 6503) where circular velocities of stars and gas are shown as a function of their distance from the
galactic centre. Here, the dotted, dashed and dash-dotted lines are the contributions of gas, disk and dark matter, respectively. Adapted from ref.~\cite{Begeman1991}.}
        \label{fig:rotation}
\end{figure}

The argument for the existence of dark matter, \ie the mass density that is not luminous and cannot be seen in telescopes, is actually very old.  Zwicky back in 1933 already reported the ``missing mass'' in the Coma cluster of galaxies by studying the motion of galaxies in the cluster and using the virial theorem~\cite{Zwicky}.  A classic strong evidence for dark matter existing in the scale of galaxies comes from the study of rotation curves in spiral galaxies by Rubin~\cite{Rubin1970,Rubin1978,Rubin1980,Rubin1985}. The observed rotation curves are not consistent with the standard theoretical assumptions unless one assumes the existence of dark matter halos surrounding all known contents of the galaxies, \ie stars and gas (for an example, see \eg \fig{fig:rotation}).

\begin{figure}[t]
\centering
\includegraphics[width=0.7\textwidth]{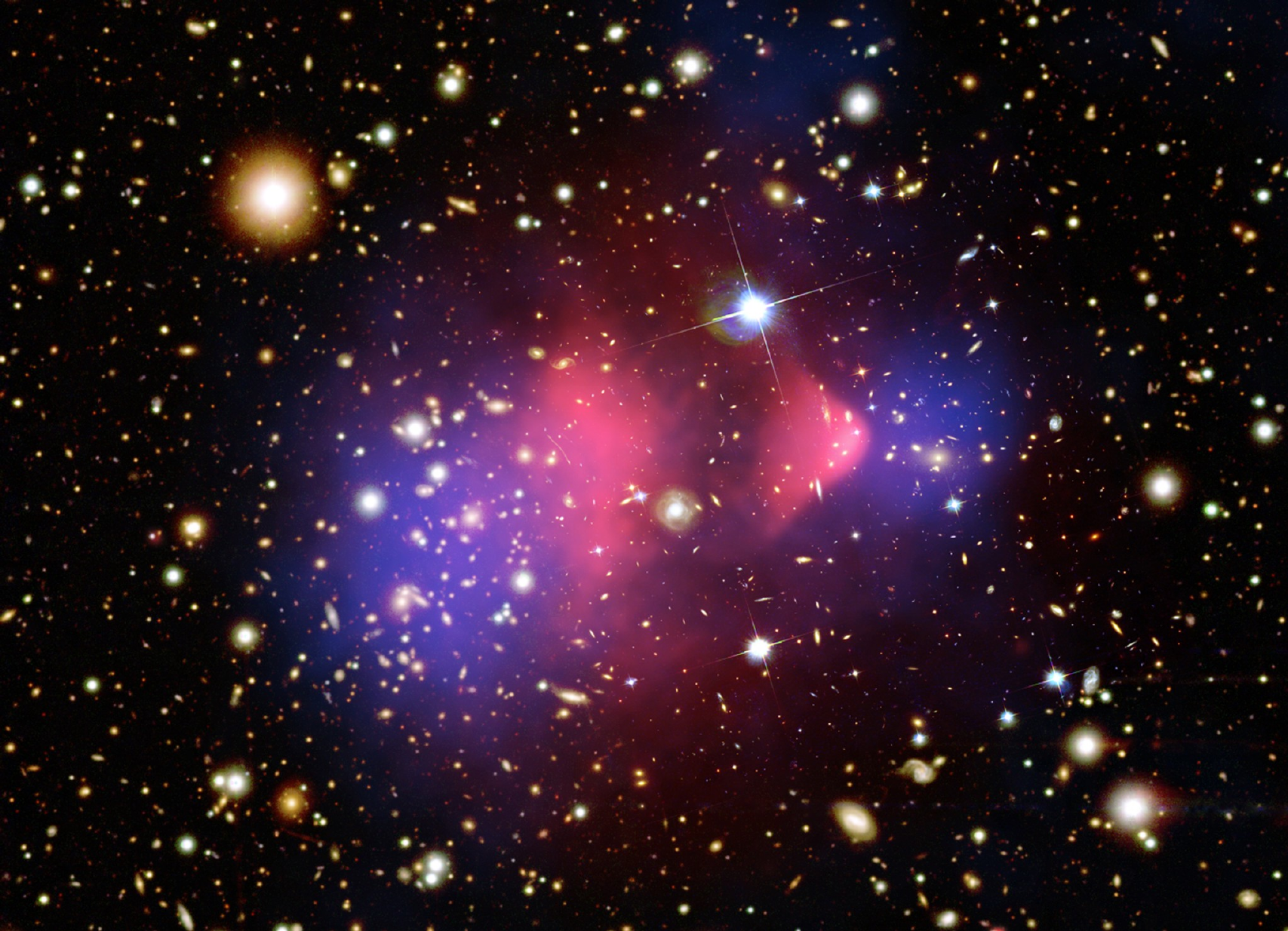}
\caption{The colliding Bullet Cluster. This is a composite image that combines the optical image of the object with a gravitational lensing map (in blue) and X-ray observations (in pink). Optical data: NASA/STScI; Magellan/U.Arizona/\cite{Clowe:2006eq}. Lensing map: NASA/STScI; ESO WFI; Magellan/U.Arizona/\cite{Clowe:2006eq}.  X-ray data: NASA/CXC/CfA/\cite{Markevitch06}.}
\label{fig:bullet}
\end{figure}

We should note here that some alternative explanations have been put forward that claim the anomalous observational data do not necessarily lead to the conclusion that dark matter exists. Some of these alternative proposals, such as the ones in the context of modified Newtonian dynamics (MOND)~\cite{Milgrom:1983ca,Bekenstein:2004ne}, have been successful in for example explaining the rotation curves of spiral galaxies (although in a rather ad hoc way). As we saw, the dark matter problem is not limited to astrophysical phenomena on particular scales and shows up in different observations from the scale of a galaxy to cosmological scales. It is in fact extremely difficult to explain all those observations without dark matter.

Perhaps the best direct evidence for the existence of dark matter is the so-called Bullet Cluster\cite{Clowe:2006eq} (see \fig{fig:bullet}). The Bullet Cluster consists of two galaxy clusters that have recently collided. \fig{fig:bullet} is a composite picture that shows (apart from the optical image) two types of observations of the cluster: gravitational lensing (in blue) and X-ray observations (in pink). Comparing these two cases evidently show that the baryonic gas component, which emits X-ray radiation, does not form the total mass of the cluster. Most of the mass, mapped by the lensing measurement, seem to come from a component that, in contract with the baryons, is collisionless: it does not interact with either baryonic gas or itself. These properties are all consistent with the assumption of dark matter.

\section{Weakly Interacting Massive Particles}
\label{sec:WIMPs}

The astrophysical/cosmological observations we discussed in the previous section all imply that dark matter probably exist. The next question we need to answer is what is the nature of dark matter, \ie what are the basic constituents of it. We have already inferred some of the properties the dark matter components should possess: (1) They must be massive otherwise we would not have seen their gravitational effects. (2) They must be dark, \ie they should not emit or absorb electromagnetic radiation (at least not noticeably), otherwise they would have already been detected by our telescopes. (3) They must be non-baryonic (confirmed by \eg the observations of CMB anisotropies and BBN). (4) They must be effectively collisionless with respect to both normal matter and themselves, otherwise they would loose energy through electromagnetic (or stronger) interactions and form dark matter disks (which contradicts the observations of galactic rotation curves). Observations of astrophysical systems like the Bullet Cluster could also not be explained in this case. (5) Dark matter must be cold(ish) (\ie almost non-relativistic), otherwise it would have not given rise to proper structure formation as we observe on cosmological scales. (6) It must be stable or at least very long-lived (compared to the age of the Universe); this is required because dark matter comprises a significant fraction of the total energy of the Universe at the present time (this fraction is given in terms of the dark matter relic abundance $\Omega_{DM}$).

Unfortunately, all attempts at finding a suitable dark matter candidate in the framework of the SM of particle physics have so far failed. This is because there are no standard particles that can satisfy all the requirements we listed above, and this means that cosmology requires new particles. This takes us to the realm of particle dark matter, namely that dark matter is composed of some new particles that have not been discovered yet. The need for particle dark matter is one of the main motivations for us to go beyond the SM (for detailed introductions to particle dark matter, see \eg refs.~\cite{Bergstrom:2000pn,Bertone:2004pz,DMBertone:2010}).

Fortunately, several viable dark matter candidates have been proposed in the literature (for a review, see \eg ref.~\cite{Bergstrom:2009ib}) and most of the interesting ones fall into the class of Weakly Interacting Massive Particles (WIMPs). WIMPs are particles that couple to the SM particles only through interactions that are of the order of the weak nuclear force (or weaker). This immediately tells us that WIMPs are electrically neutral, dark, effectively collisionless and non-baryonic. They are also massive, usually with masses within a few orders of magnitude of the electroweak scale. Having high enough masses also means that they are cold. WIMPs are also stable on cosmological timescales and this characteristic comes from a (usually imposed) discrete symmetry of the theory that gives WIMPs some conserved quantum number. This quantum number then prevents WIMPs from decaying into other particles and therefore makes them stable. In most scenarios, WIMPs are produced thermally in the early Universe~\cite{Zeldovich:1965,Chiu:1966kg,Steigman:1979kw,Scherrer:1985zt}. A generic (and highly interesting) feature of thermally-produced WIMPs is that they naturally provide the correct relic density of dark matter ($\Omega_{DM}$), \ie a value that is in excellent agreement with observations. We explain this intriguing feature in more detail below.

In the early Universe, right after the Big Bang, all the created particles (including WIMPs) are in both chemical and thermal equilibrium. Here chemical equilibrium refers to the situation where the primordial particles are created and destructed with almost equal rates and no net changes in their abundances with time. On the other hand, by thermal equilibrium (which is also called kinetic equilibrium) we mean that the particles are in thermal contact with each other without a net exchange of energy. In this latter case the temperatures associated with the particles follow the global temperature of the Universe.

Suppose that the number density associated with our hypothetical WIMP particles $\chi$ is $n_\chi$, their relative velocity is $v$ and they annihilate into lighter particles with the total annihilation cross-section $\sigma$. The equation governing the evolution of the WIMP density is the Boltzmann equation~\cite{Jungman:1995df}
\beq
\label{boltzmann}
\frac{dn_\chi}{dt}=-3Hn_\chi-\langle \sigma v \rangle (n_\chi^2-n_{\chi, eq}^2),
\eeq
where $n_{\chi, eq}$ is the equilibrium number density of the WIMPs, $H$ is the Hubble parameter and the brackets $\langle ... \rangle$ denote thermal average. For WIMPs with the mass $m_\chi$, the equilibrium number density (in the non-relativistic limit) at the temperature $T$ reads
\beq
n_{\chi, eq}=g_\chi(\frac{m_\chi T}{2\pi})^{3/2}e^{\frac{-m_\chi}{T}},
\eeq
where $g_\chi$ is the number of degrees of freedom associated with the species $\chi$.

\begin{figure}[t]
\centering
\includegraphics[width=0.7\textwidth]{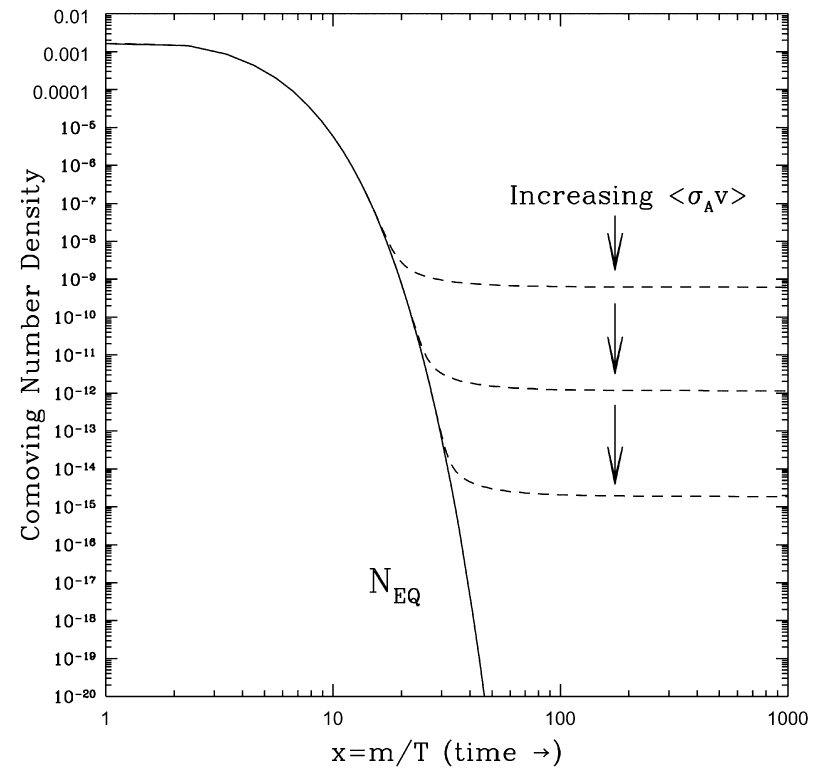}
\caption{Chemical freeze-out of WIMPs. Initially when the particles are in chemical (and thermal) equilibrium, their actual number density follows the equilibrium value $N_{EQ}$. At some later time, the particles fall out of chemical equilibrium (or freeze out) and their comoving number density becomes fixed. Adapted from ref.~\cite{Jungman:1995df}.}
\label{fig:freezeout}
\end{figure}

A direct implication of \eq{boltzmann} is that as long as the creation and annihilation of the WIMPs is larger than (or comparable with) the expansion rate of the Universe (specified by the Hubble parameter), the particles remain in chemical equilibrium. However, the Universe expands and cools, and this means that at some time and temperature, the interaction rate drops below the expansion rate and the equilibrium can no longer be maintained. This process during which the WIMPs decouple from the other particles is called chemical `freeze-out'. The number density of such thermally-produced WIMPs at the end of chemical freeze-out determines the relic density of dark matter today. Obviously, the abundance of WIMPs at freeze-out (and consequently the dark matter relic density) depends on how large the annihilation cross-section is: Larger cross-sections cause the WIMPs to remain in chemical equilibrium for a longer period and therefore generate a lower relic density (see \fig{fig:freezeout}).

Chemical freeze-out happens at a temperature $T_F$ that for WIMPs with weak-scale masses $m_\chi$ is given approximately as $T_F=m_\chi/20$~\cite{Jungman:1995df}. After chemical freeze-out, WIMPs still remain in thermal contact with the other particles for some time and kinetic freeze-out (or decoupling) happens later. The temperature of the WIMPs before this time is the same as the equilibrium temperature, and becomes fixed by kinetic decoupling afterwards. This means that the WIMPs will have a temperature lower than $T_F$ after kinetic freeze-out and this makes the WIMPs move non-relativistically up to the present moment. This characteristic is crucial for WIMPs to be `cold' dark matter.

In order to obtain the relic density of WIMPs $\Omega_\chi$, one needs to solve \eq{boltzmann} numerically. However, to a first-order approximation, it can be shown that under very general assumptions $\Omega_\chi$ does not depend explicitly on the WIMP mass and only depends on its annihilation cross-section~\cite{Bergstrom:2000pn,Jungman:1995df} in the following way:
\beq
\label{wimpmiracle}
\Omega_\chi h^2\approx\frac{3\times 10^{-27} \mbox{cm}^3 \mbox{s}^{-1}}{\langle \sigma v \rangle}.
\eeq
where $h\equiv H_0/100\mbox{km}\mbox{s}^{-1}\mbox{Mpc}^{-1}\approx 0.7$. For weakly-interacting particles with reasonable masses (\ie with values close to the scale of the electroweak symmetry breaking), the quantity $\langle \sigma v \rangle$ can be estimated as $\langle \sigma v \rangle\approx\alpha^2/m_\chi$, where $\alpha$ is the fine structure constant. Assuming a typical value of $m_\chi\sim 100$ GeV for the WIMP mass, we obtain $\langle \sigma v \rangle\approx 10^{-26}\mbox{cm}^3\mbox{s}^{-1}$. By inserting this value into \eq{wimpmiracle}, we obtain an approximate value for $\Omega_\chi$ with the right order of magnitude. This interesting `coincidence', also often referred to as `the WIMP miracle', means that, under the assumption of chemical freeze-out as the actual dark matter production mechanism occurred in the early Universe, any particles with generic properties of WIMPs can provide a dark matter relic density of the correct order. This particular characteristic of WIMPs makes them amongst the most interesting and popular dark matter candidates.

There are a large number of WIMP dark matter candidates on the market proposed in different contexts~\cite{Bergstrom:2009ib}, amongst which the lightest neutralino in supersymmetry~\cite{Jungman:1995df,Goldberg:1983nd,Ellis:1983ew}, the lightest Kaluza-Klein particle in models of Universal Extra Dimension (UED)~\cite{Servant:2002aq} and the lightest inert scalar in the Inert Doublet Model (IDM)~\cite{Barbieri:2006dq,LopezHonorez:2006gr} are the most widely-studied ones. The first one, \ie the lightest neutralino provides arguably the leading dark matter candidate with almost all desired properties. A substantial part of this thesis is devoted to the phenomenological aspects of the neutralino with particular emphasis on its implications for constraining models of weak-scale supersymmetry.

Before we end this section, let us emphasize that although WIMP dark matter proves to be an extremely powerful idea that provides extensive scope for phenomenological studies of particle dark matter, there are a number of other viable dark matter candidates that are either entirely non-WIMP or only WIMP-inspired. We do not intend to go through any of them here and just provide a list of the most interesting ones and refer the reader to the given references for detailed discussions (see also ref.~\cite{Bergstrom:2009ib} for a comprehensive review): axions~\cite{Kuster:2008zz,Hertzberg:2008wr}, gravitinos~\cite{Bailly:2009pe}, axinos~\cite{Covi:2009pq}, sterile neutrinos~\cite{Kusenko:2009up}, WIMPzillas~\cite{Kolb:1998ki}, Minimal Dark Matter~\cite{Cirelli:2005uq,Cirelli:2009uv}, Inelastic Dark Matter (iDM)~\cite{TuckerSmith:2001hy,TuckerSmith:2004jv}, eXciting Dark Matter (XDM)~\cite{Finkbeiner:2007kk}, WIMPless dark matter~\cite{Feng:2008ya,Feng:2010tg} and models with Sommerfeld enhancement~\cite{ArkaniHamed:2008qn,Nomura:2008ru}.

\end{fmffile}

\setlength{\unitlength}{1mm}
\begin{fmffile}{feyn}

\chapter{Theoretical issues with the Standard Model}
\label{sec:SMissues}

As stated earlier in chapter~\ref{sec:whyDMwhyBSM}, the Standard Model of particle physics is currently the minimal mathematical description of all known matter particles and their interactions that consistently explains various experimental observations, and holds over a wide range of energies. This includes phenomena that we observe in our everyday experiments (\ie energies of the order of a few eV), as well as the ones that can be observed only at high-energy colliders and astrophysical processes (\ie energies of $\sim 100$ GeV). The only key ingredient of this mathematical framework that still needs to be confirmed experimentally is the Higgs boson which is thought to be responsible for giving masses to the other particles. There are however alternative proposals for making the particles massive that although not excluded yet, are arguably less motivated (see \eg refs.~\cite{Farhi:1980xs,Lane:2002wv} for one of the most competitive ones). Having said that, it became relatively manifest soon after its establishment in the 1970s that for purely theoretical reasons the SM is incomplete and probably not the end of the story. It therefore has to be modified or extended beyond certain energies (which are argued to be energies higher than TeV scales).

As we discussed in the previous chapter, the need for a viable dark matter candidate is one pivotal reason for thinking about extensions of the SM. We advertised supersymmetry as one of the leading theories beyond the SM that provides such candidates. However, the nice thing with supersymmetry is that it also helps us circumvent many of the theoretical issues with the SM that are not related to the dark matter problem.

Before we introduce supersymmetry and review supersymmetric models, their properties and phenomenological implications in chapters~\ref{sec:SUSYfoundations},~\ref{sec:SUSYreal} and~\ref{sec:SUSYobs}, we remind ourselves in this chapter of some of the most notable theoretical problems in the SM and corresponding arguments in support of the physics beyond the SM, in particular supersymmetry. Clearly without describing its mathematical foundations and concrete realisations in particle physics, we cannot discuss in detail how supersymmetry helps us address these problems. We will therefore come back to some of the issues raised here in chapter~\ref{sec:SUSYreal} and explain how they can be gracefully resolved in some interesting supersymmetric models.

\section{The gauge hierarchy problem}
\label{sec:hierarchy}

In any quantum field theory, including the SM, all present parameters (such as masses and coupling constants) are affected by quantum radiative corrections. The amount of the corrections is generically a function of the cut-off scale that is used in the process of renormalising the theory or removing the divergences arising from various loop integrals. For the case of fermions (\ie particles with half-integer spin) interacting with photons, the radiative corrections to the fermion masses $m_f$ have a logarithmic dependence on the cut-off scale $\Lambda$ (here we use $\Lambda$ for a Lorentz-invariant cut-off): $\delta m_f \propto m_f \ln{(\Lambda / m_f)}$ (see \eg refs.~\cite{SUSY:Baer,SUSY:Aitchison,Martin:1997ns} for a detailed discussion). For gauge bosons (\ie particles with spin $1$ in the SM, such as photons), by using a gauge-invariant regulator (as is for example used in dimensional regularisation), one can show that the radiative corrections to the masses vanish. The reason for the absence of linear, quadratic, or higher-order corrections to the masses of fermions and gauge bosons is known and attributed to the presence of some particular symmetries of the theory: chiral symmetry in the former case and gauge invariance in the latter. Such symmetries are said to protect the particle masses from large radiative corrections.

The situation is however different for the scalar fields present in the theory, such as the Higgs boson of the SM. Restricting the discussion to the SM Higgs mass, the radiative correction to its mass from the self-interaction $H^4$ term in the SM Lagrangian reads
\beq
\delta m^2_H \propto m^2_H \left\{\Lambda^2-m^2_H \ln{\frac{\Lambda^2}{m^2_H}}+\emph{O}(\frac{1}{\Lambda^2})\right\} ,
\label{Higgscorr}
\eeq
which is quadratically divergent (\ie when $\Lambda$ increases to infinity, the term quadratic in $\Lambda$ dominates over the others and $\delta m^2_H$ becomes infinitely large). It should be noted that this is not the only quadratically divergent contribution to the radiative mass corrections for the Higgs boson: others come from gauge boson loops and fermion loops. An interesting feature of field theory is that the quadratically divergent contributions from the fermion loops have opposite signs relative to the contributions from the boson loops, an observation that, as we will argue below, plays an important role in one of our strongest motivations for extending the SM to its supersymmetric version.

Since the SM is a renormalisable theory, there is in principle no problem with the divergent radiative corrections to exist, because they can be absorbed into the so-called bare mass parameter. However, in an `effective field theory' interpretation of the SM (for an introduction, see \eg ref.~\cite{Burgess:2007pt}), it is believed that the model is a valid description of particle physics up to some particular energy scale which is characterised by the cut-off scale $\Lambda$. At energies beyond $\Lambda$, the SM may be modified by adding new degrees of freedom (\ie new fields) that are associated with some heavy particles whose effects are neglected at low energies. One example of such modifications is the assumption that the gauge group of the SM (\ie $SU(3)_C \times SU(2)_L \times U(1)_Y$) is generalised to a larger grand unification group such as $SU(5)$ or $SO(10)$. A rather trivial value for $\Lambda$ beyond which we expect new degrees of freedom to become important is the Planck scale $\MPlanck\sim10^{19}$ GeV, but $\Lambda$ can certainly be as small as TeV scales beyond which the SM has not been tested yet. In this effective field theory framework, quadratically divergent corrections pose a theoretical problem.

There are several reasons which indicate that the `physical' Higgs mass (the mass that is measured experimentally) has to be no larger than a few hundred GeV. This is the total value after adding the correction given in \eq{Higgscorr} to the bare mass parameter of the theory. If $\Lambda$ becomes very large, the quadratic term in \eq{Higgscorr} will dominate over the other terms and this effectively means that the physical mass is determined by the bare mass and the quadratic term. If one now assumes that the SM is valid below the scale of grand unification theories (GUTs) $\MGUT \sim 10^{16}$ GeV (see section~\ref{sec:unification}), the required cancellation of the two large values implies that the bare Higgs mass parameter will have to be ``fine-tuned'' to $1$ part in $10^{26}$. This becomes even worse if $\Lambda$ is as large as the obvious cut-off scale of $\MPlanck$. This fine-tuning problem is often referred to as the `gauge hierarchy problem' of the SM~\cite{hierarchy:Weinberg,hierarchy:Gildener,hierarchy:Susskind,hierarchy:tHooft}. In other words, the large quadratic corrections imply that if $\Lambda >> 1$ TeV, any predictions we make for physics at TeV energies are highly sensitive to the structure of the underlying high-energy theory with the SM being its effective incarnation at low energies.

Although such a fine-tuning of the SM structure is mathematically allowed, it has been taken as a strong hint (although not necessarily\footnote{Examples of the alternative approaches include: (1) Simply accepting that Nature is actually fine-tuned. (2) Leaving the assumption that elementary scalar fields exist in Nature, in models with composite states of bound fermions such as the idea of technicolor~\cite{Farhi:1980xs,Lane:2002wv}. (3) Assuming that the Higgs bosons interact strongly (rather than perturbatively) with themselves, gauge fields or fermions at the cut-off scale $\Lambda$~\cite{Bagger:1993zf,Bagger:1995mk}. (4) Making gravitational effects strong at energies close to TeV scales by for example assuming the existence of additional compact spatial dimensions~\cite{Csaki:2004ay,Hewett:2002hv}. (5) Assuming that the quadratic divergences only show up at multi-loop level and not necessarily at the lowest order in models such as the Little Higgs~\cite{Schmaltz:2002wx}.}) that some new degrees of freedom must exist above the electroweak scale that `naturally' cancel the problematic quadratic corrections in \eq{Higgscorr}. These new degrees of freedom should then be soon revealed by TeV-energy experiments and observations both at colliders and in high-energy astrophysical phenomena.

\begin{figure}[t]
\centering
\includegraphics[width=1\textwidth]{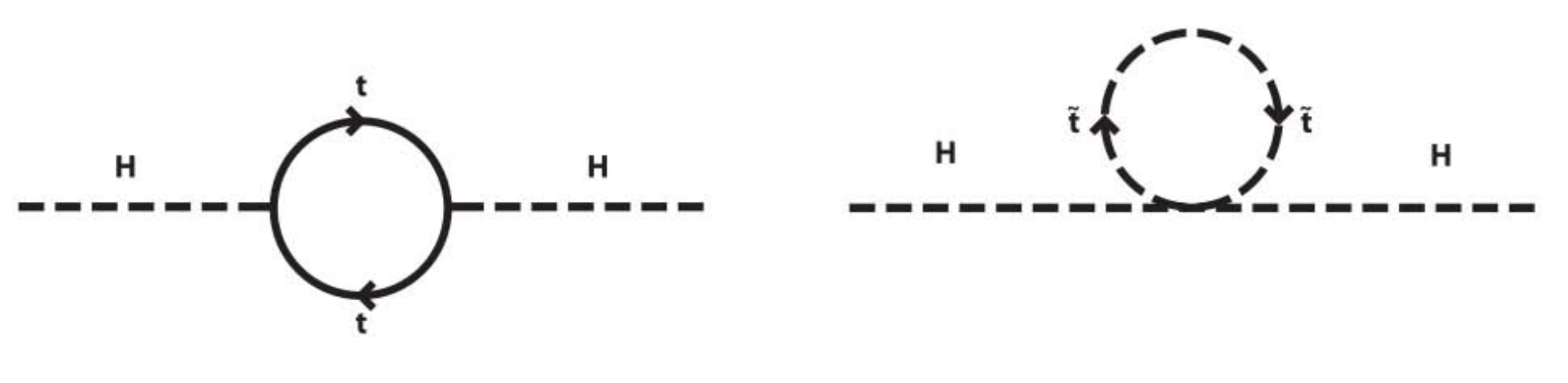}
\caption{An example of how supersymmetry solves the gauge hierarchy problem of the standard model. Quadratically divergent quantum corrections to the Higgs mass can be cancelled through the presence of equal numbers of fermion and boson loops that contribute equally but with opposite signs. Here the loop contributions are shown for the top quark $t$ and its supersymmetric partner $\tilde t$.}
\label{fig:hierarchy}
\end{figure}

Weak-scale supersymmetry is arguably the leading proposal that provides the required new degrees of freedom and solves the hierarchy problem in a simple and elegant way. We mentioned earlier in this section that the fermion and boson loops contribute to the dangerous quadratic divergences with opposite signs. This immediately suggests that in a theory with equal numbers of fermionic and bosonic degrees of freedom, the quadratic divergences will be cancelled. In order for this idea to work at any loop level, the couplings of fermions and bosons are additionally required to be related due to some symmetry. As we will see in the following chapters, both of these requirements are fulfilled in supersymmetry as a symmetry that transforms fermions to bosons and vice versa (see \eg \fig{fig:hierarchy}).

As we will argue in section~\ref{sec:SUSYbreaking}, even if supersymmetry is a correct extension of the SM, it has to be broken at least spontaneously (\ie through a mechanism similar to the Higgs mechanism of electroweak symmetry breaking). One can show that in a supersymmetric theory where supersymmetry is appropriately broken, the scalar masses all remain stabilised against radiative corrections and the hierarchy problem is still resolved~\cite{Witten:1981nf}. This observation is so remarkable that it essentially served as a watershed in the history of supersymmetry and provided one of the strongest motivations for it.

\section{Electroweak symmetry breaking}
\label{sec:EWSB}

Electroweak symmetry breaking (EWSB) is an essential ingredient in the SM. Through this process all the particles of the model acquire mass, a feature that is obviously a crucial requirement for the model to successfully describe the real world. EWSB is realised in the SM through the Higgs mechanism: The Higgs boson of the theory is believed to have acquired a vacuum expectation value (VEV) which results in the breaking of electroweak gauge symmetry. This is a `spontaneous' symmetry breaking, because the fundamental Lagrangian of the theory (\ie the SM Lagrangian) still remains symmetric while the ground state is no longer invariant under the symmetry. In order for the Higgs boson to develop an appropriate VEV, a so-called scalar potential of the theory should be minimised properly. This requires some particular parameters of the potential to acquire specific values. Strictly speaking, in order for the EWSB mechanism to work, some squared mass parameter for the Higgs boson has to be negative and this can be achieved only if some parameters of the model possess certain values. Although these values have been set experimentally, there is no explanation for such choices and again some fine-tuning seems to be necessary.

As we will discuss in sections~\ref{sec:higgssector} and~\ref{sec:CMSSM} for particular supersymmetric models, supersymmetry can naturally lead to EWSB and provide a deeper understanding of why it happens. This is mainly because in supersymmetric models one usually does not have to tune the EWSB parameters directly: The conditions of EWSB can be satisfied by setting the model parameters to some typical values that are motivated for other reasons. In models for supersymmetry with parameters that are set at some high-energy scales (such as the models of sections~\ref{sec:fundamentals} and~\ref{sec:CMSSM}), starting from a few parameters and evolving them with energy by means of the so-called renormalisation group equations (RGEs; see \eg section~\ref{sec:RGEs}) can give rise to EWSB at the electroweak scale. This process is often referred to as `radiative electroweak symmetry breaking' (REWSB).

\section{Gauge coupling unification}
\label{sec:unification}

The SM is constructed based on the gauge group $SU(3)_C \times SU(2)_L \times U(1)_Y$ and all particles are different representations of this particular symmetry group. But why is this group special? It certainly looks peculiar and there is no theoretical explanation within the framework of the SM for this particular choice. 

The three subgroups of the above gauge group (\ie $SU(3)_C$, $SU(2)_L$ and $U(1)_Y$), correspond to three forces of Nature, \ie strong, weak and electromagnetic forces, respectively. Each group has a coupling constant that determines the strength of its associated force. Experimental measurements over a wide range of energies tell us that the three forces are very different in strength and this is related to the fact that the three corresponding coupling constants have very different values. Like any other quantity in quantum field theory that in general runs with energy, the couplings are also scale-dependent. However, experiments indicate that even at energies slightly higher than the electroweak scale where the spontaneously broken (sub-)symmetry $SU(2)_L \times U(1)_Y$ becomes restored, the two associated coupling constants do not unify (see \eg the left panel of \fig{fig:unification} in section~\ref{sec:RGEs}).

The peculiar gauge structure of the SM has however important implications. For example, it prevents the occurrence of some unwanted phenomena such as proton decay and large flavour-changing neutral currents (FCNCs).  Although these characteristics are crucial for the success of the model, the way they are achieved in the SM is highly non-trivial and seems to be pure luck. In addition, the SM contains many free parameters whose values are constrained by experiments.  There is however no theoretical explanation for such experimentally favoured values. All these types of tuning problems, as well as the question about different values of gauge coupling constants find reasonable explanations through the intriguing idea of `unification'.  

In unified theories, the gauge symmetries of the SM are assumed to be extended to larger symmetries. For example in the so-called grand unified theories (GUTs), that are of particular interest in this respect, the SM symmetry group is extended to some simple Lie groups such as $SU(5)$~\cite{Bajc:2002pg} or $SO(10)$~\cite{Albright:2000dk,Aulakh:2003kg,Fukuyama:2004pb}.  This extension is largely motivated by the fact that the SM field content perfectly fits into multiplets (or representations) of these groups, \ie these larger groups include the SM group as their subgroup~\cite{Georgi:1974sy}.  This can therefore potentially explain the reason for the particular assignment of quantum numbers (such as hypercharges) in the SM (which seem to be randomly assigned). This consequently illuminates why dangerous experimental processes are forbidden in the SM.

One requirement for unification to occur is that all gauge couplings of the theory unify to a single quantity. As we mentioned above, this is not the case for the SM.  We however know that these couplings, as well as all other parameters of the model, generally evolve with energy through the RGEs.  This then gives the hope that although the gauge couplings have different values at low energies, they may unify at some high energy scale where the underlying larger symmetry group manifests itself. If this scenario is true, it provides an appropriate answer to the question why different forces of Nature have different strength: this is only a natural consequence of running of parameters with energy in quantum field theory. In addition, as a bonus, unification usually provides extra relations between various parameters of the theory and therefore gives rise to a (sometimes dramatical) reduction in the number of free parameters of the model. This alleviates the problem with the large number of free parameters in the SM. Finally, promoting the peculiar gauge group of the SM to a simple group such as $SU(5)$ or $SO(10)$ is on its own an interesting feature.  

The problem manifests itself if we now solve the RGEs for the SM gauge couplings up to very high energies: the result is that the couplings do not unify at any scale (again see \eg the left panel of \fig{fig:unification} in section~\ref{sec:RGEs}) and the idea of unification seems to be excluded. However, the unification scale (if exists) cannot be chosen arbitrarily and is determined by the particle content of the theory and measured values of different parameters at some energy scale (\eg the weak scale).  Although for the SM, with the known particle content and experimental constraints on its free parameters, the gauge couplings do not unify at any scale, a way out is to modify the particle content appropriately by adding new degrees of freedom to the model. Clearly these new particles should be heavy enough so as to remain hidden at low energies.

This is exactly where supersymmetry enters the game and turns out to be quite helpful. In most interesting versions of weak-scale supersymmetric models (as we will see in section~\ref{sec:RGEs}) the SM field content is modified such that the gauge coupling unification can be elegantly achieved. In the case of minimal supersymmetric extensions of the SM, the unification is obtained typically at a GUT scale of $\MGUT \simeq 2\times 10^{16}$ GeV with a unified gauge coupling of $\GGUT \simeq 0.7$ (see \eg the right panel of \fig{fig:unification} in section~\ref{sec:RGEs})~\cite{Dimopoulos:1981zb,Dimopoulos:1981yj,Sakai:1981gr,Ibanez:1981yh,Einhorn:1981sx}. A detailed and technical discussion of why gauge unification is achieved in concrete realisations of supersymmetry can be found in section~\ref{sec:RGEs} of this thesis. 

Before we stop our discussion here, let us note that although unification can be obtained this way, it is however highly non-trivial from a theoretical point of view. For example, no firm theoretical explanations exist for why $\GGUT$ should remain in perturbative regime, or why the GUT scale $\MGUT$ resides in a narrow energy range that is required for both suppression of proton decay and prevention of possible quantum gravitational effects. These characteristics therefore remain as accidental properties of the theory.

\section{Experimental bounds on the Higgs boson mass}
\label{sec:Higgsbounds}

In the SM, the Higgs mass is set by the quartic Higgs coupling $\lambda$ which is fairly unrestricted. Consequently, there are no strong limits on the Higgs mass and while the lower limit is set by experiments such as the Large Electron-Positron (LEP) collider to be about $114.4$ GeV~\cite{Schael:2006cr}, the mass can be as large as about $800$ GeV.

On the other hand, as we will see in section~\ref{sec:higgssector}, in the most widely-studied supersymmetric extensions of the SM, the Higgs mass is not a free parameter and is actually a prediction of the theory. In these models, the lightest Higgs scalar\footnote{As we will see later, consistency conditions require supersymmetric models to have more than one Higgs boson.} is required to be lighter than about $120-135$ GeV and this much narrower range for the Higgs mass makes the theory more falsifiable and therefore phenomenologically more interesting.

\begin{figure}[t]
\centering
\includegraphics[width=0.5\textwidth]{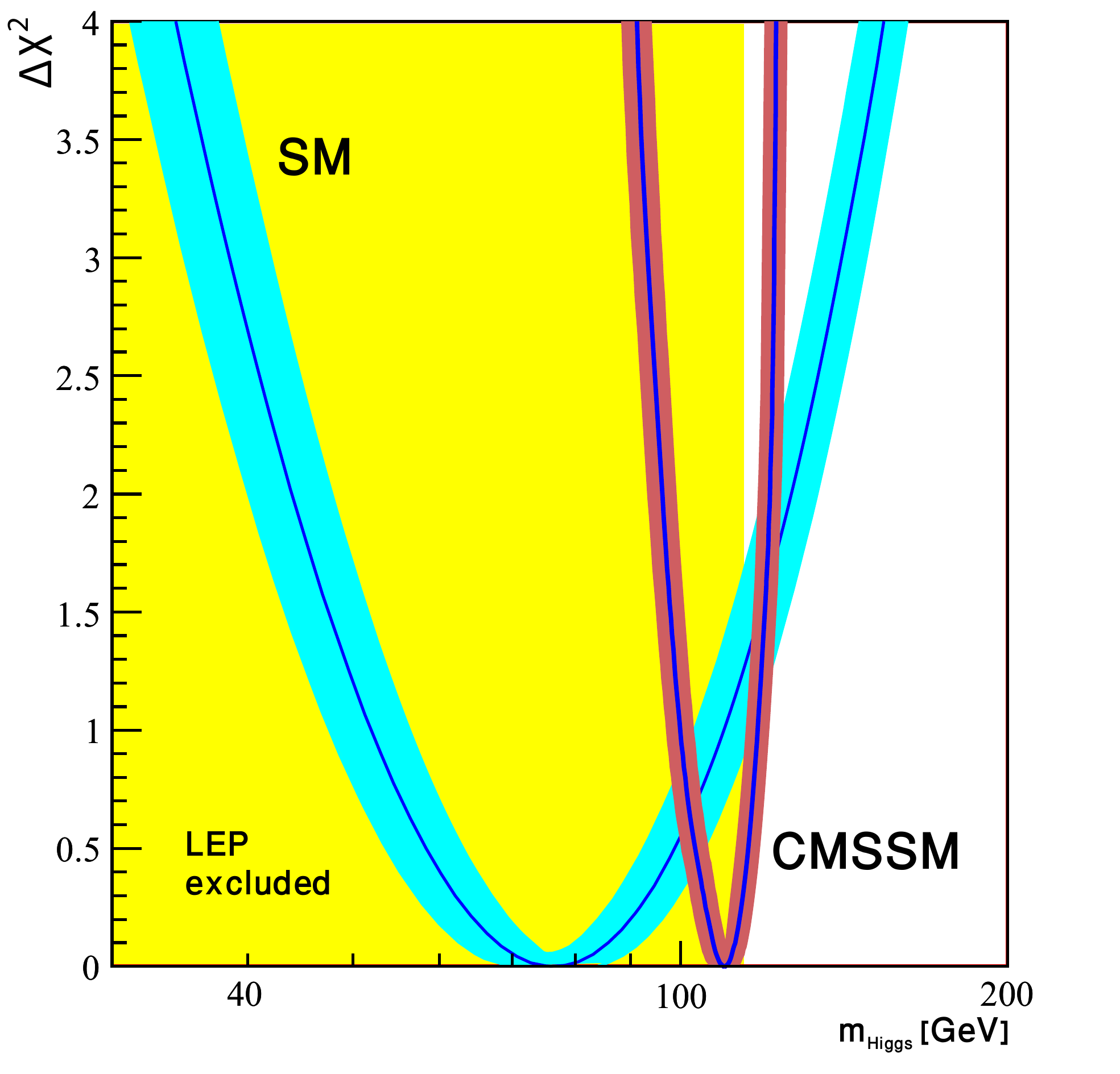}
\caption{Experimental constraints on the Higgs mass. The region excluded by direct search limits from the Large Electron-Positron (LEP) collider is shown in yellow. The blue band represents the results of fitting the Standard Model parameters to the electroweak precision data. The red band depicts a similar fit when the SM is minimally extended to its supersymmetric version. Adapted from ref.~\cite{Buchmueller:2007zk}}
\label{fig:Higgs}
\end{figure}

On the other hand, by fitting the SM parameters to the available electroweak precision data, the favoured value for the Higgs mass (\ie the minimum-$\chi^2$ point) is fairly low and well below the experimental direct limit from the LEP (see \fig{fig:Higgs}). Although this discrepancy is not statistically very significant, it definitely shows some tension. For comparison, \fig{fig:Higgs} shows also the result of a typical supersymmetric fit using one of the simplest supersymmetric extensions of the SM called the CMSSM (see section~\ref{sec:CMSSM})~\cite{Buchmueller:2007zk}. It can be seen from this example that it is possible to reconcile theoretical predictions for the Higgs mass with experimental data within the supersymmetric extensions of the SM.

\section{The need for quantum gravity}
\label{sec:quantumgravity}

As stated in chapter~\ref{sec:whyDMwhyBSM}, the SM of particle physics as a framework for describing the matter components of the Universe and their interactions, has been able to provide such a description in a mathematically consistent way only for three fundamental forces (out of four). The SM as a quantum-field theoretical framework is renormalisable only if gravitation is not included. Arguably, string theory (for an introcuction, see \eg ref.~\cite{stringtheory}) has so far been the most favoured candidate for a consistent quantum theory of gravitation which is expected to include the SM as its effective field theory valid at low energies. It is however highly difficult to build a phenomenologically successful string theory that does not require supersymmetry, and this means that supersymmetry is an essential ingredient of the best quantum description of gravitation so far. This makes supersymmetry particularly interesting.

Even if we do not believe in string theory as a valid description of high-energy phenomena, there is yet another intriguing connection between supersymmetry and gravity. As will be seen in the following chapters, the phenomenologically interesting versions of supersymmetry that we will consider are all based on a `global' symmetry. In the language of chapter~\ref{sec:SUSYfoundations} this means that the generators of supersymmetric transformations are not functions of space and time. It is however entirely justified to promote the global symmetry to a local one, in a way analogous to the gauge symmetries of the SM. Such a localisation process is shown to inevitably lead to the existence of a new spin-$2$ massless gauge field together with its supersymmetric partner, a spin-$3/2$ particle (see \eg ref.~\cite{SUSY:Baer}). The former is exactly the particle that is assumed to be responsible for gravitational interactions, and is accordingly called the graviton. The interesting characteristic of the supersymmetric graviton is that its dynamics, which is entirely fixed by local supersymmetry, contains Einstein's general relativity as our currently best classical theory of gravity. Regarding this connection with gravity, local supersymmetry is often called `supergravity' (or SUGRA). Although such a supergravity theory is not renormalisable\footnote{Strictly speaking, this statement may not be correct. There is a particular version of supergravity, called `$N=8$ supergravity' (see chapter~\ref{sec:SUSYfoundations} for the terminology), which is conjectured to be renormalisable~\cite{Bern:2009kd}. As we will point out in the next chapter, these versions of supersymmetry are however phenomenologically not very interesting.}, its natural connection to gravity should not be ignored.
 
\section{Other issues}
\label{sec:otherissues}

In addition to the issues with the SM we enumerated in the previous sections, there are a few other reasons to believe that the SM is not the complete theory of Nature. Most of these arguments are again purely theoretical (or aesthetic) in nature, but are still highly intriguing so that one cannot simply ignore them. We will not attempt to detail these other problems here and only list (or briefly introduce) a few interesting ones with some references for further reading.

The first problem comes from the observations of neutrino oscillations. These indicate that neutrinos have small but non-zero masses. In the SM, neutrinos are however massless and this directly implies that the model must be extended so as to accommodate massive neutrinos. In order to avoid a `fine-tuning' problem, this is usually done through the so-called `seesaw' mechanism that is generally implemented within the framework of grand unified theories discussed in section~\ref{sec:unification} (for a review of neutrino masses and mixing, see \eg ref.~\cite{Ray:2010rz} and references thein).

The other problem that is again related to a fine-tuning within the SM, is the `strong CP problem'. This deals with the fact that the quantum chromodynamics (QCD) sector of the SM, contrary to the electroweak sector, respects the CP-symmetry. This leads to a highly fine-tuned value for a parameter called `vacuum angle' and denoted by $\theta$~\cite{Kuster:2008zz}. The strong CP problem finds natural resolutions in models of physics beyond the SM, in particular through the introduction of new particles called axions (these are the same particles as the axions we mentioned in section~\ref{sec:WIMPs} in our list of viable dark matter candidates)~\cite{Peccei:1977hh}.

Let us end this chapter by adding to our list of issues two other theoretical speculations on the structure of the SM: (1) Why are there only three generations for matter particles, \ie for leptons and quarks? (2) What is the origin of fermion masses? These two may also find appropriate answers in theories beyond the SM.

\end{fmffile}

\setlength{\unitlength}{1mm}
\begin{fmffile}{feyn}

\chapter{Theoretical foundations of supersymmetry}
\label{sec:SUSYfoundations}

In the previous chapters, we attempted to review some answers to the question why we are interested in physics beyond the Standard Model of particle physics and in particular its supersymmetric extensions. Our discussions so far have been based on a very vague understanding of supersymmetry. Before we enter the world of concrete supersymmetric models in the next chapter and investigate various observational constraints on these models in chapter~\ref{sec:SUSYreal}, we briefly introduce supersymmetry in this chapter and review some of its fundamental properties. In addition, some formalisms are discussed and basics of supersymmetric model building are presented. This chapter is a rather technical one and the reader who is only interested in phenomenological aspects of the field can simply skip it and continue directly from chapter~\ref{sec:SUSYreal}.

\section{Supersymmetry is a symmetry}

All the known elementary particles are either bosons or fermions. Bosons are those particles (or fields) that obey Bose-Einstein statistics and this means that they can occupy the same quantum state at any given time. Fermions, on the other hand, obey Fermi-Dirac statistics and, consequently, only one fermion can occupy a particular quantum state at a time. Although the quantum mechanical distinction between matter and force is not a clear cut, fermions are often associated with matter whereas bosons are considered as carriers of forces and interactions between the fermions. According to the so-called spin-statistics theorem in quantum field theory, bosons have integer spin while fermions possess half-integer spin (for an introduction to quantum field theory, see \eg ref.~\cite{QFTPeskin}).

Elementary fermions that are known to exist in Nature, according to the Standard Model of particle physics, are categorised as quarks (6 particles $u$ (up), $d$ (down), $c$ (charm), $s$ (strange), $t$ (top), $b$ (bottom) and 6 corresponding antiparticles $\bar{u}$, $\bar{d}$, $\bar{c}$, $\bar{s}$, $\bar{t}$, $\bar{b}$) or leptons (3 charged particles $e^-$ (electron), $\mu^-$ (muon), $\tau^-$ (tau), 3 neutrinos $\nu_e$, $\nu_\mu$, $\nu_\tau$ and 6 corresponding antiparticles $e^+$ (positron), $\mu^+$, $\tau^+$, $\bar{\nu}_e$, $\bar{\nu}_\mu$, $\bar{\nu}_\tau$). The SM also contains 7 elementary bosons in total (if we include the graviton), some of which, such as the gauge bosons $\gamma$ (photon), $g$ (gluon), $W^{\pm}$ and $Z$ have already been discovered, while the other two, \ie $H^0$ (Higgs boson) and $G$ (graviton) are to be observed experimentally  (see \eg ref.~\cite{SM:Burgess} for an introduction to the SM).

As we pointed out in the previous chapters, the mathematical structure of the SM that describes its field content and various interactions between the fields, is constructed based on some particular symmetries, some of which are thought to be fundamental.

The first symmetry from the latter category is called Pioncar\'e symmetry and is a `spacetime' symmetry. The Poincar\'e group (for an introduction to group theory and its applications in particle physics, see \eg ref.~\cite{LieGroupGeorgi}) is the full symmetry of special relativity and correspondingly any relativistic field theory; the SM is no exception. This is a 10-dimensional noncompact Lie group and is the group of isometries of Minkowski spacetime. The Poincar\'e group includes the Lorentz group as a subgroup and is a semi-direct product of translations in spacetime and Lorentz transformations (\ie $R^{1,3}\times O(1,3)$, where $R^{1,3}$ stands for the former and $O(1,3)$ denotes the latter). Mathematically speaking, all elementary particles (or fields) are different `irreducible representations' of the group and are specified by two quantities: mass (or four-momentum) and spin (an intrinsic quantum number). The Poincar\'e symmetry is considered as a fundamental symmetry which every quantum-field theoretical framework that describes particles and their interactions should possess (including the SM and its potential extensions).

The other symmetry that is implemented in the SM, and has been used as a guiding principle in constructing its theoretical structure, is an `internal' symmetry, \ie a symmetry which is not obviously related to space and time. This determines how different components of a theory (\eg different fields in the SM) transform into each other. The SM is called a `gauge theory', and this is because the fundamental Lagrangian of the theory is invariant (or symmetric) under a particular non-abelian gauge symmetry: $SU(3)_C \times SU(2)_L \times U(1)_Y$. This gauge symmetry is an example of internal symmetries.

While spacetime symmetries of a quantum field theory dictate the properties of the field components and classify them into various categories of scalars, vectors, tensors and spinors, internal symmetries rather determine how different terms in the Lagrangian must be written. The two symmetries are entirely independent in the SM.

Supersymmetry or SUSY (for an introduction, see \eg refs.~\cite{SUSY:Baer,SUSY:Aitchison,Martin:1997ns,Shirman:2009mt,Bilal:2001nv}) is a symmetry that transforms fermionic degrees of freedom into bosonic ones and vice versa. In a supersymmetric theory, every fermion has a bosonic `superpartner' and every boson has a fermionic superpartner. With this definition, supersymmetry can be considered as a new internal symmetry because it gives a new way to transform some components of the theory, say fermions, to some other ones, \ie bosons. This is however not entirely true. Supersymmetry is actually also an extension of the Poincar\'e group in the sense that it extends the `Poincar\'e algebra' (and therefore special relativity) through the introduction of four anticommuting `spinor' generators. In other words, although in supersymmetry, fermions are transformed into bosons and vice versa, these transformations only modify the particles' spin and this is essentially a spacetime property.

\section{The supersymmetry algebra}
\label{SUSYalgebra}

Supersymmetry, as any other continuous symmetry, is characterised by a symmetry algebra, and as we mentioned in the previous section, this algebra is obtained by extending the Poincar\'e algebra such that it relates two types of fields (bosons and fermions) in a single algebra. The resultant algebra is called a Lie `superalgebra'.

Let us first look at the Poincar\'e algebra: Since the Poincar\'e group is a semi-direct product of the Lorentz group and the group of spacetime translations, a general Poincar\'e transformation contains both Lorentz transformations and translations. The Lorentz group has 6 generators: $J_i;~i=1,2,3$ (3 rotations) and $K_i;~i=1,2,3$ (3 boosts). One often denotes the generators of translations as $P_\mu;~\mu=0,1,2,3$. In a more covariant looking form, the Lorentz generators are usually written as $M_{\mu\nu}=-M_{\mu\nu}$, where $M_{0i}=K_i$ and $M_{ij}=\epsilon_{ijk}J_k$. In this notation, the full Poincar\'e algebra can be written as~\cite{Bilal:2001nv}
\begin{eqnarray}
\left[P_\mu,P_\nu \right] &=& 0, \nonumber \\
\left[M_{\mu\nu},M_{\rho\sigma} \right] &=& ig_{\nu\rho}M_{\mu\sigma}-ig_{\mu\rho}M_{\nu\sigma}-ig_{\nu\sigma}M_{\mu\rho}+ig_{\mu\sigma}M_{\nu\rho}, \nonumber \\
\left[M_{\mu\nu},P_\rho \right] &=& -ig_{\rho\mu}P_{\nu}+ig_{\rho\nu}P_\mu.
\end{eqnarray}

In supersymmetry, the Poincar\'e algebra is enlarged by generators that are `spinorial', \ie transform as spinors in contrast to the original `tensorial' Poincar\'e generators which transform as tensors. Such generators are often denoted by dotted and undotted spinors $Q^I_\alpha$ and $\bar{Q}^I_{\dot{\alpha}}$. These are objects that transform under the group $Sl(2,C)$ as
\begin{eqnarray}
Q^I_\alpha \rightarrow Q'^I_\alpha=M_\alpha^{\ \beta} Q^I_\beta,~~~~~~\bar{Q}^I_{\dot{\alpha}} \rightarrow \bar{Q}'^I_{\dot{\alpha}}={M^*}_{\dot{\alpha}}^{\ \dot{\beta}} \bar{Q}^I_{\dot{\beta}},
\end{eqnarray}
where $M\in Sl(2,C)$ and $\alpha,\beta=1,2$. Here, the extra index $I$ labels distinct SUSY generators in case there are more than one pair. The number of such generator pairs is usually denoted by $N$ (\ie $I=1,...,N$). The simplest case with only one pair is accordingly called `$N=1$ supersymmetry'.

Mathematically, there is no limit on $N$, but with increasing $N$ the theories contain particles of increasing spin. Since no consistent quantum field theory with spins larger than two exists, this then leads to the condition $N\leq8~\cite{Bilal:2001nv}.$\footnote{This is however the case when gravity is part of the theory, otherwise, spins cannot be larger than one and this leads to $N\leq4$.} $N=1$ supersymmetry (that is also called `unextended' supersymmetry) is of particular interest since it is the only case which permits chiral fermions. We know that chiral fermions exist, therefore from a phenomenological point of view, any supersymmetric theory of particle physics has to be of $N=1$ type, at least at low energies. We therefore restrict our discussions to $N=1$ supersymmetry. 

The extended Poincar\'e algebra (\ie the superalgebra) includes the following new commutation and anti-commutation relations:
\begin{eqnarray}
&\left[P_\mu,Q^I_\alpha\right]=0\ ,~~~ \left[P_\mu,\bar{Q}^I_{\dot{\alpha}}\right]=0 \ , \nonumber \\
&\left[M_{\mu\nu},Q^I_\alpha\right]=i (\sigma_{\mu\nu})_\alpha^{\ \beta} Q^I_\beta ,~~~
\left[M_{\mu\nu},\bar{Q}^{I\dot{\alpha}}\right]= i (\bar{\sigma}_{\mu\nu})^{\dot{\alpha}}_{\ \dot\beta} \bar{Q}^{I\dot{\beta}}, \nonumber \\
&\{Q^I_\alpha, \bar{Q}^J_{\dot{\beta}} \}=2 \sigma^\mu_{\alpha{\dot{\beta}}} P_\mu \delta^{IJ},~~~
\{Q^I_\alpha, Q^J_\beta\}= \epsilon_{\alpha\beta} Z^{IJ}\ ,~~~
\{\bar{Q}^I_{\dot{\alpha}}, \bar{Q}^J_{\dot{\beta}}\} = \epsilon_{\dot{\alpha}\dot{\beta}} (Z^{IJ})^{*} \nonumber \ .
\end{eqnarray}
Here, $Z^{IJ}=-Z^{JI}$ are the so-called `central charges' of the group, and they are the members that commute with all generators of the algebra. The $N=1$ (or unextended) SUSY algebra, is the simplest supersymmetry algebra which has no central charges. One important property of any supersymmetric theory that can be inferred from the above SUSY algebra is that the energy $P_0$ is always positive.

Any irreducible representation of the Poincar\'e algebra is associated with a particle. Since the Poincar\'e algebra is a subalgebra of the superalgebra, any representation of the latter is also a representation of the former. However in general, an irreducible representation of the superalgebra corresponds to a reducible representation of the Poincar\'e algebra, and this means that it corresponds to several particles. The particles of each SUSY representation are related to each other by the SUSY generators $Q^I_\alpha$ and $\bar{Q}^J_{\dot{\beta}}$. This means that these particles have spins that differ by units of one half, \ie some are bosons and some are fermions. The spin-statistics theorem then implies that the generators $Q^I_\alpha$ and $\bar{Q}^J_{\dot{\beta}}$ transform fermions to bosons and vice versa. The particles that are obtained via supersymmetric transformations of other particles, are called supersymmetric partners or simply `superpartners' of the original particles. 

An irreducible representation of supersymmetry that is equivalent to a set of supersymmetrically-related particle states is called a `supermultiplet'. All particles (or states) belonging to a supermultiplet have equal masses, and any supermultiplet contains an equal number of fermionic and bosonic degrees of freedom. One can show that for $N=1$ (\ie unextended) supersymmetry (with gravity), only three types of massless multiplets exist: chiral multiplets (consisting of a Weyl fermion with spin $1/2$ and a complex scalar with spin $0$), vector multiplets (consisting of a gauge boson with spin $1$ and a Weyl fermion) and graviton multiplets (consisting of a graviton with spin $2$ and a gravitino with spin $3/2$). It is tempting to also add gravitino multiplets (consisting of a gravitino and a gauge boson) to the above set of multiplets, but such a multiplet can only happen in an extended (\ie $N>1$) SUSY.

\section{The Wess-Zumino model}
\label{sec:WZ}

In this section, we briefly describe how a supersymmetric field theory can be constructed at the `action' level. Although the described model is too simple compared to the more sophisticated ones thought to be implemented in Nature (see chapter~\ref{sec:SUSYreal}), it shows the main ingredients of any supersymmetric theories including the ones that are of phenomenological interest (see also refs.~\cite{SUSY:Baer,SUSY:Aitchison,Martin:1997ns,Shirman:2009mt}).

By looking again at the possible SUSY multiplets we enumerated in the previous section, we see that the simplest representation of supersymmetric transformations which includes a chiral fermion (with spin $1/2$) is a chiral multiplet. In addition to a left-handed two-component Weyl fermion, that we denote by $\psi$, this multiplet includes a complex scalar field, say $\phi$.

Let us now try to write down a four-dimensional SUSY-invariant action that is composed of $M$ chiral multiplets with $M$ scalar fields $\phi_i$ and $M$ Weyl fermions $\psi_i$ ($i=1,...,M$). We demand the action to be supersymmetric `off-shell'. This means that the action is invariant under supersymmetry even if the classical equations of motion are not satisfied. The latter requirement leads to the addition of a set of `auxiliary' complex scalar fields $F_i$ (fields without kinetic terms) to the field content of the theory. The resultant Lagrangian density for our SUSY theory reads
\beq
\lagr_{\rm free} &=&
\BDpos \partial^\mu \phi^{*} \partial_\mu \phi
+ i {\psi}^{\dagger} \sigmabar^\mu \partial_\mu \psi
+ F^{*} F .
\label{lagrfree}
\eeq

This is a theory for massless and free fields (\ie includes no interaction terms) and was first derived by Wess and Zumino~\cite{Wess:1974tw}. The next step is obviously to add non-gauge interaction terms to the Lagrangian such that they preserve the supersymmetric property of the action. If we only retain the renormalisable interactions (\ie ones with mass dimension $\leq 4$), it can be shown that the most general Lagrangian with non-gauge interactions for chiral multiplets has to have the following form:
\begin{eqnarray}
\lagr_{\rm chiral}&=&\lagr_{\rm free} + \lagr_{\rm int} =\BDpos\partial^\mu \phi^{*i} \partial_\mu \phi_i + i \psi^{\dagger i} \sigmabar^\mu \partial_\mu \psi_i \nonumber \\
&&-\half \left (W^{ij} \psi_i \psi_j  + W^{*}_{ij} \psi^{\dagger i}
\psi^{\dagger j} \right )
- W^i W^{*}_i.
\label{noFlagr}
\end{eqnarray}
Here $W_i$ and $W_{ij}$ are defined as derivatives of the so-called `superpotential' $W$ that is a function of the scalar fields $\phi_i$:
\begin{eqnarray}
\label{superpotdef}
W &=& L^i \phi_i +
{1\over 2} M^{ij} \phi_i \phi_j + {1\over 6} y^{ijk} \phi_i \phi_j \phi_k , \nonumber \\
W^i &=& {\delta W\over \delta \phi_i} , \\
W^{ij} &=& {\delta^2 \over \delta\phi_i\delta\phi_j} W . \nonumber
\end{eqnarray}
Here, $M^{ij}$ is a symmetric mass matrix for the fermions, $y^{ijk}$ is a Yukawa coupling of a scalar and two fermions, and $L^i$ are some additional parameters that influence only the scalar potential of the Lagrangian~\cite{Martin:1997ns}. Auxiliary fields $F_i$ are eliminated from the expression using their classical equations of motion. The term $W^i W^{*}_i$ in \eq{noFlagr} is only a function of the scalar fields $\phi$ and $\phi^{*}$ and is essentially the `scalar potential' of the theory (usually denoted by $V(\phi,\phi^{*})$). The model introduced in \eq{noFlagr} is called the Wess-Zumino model~\cite{Wess:1973kz}.

\section{Supersymmetric gauge theories}
\label{sec:SUSYgauge}

In the previous section, we showed how a simple supersymmetric theory looks like for a chiral supermultiplet. This can be used in constructing a model that describes particle physics fermions (\ie leptons and quarks) and scalars (such as the Higgs boson). We however know that in reality, at least at low energies, there are other types of fields which should also be described in any supersymmetric extension of the SM: gauge fields.

As pointed out in section~\ref{SUSYalgebra}, any SUSY theory for gauge fields should include vector (or gauge) multiplets as basic ingredients. These multiplets have massless gauge bosons (that we denote by $A_\mu^a;~\mu=0,1,2,3$), as well as Weyl fermions $\lambda^a$. Here the index $a$ can take on different integer values depending on the particular gauge group of the theory (\eg $a=1,...,8$ for $SU(3)_C$, $a=1,2,3$ for $SU(2)_L$ and $a=1$ for $U(1)_Y$). As for the chiral multiplet case, one has to also add an auxiliary field to the field content of the theory. Such a field, traditionally named $D^a$, is real and bosonic, and is required for the action to be SUSY-invariant off-shell. The Lagrangian density for the gauge multiplet is shown to have the following form:
\beq
\lagr_{\rm gauge} = -{1\over 4} F_{\mu\nu}^a F^{\mu\nu a}
+ i \lambda^{\dagger a} \sigmabar^\mu D_\mu \lambda^a
+ {1\over 2} D^a D^a ,
\label{lagrgauge}
\eeq
where $F_{\mu\nu}^a$ is the Yang-Mills field strength for the gauge fields $A_\mu^a$ and $D_\mu \lambda^a$ is the covariant derivative of the $\lambda^a$ field~\cite{Martin:1997ns}.

As the final step towards constructing a general supersymmetric Largangian, one needs to consider both contributions from the chiral and gauge supermultiplets, as well as any additional interaction terms that are allowed by gauge invariance and keep the theory supersymmetric. Adding the requirement that the interaction terms should be renormalisable (\ie of mass dimension $\leq 4$ in four dimensions), our general Lagrangian density will have the following form:
\beq
\lagr & = & \lagr_{\rm chiral} + \lagr_{\rm gauge} 
\nonumber\\
        && - \sqrt{2} g 
(\phi^* T^a \psi)\lambda^a 
- \sqrt{2} g\lambda^{\dagger a} (\psi^\dagger T^a \phi)
+ g  (\phi^* T^a \phi) D^a ,
\label{gensusylagr}
\eeq
where $\lagr_{\rm chiral}$ and $\lagr_{\rm gauge}$ are defined in \eqs{noFlagr} and~\ref{lagrgauge}, respectively. The only difference is that the ordinary derivatives $\partial_\mu$ in \eq{noFlagr} for the chiral supermultiplet Lagrangian $\lagr_{\rm chiral}$ are now replaced by gauge-covariant derivatives $D_\mu$. $T^{a}$ are the generators of the gauge group that satisfy $[T^a,T^b]=i f^{abc} T^c$. Here $f^{abc}$ are the structure constants that define the gauge group and $g$ is the `gauge coupling'.

The complete scalar potential of the theory in this case is shown to be expressible purely in terms of the auxiliary fields $F_i$ and $D^a$ (which are in turn expressible only in terms of the scalar fields $\phi_i$):
\beq
V(\phi,\phi^*) = F^{*i} F_i + \half \sum_a D^a D^a .
\label{fdpot}
\eeq

The first and second terms in \eq{fdpot} are called $F$-terms and $D$-terms, respectively. The former are entirely fixed by Yukawa couplings and fermion mass terms, while the latter are fixed by the gauge interactions. In addition, the scalar potential $V(\phi,\phi^{*})$ can be shown to be bounded from below, \ie it is always greater than or equal to zero.

Finally, one should notice here that the theory defined in \eq{gensusylagr} is invariant under `global supersymmetry'. This means that the parameters of supersymmetric transformations do not depend on the spacetime co-ordinates. We however know that local symmetries also exist and some of them have played critical roles in our current description of particle physics: The best example is the gauge symmetries of the SM. `Local supersymmetry' also exists and as we pointed out in section~\ref{sec:quantumgravity} makes an interesting connection between supersymmetry and gravity (in the context of `supergravity'). The theory of supergravity is highly technical and we do not detail it in this thesis. We only briefly describe in the next chapter (mainly section~\ref{sec:shrink}) some phenomenologically interesting models that have been constructed based on supergravity assumptions. We refer the interested reader to the literature for detailed discussions (see \eg refs.~\cite{SUSY:Baer,Martin:1997ns} and references therein).

\section{Spontaneous supersymmetry breaking}
\label{sec:SUSYbreaking}

As we will argue in the next chapter, supersymmetry cannot be implemented in Nature as an exact symmetry and is required to be broken appropriately. From a theoretical point of view, arguably the most interesting way of breaking a symmetry in any quantum field theory is via a `Higgs-like' mechanism, where the symmetry is broken `spontaneously'. This idea has seemingly worked very well in the SM when the electroweak gauge symmetry is broken at TeV scales. It is therefore quite interesting to see how the same idea could work for supersymmetry (see also refs.~\cite{SUSY:Baer,SUSY:Aitchison,Martin:1997ns,Shirman:2009mt}).

Spontaneous supersymmetry breaking means that while the Lagrangian of the theory is SUSY-invariant, the vacuum state $|0 \rangle$ is not, \ie $Q_\alpha |0 \rangle \neq 0$ and $Q^{\dagger}_{\dot{\alpha}} |0 \rangle \neq 0$. In an unbroken supersymmetry, the vacuum has zero energy (since $H|0 \rangle=0$, where $H$ is the Hamiltonian operator), while in a spontaneously-broken supersymmetry the vacuum has positive energy (\ie $\langle 0|H|0 \rangle>0$). It can be shown from this that if the vacuum expectation value (or VEV) of $F_i$ and/or $D^a$ (the auxiliary fields introduced in sections~\ref{sec:WZ} and~\ref{sec:SUSYgauge}) become non-zero (\ie $\langle D^a\rangle\neq0$ and/or $\langle F_i \rangle\neq0$), supersymmetry will be spontaneously broken.

In SUSY-breaking models in which the vacuum state we live in is assumed to be the true ground state of the theory, the structure of the models usually imply that the equations $F_i=0$ and $D^a=0$ cannot be satisfied simultaneously and this breaks SUSY spontaneously. Other models exist in which we are not assumed to live in the true ground state and instead live in a metastable SUSY-breaking state with sufficiently long lifetime (comparable to the current age of the Universe) (see \eg ref.~\cite{Intriligator:2006dd}). This metastable state might have been chosen by some finite temperature effects in the early Universe.

Spontaneous SUSY breaking is usually implemented in different models either through the Fayet-Iliopoulos (or `$D$-term') mechanism~\cite{Fayet:1974jb,Fayet:1974pd} or through the O'Raifeartaigh (or `$F$-term') mechanism~\cite{O'Raifeartaigh:1975pr}.

In the $D$-term SUSY-breaking mechanism, the gauge symmetry group needs to contain a $U(1)$ subgroup with a non-zero $D$-term VEV. Supersymmetry is then broken by introducing the following little extra piece to the SUSY Lagrangian:
\beq
\lagr_{\rm Fayet-Iliopoulos} = -\kappa D,
\label{FI}
\eeq
which is a term proportional to $D$ ($\kappa$ being a constant).

In the $F$-term mechanism, SUSY breaking occurs due to the existence of a non-vanishing $F$-term VEV that comes from a particular property of the superpotential $W$, namely that there is no simultaneous solutions for the equations
\beq
F_i=-W^*_i=0,
\eeq
with $W^*_i$ defined in \eq{superpotdef} (\ie $W^*_i=\delta W^*/\delta \phi^{*i}$).

One property of all types of spontaneous global SUSY breaking models (with stable or metastable vacuum states), is the existence of a massless neutral Weyl fermion as the Nambu-Goldstone mode. This fermion is called goldstino, is denoted by $\tilde G$ and possesses the same quantum numbers as the broken symmetry generator (which in our case is the fermionic charge $Q_{\alpha}$).  The $\tilde G$ can be shown to have the form $(\langle D^a\rangle/\sqrt{2}, \langle F_i \rangle)^T$, \ie its components are proportional to the VEVs of the auxiliary fields $F_i$ and $D^a$~\cite{Martin:1997ns}.

With this brief introduction to the two aforementioned SUSY-breaking mechanisms, we stop our discussion here. We will instead come back to the discussion of supersymmetry breaking in section~\ref{sec:fundamentals} of the next chapter where we discuss various concrete scenarios in connection with phenomenologically interesting SUSY models. We will see how some of the general strategies described here can be used in constructing real-world theories.

\section{Superfield formalism}

In order to construct more complex supersymmetric Lagrangians, with larger numbers of fields and more complicated interaction terms, one needs to develop a rather general procedure that generates SUSY-invariant interactions in a systematic way. A compact and convenient way is to use `superspace' and `superfield' formalism. In $N=1$ supersymmetry, the superspace is the usual four-dimensional spacetime (labelled by the four coordinates $x^\mu;~\mu=0,1,2,3$) enlarged by adding four anticommuting `Grassmannian coordinates' $\theta_\alpha$ and $\theta_{\dot{\alpha}}$. These new coordinates are fermionic and transform as a two-component spinor and its conjugate. In general, for an extended supersymmetry with $N$ SUSY generator pairs, there are $4N$ extra fermionic coordinates. Superfields are quantum fields that differ from the usual ones in that they are defined on the superspace rather than the spacetime. Superfields are defined as single objects with components being all the different fields (fermionic, bosonic and auxiliary) that belong to a supermultiplet.

The main advantage of using superfield formalism is that the invariance under SUSY transformations remain manifest during the Lagrangian construction; this is because the Lagrangian is defined in terms of integrals over the superspace. Working with superfield formalism has also the advantage that the spacetime nature of supersymmetric transformations is more manifest. Despite all the definite benefits of working with superfields, the formalism is fairly complicated and we do not detail it here. We refer the interested reader to \eg ref.~\cite{SUSY:Baer} for a detailed introduction.

\end{fmffile}

\setlength{\unitlength}{1mm}
\begin{fmffile}{feyn}

\chapter{Supersymmetry in real life}
\label{sec:SUSYreal}

In the previous chapter, we described supersymmetry in general and discussed various properties of a supersymmetric field theory. This was done mainly through the presentation of the simplest possible SUSY models with a minimal field content, \ie the Wess-Zumino model and its gauge extension. However, these models are obviously too simple to describe the real world. It is the goal of the present chapter to discuss viable SUSY models and scenarios that may describe reality. We also argued in chapters~\ref{sec:whyDMwhyBSM},~\ref{sec:cosmology} and~\ref{sec:SMissues} why a supersymmetric extension of the SM is helpful, although our discussions were limited to rather general arguments. We may therefore want to see in a more explicit way how suprsymmetric models could address the issues discussed there. We will detail in this chapter `some' of those issues in terms of definite SUSY models. Finally, in order to examine how observations could enhance our knowledge about supersymmetry, its validity and possible implementations in Nature (which has been the primary objective of this thesis), we need to have concrete theoretical frameworks to work in. The present chapter also provides these frameworks.

\section{The Minimal Supersymmetric Standard Model}
\label{sec:MSSM}

There are various strategies in building a SUSY model that has to do with reality. In a top-down approach, one looks at some fundamentally motivated theories that accommodate supersymmetry, such as string theory. These theories are usually defined at very high energies that are not accessible by experiments. Phenomenological studies can then be carried out by extracting an effective supersymmetric field theory valid at low energies.

In an alternative bottom-up approach, one starts with the SM itself and adds all the ingredients that are required for it so as to become supersymmetric. It is important that in the latter approach one takes into account all phenomenological considerations and constraints in such a way that the emergent theory is consistent with observations as well as theoretical conditions. It must also give the SM as an effective theory valid up to certain energies since we know that the SM is an excellent description of particle physics below those energies.

The simplest phenomenologically-constructed SUSY model (\ie obtained through a bottom-up approach) is the so-called `Minimal Supersymmetric Standard Model (MSSM)'~\cite{Dimopoulos:1981zb} (see also refs.~\cite{SUSY:Baer,SUSY:Aitchison,Martin:1997ns} for comprehensive introductions to the MSSM). It is minimal in the sense that it contains the smallest number of new particles (or fields) that can be added to the SM in order to make it supersymmetric, and the theory still remains consistent with all phenomenological requirements. In this section, we describe the MSSM and its properties that are of most interest for phenomenological studies of supersymmetry. 

\subsection{Field content and superpotential}
\label{sec:fieldcontent}

In every supersymmetric model, including the MSSM, the number of degrees of freedom for particles and their corresponding supersymmetric partners (or superpartners) match. This particularly implies that some SM particles have more than one superpartner. For example, the elementary fermionic spin-$1/2$ particles with two degrees of freedom (such as leptons and quarks) need two scalar superpartners with one degree of freedom each. The superpartners of the SM fermions are called `sfermions' (sleptons for leptons and squarks for quarks) and the superpartners of the bosons are called `bosinos' (gauginos for gauge bosons and Higgsinos for Higgs bosons). We also often refer to the superpartners of the SM particles simply as `sparticles'.

In the MSSM, every known (\ie SM) particle has a spin $0$, $1/2$ or $1$ and must therefore reside, together with its superpartners, in either a chiral or gauge supermultiplet (see the previous chapter). We summarise in Tab.~\ref{MSSMparticles} all the particles and spartners in the MSSM. As we see, they are divided into two categories of chiral and gauge supermultiplets. Tab.~\ref{MSSMparticles} also shows different hypercharges associated with the particles. These correspond to the three SM gauge groups.

One interesting feature of the MSSM is that, contrary to the SM, it contains `two' Higgs doublets (shown as $(H_u^+,~H_u^0)$ and $(H_d^0,~H_d^-)$ in Tab.~\ref{MSSMparticles}) that consequently require two chiral supermultiplets. There are two main reasons for this: (1) Only one Higgs chiral supermultiplet would introduce a gauge anomaly in the electroweak gauge symmetry that would make the theory quantum-mechanically inconsistent. (2) The Higgs chiral supermultiplet that has the Yukawa couplings necessary for giving masses to the up-type quarks, has a hypercharge that is different from the hypercharge of the Higgs chiral supermultiplet that has the Yukawa couplings necessary for giving masses to the down-type quarks and the charged leptons (see \eg ref.~\cite{Martin:1997ns} for more details).

The existence of two Higgs doublets in the MSSM and the fact that every bosonic degree of freedom has a corresponding fermionic degree of freedom and vice versa, together imply that the MSSM particle content is slightly more than a doubling of the SM particle content. It is also important to note that the sparticles presented in Tab.~\ref{MSSMsuperpot} are the `interaction' (or gauge) eigenstates of the theory and the `mass' eigenstates are in general linear combinations of the gauge eigenstates (we will come back to this in section~\ref{sec:sparticlemasses}).

\begin{table}[tb]
{\footnotesize
\begin{center}
\begin{tabular}{c c c c c c c}
\toprule
\multicolumn{7}{c}{\textbf{Chiral supermultiplets}} \\
\toprule
\textbf{Name} & \textbf{Symbol} & \textbf{spin 0} & \textbf{spin 1/2} & \textbf{$SU(3)_C$} & \textbf{$SU(2)_L$} & \textbf{$U(1)_Y$} \\  \toprule
{\footnotesize (s)quarks} & $Q$ & $({\stilde u}_L\>\>\>{\stilde d}_L )$&
 $(u_L\>\>\>d_L)$ & ${\bf 3}$ & $\bf 2$ & $1\over 6$
\\
{\footnotesize $\times 3$} & $\sbar u$
&${\stilde u}^*_R$ & $u^\dagger_R$ & 
${\bf \overline 3}$ & ${\bf 1}$ & $-{2\over 3}$
\\ & $\sbar d$ &${\stilde d}^*_R$ & $d^\dagger_R$ & 
${\bf \overline 3}$ & ${\bf 1}$ & ${1\over 3}$
\\  \midrule
{\footnotesize (s)leptons} & $L$ &$({\stilde \nu}\>\>{\stilde e}_L )$&
 $(\nu\>\>\>e_L)$ & ${\bf 1}$ & ${\bf 2}$ & $-{1\over 2}$
\\
{\footnotesize $\times 3$} & $\sbar e$
&${\stilde e}^*_R$ & $e^\dagger_R$ & ${\bf 1}$ & ${\bf 1}$ & $1$
\\  \midrule
{\footnotesize Higgs(inos)} &$H_u$ &$(H_u^+\>\>\>H_u^0 )$&
$(\stilde H_u^+ \>\>\> \stilde H_u^0)$& 
${\bf 1}$ & ${\bf 2}$ & $+{1\over 2}$
\\ &$H_d$ & $(H_d^0 \>\>\> H_d^-)$ & $(\stilde H_d^0 \>\>\> \stilde H_d^-)$& 
${\bf 1}$ & ${\bf 2}$ & $-{1\over 2}$
\\  \toprule
\multicolumn{7}{c}{\textbf{Gauge supermultiplets}} \\
\toprule
\textbf{Name} & & \textbf{spin 1/2} & \textbf{spin 1} & \textbf{$SU(3)_C$} & \textbf{$SU(2)_L$} & \textbf{$U(1)_Y$}\\
\toprule
{\footnotesize gluon/gluino} & & $ \stilde g$& $g$ & ${\bf 8}$ & ${\bf 1}$ & $0$
\\
\midrule
{\footnotesize winos, W bosons} & & $ \stilde W^\pm\>\>\> \stilde W^0 $&
 $W^\pm\>\>\> W^0$ & ${\bf 1}$ & ${\bf 3}$ & $0$
\\
\midrule
{\footnotesize bino, B boson} & & $\stilde B^0$&
 $B^0$ & ${\bf 1}$ & ${\bf 1}$ & $0$
\\
\bottomrule
\end{tabular}
\caption{Chiral and gauge supermultiplets in the Minimal Supersymmetric Standard Model. The table is based on similar tables in ref.~\cite{Martin:1997ns}}\label{MSSMparticles}
\vspace{-0.6cm}
\end{center}
}
\end{table}

Like any other supersymmetric theory, the SUSY part of the MSSM Lagrangian is determined by a superpotential that is defined in terms of the chiral supermultiplets (see the previous chapter). The MSSM superpotential is~\cite{Martin:1997ns}
\beq
W_{\rm MSSM} =
\sbar u {\bf y_u} Q H_u -
\sbar d {\bf y_d} Q H_d -
\sbar e {\bf y_e} L H_d +
\mu H_u H_d \> ,
\label{MSSMsuperpot}
\eeq
where $Q$, $L$, $\sbar u$, $\sbar d$, $\sbar e$, $H_u$, and $H_d$ denote chiral superfields of the theory. It is important to notice that there are three generations (\ie three families) for quarks/squarks and lepton/sleptons and although not written explicitly in \eq{MSSMsuperpot}, the summation over the generations is understood. Similarly, all gauge indices and summations are suppressed. The presence of three generations implies that the Yukawa couplings ${\bf y_u}$, ${\bf y_d}$ and ${\bf y_e}$ in \eq{MSSMsuperpot} (which are exactly the same Yukawa couplings as those that enter the SM Lagrangian) are $3\times 3$ matrices in the family space.

The superpotential defined in \eq{MSSMsuperpot} completely determines the structure of the MSSM if SUSY is not broken (see the next section). This means that we have now obtained an exactly supersymmetrised version of the SM albeit at the cost of introducing one new parameter, \ie $\mu$.

\subsection{SUSY breaking and soft terms}

Simple phenomenological considerations imply that supersymmetry cannot be an exact symmetry of Nature (at least at low energies), and if implemented in Nature, must be broken spontaneously. In other words although the fundamental Lagrangian might be SUSY invariant, the vacuum state that Nature has chosen need not be (see section~\ref{sec:SUSYbreaking} in the previous chapter). In a fully supersymmetric theory, masses of particles and their corresponding superpartners are equal. This immediately puts the theory in trouble if it is to describe reality. For example masses of selectrons (\ie the superpartners of electrons) should be as low as the electron mass, \ie about $0.5$ MeV. A particle with such a low mass should be easily detected experimentally, as electron is, and should essentially show up in our everyday life. This all means that SUSY is a broken symmetry. In addition, all sparticle masses should be much higher than the SM masses (in order not to have been observed in low-energy experiments).

As we stated earlier, the MSSM is a phenomenological model in the sense that its general structure is not set by any fundamental high-energy theory. This clearly means that phenomenological considerations should also fix the structure of any SUSY-breaking terms that we may add to the MSSM Lagrangian (see \eg refs.~\cite{SUSY:Baer,SUSY:Aitchison,Martin:1997ns,Shirman:2009mt,Chung:2003fi,Luty:2005sn}).

One important guiding principle in determining the SUSY-breaking interactions in the MSSM comes from one of the strongest theoretical motivations for extending the SM to its SUSY version, \ie providing a solution to the gauge hierarchy problem. We argued in section~\ref{sec:hierarchy} that the quadratic divergences from the radiative corrections to the scalar masses can be cancelled out in a supersymmetric theory if fermionic fields and their bosonic partners have equal masses. This is clearly not the case in a SUSY-broken theory. It can however be shown that if the sparticles have masses not much larger than TeV scales, the cancellation of the different loop contributions does not require huge fine-tuning and therefore SUSY can still provide a solution to the hierarchy problem~\cite{Witten:1981nf}.

Supersymmetry is broken in the MSSM by adding the so-called `soft SUSY-breaking terms' to the exact supersymmetric Lagrangian described in the previous section. These are the terms that while breaking supersymmetry, satisfy four conditions: (1) They do not reintroduce quadratic divergences to the Higgs mass (\ie the gauge hierarchy remains stabilised). (2) They preserve the gauge invariance of the SM (and correspondingly the SUSY-unbroken MSSM). (3) They do not violate the renormalisability of the theory (which can be achieved by adding only mass terms and coupling parameters with positive mass dimensions). (4) They respect baryon and lepton symmetries of the SM and therefore conserve the corresponding quantum numbers $B$ and $L$. The most general soft supersymmetry-breaking Lagrangian then reads
\beq
\lagr_{\rm soft}^{\rm MSSM} &=& -\half\left ( M_3 \stilde g\stilde g
+ M_2 \stilde W \stilde W + M_1 \stilde B\stilde B 
+\conj \right )
\nonumber
\\
&&
-\left ( \stilde {\sbar u} \,{\bf a_u}\, \stilde Q H_u
- \stilde {\sbar d} \,{\bf a_d}\, \stilde Q H_d
- \stilde {\sbar e} \,{\bf a_e}\, \stilde L H_d
+ \conj \right ) 
\nonumber
\\
&&
-\stilde Q^\dagger \, {\bf m^2_{Q}}\, \stilde Q
-\stilde L^\dagger \,{\bf m^2_{L}}\,\stilde L
-\stilde {\sbar u} \,{\bf m^2_{{\sbar u}}}\, {\stilde {\sbar u}}^\dagger
-\stilde {\sbar d} \,{\bf m^2_{{\sbar d}}} \, {\stilde {\sbar d}}^\dagger
-\stilde {\sbar e} \,{\bf m^2_{{\sbar e}}}\, {\stilde {\sbar e}}^\dagger
\nonumber \\
&&
- \, m_{H_u}^2 H_u^* H_u - m_{H_d}^2 H_d^* H_d
- \left ( b H_u H_d + \conj \right ) .
\label{MSSMsoft}
\eeq
Here, $M_1$, $M_2$ and $M_3$ are bino, wino and gluino mass terms, respectively. Trilinear couplings ${\bf a_u}$, ${\bf a_d}$ and ${\bf a_e}$ are complex $3\times 3$ matrices in the family space and are in one-to-one correspondence to the Yukawa couplings ${\bf y_u}$, ${\bf y_d}$ and ${\bf y_e}$ in the superpotential (see \eq{MSSMsuperpot}). Squark and slepton mass terms ${\bf m^2_{ Q}}$, ${\bf
m^2_{{\sbar u}}}$, ${\bf m^2_{{\sbar d}}}$, ${\bf m^2_{L}}$ and ${\bf m^2_{{\sbar e}}}$ are also $3 \times 3$ (Hermitian) matrices in the family space with potentially complex entries. $m_{H_u}^2$
and $m_{H_d}^2$ are explicit real mass terms in the Higgs sector and $b$ is a complex bilinear coupling.

\subsection{Electroweak symmetry breaking and Higgs sector}
\label{sec:higgssector}

In order for the Higgs mechanism to work, the scalar potential for the Higgs scalar fields needs be minimised and the minimum then breaks electroweak symmetry. It can be shown that at the minimum, both $H_u^+$ and $H_d^-$ can be set to $0$, a property that is satisfactory. The reason for this satisfaction is that electromagnetism is not spontaneously broken at the minimum. Ignoring the terms in the potential that involve $H_u^+$ or $H_d^-$, one obtains the following expression for the Higgs scalar potential that only contains the neutral Higgs fields $H_u^0$ and $H_d^0$: 
\begin{eqnarray}
\label{Higgspot}
V_H &=& (m_{H_u}^2 + |\mu|^2) |H_u^0|^2
+ (m_{H_d}^2 + |\mu|^2) |H_d^0|^2 \nonumber \\
&-& (b H_u^0 H_d^0 + \text{c.c.}) + \frac{1}{8} (g^2 + g'^2) (|H_u^0|^2 - |H_d^0|^2)^2.
\end{eqnarray}
Here $g$ and $g'$ are the $SU(2)_L$ and $U(1)_Y$ gauge coupling constants, respectively, and $b$ is the parameter defined in \eq{MSSMsoft}.

Now, in order to break electroweak symmetry, $V_H$ is required to be minimised (with a stable minimum) and the fields $H_u^0$ and $H_d^0$ acquire real and non-zero vacuum expectation values (VEVs). We denote these VEVs by $v_u$ and $v_d$, \ie
\beq
\label{HiggsVEVs}
v_u \equiv \langle H_u^0 \rangle,~~~~~~~~v_d \equiv \langle H_d^0 \rangle.
\eeq

One can show that in order for the scalar potential $V_H$ to develop a well-defined local minimum such that electroweak symmetry is appropriately broken, the following two conditions must be satisfied~\cite{SUSY:Baer}:
\beq
\label{EWSBconds}
b^2 &>& (m_{H_u}^2 + |\mu|^2)(m_{H_d}^2 + |\mu|^2), \nonumber \\
2|b| &>& m_{H_u}^2 + m_{H_d}^2 + 2|\mu|^2.
\eeq

The existence of two Higgs doublets in the MSSM implies that the Higgs sector of the theory consists of eight degrees of
freedom (two per each field for $H_u^0$, $H_d^0$, $H_u^+$ and $H_d^-$).  As in the SM, when electroweak symmetry is broken, three of these degrees of freedom are eaten so as to make $W$
and $Z$ bosons massive. This means that five degrees of freedom remain intact and form five physical Higgs scalars. They are usually shown as
\begin{eqnarray}
\quad h, H, A, H^{\pm},
\end{eqnarray}
where $h$ and $H$ are CP-even and neutral (with $h$
lighter than $H$), $A$ is CP-odd and neutral, and $H^{\pm}$ are charged (with charges $\pm 1$).

The quantities $v_u$ and $v_d$ in \eq{HiggsVEVs} are related to the masses of the $Z$-boson and $W$-boson ($m_Z$ and $m_W$) and the gauge couplings $g$ and $g'$ in the following way:
\beq
\label{vplusv}
v_u^2 + v_d^2 = \frac{2m_Z^2}{g^2+g'^2} = \frac{2m_W^2}{g^2}.
\eeq

In addition, the ratio of the two above VEVs parametrises how the total Higgs vacuum expectation value is divided between the two neutral Higgs scalars. This is an interesting quantity and is usually denoted by $\tan\beta$, \ie
\beq
\tan\beta \equiv \frac{v_u}{v_d}.
\eeq

In most phenomenological studies of the MSSM, all the `effectively three' parameters of the superpotential $m_{H_u}^2+ |\mu|^2$, $m_{H_d}^2 + |\mu|^2$ and $b$ are usually traded for the three more physically interesting parameters (1) $v^2\equiv v_u^2 + v_d^2$, (2) $\tan\beta$ and (3) the mass of one of the physical Higgs bosons (conventionally taken to be $m_A$). Furthermore, since the value of $v^2$ is fixed by the $Z$- and/or $W$-boson masses (\eq{vplusv}), one is left with only two parameters that need to be determined experimentally: $\tan\beta$ and $m_A$.

One main reason for this new parametrisation is that the EWSB conditions given in \eq{EWSBconds} can be written as~\cite{SUSY:Baer}
\beq
b &=& \frac{(m_{H_u}^2 + m_{H_d}^2 + 2|\mu|^2)\sin 2 \beta}{2}, \nonumber \\
|\mu|^2 &=& \frac{m_{H_d}^2 - m_{H_u}^2 \tan^2{\beta}}{\tan^2{\beta}-1}-\frac{m_Z^2}{2},
\label{EWSBconds2}
\eeq
the first of which allows us to trade $b$ for $\tan\beta$, and the second can be used to fix the magnitude (but not the sign) of $\mu$ so as to obtain the measured value of $m_Z$. In addition, $m_A$ can be written in terms of the original parameters as~\cite{Martin:1997ns} $m_A^2=2b/\sin{2\beta}=m_{H_u}^2 + m_{H_d}^2 + 2|\mu|^2$.

The masses of the other physical Higgs bosons $h$, $H$ and $H^{\pm}$ can be obtained in terms of the above parameters:
\begin{eqnarray}
\label{CPevenmasses}
m^2_{\stackrel{H}{h}} \! \! &=& \! \! \! \frac{m_A^2 \! + \! m_Z^2
\! \pm \! \sqrt{(m_A^2 \! + \! m_Z^2)^2 
\! - \! 4 m_A^2 m_Z^2 (\cos 2 \beta)^2}}{2}, \nonumber \\
m_{H^{\pm}}^2 &=& m_A^2 + m_W^2.
\end{eqnarray}

These expressions are however valid only at tree level and for example large radiative corrections from the top squark/quark
loops can elevate $m_h$ to higher values (see \eg ref.~\cite{Barbieri:1991tk}).  These corrections turn out to be rather crucial for the model to survive because for example the experimental bounds on the Higgs mass indicate that $m_h$ should be larger than $m_Z$, a condition that is not consistent with what we obtain from the above relations: $m_h < m_Z |\cos 2\beta|$ (see also section~\ref{sec:expboundsHiggs}).

\subsection{Sparticles and their masses}
\label{sec:sparticlemasses}

We briefly pointed out in section~\ref{sec:fieldcontent} that in the MSSM, the sparticles of Tab.~\ref{MSSMparticles} are in general different from the mass eigenstates of the theory. This is because after SUSY breaking and electroweak symmetry breaking, some particles share quantum numbers and can consequently mix. Except for the gluino that does not have quantum numbers similar to other particles and therefore does not mix with them, the mixing happens for other gauginos, higgsinos, squarks and sleptons (gauginos and higgsinos can also mix with each other even though they belong to different groups). In this section we review very quickly different mass mixing matrices within the MSSM and their corresponding eigenstates (see \eg refs.~\cite{SUSY:Baer,SUSY:Aitchison,Martin:1997ns} for more detailed discussions).

\subsubsection{Gluinos}

The simplest case to study is the gluino. Since it is a colour octet fermion, and $SU(3)_C$ is unbroken, it does not mix with any other MSSM particle and is a mass eigenstate. This all means that the gluino mass comes solely from the corresponding soft supersymmetry-breaking term of \eq{MSSMsoft} 
\beq
-\half M_3 \stilde g\stilde g + \conj,
\eeq
and the gluino mass is therefore equal to $|M_3|$.

\subsubsection{Neutralinos}
\label{sec:neutralinos}

After electroweak symmetry breaking, the neutral higgsinos ($\stilde H_u^0$ and $\stilde H_d^0$) and the neutral gauginos ($\stilde B$ and $\stilde W^0$) mix. The four mass eigenstates corresponding to these fields are called `neutralinos'. In the gauge eigenstate basis $(\stilde B, \stilde W^0, \stilde H_d^0, \stilde H_u^0)$, the neutralino mass mixing matrix has the following form:
\beq
{\bf M}_{\stilde\chi^0} \,=\, \begin{pmatrix}
  M_1 & 0 & -g' v_d/\sqrt{2} & g' v_u/\sqrt{2} \cr
  0 & M_2 & g v_d/\sqrt{2} & -g v_u/\sqrt{2} \cr
  -g' v_d/\sqrt{2} & g v_d/\sqrt{2} & 0 & -\mu \cr
  g' v_u/\sqrt{2} & -g v_u/\sqrt{2}& -\mu & 0 \cr \end{pmatrix}.
\label{preneutralinomassmatrix}
\eeq
Here $M_1$ and $M_2$ are the mass term parameters in \eq{MSSMsoft}, $\mu$ is the SUSY higgsino mass parameter in the MSSM superpotential (\eq{MSSMsuperpot}) and $g$ and $g'$ are again the $SU(2)_L$ and $U(1)_Y$ gauge coupling constants. The relation between $g$ and $g'$ is $g'=g\tan{\theta_W}$ where $\theta_W$ is the weak mixing angle. $v_u$ and $v_d$ are the neutral Higgs VEVs we introduced in section~\ref{sec:higgssector}.

In order to find the mass eigenstates, one diagonalises the mass matrix ${\bf M}_{\stilde\chi^0}$ and the four eigenstates that are obtained from this are the four neutralinos $\tilde{\chi}^0_i$ ($i=1,2,3,4$):
\beq
\tilde{\chi}^0_i = N_{i1} \tilde{H}_u^0 + N_{i2} \tilde{H}_d^0 + N_{i3} \tilde{W}_3^0 + N_{i4} \tilde{B}^0.
\label{neutralino}
\eeq
$N$ in the above equation is the unitary matrix that diagonalises ${\bf M}_{\stilde\chi^0}$. The lightest neutralino is of particular interest (see section~\ref{sec:rparity}) and is the particle that is usually meant by people when they speak of `the neutralino'. It is often denoted by $\tilde{\chi}^0_1$ or simply $\chi$.

\subsubsection{Charginos}

Following electroweak symmetry breaking, the charged higgsinos ($\stilde H_u^+$ and $\stilde H_d^-$) and winos ($\stilde W^+$ and $\stilde W^-$) also mix and the mass eigenstates become linear combinations of these particles. We call the resultant mass eigenstates `charginos'.

The mass mixing matrix in this case and in the gauge eigenstate basis $(\stilde W^+,\, \stilde H_u^+,\, \stilde W^- ,\, \stilde H_d^- )$ reads
\beq
{\bf M}_{\stilde\chi^\pm} &=& \begin{pmatrix}M_2 & g v_u\cr
                     g v_d & \mu \cr \end{pmatrix},
\label{charginomassmatrix}
\eeq
and can be diagonalised using two unitary $2\times 2$ matrices ${\bf U}$ and ${\bf V}$. The mass eigenstates that are obtained this way are
\begin{eqnarray}
\tilde{\chi}^-_i &=& U_{i1} \stilde W^- + U_{i2}\stilde H_d^- , \nonumber \\
\tilde{\chi}^+_i &=& V_{i1} \stilde W^+ + V_{i2}\stilde H_u^+ .
\end{eqnarray}

\subsubsection{Squarks and sleptons}

We write the squark squared-mass matrices ${\bf M}_{\tilde u}^2$ and ${\bf M}_{\tilde d}^2$ in a basis where the squarks and their corresponding quarks are rotated in the same way. The squared-mass matrices for the sleptons and sneutrinos, ${\bf M}^2_{\tilde e}$ and ${\bf M}^2_{\tilde\nu}$, respectively, are obtained in a similar way. The squared-mass matrices for the squarks and sleptons then read
\begin{eqnarray}
  {\bf M}_{\tilde u}^2 &\!\!\!\!=\!\!\!\!& \left( \begin{array}{cc}
  {\bf m^2_{ Q}} + {\bf m_u}^\dagger {\bf m_u} +
      D_{LL}^{u} {\bf 1} &
    {\bf m_u}^\dagger 
        ( {\bf a_u}^\dagger - \mu^* \cot\beta ) \\
   ( {\bf a_u} - \mu \cot\beta ) {\bf m_u} &
  {\bf m^2_{{\sbar u}}} + {\bf m_u} {\bf m_u}^\dagger +
      D_{RR}^{u} {\bf 1} \\
  \end{array} \right) ,
  \label{mutilde} \nonumber \\
  {\bf M}_{\tilde d}^2 &\!\!\!\!=\!\!\!\!& \left( \begin{array}{cc}
  {{\bf V}^\dagger {\bf m^2_{ Q}} {\bf V}+
  {\bf m_d}{\bf m_d}^\dagger+D_{LL}^{d}{\bf 1}}&
  {{\bf m_d}^\dagger ( {\bf a_d}^\dagger-\mu^*\tan\beta )}\\
  {( {\bf a_d}-\mu\tan\beta ) {\bf m_d}}&
  {{\bf m^2_{{\sbar d}}}+{\bf m_d}^\dagger {\bf m_d}+
      D_{RR}^{d}{\bf 1}}\\
  \end{array} \right) ,
  \label{mdtilde} \nonumber \\
  {\bf M}^2_{\tilde e} & = &\left( \begin{array}{cc}
  {{\bf m^2_{L}}+{\bf m_e}{\bf m_e}^\dagger+
       D_{LL}^{e}{\bf 1}}&
  {{\bf m_e}^\dagger ( {\bf a_e}^\dagger-\mu^*\tan\beta )}\\
  {( {\bf a_e}-\mu\tan\beta ) {\bf m_e}}&
  {{\bf m^2_{{\sbar e}}}+{\bf m_e}^\dagger{\bf m_e}+
       D_{RR}^{e}{\bf 1}}
  \end{array} \right) , \nonumber \\
  \label{metilde}
  {\bf M}^2_{\tilde\nu} & = & {\bf m^2_{L}} + D^\nu_{LL} {\bf 1} .
\label{sfermionmasses}    
\end{eqnarray}
Here,
\begin{eqnarray}
  D^f_{LL} & = & m_Z^2\cos 2\beta(T_{3f}-e_f\sin^2\theta_W), \nonumber \\
  D^f_{RR} & = & m_Z^2\cos 2\beta e_f\sin^2\theta_W,
\end{eqnarray}
where $T_{3f}$ is the third component of the weak isospin, $e_f$ is the electric charge and $\theta_W$ is the weak mixing angle. ${\bf V}$ in \eqs{sfermionmasses} denotes the Cabibbo-Kobayashi-Maskawa (CKM) matrix. The matrices ${\bf m^2_{ Q}}$, ${\bf m^2_{{\sbar u}}}$, ${\bf m^2_{{\sbar d}}}$, ${\bf m^2_{L}}$ and ${\bf m^2_{{\sbar e}}}$ are the soft sfermion squared-mass matrices defined in \eq{MSSMsoft}, ${\bf a_u}$, ${\bf a_d}$ and ${\bf a_e}$ are trilinear couplings, and $\mu$ is the $\mu$-parameter in the MSSM superpotential (\eq{MSSMsuperpot}).  ${\bf m_u}$, ${\bf m_d}$ and ${\bf m_e}$ are diagonal matrices defined in terms of the quark and charged-lepton masses as
\begin{eqnarray}
  {\bf m_u} & = & diag(m_u,m_c,m_t) , \nonumber \\
  {\bf m_d} & = & diag(m_d,m_s,m_b) , \nonumber \\
  {\bf m_e} & = & diag(m_e,m_\mu,m_\tau).
\end{eqnarray}

\subsection{$R$-parity}
\label{sec:rparity}

We discussed earlier in this chapter that the soft SUSY-breaking terms are chosen such that they conserve both baryon ($B$) and lepton ($L$) numbers. One immediate consequence of this property is that a discrete symmetry exists between the SM particles and their superpartners. This symmetry has a corresponding multiplicative quantum number, known as $R$-parity which is conserved and can be written in terms of $B$, $L$ and the particle's spin $s$ in the following form:
\beq
R = (-1)^{3(B-L)+2s}.
\label{rparity}
\eeq
It is not difficult to show that all SM particles have $R$-parity $+1$ while all their superpartners have $R$-parity $-1$.

$R$-parity conservation in the MSSM has an extremely interesting phenomenological consequence: the lightest supersymmetric particle (LSP) is stable and does not decay into lighter SM states (clearly it also does not decay into any other SUSY-particles since by definition it is the lightest member of that group). If the LSP is also weakly-interacting and electrically neutral, it can be a viable dark matter candidate (see \eg section~\ref{sec:WIMPs} and references therein). One example is the neutralino that we introduced in section~\ref{sec:neutralinos}. It is arguably the most favoured dark matter candidate and has received the bulk of attention to date.

$B$ and $L$ conservation, and consequently $R$-parity conservation, are not fundamental assumptions in the MSSM and there are models with $R$-partity violation. However, there are good reasons to think that the assumption is not too ad hoc. For example some grand unified theories (see section~\ref{sec:unification}), which give the MSSM as their effective low-energy versions, accommodate $R$-parity conservation in their structure (see \eg ref.~\cite{Lee:1994je} for certain $SO(10)$ theories).

\subsection{Renormalisation Group Equations}
\label{sec:RGEs}

In the MSSM, analogous to any other quantum field theory, all the parameters of the model are subject to running (\ie evolving with energy scale). These include gauge coupling constants, parameters in the superpotential and soft SUSY-breaking terms. The running of the parameters can be calculated using the renormalisation group equations (RGEs).

\subsubsection{Gauge coupling constants}

Let us first look at the RGE evolution of the gauge coupling constants by introducing the commonly used couplings $\alpha_i$ ($i=1,2,3$):
\beq
\alpha_i=\frac{g_i^2}{4\pi}.
\eeq
Here, $g_1=\sqrt{5/3}g'$, $g_2=g$ and $g_3=g_s$, where $g'$, $g$ and $g_s$ are the $U(1)_Y$, $SU(2)_L$ and $SU(3)_C$ gauge coupling constants, respectively. To one-loop order, the RGEs for the couplings $\alpha_i$ have the form
\beq
{d\over dt}\alpha_i=-\frac{b_i}{2\pi}\alpha_i^2,
\label{running1}
\eeq
where $t=\ln{Q}$ and $Q$ is the energy scale of the running. $b_i$ are some quantities that are fixed by the structure of the gauge group and the matter multiplets (to which the gauge fields couple) of the theory. It can be shown that for the SM $b_1^{SM}=-41/10$, $b_2^{SM}=19/6$ and $b_3^{SM}=7$, and for the MSSM $b_1^{MSSM}=-33/5$, $b_2^{MSSM}=-1$ and $b_3^{MSSM}=3$~\cite{SUSY:Aitchison}.

\begin{figure}[t]
        \centering
\includegraphics[scale=0.5, trim = 0 0 0 0, clip=true]{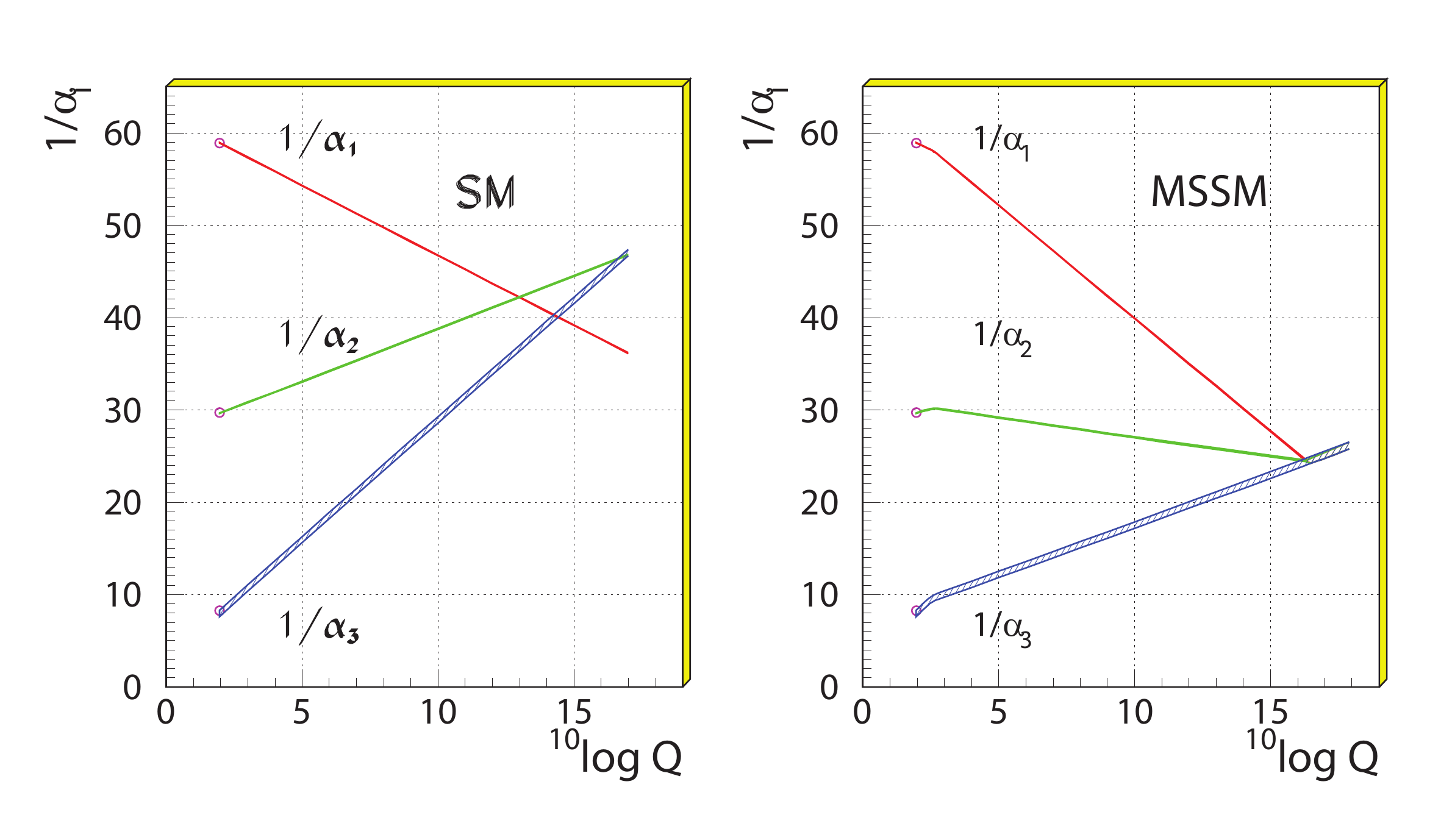}
 \caption{Running of the gauge coupling constants in both the Standard Model (SM) and its minimal supersymmetric extension (MSSM). The left panel depicts the failure of the SM gauge coupling to unify, while the right panel shows that the gauge coupling unification can happen in the MSSM. Adapted from ref.~\cite{Kazakov:2000ra}.}
        \label{fig:unification}
\end{figure}

\eqs{running1} can be integrated to give the following equations in terms of the inverse of the couplings $\alpha_i$:
\beq
\alpha_i^{-1}(Q)=\alpha_i^{-1}(Q_0)+\frac{b_i}{2\pi}\ln{\frac{Q}{Q_0}}.
\label{running2}
\eeq
Here $Q_0$ is the energy scale at which we start the running, and is usually taken to be $m_Z$ since the couplings are well measured at the energy scale $m_Z$. Taking into account the effects of two-loop corrections, and other subtleties, one can solve \eqs{running2} and the results are shown in \fig{fig:unification} for both the SM and the MSSM.

It is interesting to see how gauge coupling unification can occur in the MSSM while it fails in the SM. This provides another explicit example of our general discussions in chapter~\ref{sec:SMissues}, namely that extending the SM to its supersymmetric version provides appropriate solutions to some issues with the SM (see section~\ref{sec:unification} for details). The other interesting observation is the scale $\MGUT$ at which the unification happens: it is about $2.2\times 10^{16}$ GeV, a value that is predicted as the unification scale in many grand unified theories but with entirely different motivations (again see section~\ref{sec:unification} and references therein).

\subsubsection{Superpotential parameters}

The parameters in the MSSM superpotential include the Yukawa coupling parameters ${\bf y_u}$, ${\bf y_d}$, ${\bf y_e}$, and the $\mu$ parameter (see \eq{MSSMsuperpot}). In order to calculate the RGEs for the Yukawa couplings, one usually makes an approximation that only the third-family components are important. This is because the third-family particles, \ie the top quark, bottom quark and tau lepton, are the heaviest fermions in the SM. In other words, one assumes that the Yukawa couplings have the diagonal forms ${\bf y_u}=diag(0,0,y_t)$, ${\bf y_d}=diag(0,0,y_b)$ and ${\bf y_e}=diag(0,0,y_\tau)$. The one-loop RGEs for the parameters $y_t$, $y_b$ and $y_\tau$, as well as the parameter $\mu$ are~\cite{Martin:1997ns}:
\beq
{d\over dt} y_t \!&=&\! {y_t \over 16 \pi^2} \Bigl [ 6 y_t^* y_t + y_b^* y_b
- {16\over 3} g_3^2 - 3 g_2^2 - {13\over 15} g_1^2 \Bigr ],
\label{eq:betayt}
\\
{d\over dt} y_b \!&=&\! {y_b \over 16 \pi^2} 
\Bigl [ 6 y_b^* y_b + y_t^* y_t + 
y_\tau^* y_\tau - {16\over 3} g_3^2 - 3 g_2^2 - {7\over 15} g_1^2 \Bigr ],
\\
{d\over dt} y_\tau \!&=&\! {y_\tau \over 16 \pi^2} 
\Bigl [ 4 y_\tau^* y_\tau 
+ 3 y_b^* y_b - 3 g_2^2 - {9\over 5} g_1^2 \Bigr ],
\\
{d\over dt} \mu \!&=&\! {\mu \over 16 \pi^2} 
\Bigl [ 3 y_t^* y_t + 3 y_b^* y_b
+ y_\tau^* y_\tau - 3 g_2^2 - {3\over 5} g_1^2 \Bigr ].
\label{eq:betamu}
\eeq

\subsubsection{Soft SUSY-breaking parameters}

We now look at the RGEs for the parameters in the soft SUSY-breaking sector of the MSSM (\ie \eq{MSSMsoft}). These parameters are divided into four categories: the gaugino masses $M_1$, $M_2$ and $M_3$, the trilinear couplings ${\bf a_u}$, ${\bf a_d}$ and ${\bf a_e}$, the squark and slepton squared-masses ${\bf m^2_{ Q}}$, ${\bf m^2_{L}}$, ${\bf m^2_{{\sbar u}}}$, ${\bf m^2_{{\sbar d}}}$ and ${\bf m^2_{{\sbar e}}}$, and the Higgs squared-mass parameters $m_{H_u}^2$, $m_{H_d}^2$ and $b$.

For the gaugino masses $M_i$ ($i=1,2,3$), the one-loop RGEs read (in terms of the quantities $b_i$ and $\alpha_i$)
\beq
{d\over dt} M_i \,=\, -\frac{b_i}{2\pi} \alpha_i M_i~~~\Longrightarrow~~~{d\over dt}(M_i/\alpha_i)=0.
\label{gauginomassrge}
\eeq
This shows that the three ratios $M_i/\alpha_i$ are scale independent, \ie do not change with renormalisation scale. In addition, we observed that the gauge couplings unify at the GUT scale $\MGUT$. This suggests to assume that the gaugino masses also unify at $\MGUT$. This is a popular assumption and the unified value is usually denoted as $m_{1/2}$ (see also section~\ref{sec:shrink}).

For the trilinear couplings ${\bf a_u}$, ${\bf a_d}$ and ${\bf a_e}$, it is common to assume that they are proportional to the Yukawa couplings ${\bf y_u}$, ${\bf y_d}$, ${\bf y_e}$, and accordingly have the diagonal forms ${\bf a_u}=diag(0,0,a_t)$, ${\bf a_d}=diag(0,0,a_b)$ and ${\bf a_e}=diag(0,0,a_\tau)$. In this case, the one-loop RGEs for the parameters $a_t$, $a_b$ and $a_\tau$ have the following form~\cite{Martin:1997ns}:
\beq
{d\over dt} a_t \!&=&\! \frac{1}{16\pi^2} \Bigl [ a_t ( 18 y_t^* y_t + y_b^* y_b
- {16\over 3} g_3^2 - 3 g_2^2 - {13\over 15} g_1^2 )
\nonumber\\ &&
\!+ 2 a_b y_b^* y_t + y_t ( {32\over 3} g_3^2 M_3 + 6 g_2^2 M_2 + {26\over 15} g_1^2 M_1 ) \Bigl ],
\label{atrge}
\nonumber \\
{d\over dt} a_b \!&=&\! \frac{1}{16\pi^2} \Bigl [ a_b ( 18 y_b^* y_b + y_t^* y_t +
y_\tau^* y_\tau - {16\over 3} g_3^2 - 3 g_2^2 - {7\over 15} g_1^2 )
\nonumber\\ &&
\!+ 2 a_t y_t^* y_b + 2 a_\tau y_\tau^* y_b + y_b ( {32\over 3} g_3^2 M_3 + 6 g_2^2 M_2 + {14 \over 15} g_1^2 M_1) \Bigl ],
\nonumber \\
{d\over dt} a_\tau \!&=&\! \frac{1}{16\pi^2} \Bigl [ a_\tau ( 12 y_\tau^* y_\tau 
+ 3 y_b^* y_b - 3 g_2^2 - {9\over 5} g_1^2 )
\nonumber\\ &&
\!+ 6 a_b y_b^* y_\tau
+ y_\tau ( 6 g_2^2 M_2 + {18\over 5} g_1^2 M_1 ) \Bigr ].
\eeq

The squark and slepton squared-masses, that are $3\times 3$ matrices, are usually assumed to have diagonal forms (in order for the potentially dangerous flavour-changing and CP-violating effects in the MSSM to be suppressed). For example for ${\bf m^2_{ Q}}$ this means that ${\bf m^2_{ Q}}=diag(m^2_{Q_1}, m^2_{Q_2}, m^2_{Q_3})$. Analogous forms are assumed for the other squarks and sleptons.

The one-loop RGE expressions for these squared-masses are relatively lengthy and we therefore give here, as an example, the expressions only for the third-family quantities~\cite{Martin:1997ns}:
\beq 
{d\over dt} m_{Q_3}^2 \!&=&\!
\frac{1}{16\pi^2} \Bigl [ X_t +X_b-{32\over 3} g_3^2 |M_3|^2 -6 g_2^2 |M_2|^2 -{2\over 15} g_1^2 |M_1|^2
+ \frac{1}{5} g^{2}_1 S \Bigl ],
\label{mq3rge} 
\nonumber \\
{d\over dt} m_{\sbar u_3}^2 \!&=&\!
\frac{1}{16\pi^2} \Bigl [ 2 X_t - {32\over 3} g_3^2 |M_3|^2 - {32\over 15} g_1^2|M_1|^2
- \frac{4}{5} g^{2}_1 S \Bigl ],
\label{mtbarrge}
\nonumber \\
{d\over dt} m_{\sbar d_3}^2 \!&=&\!
\frac{1}{16\pi^2} \Bigl [ 2 X_b - {32\over 3} g_3^2 |M_3|^2 - {8\over 15} g_1^2|M_1|^2 
+ \frac{2}{5} g^{2}_1 S \Bigl ],
\label{md3rge}
\nonumber \\
{d\over dt} m_{L_3}^2 \!&=&\!
\frac{1}{16\pi^2} \Bigl [ X_\tau  - 6 g_2^2 |M_2|^2 - {6\over 5} g_1^2 |M_1|^2 - \frac{3}{5} g^{2}_1 S \Bigl ],
\nonumber \\
{d\over dt} m_{\sbar e_3}^2 \!&=&\!
\frac{1}{16\pi^2} \Bigl [ 2 X_\tau - {24\over 5} g_1^2 |M_1|^2  + \frac{6}{5} g^{2}_1 S \Bigl ],
\label{mstaubarrge}
\eeq
where
\beq
S \!&=& \! m_{H_u}^2 - m_{H_d}^2 + {\rm Tr}[
{\bf m^2_Q} - {\bf m^2_L} - 2 {\bf m^2_{\overline u}}
+ {\bf m^2_{\overline d}} + {\bf m^2_{\overline e}}], \nonumber \\
X_t \!&=&\!  2 |y_t|^2 (m_{H_u}^2 + m_{Q_3}^2 + m_{\sbar u_3}^2) +2 |a_t|^2,
\nonumber \\
X_b \!&=& \! 2 |y_b|^2 (m_{H_d}^2 + m_{Q_3}^2 + m_{\sbar d_3}^2) +2 |a_b|^2,
\nonumber \\
X_\tau\! &=&\!  2 |y_\tau|^2 (m_{H_d}^2 + m_{L_3}^2 + m_{\sbar e_3}^2)
+ 2 |a_\tau|^2.
\label{SandXs}
\eeq

Finally, let us look at the RGEs for the Higgs squared-mass parameters $m_{H_u}^2$, $m_{H_d}^2$ and $b$. They are of the following forms~\cite{Martin:1997ns}:
\beq
{d\over dt} m_{H_u}^2 \!&=&\!
\frac{1}{16\pi^2} \Bigl [ 3 X_t - 6 g_2^2 |M_2|^2 - {6\over 5} g_1^2 |M_1|^2 + \frac{3}{5} g^{2}_1 S \Bigl ],
\label{mhurge}
\nonumber \\
{d\over dt} m_{H_d}^2 \!&=&\!
\frac{1}{16\pi^2} \Bigl [ 3 X_b + X_\tau - 6 g_2^2 |M_2|^2 - {6\over 5} g_1^2 |M_1|^2 - \frac{3}{5} g^{2}_1 S \Bigl ],
\nonumber \\
{d\over dt} b \!&=&\! \frac{1}{16\pi^2} \Bigl [ b ( 3 y_t^* y_t + 3 y_b^* y_b
+ y_\tau^* y_\tau - 3 g_2^2 - {3\over 5} g_1^2 )
\nonumber \\ && 
\!+ \mu ( 6 a_t y_t^* + 6 a_b y_b^* + 2 a_\tau y_\tau^* +
6 g_2^2 M_2 + {6\over 5} g_1^2 M_1 ) \Bigl ] ,
\label{mhdrge}
\eeq
with $X_t$, $X_b$, $X_\tau$ and $S$ defined in Eqs.~\ref{SandXs}.

\subsection{Parameter space}
\label{sec:paramspace}

Looking at the full Lagrangian of the MSSM, including soft terms (\eq{MSSMsoft}), one realises that the model, compared to the SM, possesses a much larger number of free parameters, most of which come from the SUSY-breaking sector, \ie soft terms. Indeed only one of the new parameters, $\mu$, belongs to the SUSY invariant sector.

Let us try to count the parameters in the MSSM: In the fermion sector, the theory has $5$ Hermitian $3 \times 3$ mass-squared matrices plus $9$ complex $3 \times 3$ trilinear coupling matrices. These give $5\times 3 \times 3+6\times3\times 3 \times 2 = 153$ real parameters. Due to some field redefinitions, this number is reduced to $110$. The gauge sector of the MSSM has $3$ usual gauge couplings $g'$, $g$ and $g_s$ plus the QCD vacuum angle $\theta$ (see section~\ref{sec:otherissues} and references therein). If we add to these the $6$ gaugino masses, this gives $3+1+6=10$. Using a field transformation, one of the CP-violating masses can be removed, reducing the number by $1$ and giving rise to $9$ parameters in the gauge sector. The Higgs terms of the soft SUSY-breaking sector contribute by $2$ real squared masses $m^2_{H_u}$ and $m^2_{H_d}$, and $1$ complex coefficient $b$, and this gives $1+1+2=4$ parameters. The SUSY invariant sector contributes with the parameter $\mu$ (being complex). A field definition helps reduce $1$ parameter and we are left with $1$. We can now count the total number of free parameters in the MSSM: It is $110+9+4+1=124$.  The number of free parameters in the SM is $19$, meaning that the full MSSM has $105$ more parameters than the SM.

\section{... and beyond}

We introduced in section~\ref{sec:MSSM} the MSSM as a phenomenological supersymmetric extension of the SM that contains the minimum number of new fields needed for the supersymmetrisation process. Respecting the renormalisability, gauge symmetry and $B$ \& $L$ conservation in the SM, the MSSM allows for the most general terms in the soft SUSY-breaking sector of the theory. Although there are models with less number of free parameters (coming either from particular SUSY-breaking mechanisms or from purely phenomenological assumptions, as we will see in section~\ref{sec:shrink}), most of them have usually the same field content as the MSSM while defined over particular parts of the MSSM parameter space. The MSSM is therefore considered by almost all particle physicists as the most interesting framework to study supersymmetric extensions of the SM and provides extensive scope for various phenomenological investigations of such extensions. On the other hand, as we will see in section~\ref{sec:shrink}, the MSSM parameter space is already too large and cannot be fully analysed. Although these all indicate that the main focus of SUSY phenomenologists will, for the next few years, be the MSSM and its subclasses, one should however notice that attempts for going even beyond the MSSM have already begun. This is mainly because the MSSM itself is not flawless and in fact possesses several problems. In what follows, we give two examples of SUSY models beyond the MSSM (and the problems they aim to solve) that have attracted much attention in the past few years.

\subsection{BMSSM}

In the MSSM, the quartic Higgs coupling constant is given entirely in terms of the electroweak gauge couplings $g$ and $g'$. This can be seen \eg from \eq{Higgspot} where this quartic coupling (that we denote by $\lambda$) is expressed as $\lambda=(g^2+g'^2)/8$. As a result, the value of $\lambda$ becomes small ($\lambda \approx 0.07$). This smallness poses a problem.

As we discussed in section~\ref{sec:higgssector}, the theoretical tree-level lightest Higgs mass $m_h$ is too low to be consistent with experimental constraints. One can show that the value of this tree-level mass is related to the value of the quartic Higgs coupling constant $\lambda$ and the smallness of the latter implies the same for the former. This all means that in the MSSM, large loop corrections are needed to bring the Higgs mass above the experimental bounds. The experimental lower bound on the Higgs mass (\ie $114$ GeV) excludes only a small part of the SM parameter space. On the contrary, the requirement of fairly large loop corrections excludes most of the natural values for the MSSM parameters. This is called the `little hierarchy problem' of the MSSM (see \eg ref.~\cite{Haber:1990aw}).

Many possible extensions of the MSSM have been proposed so far in attempt to address the little hierarchy problem (for a review, see \eg ref.~\cite{Giudice:2006sn} and references in ref.~\cite{Bae:2010cd}). One example is called `beyond the MSSM' (or BMSSM)~\cite{Dine:2007xi} (see also refs.~\cite{Berg:2009mq,Bernal:2009hd,Bernal:2009jc,Carena:2010cs} and references therein for more phenomenological studies of the model).

In this model, with an effective field theory approach (see \eg ref.~\cite{Burgess:2007pt}), the field content of the MSSM remains intact, while the quartic Higgs couplings are modified by the effects of some new physics that might exist at high energies. In the MSSM, the quartic Higgs terms are exactly supersymmetric, because in the MSSM the only SUSY-breaking terms that are allowed are those with mass dimension three or lower (soft terms). As we discussed in the previous section, this condition is imposed so as to prevent the reintroduction of quadratic divergences. In an effective field theory approach however, the effects of new physics can be parametrised by non-renormalisable terms.

In the simplest version of BMSSM only two leading terms with particular properties are added to the MSSM. One is a dimension-five operator $W_5$ that is added to the MSSM superpotential $W_{MSSM}$. $W_5$ has the following form:
\beq
W_5=-\frac{\epsilon_1}{\mu^*}(H_uH_d)^2.
\eeq
Here $\epsilon_1$ is a new effective real parameter of the model that is free and should be determined experimentally.

The other new operator of the theory is another dimension-five operator that is added to the soft SUSY-breaking sector of the MSSM. This term is
\beq
\epsilon_2(H_uH_d)^2+h.c.
\eeq 
with $\epsilon_2$ being another new effective real parameter.

After introducing the new parameters $\epsilon_1$ and $\epsilon_2$, we get the following new terms added to the full MSSM Lagrangian:  
\begin{eqnarray}
\label{expllag}
&\delta\lagr = - 2\epsilon_1 (\hu\hd)(\hu^\dagger \hu + \hd^\dagger \hd) - \epsilon_2 (\hu\hd)^2 + \frac{\epsilon_1}{\mu^*} \Bigl [ 2 (\thu \thd) (\hu \hd) \nonumber \\
&+ 2 (\hu \thd) (\thu \hd) + (\thu \hd)^2 + (\hu \thd)^2 \Bigl ] + \hc
\end{eqnarray}

The above correction terms affect the Higgs potential, neutralino masses and mixings, and chargino masses and mixings~\cite{Berg:2009mq}. The first effect is helpful in alleviating the little hierarchy problem of the MSSM since the Higgs quartic self-coupling now receives new contributions with parameters that are not fixed by the theory. BMSSM also has interesting implications for cosmology, in particular dark matter phenomenology and baryogenesis that are studied for example in refs.~\cite{Bernal:2009hd,Berg:2009mq}.

\subsection{NMSSM}

One other issue that cannot be explained in the MSSM is the so-called $\mu$-problem~\cite{Kim:1983dt}, which has to do with the $\mu$-term in the MSSM superpotential (\eq{MSSMsuperpot}). The $\mu$ parameter is dimensionful (\ie with positive mass dimension) but its value can be arbitrary and is not associated with any particular scale of the theory such as the SUSY-breaking scale. Phenomenologically, $\mu$ is required to have a value close to the EWSB scale, and this scale is not `natural'.

The $\mu$-problem has served as the primary motivation for proposing an extension of the MSSM that is called `Next-to-Minimal' Supersymmetric Standard Model (or NMSSM) (for a review, see \eg refs.~\cite{Maniatis:2009re,Ellwanger:2009dp}).

In the NMSSM, a new gauge singlet is added to the MSSM and consequently an effective $\mu$-term is generated dynamically (\ie spontaneously). The introduction of the additional gauge-singlet superfield is essentially the price to pay in order to solve the $\mu$-problem. The NMSSM however provides explanation for the other problems of the MSSM, including the little hierarchy problem described in the previous section. This is because the Higgs-boson sector of the theory is much less restricted compared to the MSSM and the predicted lower mass bound on the Higgs mass is in general substantially shifted. The NMSSM contains two additional Higgs bosons and one additional neutralino (that is called singlino). The model generally offers different Higgs-boson phenomenology, compared to what we expect from the MSSM, with interesting implications for collider searches.

\section{Shrinking the parameter space}
\label{sec:shrink}

We saw in section~\ref{sec:paramspace} that the full MSSM Lagrangian possesses $124$ free parameters that should be determined experimentally. This huge parameter space makes the phenomenological studies of the model practically difficult. In addition, the structure of the theory is such that most combinations of the parameters give experimental predictions that are excluded. For example many non-diagonal terms in the MSSM Lagrangian generate too large FCNCs, at levels that are experimentally excluded. The parameter space is therefore highly porous and has a very non-trivial structure. It is therefore rather crucial to work with sub-models of the MSSM with substantially less numbers of free parameters. In this section, we briefly review some approaches to this problem and different strategies in reducing the size of the parameter space.

\subsection{Connections with fundamentals}
\label{sec:fundamentals}

As we discussed earlier in this chapter, all the new parameters in the MSSM, except one, come from the SUSY-breaking sector, \ie the soft SUSY-breaking terms of \eq{MSSMsoft}. We argued that the soft terms are introduced based on some theoretically and/or phenomenologically motivated properties that one expects the MSSM to possess at low energies. Although such terms break supersymmetry explicitly, they are widely considered as useful low-energy terms that parametrise our ignorance of some underlying mechanism that breaks SUSY spontaneously (see section~\ref{sec:SUSYbreaking}).

Various SUSY-breaking mechanisms have so far been proposed, each of which imposes its own set of relations between different soft terms and corresponding parameters. For example in many SUSY-breaking scenarios some generally non-zero or complex parameters of the MSSM are predicted to be vanishing (or extremely small) or real, and universality conditions are imposed on some otherwise unrelated parameters. Usually in these scenarios, the parameter space is significantly contracted and the analysis of the model predictions becomes considerably easier.

In section~\ref{sec:SUSYbreaking}, we discussed general strategies one can use for breaking supersymmetry spontaneously. They were classified into two categories of $F$ or $D$-term SUSY-breaking mechanisms.

Unfortunately, the structure of the MSSM does not allow any of the two above strategies for breaking supersymmetry to be realised without extending the field content of the theory~\cite{Martin:1997ns}: (1) Giving a VEV to the $D$-term associated with the $U(1)_Y$ part of the MSSM gauge structure has turned out to give rise to an inappropriate mass spectrum. (2) The MSSM does not contain any gauge singlet with a corresponding $F$-term that develops a VEV.

\begin{figure}
        \centering
\includegraphics[scale=0.75, trim = 0 0 0 0, clip=true]{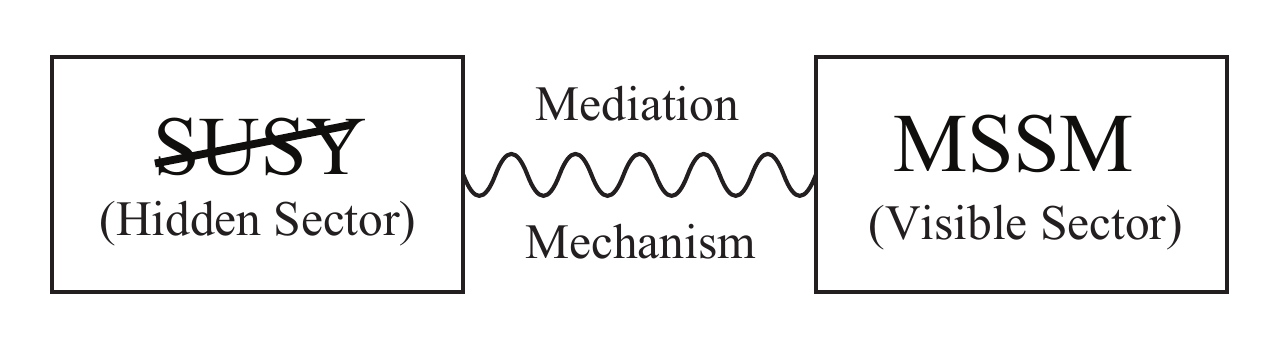}
 \caption{Schematic description of supersymmetry breaking mechanisms.}
        \label{fig:hiddensector}
\end{figure}

Even if we extend the MSSM in such a way that it includes new supermultiplets (including gauge singlets necessary for $F$-term SUSY-breaking), it is highly difficult to achieve phenomenologically viable spontaneous supersymmetry breaking only using renormalisable interactions at tree level. For example, due to the existence of particular `sum rules' for the tree-level mass terms, some squarks and sleptons in the MSSM have to have experimentally excluded low masses.

These types of problems can be evaded if one assumes that supersymmetry is broken in a different sector which communicates with the MSSM particles only indirectly \ie via either `non-renormalisable' interactions or couplings at `loop level'. The new SUSY-breaking sector is accordingly referred to as ``the hidden sector'' as opposed to the MSSM sector being the visible sector. The MSSM soft terms are then thought of as being the results of the `mediation' of SUSY-breaking from the hidden sector to the visible sector (see \fig{fig:hiddensector} for a schematic picture of viable SUSY-breaking mechanisms).

In order to understand a SUSY-breaking mechanism in full details, one needs to know exactly which theory governs the laws of physics at high energies. Such an understanding is still lacking and the usual approach is therefore to propose models of SUSY-breaking based on rather general frameworks and assumptions that capture interesting features of more fundamental theories.

Existing SUSY-breaking models can be categorised into two general classes: either they are based on local supersymmetry (or supergravity; see sections~\ref{sec:quantumgravity} and~\ref{sec:SUSYgauge}) in which SUSY breaking is mediated through gravitational non-renormalisable interactions, or the soft SUSY-breaking terms are generated only at loop level~\cite{Martin:1997ns}. As we stated earlier, in both classes the problematic sum rules are circumvented and supersymmetry is broken spontaneously.

In the following sections we briefly review three main SUSY-breaking scenarios that have been proposed in the literature and have as of yet received most of the attention. We describe their theoretical foundations only in a few words and pay most of our attention to the simplest (but practically the most interesting) models within each framework and the corresponding phenomenological aspects (for detailed introduction to SUSY-breaking mechanisms, see \eg refs.~\cite{SUSY:Baer,Martin:1997ns}).

\subsubsection{Planck-scale-mediated SUSY breaking: mSUGRA}

In this class of models (which has been historically the most popular one) supersymmetry is broken in a hidden sector which interacts with the visible sector only through gravitational effects near the Planck scale, hence the names ``gravity-mediated'' SUSY-breaking or ``Planck-scale-mediated'' SUSY-breaking (PMSB) mechanism~\cite{MSUGRA1,MSUGRA2,MSUGRA3,MSUGRA4,MSUGRA5,MSUGRA6,MSUGRA7}.

The presence of gravitational interactions in these models requires a supersymmetric theory that includes gravity. We mentioned earlier in sections~\ref{sec:quantumgravity} and~\ref{sec:SUSYgauge} that such a theory can be provided if the global supersymmetry is upgraded to a local one, \ie the parameters of supersymmetry transformations become space and time dependent. Supersymmetry is a spacetime symmetry and its local version automatically involves gravitation and is therefore also referred to as supergravity or SUGRA. The phenomenologically interesting versions of SUGRA, like any other known quantum field theory that contains gravity has turned out to be non-renormalisable.

The non-renormalizable terms in the SUGRA effective Lagrangian are fortunately suppressed by powers of $1/\MPlanck$ and their effects are consequently negligible for all phenomomenological studies that normally probe energies below TeV scales. Such terms can however provide a mechanism through which the hidden and visible sectors communicate and SUSY breaking is mediated from the former to the latter.

The supergravity Lagrangian, amongst other terms, usually contains non-renormalisable (NR) terms that look like
\beq
\lagr_{\rm NR} \!  &=& \!
-{1\over \MPlanck} \FX\, \mathcal{F}(\lambda^a, \phi_i)
- {1\over \MPlanck^2} \FX \FX^*\,
\mathcal{G}(\phi^i).
\label{hiddengrav}
\eeq
Here, $\phi_i$ and $\lambda_a$ are the scalar and gaugino fields of the MSSM and $\mathcal{F}$ and $\mathcal{G}$ are two functions of the fields whose exact forms are not important for our discussion here (see \eg ref.~\cite{Martin:1997ns} for details). $F$ denotes the auxiliary field corresponding to a chiral multiplet that is assumed to exist in the hidden sector. By choosing the right functional forms for $\mathcal{F}$ and $\mathcal{G}$, one can show that \eq{hiddengrav} is equivalent to \eq{MSSMsoft} for the MSSM soft terms with a mass scale of
\beq
m_{\mathrm{soft}}\sim\frac{\langle F \rangle}{\MPlanck}.
\eeq

The SUSY-breaking scale in the hidden sector $M_{\cancel{\mathrm{SUSY}}}$ is determined by the VEV $\langle F \rangle$ such that
\beq
M_{\cancel{\mathrm{SUSY}}}\sim\sqrt{\langle F \rangle}.
\eeq
This means that in order to have $m_{\mathrm{soft}}\sim 1$ TeV (as we require for weak-scale supersymmetry), $M_{\cancel{\mathrm{SUSY}}}$ should be about $10^{11}$ GeV.

Functions $\mathcal{F}$ and $\mathcal{G}$ in \eq{hiddengrav} in general contain many free parameters whose values are to be determined by the exact form of the underlying supergravity theory. One popular way of simplifying the PMSB scenario is to work in the framework of ``minimal supergravity'' or mSUGRA (also called the `supergravity-inspired scenario'). In this case all the soft parameters are fully determined by only four parameters
\beq
m_{1/2}, m_0, A_0, B,
\eeq
plus the $\mu$-parameter of the MSSM superpotential. The soft parameters of \eq{MSSMsoft} in this case read
\beq
&&M_3 = M_2 = M_1 = m_{1/2}, \nonumber \\
&&{\bf m^2_{Q}} =
{\bf m^2_{{\sbar u}}} =
{\bf m^2_{{\sbar d}}} =
{\bf m^2_{ L}} =
{\bf m^2_{{\sbar e}}} =
m_0^2\, {\bf 1}, \nonumber \\
&&m_{H_u}^2 = m^2_{H_d} = m_0^2, \nonumber \\
&&{\bf a_u} = A_0 {\bf y_u},~~
{\bf a_d} = A_0 {\bf y_d},~~
{\bf a_e} = A_0 {\bf y_e}, \nonumber \\
&&b = B \mu .
\label{mSUGRA}
\eeq

It is important to notice that the above unification relations are defined at the natural scale for gravity \ie $\MPlanck$. One then needs to use renormalisation group equations (as we discussed in section~\ref{sec:RGEs}) and evolve the soft parameters down to the electroweak scale so as to obtain the interesting low-energy quantities such as the MSSM mass spectrum. From a practical point of view however, it has become common to assume that the relations (\ref{mSUGRA}) are valid at the grand unification scale $\MGUT$ and then start the RGEs from that scale instead of $\MPlanck$. This is mostly because we do not know much about how the RGEs behave at scales between $\MGUT$ and $\MPlanck$ whereas the observed unification pattern for the MSSM gauge couplings (see section~\ref{sec:RGEs}) indicates that our understanding of the RGEs at scales below $\MGUT$ may not be far from reality. It has been shown that the effects neglected by using $\MGUT$ instead of $\MPlanck$ in the RGEs, are probably not significant. They may however cause additional important effects~\cite{PlancktoGUT1,PlancktoGUT2,PlancktoGUT3,PlancktoGUT4}.

The much simpler parameter space of mSUGRA (with only $5$ new parameters compared to the SM) has made it one of the most popular models for SUSY breaking and after some small modifications (as we will see in section~\ref{sec:phenassump}) the model has received the bulk of attention in phenomenological studies of supersymmetry. According to some authors~\cite{Ellis:2011jb}, the parameter space of mSUGRA is even simpler than what we described here. This is achieved by imposing the extra relation $A_0=B+m_0$ which eliminates one more free parameter from the MSSM parameter space; this parameter is usually taken to be $B$.

\subsubsection{Gauge-mediated SUSY breaking: mGMSB}

In gauge-mediated SUSY breaking (GMSB)~\cite{newgmsb1,newgmsb2,newgmsb3,Giudice:1998bp}, in contrast to the PMSB scenario, the effects of SUSY breaking are not mediated from the hidden sector to the observable sector (\ie the MSSM) using non-renormalisable interactions. GMSB models circumvent the problems with the sum rules in a different way, and that is to generate the soft terms through `radiative' interactions, \ie couplings at `loop level' rather than tree level.

The interactions responsible for generating soft terms in a GMSB setup are purely gauge, similar to the gauge interactions in the SM or the MSSM. The idea is the following: Assume that there are some new chiral supermultiplets that, on the one hand, couple to the source of supersymmetry breaking in the hidden sector, and on the other hand, couple to the MSSM particles through the SM gauge bosons and their superpartners gauginos. These new chiral supermultiplets are called ``messengers''. Now, when SUSY is broken in the hidden sector, the effects are first conveyed to the so-called `messenger sector' and then communicated to the visible sector through gauge interactions and `only radiatively' (\ie only at loop level).

One should notice that even in the GMSB scenarios gravitational interactions exist between the fields in the hidden, messenger and observable sectors. However, such gravitational effects are much weaker than the gauge effects. In other words, the gravitational effects are dominated by the gauge effects and consequently do not play any important role in breaking supersymmetry.

In the simplest GMSB model, called `minimal GMSB' (or mGMSB), the messenger fields couple to a gauge-singlet chiral supermultiplet.  The scalar component of the supermultiplet (denoted by $S$) and its corresponding auxiliary field $F_S$ develop VEVs $\langle S \rangle$ and $\langle F_S \rangle$, respectively, and therefore break supersymmetry. The scale of SUSY breaking in the messenger sector is $M_{\cancel{\mathrm{SUSY}}}^{\mathrm{mess}}\sim\sqrt{\langle F_S \rangle}$ and this should be distinguished from the SUSY-breaking scale in the hidden sector $M_{\cancel{\mathrm{SUSY}}}^{\mathrm{hidd}}$ that is associated with its own auxiliary field $F$.

The mGMSB contains $6$ free parameters~\cite{SUSY:Baer}
\beq
\Lambda, M, n_5, \tan{\beta}, \sgn{\mu}, C_{\mathrm{grav}}.
\eeq
Here, the two parameters $\Lambda$ (that sets the mass scale of the MSSM sparticles) and $C_{\mathrm{grav}}$ (which is called gravitino mass parameter) are defined as
\beq
\Lambda\equiv\frac{\langle F_S \rangle}{\langle S \rangle}=\frac{{(M_{\cancel{\mathrm{SUSY}}}^{\mathrm{mess}})}^2}{\langle S \rangle}, \nonumber \\
C_{\mathrm{grav}}\equiv\frac{\langle F \rangle}{\lambda \langle F_S \rangle}=\frac{{(M_{\cancel{\mathrm{SUSY}}}^{\mathrm{hidd}})}^2}{\lambda {(M_{\cancel{\mathrm{SUSY}}}^{\mathrm{mess}})}^2},
\eeq
where $\lambda$ is the common messenger-sector Yukawa coupling. The parameters $n_5$ and $M$ ($M>\Lambda$) set the number of messenger multiplets and the mass-scale associated with the messenger sector, respectively. All the soft SUSY-breaking terms of the MSSM are obtained in mGMSB by means of the RGEs that are evolved from the scale of $M$ down to the electroweak scale. $\tan{\beta}$ is the ratio of up-type to down-type Higgs VEVs at the electroweak scale (see section~\ref{sec:higgssector}). $\mu$ can in principle be treated as a free parameter like in mSUGRA, but its magnitude is usually fixed in this model by imposing `radiative electroweak symmetry breaking' (REWSB) condition at the weak scale, while its sign is to be determined experimentally (see section~\ref{sec:CMSSM} for more details about REWSB and a similar assumption for a different model). An interesting feature of mGMSB is that the trilinear coupling parameters of~\ref{MSSMsoft}, \ie ${\bf a_u}$, ${\bf a_d}$ and ${\bf a_e}$ only arise at two-loop level and are therefore very small; they are frequently assumed to be vanishing~\cite{Martin:1997ns}.

\subsubsection{Extra-dimensional-mediated SUSY-breaking: mAMSB}

In this class of SUSY-breaking models, the central idea is that the observable and hidden sectors correspond to two different spacetime manifolds that are hovering in a bulk and separated physically (for detailed discussion, see \eg refs.~\cite{SUSY:Baer,Martin:1997ns}). In the simplest models of extra-dimensional-mediated supersymmetry breaking (XMSB), it is assumed that each of the two sectors of the theory is confined to a 4-dimensional brane with a 5-dimensional bulk spacetime between them. The hidden and observable sectors then communicate in a manner that depends on whether the MSSM gauge supermultiplets are allowed to propagate in the bulk or not. If so, they can mediate supersymmetry breaking. It has been shown that in the simplest version of such models soft SUSY-breaking is dominated by gaugino masses; this scenario is therefore usually called ``gaugino mediation'' (see \eg refs.~\cite{gauginomediation1,gauginomediation2,gauginomediation3}).

One other possibility is that the gauge supermultiples are not allowed to freely travel in the bulk and, analogously to the chiral multiplets, are confined to the visible brane (\ie the MSSM sector). This means that like in the case of the gravity-mediated SUSY-breaking, some supergravity effects should be responsible for the transmission of SUSY-breaking from the hidden to the visible sector. In an interesting class of these models, which is called ``anomaly-mediated supersymmetry breaking'' (or AMSB)~\cite{AMSB1,AMSB2}, the MSSM soft terms are generated at loop level due to an anomalous violation of a particular symmetry called `local superconformal invariance'. Again, as in the GMSB scenario, gravity-mediation is present here but its effects are dominated by the anomaly-mediation interactions.

The original AMSB scenario has many unique properties that make the scenario quite interesting. For example all the soft terms generated from this model can be written in terms of only one free parameter which is the gravitino mass $m_{3/2}$. The model is however not viable: it can be shown that the sleptons have negative squared-masses (\ie the particles are tachyonic). There have been various proposals for modifying the theory to circumvent the tachyonic mass problem (see \eg ref.~\cite{Martin:1997ns} and references therein). One phenomenologically motivated approach has been to add a new parameter $m_0$ (usually set at the GUT scale) to the model which provides large contributions to the slepton squared-masses and makes them positive. This so-called `minimal AMSB' (or mAMSB) is then characterised by four free parameters~\cite{SUSY:Baer}
\beq
m_0, m_{3/2}, \tan{\beta}, \sgn{\mu}.
\eeq

\subsection{Phenomenological assumptions}

An entirely orthogonal approach to handle the large MSSM parameter space, is to simply impose phenomenologically justified assumptions and simplifications to the parameters without relying on any particular underlying supersymmetry-breaking mechanism. We briefly review some of these simplified models in this section.

\label{sec:phenassump}
\subsubsection{Low-energy models: MSSM-7,8,10,11,18,19,24}

It is already known that extensive regions of the MSSM parameter space are excluded experimentally. This is mainly due to the fact that many of the soft terms in the SUSY-breaking sector of the MSSM can introduce large flavour-changing neutral currents (FCNCs) or new sources of CP-violation that are strongly constrained by experiments. These dangerous terms should therefore be suppressed. Examples of soft parameters that typically produce large FCNCs include off-diagonal entries in the trilinear coupling matrices ${\bf a_u}$, ${\bf a_d}$ and ${\bf a_e}$ and sfermion mass matrices ${\bf m^2_{ Q}}$, ${\bf m^2_{{\sbar u}}}$, ${\bf m^2_{{\sbar d}}}$, ${\bf m^2_{L}}$ and ${\bf m^2_{{\sbar e}}}$. It is common in the SUSY-phenomenology community to accordingly approximate many of these off-diagonal parameters to zero (see \eg ref.~\cite{Gondolo:2004sc}), although this is certainly not theoretically motivated at this level.

Let us emphasise here that many of the SUSY-breaking scenarios we discussed in the previous section, naturally lead to the suppression of dangerous off-diagonal terms; an example is mSUGRA. Inspired by mSUGRA assumptions, in constructing some phenomenological sub-models of the MSSM, one can assume additional properties for the mass and trilinear coupling matrices such as `reality' and `universality'. This contraction of the parameter space has turned out to be so helpful that in some popular cases the number of new free parameters of the MSSM has seen dramatic reductions, \eg from $105$ to $7$ (see section~\ref{sec:paramspace}).

The phenomenological version of the MSSM with the largest number of free parameters that has been analysed phenomenologically to date, has $24$ non-SM parameters (compared to $105$ for the full MSSM) and is called MSSM-24 or simply `phenomenological MSSM' (see \eg refs.~\cite{AbdusSalam:2009qd,Bertone:2010rv}). In one version of MSSM-24 the non-SM parameters are $3$ trilinear couplings $a_t$, $a_b$ and $a_\tau$ for the third-family, $5\times 3=15$ diagonal entries for sfermion squared-mass matrices ${\bf m^2_{ Q}}$, ${\bf m^2_{{\sbar u}}}$, ${\bf m^2_{{\sbar d}}}$, ${\bf m^2_{L}}$ and ${\bf m^2_{{\sbar e}}}$ corresponding to the three different families (no universality is imposed), $3$ gaugino masses $M_1$, $M_2$ and $M_3$, the ratio of up-type to down-type Higgs VEVs $\tan{\beta}$, the mass of the pseudoscalar Higgs $m_A$ and the $\mu$ parameter of the superpotential. All these parameters are defined at the electroweak scale.

By imposing more and more simplifications, one can decrease the number of free parameters and this has been done in the literature in various ways. The ones that have been widely used so far have $19$, $18$, $11$, $10$, $8$ or $7$ parameters.

As one example of the low-dimensional models, let us briefly describe MSSM-7. In this model only the trilinear parameters $a_t$ and $a_b$ of MSSM-24 are taken to be generally non-zero ($a_\tau$ is set to zero), all $15$ diagonal sfermion squared-mass parameters are assumed to be equal (with the universal mass parameter $m_0$), and gaugino mass parameters $M_1$, $M_2$ and $M_3$ are related in the following way:
\beq
\label{MimSUGRA}
M_3=\frac{\alpha_s}{\alpha}\sin^2\theta_W M_2=\frac{3}{5}\frac{\alpha_s}{\alpha}\cos^2\theta_W M_1.
\eeq
The above relations hold at the electroweak scale and are inspired by mSUGRA model described in section~\ref{sec:fundamentals}. These relations are the results of the RGEs evolved from the GUT scale down to the electroweak scale (for mSUGRA).

So far we have $4$ parameters $a_t$, $a_b$, $m_0$ and, say, $M_2$. If we add $\mu$, $m_A$ and $\tan{\beta}$ to these, we obtain the full set of free parameters for MSSM-7.

\subsubsection{High-energy models: CMSSM and NUHM}
\label{sec:CMSSM}
\begin{figure}
        \centering
\includegraphics[scale=0.25, trim = 0 0 0 0, clip=true]{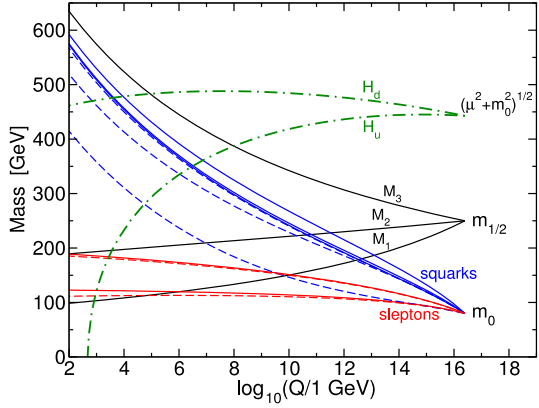}
 \caption{Renormalisation group evolution of scalar and gaugino mass parameters $m_{1/2}$ and $m_0$ in the constrained MSSM. The quantity $|\mu|^2+m_{H_u}^2$ runs negative and breaks electroweak symmetry. Squark and slepton squared-mass parameters remain positive. Adapted from ref.~\cite{Martin:1997ns}.}
        \label{fig:CMSSMrun}
\end{figure}

All phenomenological assumptions we made in the previous subsection were imposed on the MSSM parameters at `low energies', \ie directly at the electroweak scale. They are therefore different from the relations imposed by certain SUSY-breaking scenarios of section~\ref{sec:fundamentals} in that the latter are usually applied at `high energies', in particular the GUT scale. The hybrid approach is then to assume purely phenomenologically interesting simplifications and relations between the model parameters at high energy scales. The so-called ``constrained MSSM'' (or CMSSM) is an example~\cite{Kane:1993td}.

In the CMSSM, inspired by mSUGRA model, various universality assumptions are imposed on gaugino and scalar mass parameters, as well as trilinear ${\bf a}$-term couplings. One central assumption in the CMSSM that makes it different from mSUGRA is that the condition of `radiative electroweak symmetry breaking' (REWSB) is required to be fulfilled when RGEs evolve the parameters from the GUT to the electroweak scale.

In section~\ref{sec:higgssector}, we described the Higgs sector of the MSSM, as well as the conditions that must be fulfilled in order for the Higgs mechanism to break electroweak symmetry spontaneously. In fact, EWSB can be achieved in the MSSM if any of the Higgs squared-mass parameters
$m_{H_u}^2$ and $m_{H_d}^2$ in the soft SUSY-breaking sector (see \eq{MSSMsoft}) becomes negative. This can be certainly done in the MSSM by hand, namely by setting the parameters directly at the electroweak scale (of course with the condition that the relations~\ref{EWSBconds} must be satisfied). The process so far has been very similar to the SM case, with the difference that the latter has only one squared-mass parameter to be tuned.

The nice feature of the MSSM that makes it remarkably interesting from the EWSB point of view, is the observation that even if one gives positive values to the Higgs squared-masses $m_{H_u}^2$ and $m_{H_d}^2$ at some high energy, the renormalisation group equations can make $m_{H_u}^2$ negative at the weak scale. This includes models with universal mass assumptions, such as mSUGRA where all scalar mass parameters are unified at the GUT scale  (see \fig{fig:CMSSMrun}). In this latter case, $m_{H_u}^2$ which has the same value as the other scalar squared-mass parameters of \eq{MSSMsoft} (see \eqs{mSUGRA}), runs negative and makes the quantity $|\mu|^2 + m_{H_u}^2$ also negative. The latter condition then breaks electroweak symmetry. It can be observed from \fig{fig:CMSSMrun} that the squark and slepton squared-masses remain positive at all scales. This mechanism through which $m_{H_u}^2$ turns negative via the renormalisation group evolution, is called radiative electroweak symmetry breaking~\cite{SUSY:Baer}. It has been shown that REWSB can occur over a wide range of MSSM parameters if the top quark mass $m_t$ has a value between about $100$ and $200$ GeV, which is obviously the case ($m_t\simeq 172$ GeV)~\cite{SUSY:Baer}. This observation has provided one of the strong motivations for supersymmetrising the SM (see also section~\ref{sec:EWSB}).

The first equation in \eqs{EWSBconds2} of section~\ref{sec:higgssector} implies that if electroweak symmetry is broken appropriately, the value of the mSUGRA parameter $B$ can be fully determined in terms of $\mu$, the quantity $\tan\beta$ and the mass parameters of the theory that their weak-scale values are calculated by the RGEs (remember that $b=B\mu$). In addition, the second equation fixes the magnitude of the $\mu$ parameter in terms of $\tan\beta$ and the masses only leaving its sign to be determined experimentally. 
 
In the CMSSM, assuming that \eqs{EWSBconds2} hold, the high-scale parameter $B$ is eliminated in favour of $\tan{\beta}$. This leads to the following set of free parameters for the model (four continuous and one discrete):
\beq
m_0, m_{1/2}, A_0, \tan{\beta}, \sgn{\mu}.
\eeq

One concrete prediction of the CMSSM, as well as mSUGRA, is that (as we have already indicated in \eq{MimSUGRA}) the three gaugino mass parameters $M_1$, $M_2$ and $M_3$ are in a particular ratio. This is often shown approximately as $M_1 : M_2 : M_3 \simeq 1 : 2 : 7$ (according to \eq{MimSUGRA}).

We mentioned earlier (when discussing the RGEs in section~\ref{sec:RGEs}) that one reason why the gaugino masses are assumed to unify at the GUT scale is the intriguing fact that their ratios to the corresponding gauge couplings (which are unified at the GUT scale) do not evolve with energy. Such a motivation does not exist for the scalar masses and the unification assumption on their GUT values seems to be a rather arbitrary choice. This means that one can make other equally reasonable choices.

One popular example is the so-called ``non-universal Higgs model (NUHM)'' (see \eg ref.~\cite{Baer:2005bu}). In this slightly less restrictive framework, only squak and slepton masses are given a universal mass $m_0$ at the GUT scale whereas $m_{H_u}$ and $m_{H_d}$ are treated as being independent parameters. This seems to be a reasonable relaxation of the universality condition because even in grand unified theories the Higgs scalars do not necessarily belong to the same multiplet as the sfermions and there is therefore no reason for treating them on the same footing. The relaxation made in the NUHM model, introduces two new parameters $m_{H_u}$ and $m_{H_d}$ which are commonly traded for the more phenomenologically interesting weak-scale quantities $\mu$ and $m_A$. This gives the model a total number of $7$ free parameters.\footnote{Strictly speaking, the model we described here is the so-called NUHM2 model. The NUHM1 model is slightly different \eg in that it has only $1$ more free parameter than the CMSSM.}

\end{fmffile}

\setlength{\unitlength}{1mm}
\begin{fmffile}{feyn}

\chapter{Observational constraints on supersymmetry}
\label{sec:SUSYobs}

In the previous chapters, we gave a review of supersymmetry in general, as well as various supersymmetric models that have been put forward as solutions to different problems in particle physics and cosmology. Supersymmetry, as any other theories, has to be tested observationally. In fact, questions such as ``whether supersymmetry is a correct description of Nature at high energies'' and if so ``which supersymmetric model provides the best such description'' all need comparison of SUSY predictions with real experimental data.  In addition, we noticed that even within the framework of each supersymmetric model, one usually encounters a large number of free parameters that need to be determined experimentally. One of the main objectives of the present thesis has been to provide powerful tools and techniques for SUSY parameter estimation when its predictions face different observational data. We therefore in this chapter briefly review some of the most important observational strategies in testing supersymmetric models and constraining their parameter spaces.

\section{Supersymmetric WIMPs}

We discussed in section~\ref{sec:WIMPs} that weakly interacting massive particles (WIMPs) provide an elegant solution to the dark matter problem. We also mentioned that supersymmetric theories contain viable dark matter candidates such as the lightest neutralino (that we will simply call ``the neutralino'' from now on), gravitino and axino, of which the neutralino has been the most popular WIMP dark matter candidate to date (see also section~\ref{sec:neutralinos}).

A copious number of experiments are now looking for WIMPs and various observational constraints on their properties can naturally provide potential tests of supersymmetric models that contain WIMPs. In this section, we give an overview of the main strategies and techniques used to search for WIMPs, as well as the major existing observational constraints on their properties. Since the neutralino has been the only dark matter candidate we have studied in all analyses done in this thesis, we therefore restrict our discussions to its properties whenever we speak of a particular type of WIMPs.

\subsection{Cosmological relic density}
\label{sec:relic}

One of the accurately measured observables that significantly constrain the parameter spaces of most SUSY models, is the present amount of dark matter in the Universe, \ie $\Omega_{DM}\equiv\Omega_m-\Omega_b$ introduced in section~\ref{sec:cosmology}. We discussed in section~\ref{sec:WIMPs} that if the thermally-produced-WIMP scenario is correct, the present dark matter particles should have been created at the freeze-out period in the early Universe when the expansion rate of the Universe hits the WIMP interaction rate and WIMPs fall out of chemical equilibrium (and decouple). We also gave a brief exposition of some of the basic principles in calculating the dark matter relic density from WIMPs.

\begin{figure}[t]
        \centering
\includegraphics[scale=0.35, trim = 0 0 0 0, clip=true]{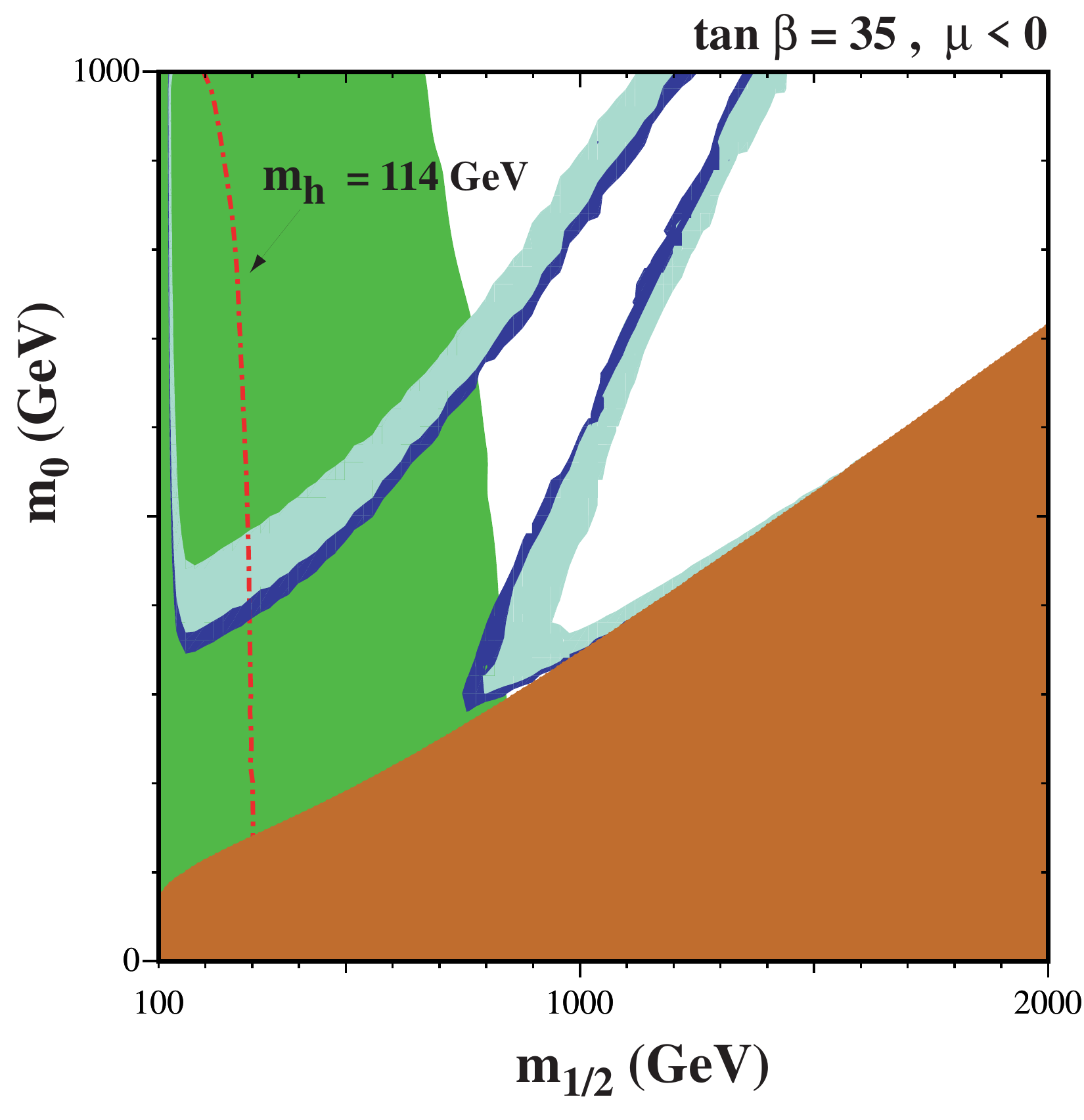}
 \caption{The $m_{1/2}$-$m_0$ planes for the CMSSM with $\tan\beta = 35$ and assuming $\mu < 0$. The region with cyan shading shows the allowed values of $m_{1/2}$ and $m_0$ when the cosmological constraint $0.1 \le \Omega_\chi h^2 \le 0.3$ has been imposed, and the region in cyan displays the allowed values when a tighter constraint ($0.094 \le \Omega_\chi h^2 \le 0.129$) has 
been imposed. Adapted from ref.~\cite{Ellis:2003cw}.}
        \label{fig:CMSSMrelic}
\end{figure}

The value for the relic density we presented in \eq{wimpmiracle} was however only approximate and one needs to solve the Boltzmann equation~\ref{boltzmann} so as to obtain the exact value for a particular WIMP candidate. The relic density of a WIMP in general depends on both its mass and annihilation cross-section which in turn contain information from the particle nature of the WIMP. These provide a connection between the actually observed relic density and the theoretical value predicted for example by a particular set of model parameters. \fig{fig:CMSSMrelic} shows an example analysis of the CMSSM (see section~\ref{sec:CMSSM}) where all but two parameters $m_0$ and $m_{1/2}$ are fixed and only the relic density constraint is imposed~\cite{Ellis:2003cw}. It is interesting to see that this observable provides such a tight constraint on the CMSSM parameter space (cyan and dark blue strips) even if no other constraints are used.

It is important to note here that the processes that give rise to the final value of the relic density for WIMPs are in most cases much more sophisticated than the simple case we have discussed so far. 
This means that for example the Boltzmann equation~\ref{boltzmann} has to be modified. One example is that the relic density of dark matter depends upon the history of the expansion rate in the early Universe before and during freeze-out. Strong modifications of that epoch can substantially impact the value of the relic density today. One other important example is when at the time of freeze-out some other particles exist that have masses close to the WIMPs and also share a quantum number with them. These particles interact with the WIMPs and in many cases can enhance the annihilation process and therefore change the relic density dramatically. These effects are usually referred to as `coannihilations'~\cite{Griest:1990kh}. The effects of including coannihilations in the calculations of relic density has been extensively studied in the literature (see \eg refs.~\cite{Edsjo:1997bg,Edsjo:2003us}). The presence of coannihilations make the relic density calculations quite difficult and they have to be done numerically. For example, for the case of the supersymmetric neutralino WIMPs the relic density is usually calculated using advanced computer packages that take various effects into account, including coannihilations. The most popular publicly available numerical codes that calculate relic densities (amongst other things) are \textsf{DarkSUSY}~\cite{Gondolo:2004sc} (available from ref.~\cite{DarkSUSY:web}) and \textsf{micrOMEGAs}~\cite{Belanger:2004yn} (available from ref.~\cite{micrOMEGAs:web}).

Finally let us mention here that our best estimation of the value of the dark matter relic density at the moment comes from the observations of the cosmic microwave background (CMB) by the WMAP satellite~\cite{Komatsu:2010fb}. The value is $\Omega_{DM}=0.222\pm 0.026$. We have used a similar value in our analyses of this thesis. A tighter constraint is expected to be provided by \eg the Planck satellite~\cite{:2011ah}.

\subsection{Direct detection}

\begin{figure}
        \centering
\includegraphics[scale=0.8, trim = 130 350 50 0, clip=true]{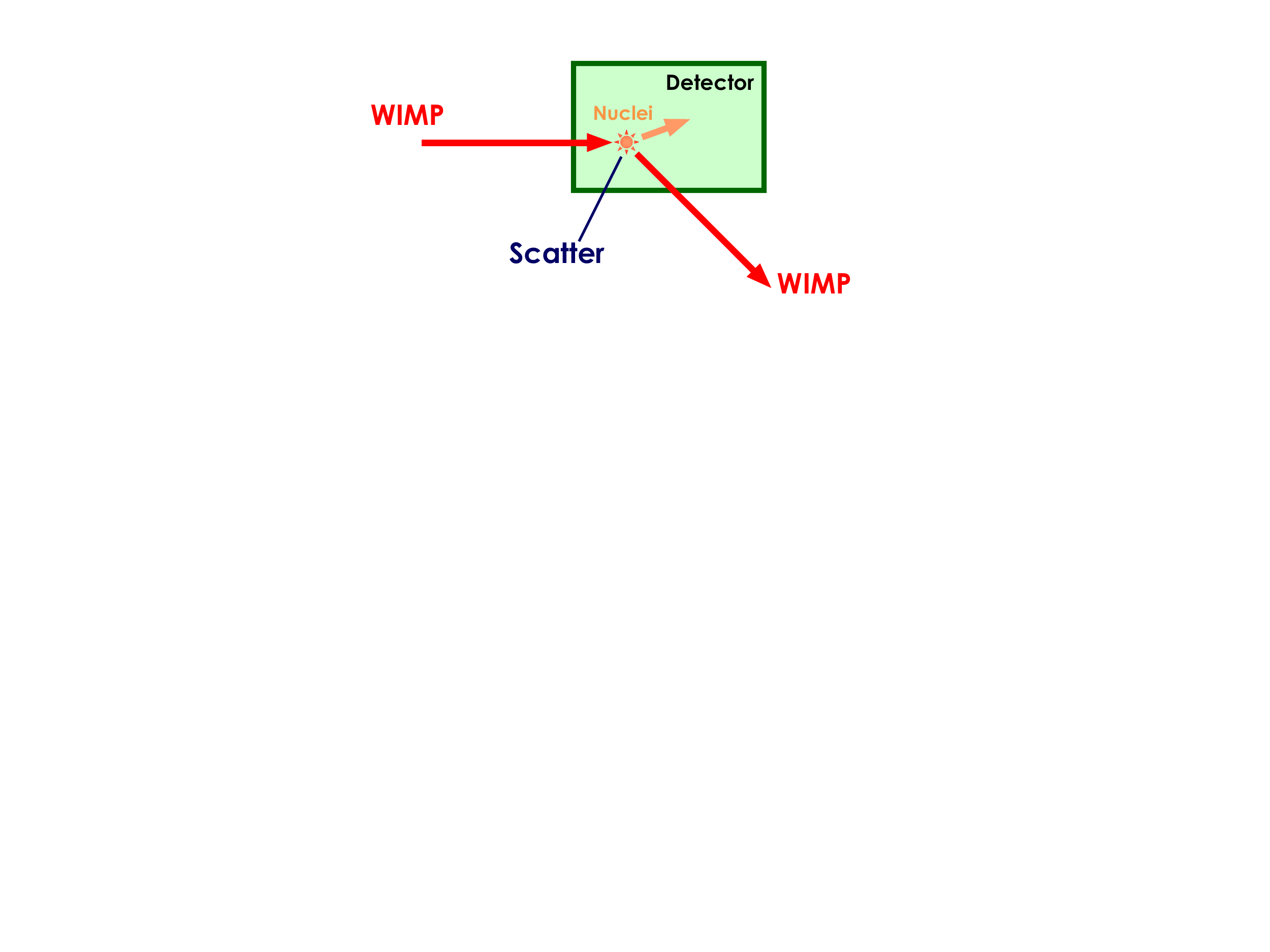}
 \caption{A WIMP scattering off normal nuclei in a dark matter direct detection experiment.}
        \label{fig:WIMPsDD}
\end{figure}

One important property of WIMPs is that they interact with the SM particles, although very weakly. This property can be used to directly search for WIMPs, \ie by looking for any interactions between them (that are supposed to fill our galaxy if they form dark matter) and some normal matter particles on Earth~\cite{Goodman:1984dc,Jungman:1995df}. This simple idea, as one of the most promising search strategies, has stimulated many experimental groups to build different small and large scale detectors looking for WIMP-SM interaction signals. Such interactions are sought for in `direct detection' experiments by recoding nuclear recoils when WIMPs scatter off the detector nuclei (see \eg \fig{fig:WIMPsDD}). This is usually done in the detectors in different ways that are usually various combinations of three different detection techniques: `ionisation', in which the atoms of the target material becomes ionised by the transferred `recoil energy'; `scintillation', in which particular materials known as scintillators are used to measure the fluorescent radiation produced by electrons in the target material when they decay after getting excited by the transferred recoil energy; and measurement of `phonon' excitations generated by nuclear recoils.

\begin{figure}[t]
        \centering
\includegraphics[scale=0.5, trim = 0 0 0 0, clip=true]{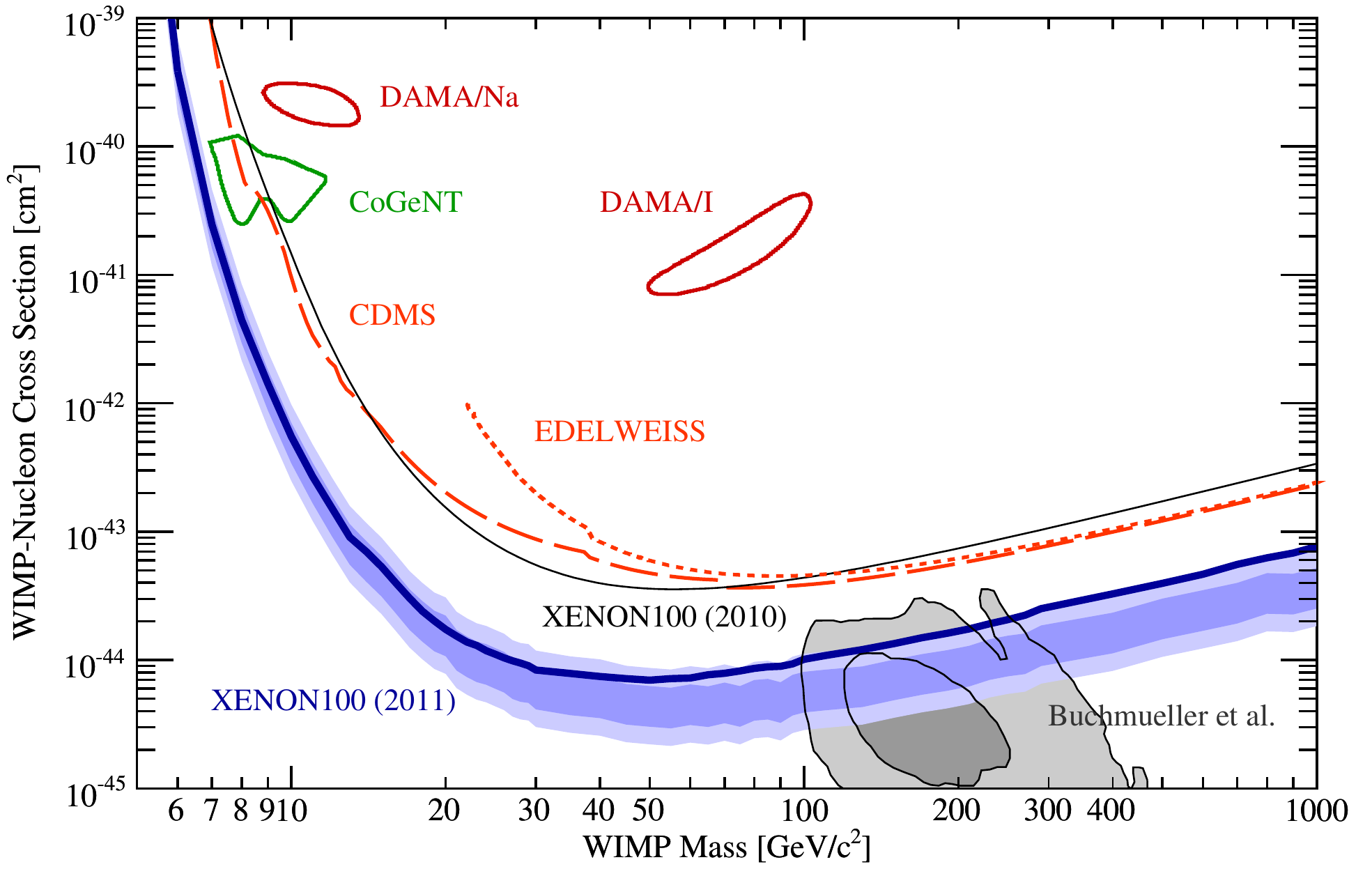}
 \caption{Existing strongest exclusion limits on the spin-independent elastic WIMP-nucleon cross-section $\sigma_{SI}$ versus the WIMP mass $m_{\chi}$, provided by XENON100~\cite{Aprile:2011hi}. The thick blue curve shows the XENON100 limit at $90\%$ C.L. and the dark and light shaded areas depict the $1\sigma$ and $2\sigma$ sensitivity of the experiment. The limit is derived with the Profile Likelihood method (see \eg section~\ref{sec:PL} of the present thesis) where different systematic uncertainties are taken into account. For comparison, limits from some other experiments or analyses are also given: XENON100 (2010)~\cite{Aprile:2010um}, EDELWEISS~\cite{:2011cy} and CDMS~\cite{Ahmed:2009zw}. The $90\%$ C.L. regions favored by the two experiments CoGeNT~\cite{Aalseth:2010vx} and DAMA~\cite{Savage:2008er} are also provided. Expected interesting $68\%$ and $95\%$ C.L. regions of the CMSSM are shown in shaded gray~\cite{Buchmueller:2011aa}. Adapted from ref.~\cite{Aprile:2011hi}.}
        \label{fig:WIMPsDDlimits}
\end{figure}

Constraints provided by direct detection experiments are usually presented in terms of exclusion limits on either `spin-dependent' or `spin-independent' cross-sections of WIMPs and normal nuclei, $\sigma^{SD}$ and $\sigma^{SI}$, respectively, versus the WIMP mass $m_\chi$. There are currently a large number of direct detection experiments looking for WIMP signals and many others are under construction or planned for construction in the near future. The strongest available limits so far have been provided by the XENON100 experiment~\cite{Aprile:2011hi}, and, as can be seen in \fig{fig:WIMPsDDlimits}, they have just started probing interesting regions of SUSY parameter space.

No positive signal from WIMPs has been detected so far, except for the detection of an annual modulation signal by the DAMA/LIBRA experiment~\cite{Bernabei:2000qi,Bernabei:2008yi} that has not been confirmed by other experiments yet. With the upcoming direct detection experiments however, it is expected that a substantial fraction of the parameter space for various dark matter models (including SUSY models such as the CMSSM) will be tested. This was the main motivation for us in writing \AkramiDD~that examines the prospects for constraining SUSY models (in the context of the CMSSM) with future ton-scale direct detection experiments. \AkramiDD~is one of the few works that has compared direct detection data (\ie the number of observed events and corresponding recoil energies) directly with the model predictions in a full likelihood setup instead of just using available exclusion bounds.

The theory and phenomenology of dark matter direct detection, has been reviewed in great detail in \AkramiDD, we therefore do not discuss those here. Various experimental issues that should be considered for correctly interpreting experimental results, as well as uncertainties in different nuisance parameters have all been discussed in \AkramiDD. Direct detection has also been our particular case of study in \AkramiCOV~where certain statistical issues in SUSY parameter estimation (as we will discuss in the next chapter) have been discussed.

\subsection{Indirect detection}

Let us look at the neutralino, our favourite supersymmetric WIMP (see section~\ref{sec:neutralinos}). Neutralinos are `Majorana' fermions meaning that they are identical with their antiparticles. This means that two neutralinos can interact and annihilate into other particles including photons, neutrinos, antimatter and other types of cosmic rays, as primary or secondary products. With `indirect detection' methods one aims to detect such products that we receive from the self-annihilation processes (for an introduction, see \eg ref.~\cite{Bergstrom:2000pn,Bertone:2004pz,DMBertone:2010}).

Similar to the direct detection case, indirect searches usually provide their constraints on the WIMP properties in terms of limits on the annihilation cross-sections as a function of the WIMP mass. The annihilation rate is proportional to the square of the WIMP density $\rho_\chi^2$, leading to that the best targets for indirect searches are the ones with the highest concentration of WIMPs, such as the Galactic Centre or dwarf spheroidal galaxies.

\begin{figure}[t]
        \centering
\includegraphics[scale=0.35, trim = 0 0 0 0, clip=true]{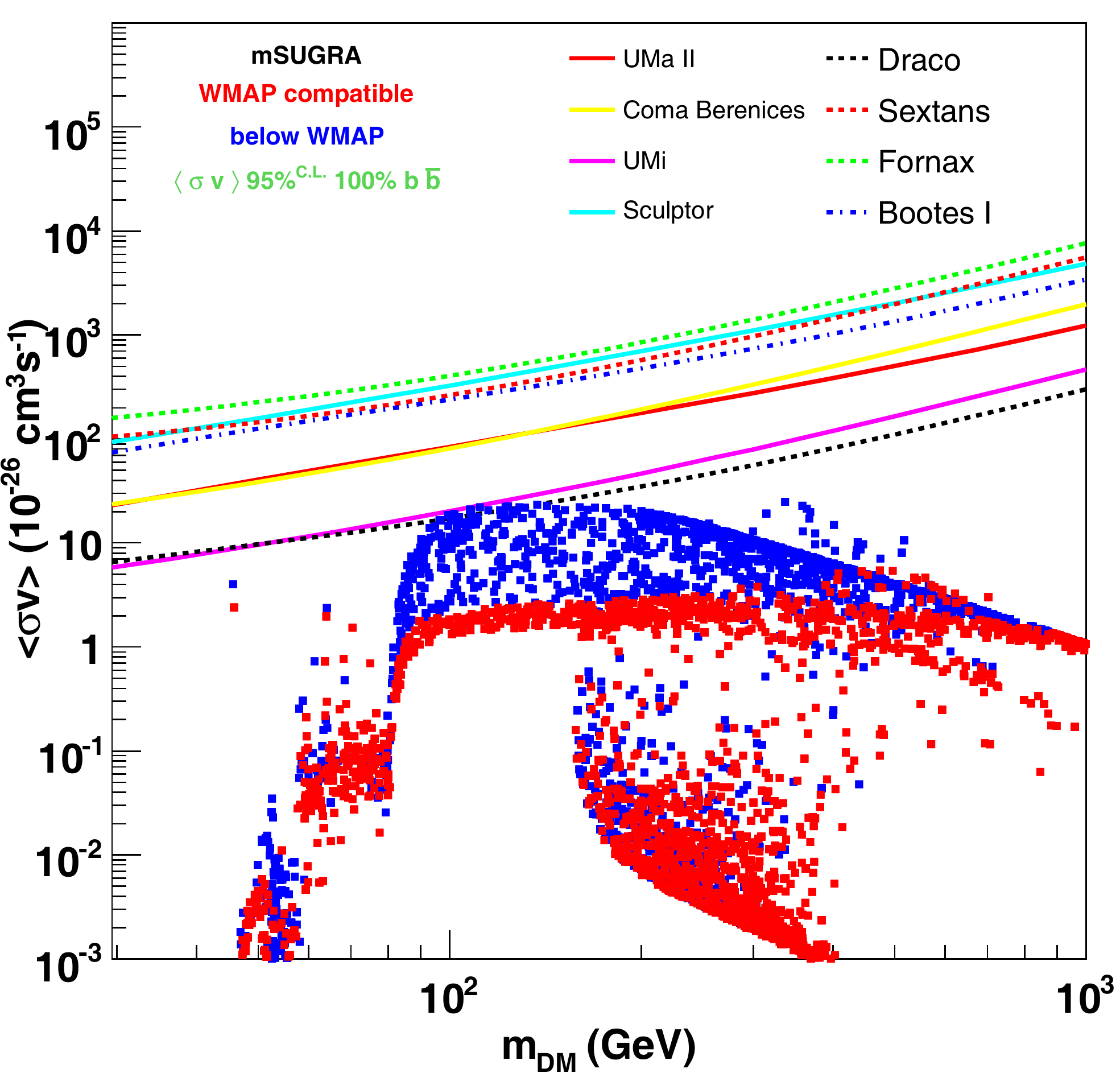}
 \caption{Exclusion limits on the velocity-averaged annihilation cross-section $\langle \sigma v \rangle$ versus the WIMP mass $m_{\chi}$, provided by \emph{Fermi} observations of Milky Way dwarf spheroidal galaxies. The red and blue points show mSUGRA models that are consistent with all accelerator constraints. The red points are models that also give a cosmological dark matter relic density equal to the observed value, while the blue ones are models with a lower thermal relic density. In the latter case, the neutralinos are assumed to be also produced non-thermally and still constitute all of the dark matter. The lines show the \emph{Fermi} $95\%$ upper limits. Adapted from ref.~\cite{Abdo:2010ex}.}
        \label{fig:FermiwarfsmSUGRA}
\end{figure}

The Galactic Centre could be one of the best choices, because it is nearby and potentially contains a large amount of WIMPs. The problem is however that it has a highly complex and poorly-understood structure with many different components~\cite{Vitale:2009hr,Collaboration:2009tm}. The dark matter profile of the Galactic Centre is also not entirely known~\cite{Stoehr:2003hf,Merritt:2010yu}. Dwarf galaxies, on the other hand, are interesting targets because of their high mass-to-light ratios, a property that reduces the astrophysical background. The problem is however that the flux coming from dwarfs is significantly lower than the Galactic Centre~\cite{Bringmann:2009vf,Pieri:2008nb,Martinez:2009jh}.

The other potentially very interesting targets for indirect detection are the so-called ``unidentified sources''. These include all sources of radiation in the sky whose astrophysical properties are not identified yet.  Some of these objects might be small clumps of dark matter whose existence is predicted theoretically in some models of the early Universe~\cite{Green:2003un,Green:2005fa,Kuhlen:2008aw,Bringmann:2009vf,Scott:2009tu}.  The problem with these objects is that their predicted properties (such as their number, mass and distance to Earth) can substantially vary in different theories.

Finally, one can go even further and look for WIMP signals coming from extragalactic sources such as clusters of galaxies~\cite{Pinzke:2009cp} or the extragalactic diffuse background (of for example gamma rays)~\cite{Abdo:2010dk}.

An indirect detection experiment usually looks for WIMP self- annihilation signals in a particular annihilation channel, \ie with particular annihilation products. In some cases, for example when photons (that are electrically neutral) are being observed, they directly point toward their sources, whereas the trajectories of electrons or antimatter particles (such as positrons and antiprotons) are easily affected by magnetic fields and consequently they do not give us information about the position of their sources.

Currently, various indirect detection experiments are observing potential dark matter self-annihilation products coming from different sources in the sky through different annihilation channels. Frontiers are: for photons, the Large Area Telescope (LAT), aboard the \emph{Fermi} gamma-ray space telescope~\cite{Atwood:2009ez} and several ground-based large air \v{C}erenkov gamma-ray telescopes (ACTs) (such as VERITAS~\cite{Holder:2008ux}, MAGIC~\cite{magic} and H.E.S.S.~\cite{Aharonian:2008aa}); for electrons, positrons and antiprotons, the PAMELA satellite~\cite{Adriani:2008zr}, \emph{Fermi}, H.E.S.S. and some balloon missions such as ATIC \cite{:2008zzr}; and for neutrinos, IceCube~\cite{Ahrens:2002dv}.

So far however, as in the direct detection case, no major signal excess has been observed by any of these experiments, and they have therefore been able to only provide some exclusion limits on the annihilation cross-section versus the neutralino mass. An example of such limits is depicted in \fig{fig:FermiwarfsmSUGRA} where limits from observations of Milky Way dwarf spheroidal galaxies by \emph{Fermi} are shown. This shows that like the previous case of direct detection, indirect detection experiments are also approaching the interesting regions of the SUSY parameter space.

In \Scott~we used \emph{Fermi} gamma-ray data obtained from observations of a particular (and arguably the most interesting) dwarf galaxy, Segue 1, to place constraints on the parameter space of the CMSSM. The main difference between our approach and the other popular approaches is that, like our previously mentioned work in \AkramiDD~for direct searches, we compared the model predictions with real data directly and in a full likelihood setup (see the next chapter). This was arguably the first paper that applied such a methodology to indirect detection analyses. In addition, we included a full treatment of the instrument response function (IRF) and its related uncertainties, as well as detailed background models.

\section{Collider constraints}

If weak-scale supersymmetry is a correct extension of the SM at high energies, it is expected to show up at current and future TeV colliders (or accelerators) such as the Large Hadron Collider (LHC) or the proposed International Linear Collider (ILC). Supersymmetric particles are expected to be discovered at these colliders and any positive or negative results can place tight constraints on interesting SUSY models. Currently, no sparticles have been found and therefore only lower bounds exist on their masses that can be used to exclude parts of the SUSY parameter spaces. In addition, SUSY contributions can indirectly affect other measured particle physics quantities including the so-called electroweak precision observables (EWPOs) and observables related to some rare processes such as $B$-physics observables (BPOs). We review in this section some of the most important collider constraints that are widely used in SUSY parameter estimation and model selection, including our analyses in this thesis.

\subsection{Electroweak precision observables (EWPOs)}
\label{sec:EWPOs}

The existence of any new physics, including supersymmetry, at close-to-electroweak energy scales can indirectly affect various precision observables at those energies, and such effects can be probed if high-precision experimental data are available~\cite{Heinemeyer:2004gx}. Some of these effects can be large such that the absence of significant deviations from the SM predictions place strong constraints on models of physics beyond the SM. As far as the supersymmetric extensions of the SM are concerned, electroweak precision observables have turned out to serve as powerful tools for testing the SM and its SUSY extensions by probing indirect effects of SUSY particles on those observables. Current precision experimental data, in all but only a few exceptional cases, fit the SM predictions very well and this implies that the data only put lower bounds on SUSY masses. The small deviations from the SM predictions can however be used to favour some regions of the SUSY parameter space which explain such deviations by higher order corrections that are caused by SUSY particles.

In addition to the EWPOs that are affected by contributions from SUSY particles, the virtual effects of these particles can be detected also by measurements of some `rare' processes. For example, there are some decays that represent flavour-changing neutral currents (FCNCs) and occur in the SM only at loop level~\cite{Hurth:2003vb}. This virtually means that such processes happen only very rarely. Examples of these rare decays are some $B$ decays such as $B\rightarrow X_s\gamma$ and $B_s\rightarrow\mu^+\mu^-$. In many SUSY models however, loop contributions from SUSY particles can be large (compared to the SM) and therefore make these processes happen more frequently, \ie with higher rates. The experimental tight constraints on their decay rates in these cases can therefore constrain the SUSY models. Obviously, the measurements of the observables associated with rare processes can also be considered as EWPOs if those measurements are of high precision. We however follow the conventions of some authors~\cite{Heinemeyer:2004gx} and discuss them under the name $B$-physics observables (BPOs) in section~\ref{sec:BPOs}. In addition, an observable like the anomalous magnetic moment of the muon (see below) corresponds both to an EWPO since it has been measured very accurately, and to a rare process. We again follow some conventions and consider it as an EWPO. Furthermore, quantities like the mass of the Higgs boson (if it exists) will be measured with high accuracy in the near future. It can therefore be also considered as an EWPOs, but regarding the existing limits on its value and the fact that the Higgs has not been measured yet, we discuss its constraints on SUSY models separately in section~\ref{sec:expboundsHiggs}.

Amongst different EWPOs, perhaps the most important ones that have been widely used in constraining SUSY models, are the $W$-boson mass $m_W$, the effective leptonic weak mixing angle $\sin^2{\theta_{\mbox{eff}}}$, the total $Z$-boson decay width $\Gamma_Z$, and the anomalous magnetic moment of the muon $a_\mu$. In the rest of this subsection, we briefly describe each quantity, its current experimentally measured value and its relevance for SUSY parameter estimation.

\subsubsection{$W$-boson mass $m_W$}

The theoretical prediction for $m_W$ can be expressed as~\cite{Heinemeyer:2004gx}
\beq
m_W^2(1-\frac{m_W^2}{m_Z^2})=\frac{\pi \alpha}{\sqrt{2}G_F}(1+\Delta r),
\eeq
where $\alpha$ and $G_F$ are the fine structure constant (calculated at the $m_Z$ renormalisation energy scale) and the Fermi weak coupling constant, respectively. $m_Z$ is the mass of the $Z$-boson. The quantity $\Delta r$ in this expression stands for all radiative corrections from new physics effects, including SUSY (see \eg refs.~\cite{Sirlin:1980nh,Marciano:1980pb} and references therein)).

Currently the best measurement of the $W$-boson mass has been provided by LEP~\cite{Alcaraz:2007ri} as
\beq
m_W=80.398 \mbox{GeV} \pm 25 \mbox{MeV}.
\eeq

In the case of supersymmetric extensions, the above experimental value is so accurate that even two-loop corrections from the superparticle effects can be probed. This means that it is quite important that we have an accurate theoretical prediction for $m_W$ for a given set of SUSY parameters. Currently there are some numerical tools for calculating this quantity that include two-loop corrections from the SM as well as the MSSM. An example is \textsf{SUSYPOPE}~\cite{Heinemeyer:2007bw,Heinemeyer:2006px} (see also ref.~\cite{Awramik:2003rn}). In order to compensate various higher-order approximations made in calculating the predicted value of $m_W$, one usually considers a theoretical uncertainty of about $10-15$ MeV in addition to the experimental uncertainty given above. These types of uncertainties are often used in fitting SUSY models to different EWPOs.

\subsubsection{Effective leptonic weak mixing angle $\sin^2{\theta_{\mbox{eff}}}$}

One important group of EWPOs are the $Z$-boson observables of which we briefly discuss here one of the most important ones for SUSY phenomenology, namely the `effective leptonic mixing angle' at the $Z$-boson resonance $\sin^2{\theta_{\mbox{eff}}}$. $\sin^2{\theta_{\mbox{eff}}}$ is a precision observable with high sensitivity for probing physics at electroweak scales. It is defined in terms of the ratio of the effective vector and axial vector couplings $g_V$ and $g_A$, when the $Z$-boson couples to leptons through the vertex $i\bar{l}\gamma^\mu(g_V-g_A\gamma_5)l Z_\mu$, where $l$ denote leptons~\cite{Bardin:1997xq}. The expression for $\sin^2{\theta_{\mbox{eff}}}$ then reads
\beq
\sin^2{\theta_{\mbox{eff}}}=\frac{1}{4}(1-Re\frac{g_V}{g_A}).
\eeq

It can be shown that $\sin^2{\theta_{\mbox{eff}}}$, if calculated at tree level, is approximately equal to $\sin^2{\Theta_W}$, where $\theta_W$ is the weak mixing angle with $\sin^2{\Theta_W}=1-m^2_W/m^2_Z$ in the on-shell renormalisation scheme. The higher-order (loop) corrections (from the SM or beyond) are all accommodated through the couplings $g_V$ and $g_A$~\cite{Hollik:2006ma,Heinemeyer:2004gx}.

The experimental value of $\sin^2{\theta_{\mbox{eff}}}$ can be measured at the electron-positron colliders through the measurements of different assymmetries around the $Z$-boson peak when QCD effects have been removed~\cite{:2005ema}. Currently the best estimate of the actual value of $\sin^2{\theta_{\mbox{eff}}}$ is~\cite{Alcaraz:2007ri}
\beq
\sin^2{\theta_{\mbox{eff}}}=0.2324\pm 0.0012.
\eeq

\subsubsection{Total $Z$-boson decay width $\Gamma_Z$}

Another important $Z$-boson observable is the `total decay width' of the $Z$-boson when it decays to different fermions, \ie in processes $Z\rightarrow f\bar{f}$. In this case $\Gamma_Z$ is defined as $\Gamma_Z=\sum_f\Gamma_f$, where $\Gamma_f$ are the `partial decay widths'. Here only decays to fermions are considered because other decay channels are relatively insignificant. The partial decay width $\Gamma_f$ can be expressed as~\cite{Heinemeyer:2004gx}
\beq
\Gamma_f=N_C^f\frac{\sqrt{2}G_F m_Z^3}{12\pi} \Bigl [|g_V^f|^2+|g_A^f|^2 (1-\frac{6m_f^2}{m_Z^2}) \Bigl ] (1+\delta_QED)+\Delta\Gamma_{QCD}^f, \nonumber
\eeq
where $g_V$ and $g_A$ are the effective coupling constants defined in the previous subsection, $N_C^f$ is the colour factor being $1$ for leptons and $3$ for quarks, $m_f$ is the mass of the fermion into which the $Z$-boson decays. $\delta_{QED}$ is some photonic QED correction, and $\Delta\Gamma_{QCD}^f$ denotes the standard gluonic QCD corrections plus possibly supersymmetric QCD corrections that involve virtual gluinos and quarks.

Another way of modifying the standard decay width $\Gamma_Z$ by SUSY processes is when $Z$-boson can decay `invisibly' into some new sparticles such as neutralinos (if they are sufficiently light). The partial decay width of the $Z$-boson to the dark matter particles with non-zero branching fraction is called `invisible $Z$ width'. In this case, a large deviation of the total width $\Gamma_Z$ from the standard value is expected.

The currently best experimental estimation for the total decay width $\Gamma_Z$ is~\cite{Alcaraz:2007ri}
\beq
\Gamma_Z=2.4952 \mbox{GeV} \pm 2.3 \mbox{MeV}.
\eeq

\subsubsection{Muon anomalous magnetic moment $a_\mu$}

The `anomalous magnetic moment of the muon' $a_\mu\equiv(g-2)_\mu/2$ is defined theoretically through the photon-muon vertex function $\Gamma_{\mu\bar\mu A^\rho}$ in the expression~\cite{Heinemeyer:2004gx}
\beq
\bar{u}(p')\Gamma_{\mu\bar\mu A^\rho}(p,-p',q) u(p)= \bar{u}(p')\left[\gamma_\rho F_V(q^2) + (p+p')_\rho F_M(q^2) + \ldots\right] u(p), \nonumber
\eeq
where $a_\mu \equiv -2m_\mu F_M(0)$. At three level, $F_M(q^2)$ is vanishing, and non-vanishing values are induced by quantum corrections at loop level. Currently the best experimental estimation of the actual value of $a_\mu$ (to eight significant figures) has been provided by the Muon G-2 collaboration through an experiment at Brookhaven National Laboratory~\cite{Bennett:2006fi}:
\beq
a_\mu^{exp}=1165920.80\pm 0.63 \times 10^{-9}.
\eeq

This measured value shows an about $3\sigma$ deviation from the SM prediction $a_\mu^{SM}=1165917.85\pm 0.61 \times 10^{-9}$~\cite{Miller:2007kk}. It is common to speak of the observed deviation in terms of the quantity $\delta a_\mu$:
\beq
\delta a_\mu \equiv a_\mu^{exp} - a_\mu^{SM}=29.5\pm8.8 \times 10^{-10}.
\eeq

Supersymmetric models typically well explain the above deviation. To show this, let us just present here a simple example where only one-loop corrections are considered and also all SUSY particles are assumed to have equal masses of the value $M_{\mathrm{SUSY}}$. The SUSY contribution to the muon anomalous magnetic moment in this case reads~\cite{Stockinger:2006zn}
\beq
a_\mu^{\mathrm{SUSY},1L}=13\times 10^{-10}(\frac{100 \mbox{GeV}}{M_{\mathrm{SUSY}}})^2\tan{\beta}\sgn{\mu},
\eeq
with $\tan{\beta}$ and $sign{\mu}$ defined for example as the CMSSM free parameters in section~\ref{sec:CMSSM}. Assuming $\mu > 0$, and $100 \mbox{GeV} \lesssim M_{\mathrm{SUSY}} \lesssim 600 \mbox{GeV}$ (depending on the value for $\tan{\beta}$), it can be seen that SUSY effects can easily provide the required contribution to the theoretical value of $a_\mu$ in order to reconcile it with observations. It is interesting to also notice that the SUSY contributions in some cases may instead deteriorate the situation by generating very large or even negative values for $a_\mu^{\mathrm{SUSY}}$. This should however be considered a plus for the muon anomalous magnetic moment as an observable because it can place strong constraints on the SUSY parameter space.

\subsection{Experimental bounds on the Higgs mass}
\label{sec:expboundsHiggs}

We saw in section~\ref{sec:higgssector} that the predicted value for the MSSM lightest Higgs boson $h$ satisfies the condition $m_h < m_Z \sim 91$ GeV at tree level. This is much lower than the currently strongest lower bound on the Higgs mass $114$ GeV provided by the LEP direct search~\cite{Schael:2006cr}. Loop corrections can however increase $m_h$ up to $135$ GeV.

Imposing the lower limit $114$ GeV can consequently exclude regions of SUSY parameter space that cannot provide the required loop corrections to avoid the limit. To be more precise, searches for the MSSM Higgs bosons at LEP have been performed using two production processes $e^+ e^- \rightarrow h Z$ and $e^+ e^- \rightarrow h A$ where $A$ is the CP-odd Higgs boson. Both processes are mediated by $s$-channel $Z$-boson exchange~\cite{Feng:2009te} and the cross-sections are proportional to the quantities $\sin^2(\beta-\alpha)$ and $\cos^2(\beta-\alpha)$, respectively. Here $\beta$ is as usual the ratio of the up-type to down-type Higgs VEVs and $\alpha$ is the mixing angle between the two CP-even Higgs bosons $h$ and $H$ and the Higgs interaction eigenstates $H_u^0$ and $H_d^0$. One can therefore be more stringent and impose the experimental bounds on SUSY models depending on the values for $sin^2{\beta-\alpha}$ (see also Tab.~\ref{massLLs}).

In addition to the LEP lower bounds on the Higgs mass, the CDF and D0 collaborations have recently excluded an additional mass range for the Higgs, but at higher values, using $p\bar{p}$ collisions at the Fermilab Tevatron at $\sqrt{s}=1.96 TeV$. Their combined results exclude the region $158 \mbox{GeV} <m_h<173 \mbox{GeV}$ at $95\%$ C.L.~\cite{Aaltonen:2011gs}. These semi-upper bounds however do not really put any new constraints on the MSSM parameters becasue such high Higgs masses cannot be acheived within the MSSM.

\subsection{Experimental bounds on sparticle masses}
\label{sec:expboundsSUSY}

\begin{table}[t] 
\linespread{1.5}
\begin{center}
\begin{tabular}{l l l} \toprule
\textbf{Particle mass} & \textbf{Commonly-used LLs} & \textbf{Conservative LLs} \\ \toprule
$\mhl$  & $>114\gev$\ & $>89.7\gev$ \\
$m_{\tilde\chi_1^0}$   & $>50\gev$ & - \\
$m_{\tilde\chi^\pm_1}$& $>103.5\gev$ & $>92.4\gev$ \\
$m_{\tilde{e}_R}$ & $>100\gev$ & $>73\gev$ \\
$m_{\tilde{\mu}_R}$ & $>95\gev$ & $>73\gev$ \\
$m_{\tilde{\tau}_1}$ & $>87\gev$ & $>73\gev$ \\
$m_{\tilde{\nu}}$ & $>94\gev$ & $>43\gev$ \\
$m_{\tilde{t}_1}$ & $>95\gev$ & $>65\gev$ \\
$m_{\tilde{b}_1}$ & $>95\gev$ & $>59\gev$ \\
$m_{\tilde{q}}$ & $>375\gev$ & - \\
$m_{\tilde{g}}$ & $>289\gev$ & - \\
\bottomrule
\end{tabular}
\end{center}
\caption{Experimental lower bounds on the Higgs and supersymmetric particles~\cite{PDG:Nakamura}.}
\label{massLLs}
\end{table}

As in the case of the Higgs boson, negative results from the current collider searches for SUSY particles places lower limits on their masses. These limits can then be used to exclude the points in the supersymmetric parameter space that predict masses violating the lower bounds.

Currently the best lower limits on the sparticle masses have been provided by the Particle Data Group~\cite{PDG:Nakamura}. These limits are however not completely model-independent and particular (usually CMSSM-like) SUSY models are often assumed in extracting the limits. In addition, the constraints on each sparticle mass is often dependent on its difference with the mass of the lightest supersymmetryc particle (LSP) which for the case of the Particle Data Group limits is assumed to be the neutralino $\tilde{\chi}^0_1$. Commonly used limits on the Higgs and sparticle masses (and the more conservative ones applied under specific conditions) are summarised in Tab.~\ref{massLLs}.

\subsection{$B$-physics observables (BPOs)}
\label{sec:BPOs}

In this section we describe four major observables involving $B$-mesons~\cite{Hurth:2003vb} and their decays into other particles that are widely used in the global fits of SUSY models to the experimental data. Those are the branching ratios $BR(b\rightarrow s\gamma)$, $B_s\rightarrow\mu^+\mu^-$, $B_u\rightarrow\tau\nu$ and the mass difference between $B_s$ and $\bar{B}_s$. There are $B$-physics observables other than these four, but they are of less interest and we do not consider them in this thesis.

\subsubsection{Branching ratio for $B\rightarrow X_s\gamma$}

Perhaps the most important $B$-physics observable is the experimentally measured value of the decay rate for the flavour changing process $B\rightarrow X_s\gamma$. In the SM, the main loop contributions are from the $W$-boson and top quark, and the SM prediction for the branching ratio of the process, that we denote by $BR(b\rightarrow s\gamma)$, is in excellent agreement with experiment. Additionally, in the MSSM, mainly chargino/stop and charged Higgs/stop loops contribute to the branching ratio~\cite{Grunewald:2003ij} and the contributions from neutralino loops are comparatively small~\cite{Wick:2008sz}. The two types of SUSY contributions can be large individually, but they can interfere destructively with each other giving rise to a value that is not significantly different from the SM prediction. This means that the observable $BR(b\rightarrow s\gamma)$ can place stringent constraints on the SUSY parameter space. The contributions from SUSY are particularly large when charged Higgs bosons are light and the parameters $\mu$ or $\tan{\beta}$ have large values.

The current world-average experimentally measured value for the branching ratio is given by the Heavy Flavor Averaging Group (HFAG) as~\cite{Barberio:2007cr}
\beq
BR(b\rightarrow s\gamma)_{exp}=(3.55\pm 0.22_{-0.10}^{+0.09}\pm 0.03)\times 10^{-4},
\eeq
whereas the SM contribution to the process at next-to-next-to leading order in QCD is theoretically predicted to be~\cite{Misiak:2006zs,Misiak:2006bw,Misiak:2006ab,Gambino:2008fj}
\beq
BR(b\rightarrow s\gamma)_{SM}=(3.28\pm 0.23)\times 10^{-4},
\eeq
which shows only a $1\sigma$ difference from the experimental value.

\subsubsection{Branching ratio for $B_s\rightarrow\mu^+\mu^-$}

Another important $B$-physics observable in SUSY phenomenology is the branching ratio for the flavour changing process $B_s\rightarrow\mu^+\mu^-$. The SM prediction for this observable is~\cite{Buras:2003td}
\beq
BR(B_s\rightarrow\mu^+\mu^-)_{SM}=(3.42\pm0.54)\times 10^{-9}
\eeq
which is well below the experimental upper bound
\beq
BR(B_s\rightarrow\mu^+\mu^-)_{exp}<5.8\times 10^{-8},
\eeq
a value given by CDF II data at $95\%$ C.L.~\cite{:2007kv}.

The MSSM however predicts that if $\tan{\beta}$ is large, neutral Higgs bosons can contribute to the branching ratio and enhance it by several orders of magnitude. This can easily violate the above experimental upper bound and therefore, the observable $BR(B_s\rightarrow\mu^+\mu^-)$ can impose important constraints on the MSSM parameter space.

\subsubsection{Branching ratio for $B_u\rightarrow\tau\nu$}

The last $B$-physics process that we discuss here is $B_u\rightarrow\tau\nu$, for which the SM prediction reads~\cite{Isidori:2006pk}
\beq
BR(B_u\rightarrow\tau\nu)_{SM}=\frac{G_F^2 m_B m_\tau^2}{8\pi}(1-\frac{m_\tau^2}{m_B^2})^2 f_B^2|V_{ub}|^2\tau_B.
\eeq 
Here $G_F$ is the Fermi weak coupling constant, $m_B$ and $m_\tau$ are the $B$-meson and $\tau$ masses, respectively, $f_B$ is the so-called $B$-meson decay constant, $V_{ub}$ is the $ub$-element of the Cabibbo-Kobayashi-Maskawa quark mixing matrix and $\tau_B$ is the $B$-meson lifetime.

The experimental value of the branching ratio for this process has been mainly measured by the two experiments BELLE~\cite{Ikado:2006un} and BABAR~\cite{Aubert:2006fk} and an often-used world-average value is~\cite{Ellis:2007fu}
\beq
BR(B_u\rightarrow\tau\nu)_{exp}=(1.31\times 0.49)\times 10^{-4}.
\eeq

This value does not quite agree with the SM prediction and the deviation can be expressed as
\beq
\frac{BR(B_u\rightarrow\tau\nu)_{exp}}{BR(B_u\rightarrow\tau\nu)_{SM}}=0.93\pm0.41.
\eeq

In the MSSM, the main contributions to $BR(B_u\rightarrow\tau\nu)$ are from the direct-exchange of a virtual Higgs boson that decay into $\tau\nu$. In scenarios with minimal flavour violation (such as the CMSSM and NUHM) one can show that~\cite{Ellis:2007fu}
\beq
\frac{BR(B_u\rightarrow\tau\nu)_{MSSM}}{BR(B_u\rightarrow\tau\nu)_{SM}}=\Bigl [ 1 - (\frac{m_{B_u}^2}{m_{H^\pm}^2})\frac{\tan^2{\beta}}{1+\epsilon_0\tan{\beta}} \Bigl ]^2,
\eeq
where $\epsilon_0$ is the effective coupling that parametrises the correction to the down-type Yukawa coupling from the gluino exchange (see below for its mathematical expression), $m_{B_u}$ is the $B$-meson mass and $m_{H^\pm}$ is the mass of the charged higgs boson.

\subsubsection{$B_s-\bar{B}_s$ mass difference}

Finally, another interesting $B$-physics quantity that is used in comparing new physics predictions (including SUSY) with experimental data is the mass difference $B_s-\bar{B}_s$ that is often denoted by $\Delta M_{B_s}$. This quantity is interesting because the frequency of oscillation between particle and antiparticle states of the neutral $B$-meson (which is measurable experimentally) is proportional to $\Delta M_{B_s}$. In addition, this quantity can be predicted theoretically both in the SM and the MSSM and therefore be used to constrain SUSY parameter spaces.

$\Delta M_{B_s}$ has been measured experimentally by the CDF collaboration to have the value~\cite{Abulencia:2006ze}
\beq
(\Delta M_{B_s})_{exp}=17.77\pm 0.12 ps^{-1}.
\eeq

The deviation from the SM prediction is~\cite{AbdusSalam:2009qd}
\beq
\frac{(\Delta M_{B_s})_{exp}}{(\Delta M_{B_s})_{SM}}=0.85\pm0.11.
\eeq

In the MSSM, the major additional contribution to $\Delta M_{B_s}$ comes from the exchange of neutral Higgs bosons~\cite{Ellis:2007fu}. The full MSSM prediction compared to the SM one is expressed as~\cite{Isidori:2006pk}
\beq
\frac{(\Delta M_{B_s})_{MSSM}}{(\Delta M_{B_s})_{SM}}=1-\frac{64\pi\sin^{\theta_w}}{\alpha m_A^2 S_0(m_t^2/m_W^2)}\frac{m_b{m_b}m_s{m_b}(\epsilon_Y \tan^2{\beta})^2}{[1+(\epsilon_0+\epsilon_Y)\tan{\beta}]^2[1+\epsilon_0\tan{\beta}]^2}. \nonumber
\eeq
Here the masses of the bottom and strange quarks $m_b$ and $m_s$ are calculated in the modified minimal subtraction renormalisation scheme $\bar{MS}$ and at the renormalisation scale $m_b$. $S_0$ is a function with the form
\beq
S_0(x) = \frac{4x-11x^2+x^3}{4(1-x)^2} - \frac{3x^3 \log
x}{2(1-x)^3}.
\eeq

$\epsilon_0$ and $\epsilon_Y$ have the forms
\beq
\epsilon_0 = -\frac{2\alpha_s \mu}{3 \pi m_{\tilde{g}}}\,H_2\left(
\frac{m_{\tilde{q}L}^2}{m_{\tilde{g}}^2},
\frac{m_{\tilde{d}R}^2}{m_{\tilde{g}}^2}\right), \quad
\epsilon_Y = -\frac{A_t y_t^2}{16 \pi^2 \mu}\,H_2\left(
\frac{m_{\tilde{q}L}^2}{\mu^2}, \frac{m_{\tilde{u}R}^2}{\mu^2}\right), \nonumber
\eeq
and are the effective couplings parametrising the corrections to the down-type Yukawa couplings. $\mu$ is the usual supersymmetric Higgs mass term, and $y_t$ and $a_t$ are the Yukawa and soft trilinear couplings that involve stopts. $H_2$ is a function with the following form:
\beq
H_2(x,y)= \frac{x \log x}{(1-x)(x-y)} + \frac{y \log y}{(1-y)(y-x)}.
\eeq
\end{fmffile}

\setlength{\unitlength}{1mm}
\begin{fmffile}{feyn}

\chapter{Statistical subtleties}
\label{sec:SUSYstat}

One principal objective of the present thesis has been to provide some additional strong support for the fact that SUSY models (even in cases where the number of free parameters is dramatically reduced) exhibit rather complex structures in their parameter spaces. In \AkramiGA~and \AkramiCOV~we have essentially tried to further demonstrate that constraining models using existing experimental data is under no circumstances an easy or straightforward task, and therefore care must be taken when particular statistical frameworks are being used in such analyses. Indeed, we have shown that current data do not sufficiently constrain the model parameters in a way completely independent of the employed statistical framework and scanning techniques.  We have extensively discussed some of the important statistical issues in the papers, concluding for example that the current scanning techniques may yet have some distance to go in this respect. The statistical frameworks and scanning strategies are also discussed in detail in the papers, in particular \AkramiGA~and \AkramiCOV. In this chapter we only go into some general descriptions of the frameworks and algorithms and refer the reader to the papers for more details.

\section{Statistical frameworks}
\label{sec:statframeworks}

In order to make any meaningful statistical inference about a theoretical setup, such as our favourite SUSY models, when its predictions are compared with experimental data, one needs to first make a decision about the statistical framework and formalism that should be used. This for example tells us how to make statistically significant statements about the model parameters. 

There are two commonly-used but fundamentally different
approaches to statistics that are based on entirely different interpretations of `probability' (for a detailed discussion, see \eg ref.~\cite{Cowan:1998}): the so-called ``frequentist'' and ``Bayesian'' statistics (or inference). The first approach deals with relative frequencies, while the second talks about subjective probabilities. Usually, the former is employed for assigning statistical errors to measurements, whereas the latter is used to also quantify systematic uncertainties. The basic difference between the two frameworks is that in
Bayesian inference one is interested in probabilities of some model parameters to have particular values when some data are given, whereas in frequentist statistics
the only meaningful quantity to work with is the probability of observing particular data when a specific set of model parameters is assumed. In this section we look into the definitions and fundamental ingredients of the two statistics and introduce some statistical measures that we used in the papers.
         
\subsection{Frequentist inference: profile likelihood} \label{sec:PL}

A frequentist accepts the most common interpretation of probability that defines it as a ``limiting relative frequency''. Let us assume for example that we measure a statistical variable through a process that is repeatable. In addition, suppose that we are interested in a particular outcome of the measurement that we denote by $\mathbb{O}$. The frequentist then defines the probability corresponding to the outcome $\mathbb{O}$, denoted by $P(\mathbb{O})$, as the fraction of times that
$\mathbb{O}$ occurs when we repeat the measurement procedure an infinite number of times. In reality however, no experiments can be repeated `an infinite number of times', and consequently the probabilities defined this way always remain hypothetical. One of the main tasks of a frequentist statistician is then to provide an estimation of the exact probabilities based on a finite
set of observed data. Having this given, the next step would then be to establish an appropriate method to investigate the compatibility of a particular theoretical model, which predicts the probabilities, with the experimental data and also place some constraints upon the model parameter space.

Let us see how this is usually performed in a frequentist setup by assuming that $n$ measurements of
a random variable $\mathbb{O}$ are made. We additionally suppose that the probability density function
(PDF) for the variable $\mathbb{O}$ (denoted by $p(\mathbb{O})$) is not known a priori. Our aim is now to `infer' properties of $p(\mathbb{O})$ based on the set of the $n$ observed data points
$d_1,...,d_n$ (that we denote by $D$). One may in particular be interested in constructing some functions
of $D$ so as to appraise different characteristics of $p(\mathbb{O})$. In most cases, a hypothetical form of the function $p(\mathbb{O})$ is available in terms of some undetermined parameters $\Theta=(\theta_1,...,\theta_m)$. This means that the PDF can now be shown as $p(\mathbb{O};\Theta)$.  The objective is then to estimate the values of the parameters $\Theta$, and this is done by constructing particular functions of the data points $D$. Such functions are called `estimators' and are often denoted for the parameter set $\Theta$ by $\hat{\Theta}$. One requirement for the estimator $\hat{\Theta}$ is that it converges to the `true' $\Theta$ (whose
actual values are, and may forever remain, unknown) when the number of data points $n$ becomes infinitely large. The entire process through which the actual model parameters $\Theta$ are being estimated from the empirical data points $D$ is called parameter estimation or parameter fitting.

The estimators $\hat{\Theta}(d_1,...,d_n)$ are themselves new random variables.  The reason is that the estimators are functions of the
measured values $D=(d_1,...,d_n)$, and obviously, if we repeat the experiment several times, each time we obtain in general different values for the measured quantities.  As a result, the estimators
$\hat{\Theta}(D)$ also receive different values in different repetitions of the measurement. One can now define another PDF, say $q(\hat{\Theta};\Theta)$, that corresponds to the statistical distribution of the estimators $\hat{\Theta}$; this PDF is called sampling distribution.  Studying various statistical properties
of estimators using the corresponding sampling distributions is another major task
of any statistical analysis.  This in particular includes estimating different experimental errors and theoretical uncertainties associated with the estimators.

One important point about the estimators is that there are various ways to construct them for a particular set of model parameters. There are however some properties that are said to be desirable for an estimator. This for example include small (ideally zero) biases and variances.  Unfortunately one cannot always achieve all the desired properties for an estimator simultaneously (an example is the trade-off between bias and variance~\cite{Cowan:1998}).  There are however some methods that provide estimators with optimal properties (\eg with reasonably small bias and variance). Examples are the methods of `maximum likelihood' (ML) and `least squares' (LS) that are arguably the most popular ones in various fields of scientific data analysis. Here we are in particular interested in the ML method and continue our discussions with describing the method and some of its interesting properties.

The central ingredient of the ML method (as it is clear from its name) is the likelihood function $\mathcal{L}$ that is defined as
\beq
\label{eq:like}
\mathcal{L}(\Theta)=\prod_{i=1}^{n}p(d_i;\Theta),
\eeq
where $p(\mathbb{O};\Theta)$ is the PDF according to which our random variable (previously shown as $\mathbb{O}$) is distributed, and $d_1,...,d_n$ are the corresponding values we have obtained in $n$ repetitions of the experiment. We additionally assume that the functional form of $p(\mathbb{O};\Theta)$ is determined although the values of the parameters $\Theta=(\theta_1,...,\theta_m)$ are not known. $\mathcal{L}$, if considered as a function of the data points $D=d_1,...,d_n$, is precisely the joint PDF for $D$. However, one can consider this quantity instead as a function of the parameters $\Theta$; the variables $D$ are assumed to be fixed in this case. Having the likelihood function defined, the ML estimators $\hat{\Theta}=\hat{\theta}_1,...,\hat{\theta}_m$ corresponding to the parameters $\Theta=\theta_1,...\theta_m$ are then defined as the values of the parameters at which $\mathcal{L}(\Theta)$ is maximised.

As we mentioned earlier, the ML estimator for a parameter of the
model is itself a random variable and this brings us to the point
where we should find an appropriate way of quantifying uncertainties in the fitted value of the parameter.  The simplest
way is to calculate and report the variance (or the standard
deviation) of the estimator, \eg by simulating several
experiments with the same number of samples in each as we had in
the actual measurement and then calculate the best-fit values
(\ie ML estimators) in each of those experiments. Then it
would be possible to work out an estimator for the variance of the
obtained parameter estimators. This procedure however, is not always adequate enough, and the
statistical uncertainty of a measurement must be given in some other more sophisticated ways.  This is
commonly performed by introducing the `confidence intervals'
for the parameters in the following way:  Assume that the experiment is repeated a large number of times, and an interval $[a,b]$ contains the true value of a parameter in a particular fraction of times, say, $\gamma$. The interval $[a,b]$ is then said to be a confidence interval at a confidence
level (C.L.) $\gamma$~\cite{Cowan:1998}. 

For the case of one parameter, the definition above is fine.  Now the question is: `can one generalise the concept of confidence interval to the case of $m$ parameters
$\Theta=(\theta_1,...,\theta_m)$ in a straightforward way?'  One way of doing this might be to introduce an
$m$-dimensional confidence interval $[\textbf{a},\textbf{b}]$ in which $\textbf{a}=(a_1,...,a_m)$ and $\textbf{b}=(b_1,...,b_m)$ in such a way that each $[a_i,b_i]$ is defined separately as a confidence interval for the $i^{th}$ parameter $\theta_i$.  This recipe is however not what is often used in the statistical inference (it is computationally difficult~\cite{Cowan:1998}).  One instead constructs a so-called `confidence region'
in the parameter space.  This is defined as the region which contains with a particular probability a point in the parameter space corresponding to the true values of the parameters $\Theta$.  Clearly, the form of this region is in general different from the aforementioned $n$-dimensional confidence interval $[\textbf{a},\textbf{b}]$.  It turns out that if $n\to\infty$, this region approaches an ellipse for $m=2$ parameters and an $m$-dimensional hyperellipsoid in general~\cite{Cowan:1998}.

Now, the question is `how to find in practice such confidence
regions for a given model at hand with unknown parameters for
which we have found ML estimators'.  One can show that both of the
joint PDF $q(\hat{\Theta}|\Theta)$ for the estimator
$\hat{\Theta}=(\hat{\theta}_1,...,\hat{\theta}_m)$ and the
likelihood function $\mathcal{L}(\Theta)$ become Gaussian in the large
sample limit (\ie when $n\to\infty$).  That is, contours of constant $q(\hat{\Theta}|\Theta)$ and
$\mathcal{L}(\Theta)$ are hyperellipsoids (ellipses in two dimensions) in
$\hat{\Theta}$-space and $\Theta$-space respectively, centred
about the true parameters $\Theta$ and the ML estimators
$\hat{\Theta}$, correspondingly~\cite{Cowan:1998}.  It can be shown that for this
likelihood function of Gaussian form, the regions in the
$\Theta$-space specified by the contours
\beq
\label{eq:confreg}
\ln \mathcal{L}(\Theta)=\ln
\mathcal{L}_{max}-\frac{Q_{\gamma,m}}{2}
\eeq
of the log-likelihood function $\ln \mathcal{L}(\Theta)$ (or
$-\frac{\chi^2}{2}$) are nothing but the previously defined
confidence regions with confidence levels $\gamma$.  The
quantities $Q_{\gamma,m}$ are the `quantiles' of orders $\gamma$
of the ${\chi}^2$ distribution, \ie
\beq
\label{eq:quantile}
\int_0^{Q_{\gamma,m}}f_{\chi^2}(z;m)dz=\gamma,
\eeq
where $f_{\chi^2}(z;m)$ indicates the $\chi^2$ distribution for $m$ degrees of
freedom.  Values of the $Q_{\gamma,m}$, for $1\sigma$ and
$2\sigma$ C.L.s (\ie $\gamma=68.3\%$ and $95.4\%$)
and $m=1,2,3,4,5$ fitted parameters, are given
in Tab.~\ref{tab:quantile}.
\begin{table}
\centering
\begin{tabular}{cccccc}
  \toprule
  \multirow{2}{*}{\textbf{$\gamma$}} &  &  & \textbf{$Q_\gamma$} &  &  \\
  \cline{2-6}
  & \textbf{$m=1$} & \textbf{$m=2$}& \textbf{$m=3$}& \textbf{$m=4$}& \textbf{$m=5$}\\
  \toprule
  $68.3\%$ ($1\sigma$)& 1.00 & 2.30 & 3.53 & 4.72 & 5.89\\
  \midrule
  $95.4\%$ ($2\sigma$)& 4.00 & 6.17 & 8.02 & 9.70 & 11.30\\
  \bottomrule
\end{tabular}
\caption[aa]{\footnotesize{Values of the quantile $Q_{\gamma,m}$ for
$68.3\%$ and $95.4\%$ (\ie $1\sigma$ and $2\sigma$) confidence
levels and for $m=1,2,3,4,5$ fitted parameters.}}
\label{tab:quantile}
\end{table}

The prescription given by \eq{eq:confreg} for determining
the confidence regions is completely true only in the case of a
Gaussian likelihood.  However, this method is also employed
for non-Gaussian functions (including the SUSY likelihood function) as a reasonably appropriate approximation to the actual confidence regions. In \AkramiCOV, we discuss one main issue with this approximation for the cases where the model parameter space is large and highly complex, namely the `statistical coverage' problem. We do not discuss the issue here. 

Now, all one needs to do in order to determine the best-fit
points and the associated uncertainties is to map the likelihood
function of the model given the experimental data.  From the
plotting point of view, it is a very straightforward procedure if
there are only one or two fitted parameters in the model
with one-dimensional (1D) and two-dimensional (2D) likelihood functions, respectively.  However, for the
higher-dimensional functions, one should have a good recipe for
summarising the statistical inference (\ie the best-fit points
and the errors) for each parameter separately, or in a 2D plane
for one parameter versus the other. A nice way of doing this is to make use of the ``profile likelihood'' (see \eg ref.~\cite{profilelike} and references therein) that is defined \eg for one parameter $\theta_i$ as
\beq
\label{eq:proflike}
\mathbb{L}(\theta_i)\equiv\max_{\theta_1,...,\theta_{i-1},\theta_{i+1},...,\theta_m}\mathcal{L}(\Theta),
\eeq
where $\mathcal{L}(\Theta)$ is the full likelihood function. This definition can easily be generalised if one is interested in two-, three-, or higher dimensional profile likelihoods, although it is often sufficient to calculate 1D and 2D profile likelihoods only.

This definition simply means that a frequentist eliminates unwanted parameters by maximising the
likelihood along the hidden dimensions. In other words, the profile likelihood is
nothing but the likelihood function of the reduced set of
parameters with the unwanted parameters at their conditional
ML estimates.  Now the interesting point about the
profile likelihood recipe is that, the approximate confidence regions can be set
using exactly the same prescription of \eq{eq:confreg} just as in
a standard $\chi^2$ fit, but now with the quantiles
$Q_{\gamma,1}$ and $Q_{\gamma,2}$ for the 1D and 2D plots,
respectively.

\subsection{Bayesian inference: marginal posterior} \label{sec:PPDF}

A fundamentally different way of interpreting probabilities is that of Bayesian or subjective statistics.  Contrary to the previous framework of frequentism where possible outcomes of a measurement constitute the sample space, here one instead talks about the hypothesis space, which consists of hypotheses.  These are
statements that are either true or false~\cite{Cowan:1998}.  Therefore, one of the
crucial features of the Bayesian framework is that a probability can be
assigned to a hypothesis, say $\mathbb{H}$.  This is not possible under the
frequentist framework, where a hypothesis can only be rejected or not
rejected.  This probability $P(\mathbb{H})$, in the Bayesian context, is interpreted as the degree of belief that the hypothesis $\mathbb{H}$ is true.  This means that, the Bayesian
probability can in particular be associated with the values of unknown parameters in a theoretical model of interest where these
parameters themselves are considered as random variables.  These subjective probabilities reflect our degree of confidence that the parameter values reside in certain intervals. Again, a probability for an unknown parameter is not meaningful within the context of the frequentist statistics.

The above description of the Bayesian inference then leads to a more practical definition of it, namely as a statistical inference in which
evidence or observations are used to update or to newly infer the
probability that a hypothesis may be true (for an introduction to general applications of Bayesian inference in physics,
see \eg ref.~\cite{D'Agostini:1995fv}, and for reviews of its applications in cosmology, see \eg refs.~\cite{Trotta:2005ar,Trotta:2008qt,Liddle:2009xe,Hobson:2010}.  The name Bayesian
comes from the frequent use of ``Bayes' theorem'' in the
inference process, which in fact forms the basis of Bayesian statistics.

Assume again (analogous to the previous subsection) that the model at hand is parametrised by $m$
unknown parameters $\Theta=(\theta_1,...,\theta_m)$ and there are
some experimentally provided data $D=(d_1,...,d_n)$ which are
supposed to be used for constraining the parameter values. Bayes'
theorem then reads
\beq
\label{eq:bayestheorem}
p(\Theta|D)=\frac{p(D|\Theta)p(\Theta)}{p(D)}.
\eeq
Here $p(\Theta)$, the so-called `prior PDF' (or simply the `prior') and usually shown as $\pi(\Theta)$, represents our
degree of belief (or the state of knowledge) that the parameters $\Theta$ are the true values
before the consideration of the data, $p(D|\Theta)$ is nothing but the
previously introduced likelihood function $\mathcal{L}(\Theta)$, \ie the
probability under the assumption of the specific values of
$\Theta$ for the model parameters, to observe the data $D$. $p(\Theta|D)$ is called the `posterior
PDF' (or simply the posterior), reflecting the
probability of $\Theta$ as being the true values after seeing the data $D$. Finally, the
quantity in the denominator, \ie $p(D)$, is called the `Bayesian evidence' (or simply the evidence), which for the
purpose of constraining parameters of a model, is in fact
nothing but a normalisation constant, \ie independent of
$\Theta$, and is often simply dropped.

It is important to realise that Bayesian statistics does not offer any fundamental rule for choosing priors for a model's parameters; this should be done based on other theoretical considerations. After choosing a particular prior, one can see how the degree of belief (or the state of knowledge) changes, \ie how the prior is updated to the posterior, when the data are used. This upgrading information is provided by the model's likelihood.

In Bayesian statistics, all of our knowledge about the parameters
$\Theta$ is contained in the posterior PDF $p(\Theta|D)$.  Very often however the parameter space of the model at hand is multidimensional and one needs to summarise the important characteristics of the multidimensional posterior in a practically appropriate way.  One usually starts with introducing an estimator, which is often taken to be either the expectation
values of the parameters $E[\Theta]$ corresponding to the
posterior PDF (\ie the `posterior mean'), or the values of
$\Theta$ that maximise it (\ie the `posterior mode');
the latter coincides with the ML estimator discussed previously if the prior $\pi(\Theta)$ is taken to be a
constant (\ie for a flat or linear prior). In this case the posterior $p(\Theta|D)$ is
proportional to the likelihood $\mathcal{L}(\Theta)$.  As the next step in
making statistical conclusions about the model parameters based on the posterior PDF, one should find an appropriate way of
quantifying uncertainties, similar to what we did by
introducing the confidence intervals and regions in a frequentist
framework. Perhaps the most natural way is then to construct a
so-called `credible region' in the parameter space that contains a certain
fraction of the total probability given by integrating the
posterior over the whole space. This can be simply done for
example for a 2-dimensional parameter space by drawing a contour
such that, say, $68.3\%$ of the total posterior falls inside the contour. This contour then describes a
credible region at a confidence level of $68.3\%$ (or
$1\sigma$).

Again, like what we did in the previous case of frequentist statistics, we should have a
good recipe for calculating and showing uncertainties about
the estimated value of one of the fitted parameters, say
$\theta_i$, or drawing the relevant contours in the 2-dimensional
plane of one parameter, say $\theta_i$, versus the other, say $\theta_j$.
This can be performed in a natural way if the corresponding 1D and
2D posterior probability density functions for those parameters
are available in some manner. Paying attention to the fact that the full
posterior $p(\Theta|D)$ is nothing but the joint PDF of all the
parameters (which are treated just as random variables in this
framework), the probability densities for a fewer number of
parameters are just `marginal' densities, defined as the
ones obtained by marginalising (\ie integrating over) the
unwanted (or hidden) parameters. That is, for a 1D posterior PDF of the parameter $\theta_i$, this procedure gives
a ``marginal posterior'' as
\beq \label{eq:margpdf} p(\theta_i|D)=\int
p(\Theta|D)d\theta_1...d\theta_{i-1}d\theta_{i+1}...d\theta_m.
\eeq
A 2D posterior is defined in an analogous manner.

Now that we have constructed these reduced PDFs, credible
intervals and regions can be defined in the same way as if we have
had only one or two fitted parameters from the beginning.

By looking at the mentioned characteristics of the two frameworks of Bayesian and frequentist statistics, we realise that, from a practical point of view, they are rather different in (1) the ways one defines estimators and associated uncertainties and (2) the recipes for discarding unwanted parameters in the statistical
inference (\ie marginalization of the posterior PDF
$p(\Theta|D)$ in the marginal posterior prescription, and maximisation of the likelihood function $\mathbb{L}(\Theta)$ in the profile likelihood prescription).  It is therefore quite clear that the results of the two inferences might not
coincide in general, even if the prior PDF $\pi(\Theta)$ is
taken to be constant, \ie when $p(\Theta|D)$ is proportional to
$\mathbb{L}(\Theta)$. This is especially true if the model likelihood has a
complex dependence on the parameters (\ie not just a simple
Gaussian form) and experimental data are not sufficiently available. Of course, in the case of the large sample limit, they
(must) give similar results (since both the statistical measures
become almost Gaussian in this case); this is why both methods are
commonly used in data analysis.

Finally, it is worth stating some aspects of perhaps the cornerstone of Bayesian statistics, \ie including prior functions in the final inference. One practically interesting
consequence of this is that it gives a powerful way of estimating
how robust a fit is.  That is, strong dependence of the posterior on different priors actually means that the data are not
sufficient or accurate enough to constrain the model parameters.
In other words, the posterior in this case is dominated by the
prior rather than the likelihood function. It can be shown also
that if a fit is robust in this language, the Bayesian and
frequentist methods should identify similar regions of
the parameter space at any particular confidence level. This robustness issues are for example
investigated in~\cite{Trotta:2008bp} for the CMSSM with
different choices of priors, where it is shown that the currently
available data are not yet sufficiently constraining to determine
the best-fit values of the parameters independent of the
priors. Therefore, one should inevitably decide which approach to use in the statistical analysis of the model.

In presenting and interpreting our direct and indirect dark matter detection results in \Scott~and \AkramiDD, we have employed both Bayesian and frequentist approaches and constructed the corresponding confidence and credible regions. In \AkramiGA~we have been particularly interested in frequentist statistics and the ability of existing scanning techniques in correctly mapping profile likelihoods in SUSY global fits. Finally, in our \AkramiCOV, we have focused on the issue of statistical coverage in SUSY parameter estimation which directly affects frequentist inference; we have however analysed the coverage for both profile likelihoods and marginal posteriors.

\section{Scanning algorithms}
\label{sec:algorithms}

Even if one decides on a particular statistical framework to work in, and also chooses the appropriate statistical measures that match the framework, there are other related issues that need to be addressed in any statistical analysis of a theoretical model; SUSY models are no exception.

One of these issues is the limitations of the available computational methods in correctly sampling a large and complex parameter space. It is for example crucial within the frequentist framework to map the profile likelihoods appropriately so as to be able to construct correct confidence regions and intervals. In addition, having a good estimate of the globally maximum value for the likelihood function is of extreme importance in this framework, because the results of the statistical inference strongly depend on that value. In a large, complex and poorly-known parameter space however, usual scanning methods such as grid or purely random scans can easily fail or otherwise take an extremely long time to generate useful results. This has therefore led many phenomenologists to employ more sophisticated scanning algorithms in their analyses.

Most of the existing advanced scanning techniques that are also widely used in SUSY phenomenology, are based on `Markov Chain Monte Carlo' (MCMC) algorithms (see \eg ref.~\cite{deAustri:2006pe}) or their different variations such as~\MN~\cite{Feroz:2007kg,Feroz:2008xx}, which is an algorithm based on nested sampling~\cite{SkillingNS1,SkillingNS2} (We do not describe these methods here and instead refer the interested readers to the given references for introductions and detailed descriptions). The structures of these algorithms are such that they exceptionally match the requirements for an appropriate scanning technique that is optimised for Bayesian statistics. These techniques have however been also used in frequentist inference, for example to map profile likelihoods and corresponding confidence regions. Even though there are currently various ongoing efforts in improving such techniques for these purposes, there are still some severe issues that have not been properly answered by them yet.

In \AkramiGA~and \AkramiCOV, we have studied some of these issues with the statistical analysis of SUSY models (in the context of the CMSSM) that we think stem from the imperfection of the utilised statistical scanning algorithms in particular cases. In \AkramiGA, we in addition propose an entirely different type of scanning techniques for SUSY parameter estimation, based on Generic Algorithms~\cite{Holland92} (GAs; for a classic introduction, see \eg ref.~\cite{Goldberg:1989}; for recent introductions, see \eg refs.~\cite{GArecrev1,GArecrev2,GArecrev3}; for a modern treatment, see \eg ref.~\cite{Goldberg:2010}), that can help in a complementary way the state-of-the-art and powerful Bayesian methods such as~\MN. The specific version of GAs that we use in our analysis is described in detail in \AkramiGA~and we therefore do not detail it here.

\end{fmffile}

\part{Summary and outlook}
\label{summaryoutlook}

\setlength{\unitlength}{1mm}
\begin{fmffile}{feyn}

\chapter{Summary of results}
\label{sec:summary}

We have so far given an introduction to the field of supersymmetry phenomenology with an emphasis on different strategies in comparing various viable supersymmetric models with observational data. This in particular includes advanced statistical techniques and innovative numerical algorithms in analysing the complex parameter spaces of SUSY models. In this chapter, we give a summary review of our contributions to the field that have been presented in the papers included in the thesis (see Part~\ref{papers}). We also discuss some future prospects for constraining supersymmetric models and parameters.

\section{Major achievements}

\Scott~\cite{Scott:2009jn} and \AkramiDD~\cite{Akrami:2010dn} deal with the observational constraints that direct and indirect searches for particle dark matter can place upon the parameter spaces of minimal SUSY extensions of the SM. We restrict our investigations to the CMSSM.

In \Scott, we analyse the CMSSM parameter space when \emph{Fermi} gamma-ray observations of dwarf galaxy Segue 1 are considered. We perform this in a global-fit framework where likelihoods from \emph{Fermi} data are combined with those from the cosmological measurements of dark matter relic density (section~\ref{sec:relic}), electroweak precision observables including the anomalous magnetic moment of the muon (section~\ref{sec:EWPOs}), collider bounds on the masses of SUSY and Higgs particles (sections~\ref{sec:expboundsHiggs} and~\ref{sec:expboundsSUSY}), and $B$-physics observables (section~\ref{sec:BPOs}). Our objective is to know which regions in the CMSSM parameter space are favoured by \emph{Fermi} data alone or in the presence of other existing constraints. Our employed \emph{Fermi} likelihoods are constructed from both the observed energy spectrum of gamma rays from Segue 1 and the spatial distribution of the gamma-ray photons at and around Segue 1. We also convolve the theoretical spectrum with the instrumental energy dispersion and point spread function. We use the \emph{Fermi} data from $9$ months observations of Segue 1 and also extrapolate our analysis to $5$ years of observations assuming that no excess events will have been observed after this time (similar to the actual $9$ months data).

We scan over the CMSSM parameter space using \MN~as implemented in \textsf{SuperBayeS}~\cite{Roszkowski:2007id,Roszkowski:2007fd,Roszkowski:2006mi,Trotta:2006ew,deAustri:2006pe,Trotta:2008bp,Trotta:2009gr} (available from ref.~\cite{SuperBayeS:web}). \textsf{SuperBayeS} is a numerical global-fit package for exploring the parameter space of the CMSSM and finding both Bayesian credible and frequentist confidence regions when the model's predictions are compared with different types of experimental data. It reconstructs the marginalised posterior PDFs, as well as the profile likelihoods although the employed scanning techniques (\ie MCMCs and~\MN) are both optimised for Bayesian statistics. For physics calculations, it uses different packages: \textsf{SOFTSUSY}~\cite{softsusy} (available from ref.~\cite{softsusy:web}), \textsf{DarkSUSY}~\cite{Gondolo:2004sc} (available from ref.~\cite{DarkSUSY:web}), \textsf{FeynHiggs}~\cite{feynhiggs:99,feynhiggs:00,feynhiggs:03,feynhiggs:06} (available from ref.~\cite{feynhiggs:web}), \textsf{Bdecay} and \textsf{micrOMEGAs}~\cite{Belanger:2004yn} (available from ref.~\cite{micrOMEGAs:web}).

\begin{figure}
\begin{center}
\subfigure{\includegraphics[scale=0.25, trim = 40 172 70 220, clip=true]{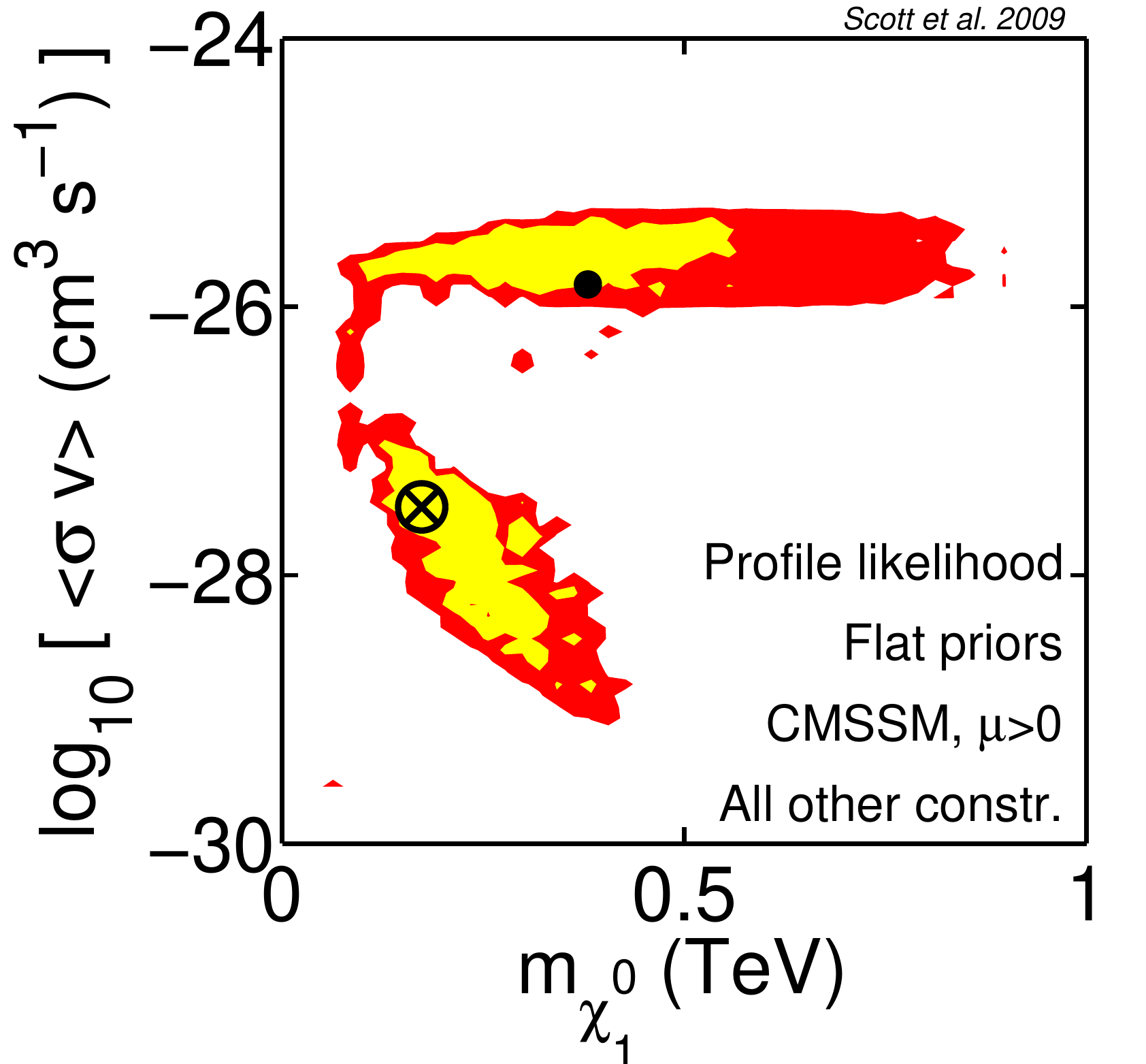}}
\subfigure{\includegraphics[scale=0.25, trim = 40 172 35 220, clip=true]{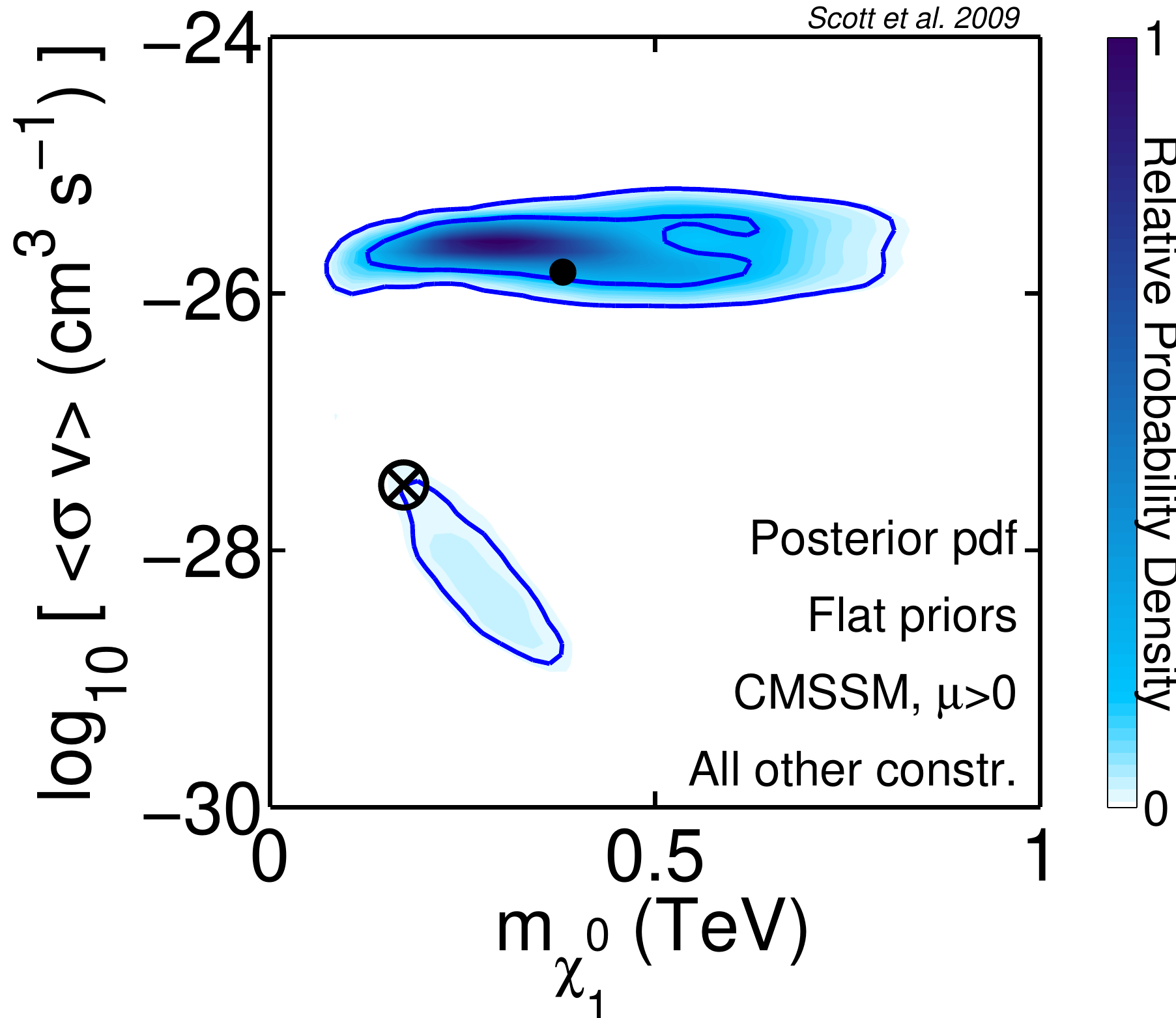}}\\
\subfigure{\includegraphics[scale=0.25, trim = 40 172 70 220, clip=true]{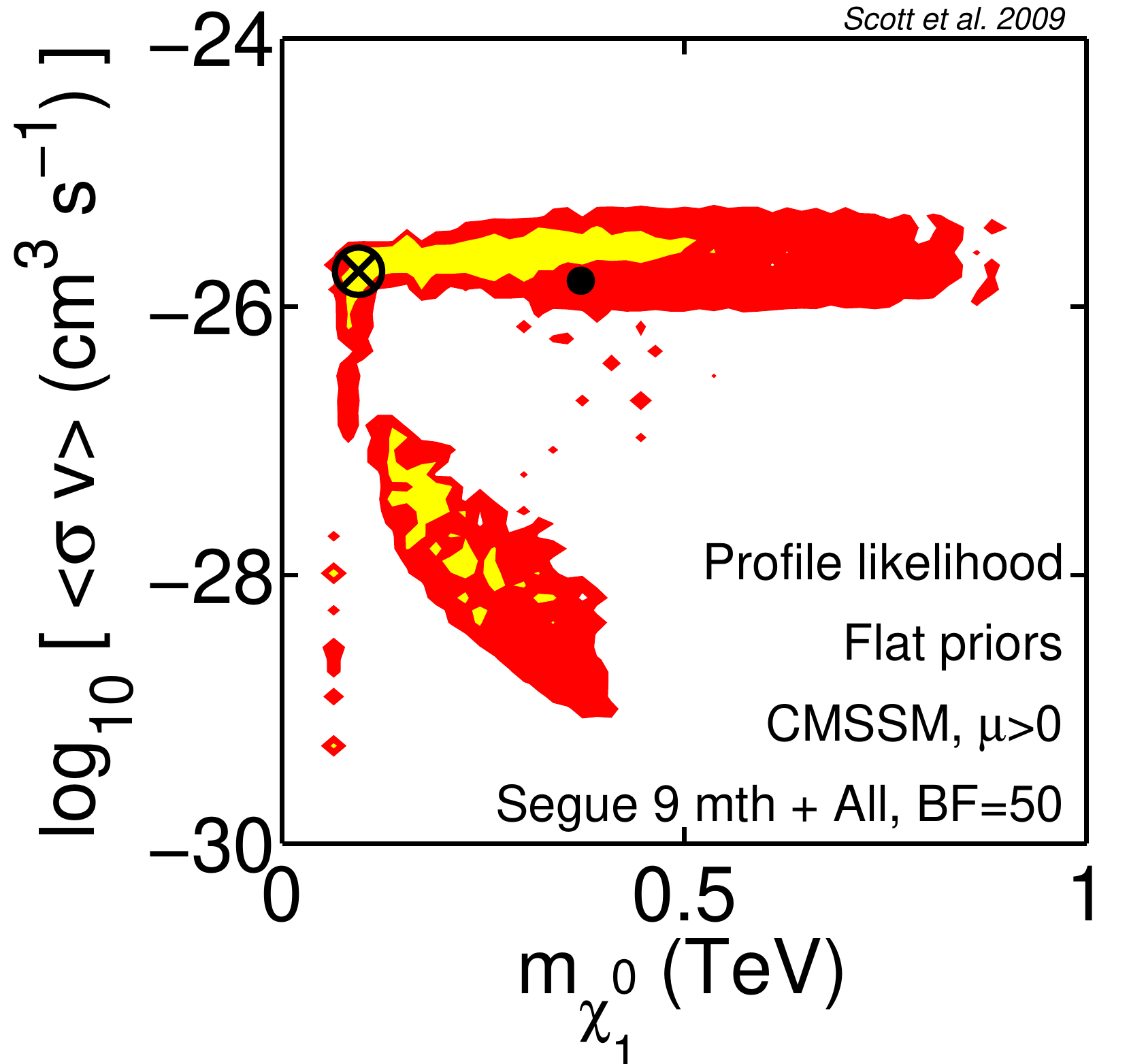}}
\subfigure{\includegraphics[scale=0.25, trim = 40 172 35 220, clip=true]{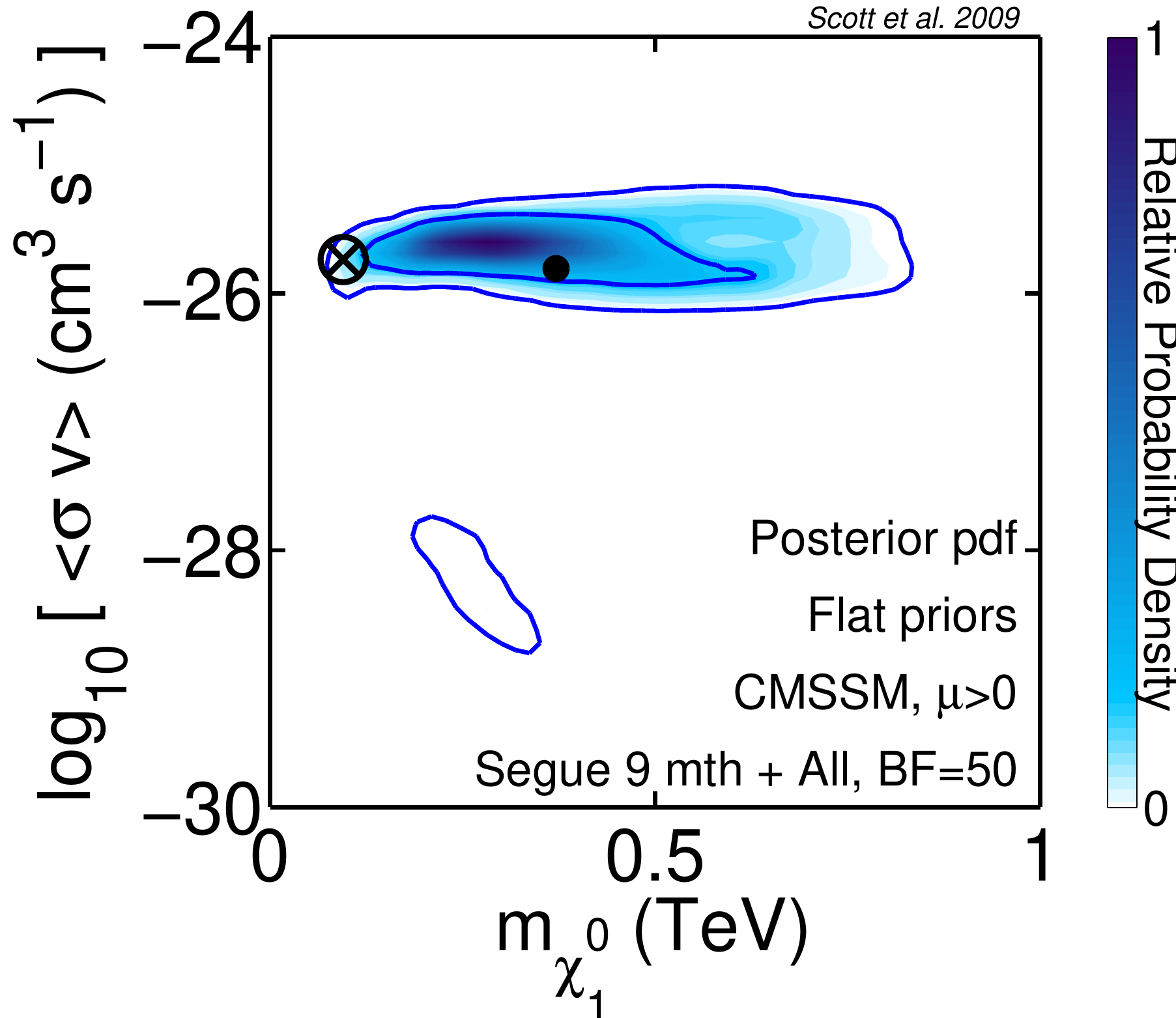}}\\
\subfigure{\includegraphics[scale=0.25, trim = 40 172 70 220, clip=true]{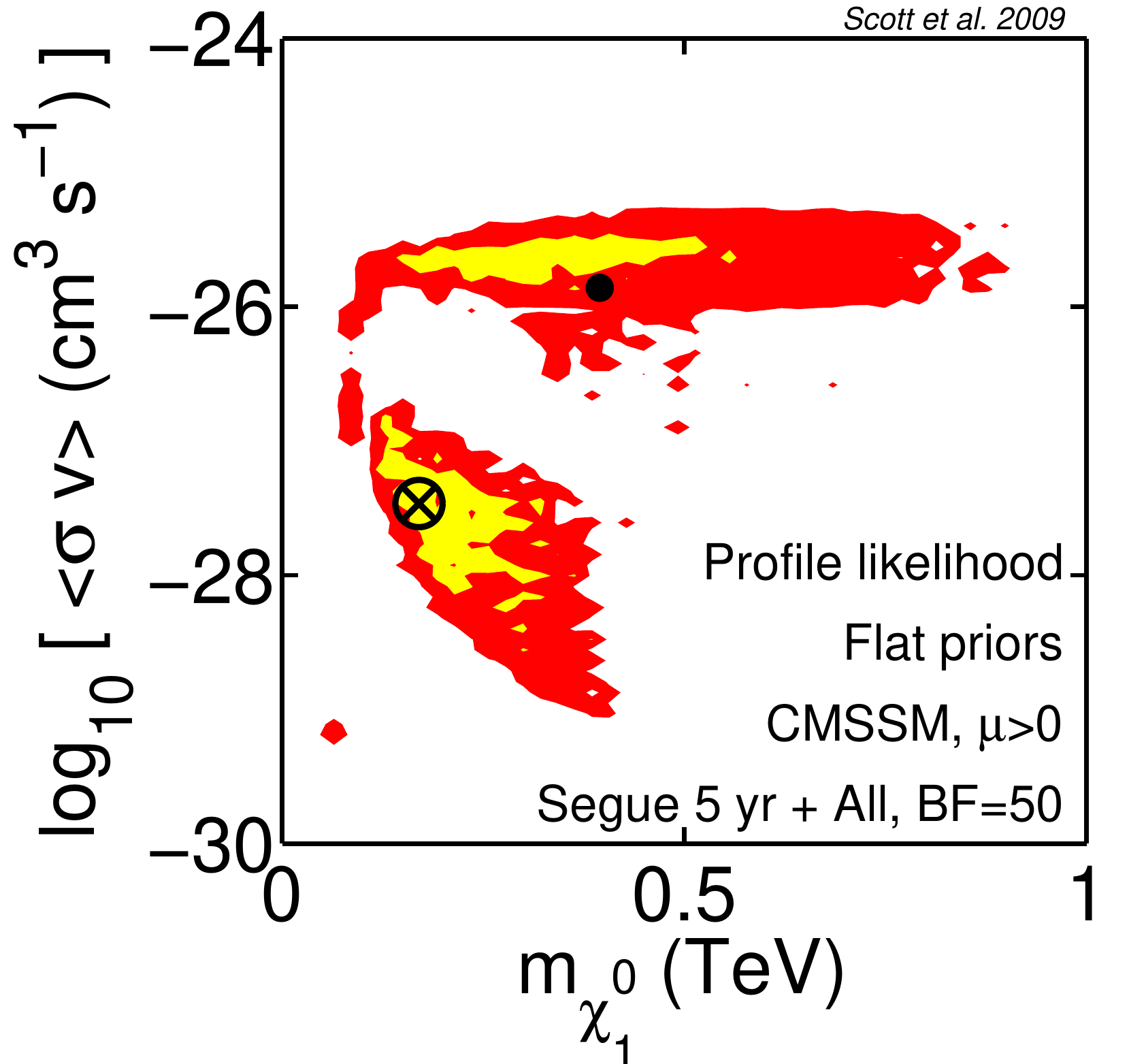}}
\subfigure{\includegraphics[scale=0.25, trim = 40 172 35 220, clip=true]{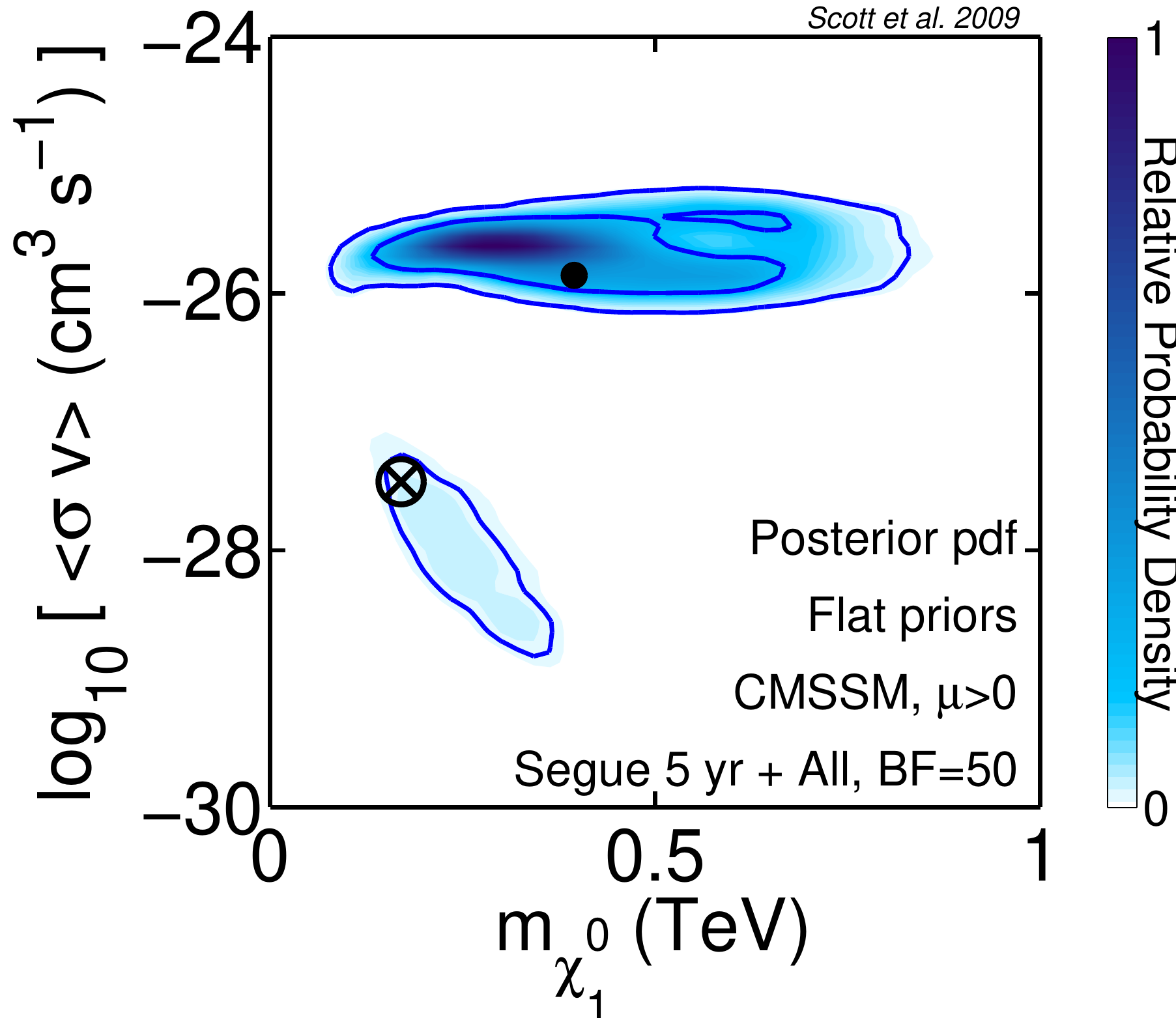}}\\
\end{center}
\caption[aa]{\footnotesize{Favoured values of the neutralino annihilation cross-section and mass for the CMSSM, when: (1) only existing constraints from the dark matter relic density, electroweak precision observables, collider bounds on the Higgs and sparticle masses, and $B$-physics observables are included (upper panels), (2) $9$ months of \emph{Fermi} observations of the dwarf galaxy Segue 1 are added (central panels), and (3) no excess events are observed from Segue 1 after $5$ years of \emph{Fermi} observations (lower panels). The left panels show frequentist profile likelihoods (where yellow and red indicate $68\%$ and $95\%$ confidence regions, respectively) and the right panels give Bayesian marginalised posterior PDFs (where solid blue contours depict $68\%$ and $95\%$ credible regions). Best-fit points and posterior means are indicated by black crosses and dots, respectively. Adapted from \Scott.}}\label{fig:FermiSegue1}
\end{figure}

The number of total CMSSM points in our final set of samples is $\sim 5.5\times 10^4$. Our results show that some points in the CMSSM parameter space are ruled out when only \emph{Fermi} data are used. These are models that give very large annihilation cross-sections ($\langle \sigma v \rangle \gtrsim 10^{-25} \mbox{cm}^3 \mbox{s}^{-1}$ in the best case where $5$ years data are used and a large boost factor at the source is assumed) and low neutralino masses. These models are excluded because no gamma-ray excess signal is detected from Segue 1. When we combine \emph{Fermi} observations with other constraints, in particular the observed value of dark matter relic density, we see that most of the CMSSM points excluded by \emph{Fermi} are already disfavoured by those other constraints (see \fig{fig:FermiSegue1}).

The analysis of \Scott~presents the first direct inclusion of constraints from indirect dark matter searches in global fits of supersymmetric models to experimental data. It is based on the full likelihood construction for the experiment, \ie the \emph{Fermi} data points are directly used in the statistical analysis instead of the usual use of exclusion limits provided by the experiment. This makes the paper a significant contribution to the field and gives a concrete example of what should be done for all types of experimental data in order to provide statistically consistent constraints on SUSY and other beyond-the-SM models. In addition, the paper is of particular interest because it was the first work in which \emph{Fermi} gamma-ray data have been used in searches for dark matter signals.

In \AkramiDD, we perform a similar analysis where data from direct dark matter searches are used in constraining SUSY parameters. Here, instead of working with real data, we are mainly interested in constraints provided by the future generation of direct detection experiments, namely the ones with ton-scale target materials. We want to know how well those experiments can detect a supersymmetric WIMP with particular masses and cross-sections and in each case, what constraints they put upon the fundamental properties of the underlying theory.

We investigate these by choosing a few benchmarks in the CMSSM parameter space with masses and cross-sections that cover an interesting range of values. We generate some synthetic data (\ie the number of events detected by an experiment and the corresponding recoil energies) and use those data to scan the CMSSM parameter space and obtain the favoured regions.

The experiments considered in this work are ton-scale extrapolations of three existing experiments: CDMS (Cryogenic Dark Matter Search) \cite{CDMS:web}, XENON~\cite{XENON:web} and COUPP (Chicagoland Observatory for Underground Particle Physics)~\cite{COUPP:web}. These experiments are expected to cooperatively explore large fractions of WIMP masses and scattering cross-sections, both spin-independent (SI) and spin-dependent (SD). The two experiments CDMS and XENON have the capability of measuring the recoil energies produced in the detector by WIMPs that interact with target nuclei. These experiments are however not very sensitive to SD scattering. COUPP on the other hand, although only measures the number of events above a threshold nuclear recoil energy and does not measure event energies associated with the event, has a significant sensitivity to SD scattering. COUPP is therefore expected to break some degeneracy in the CMSSM parameter space by providing both SI and SD measurements. Obviously this ability increases when the number of observed (SD) events becomes larger. For each experiment, we assume $1000$ kg-years of raw exposure and energy resolution, energy range and efficiency similar to the present day version of the experiment. We also assume backgrounds at target levels. In a similar way as in \Scott, our analysis is based on reconstructing full likelihoods for the experiments, namely, we use the experimental data points (\ie the number of events and recoil energies) directly in the likelihood constructions rather than only using the exclusion limits.

For each benchmark point, we scan over the CMSSM parameter space by means of \MN~again as implemented in \textsf{SuperBayeS}. We however do not combine the direct detection likelihoods with other constraints since we are only interested in what one can gain from direct detection data. Another important constituent of the paper is that we take into account existing uncertainties on hadronic matrix elements for neutralino-quark couplings (important in calculating theoretical cross-sections), as well as on halo model parameters (important in calculating differential recoil rates). We marginalise (maximise) over these nuisance parameters in order to find marginal posteriors (profile likelihoods).

\begin{figure}
\begin{center}
\subfigure{\includegraphics[scale=0.2, trim = 40 230 130 123, clip=true]{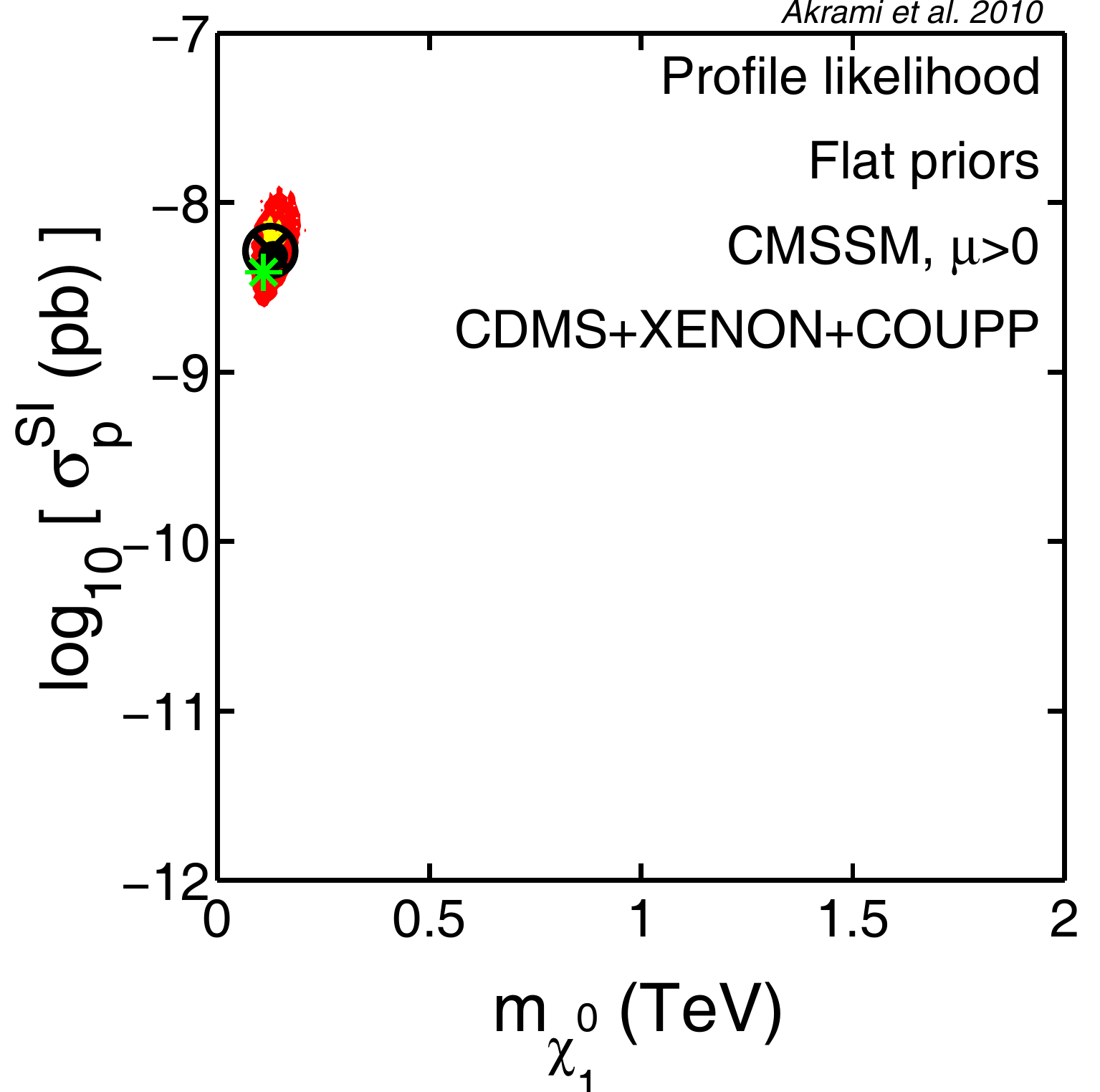}}
\subfigure{\includegraphics[scale=0.2, trim = 40 230 60 123, clip=true]{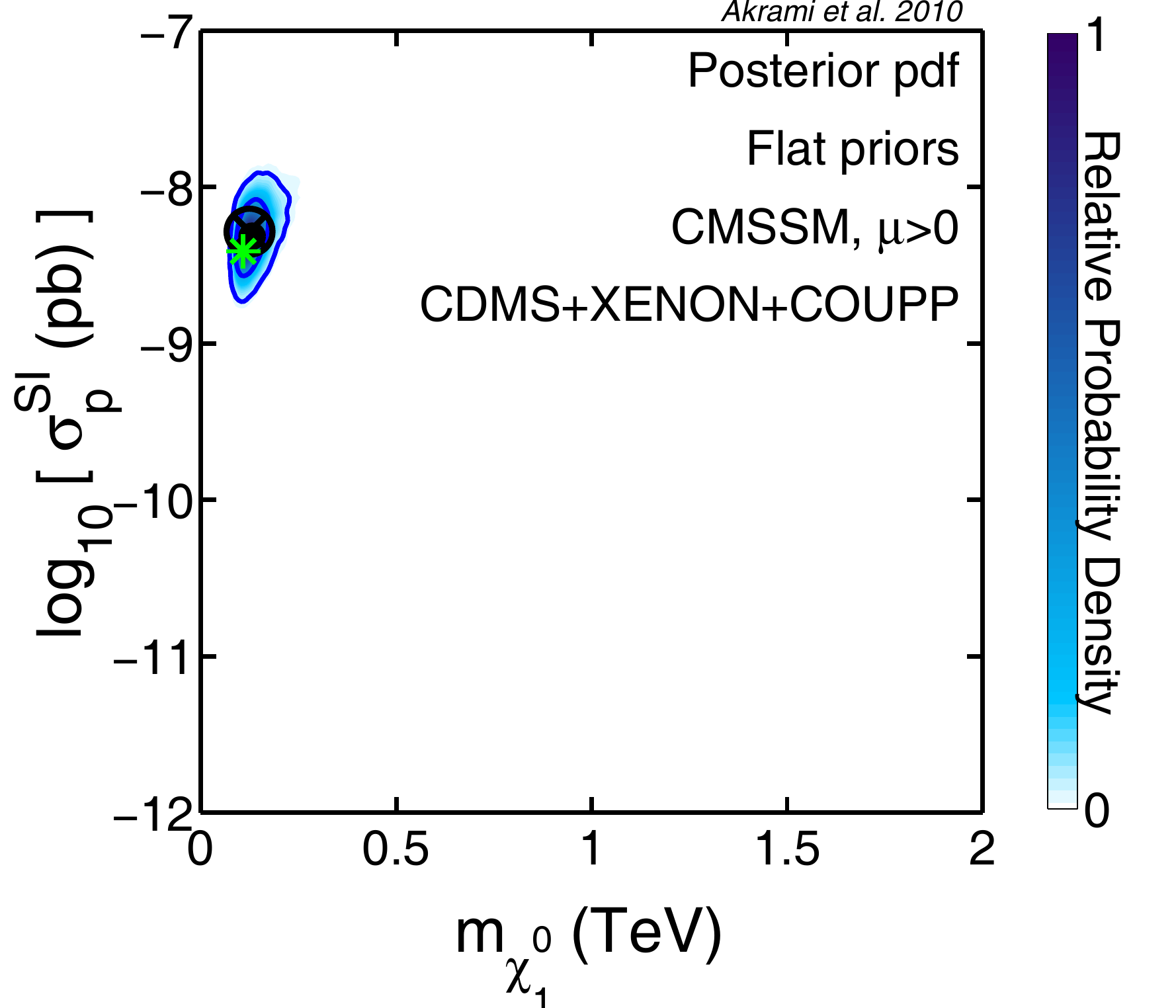}}\\
\subfigure{\includegraphics[scale=0.2, trim = 40 230 130 123, clip=true]{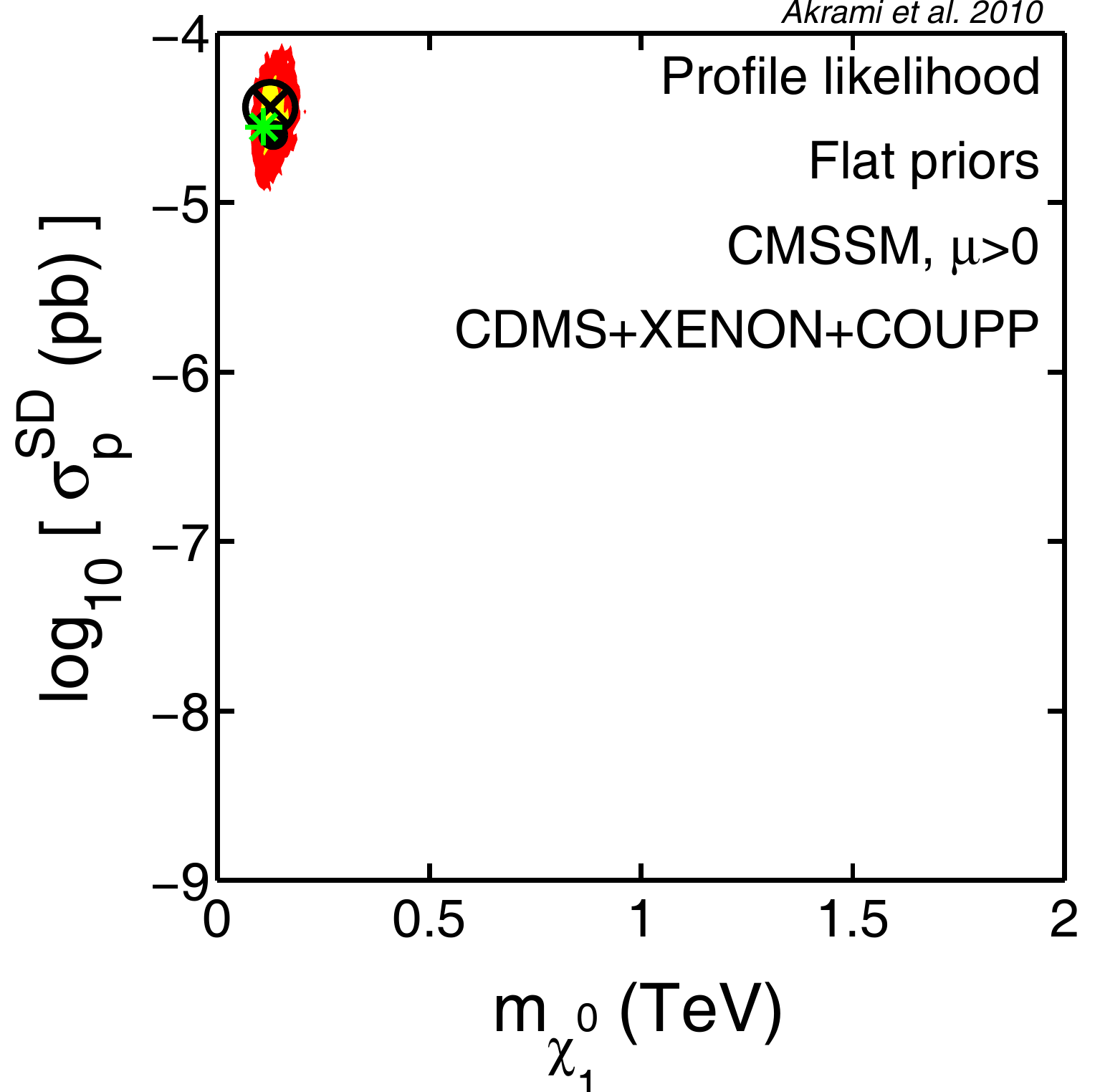}}
\subfigure{\includegraphics[scale=0.2, trim = 40 230 60 123, clip=true]{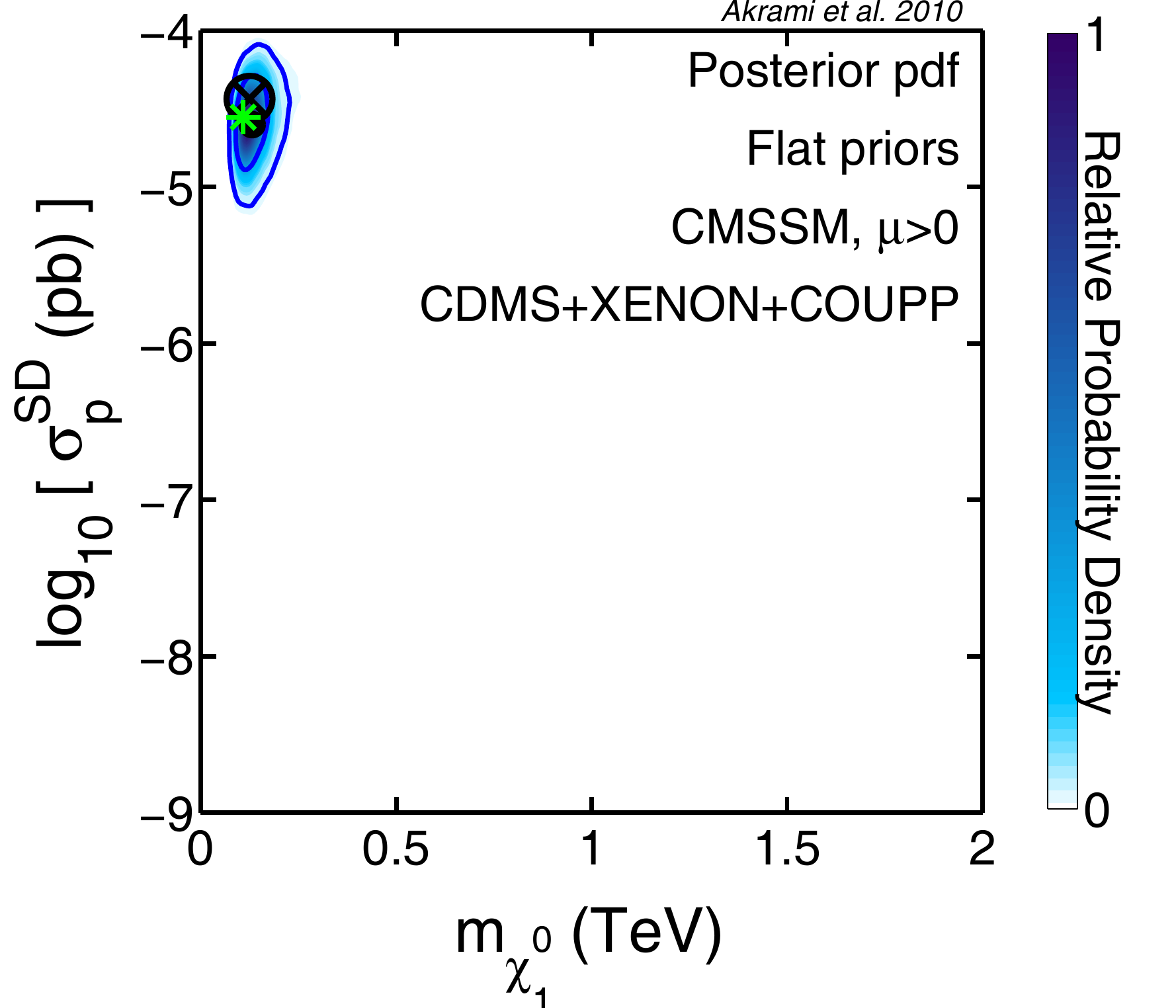}}\\
\subfigure{\includegraphics[scale=0.2, trim = 40 230 130 123, clip=true]{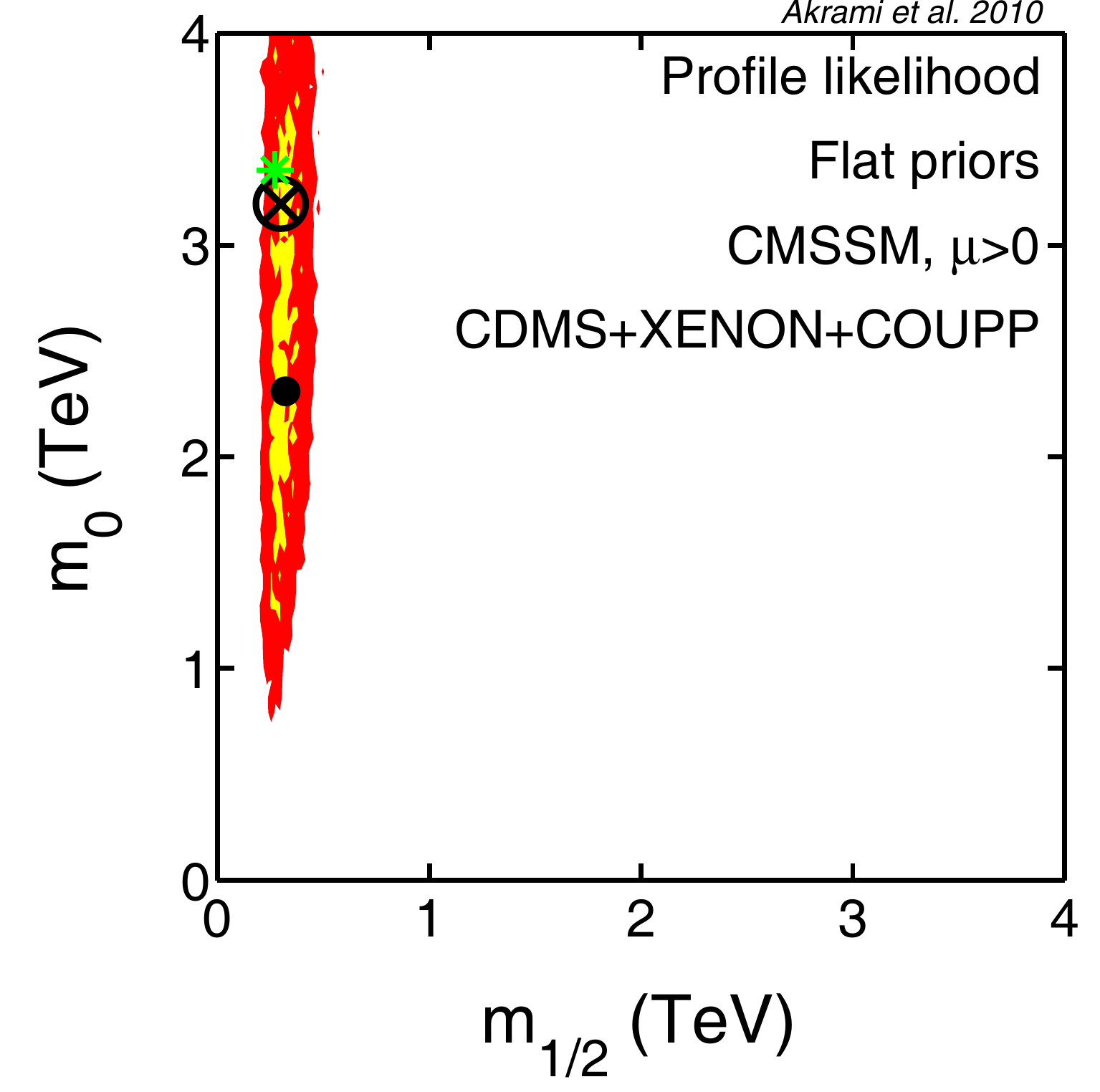}}
\subfigure{\includegraphics[scale=0.2, trim = 40 230 60 123, clip=true]{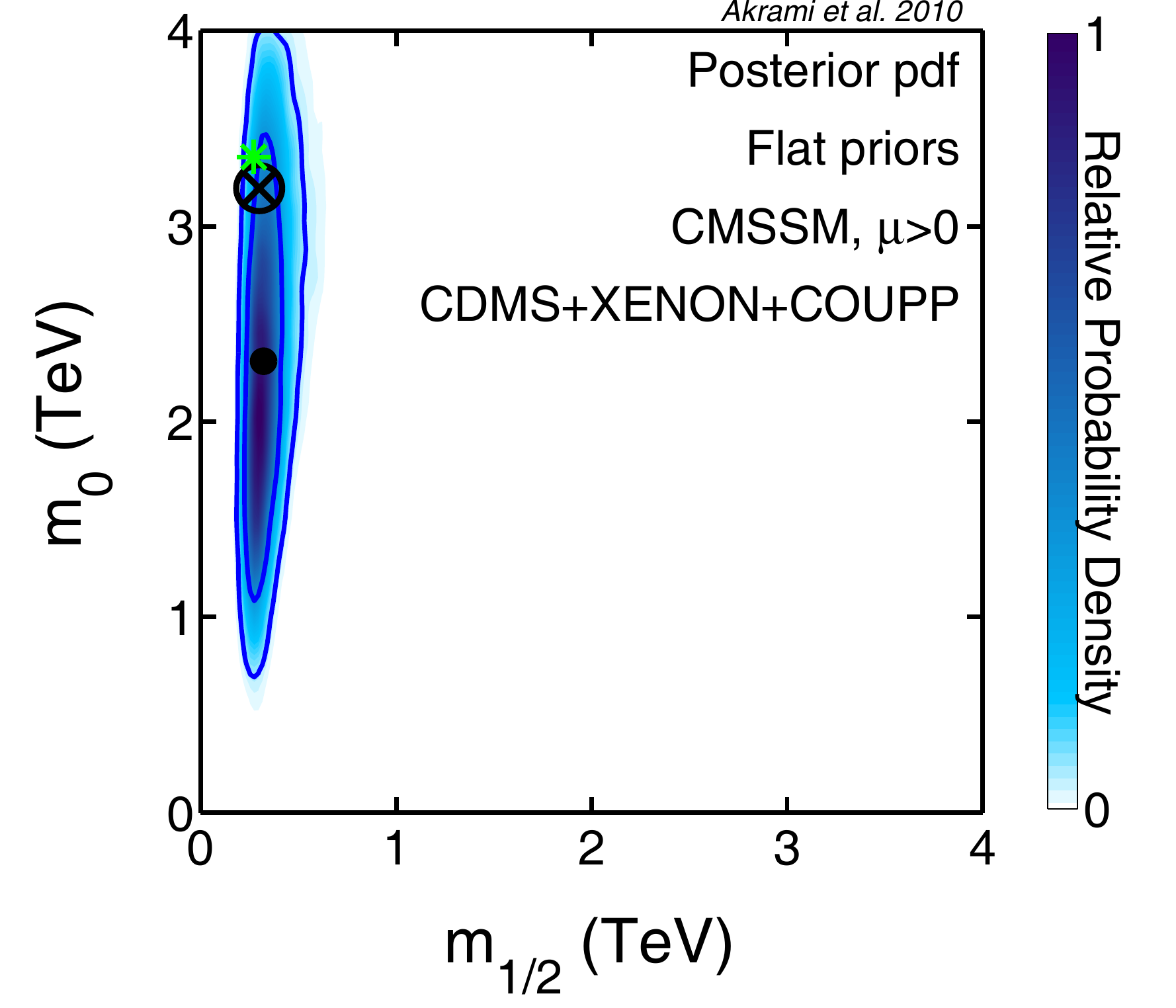}}\\
\subfigure{\includegraphics[scale=0.2, trim = 40 230 130 123, clip=true]{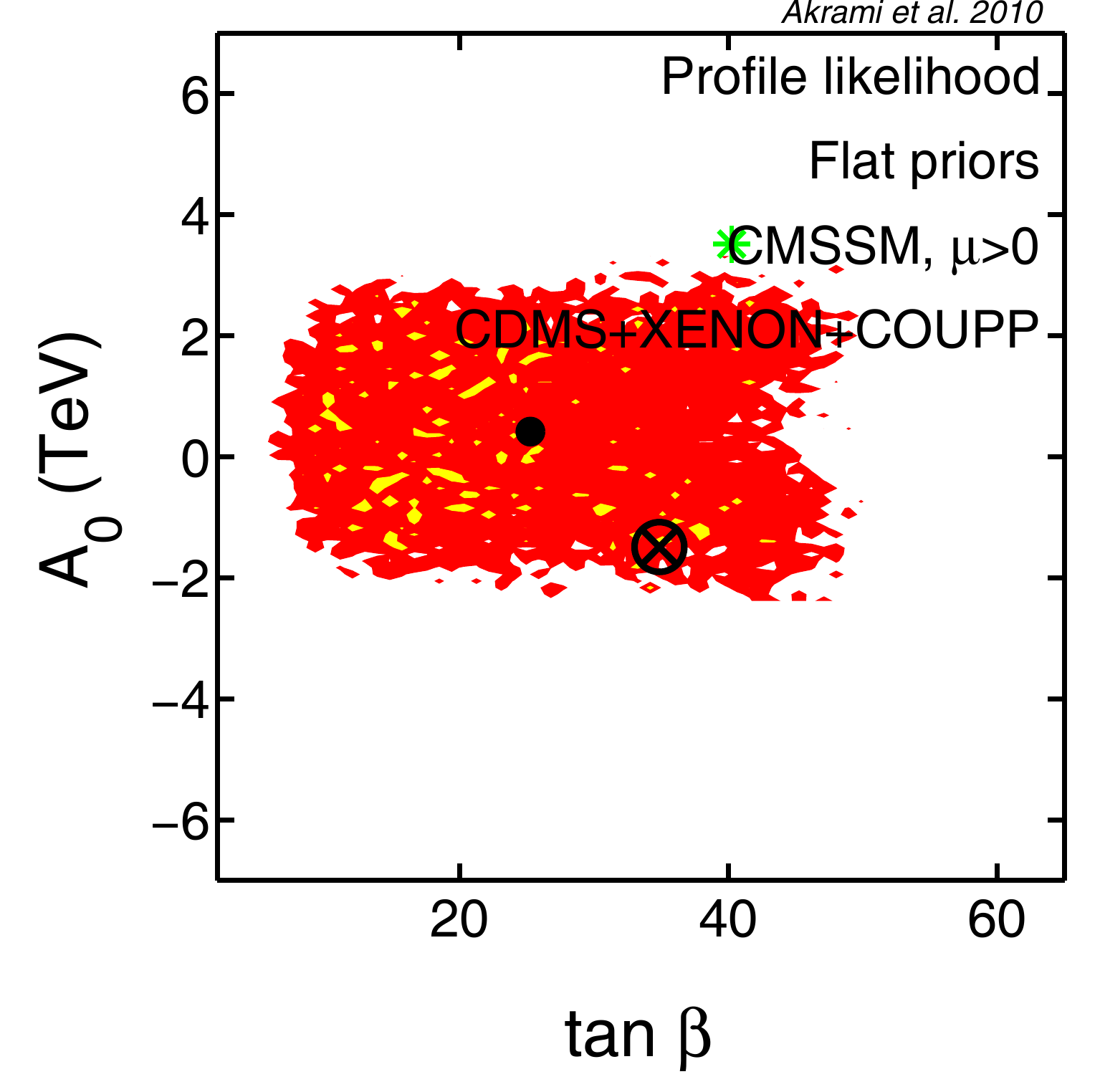}}
\subfigure{\includegraphics[scale=0.2, trim = 40 230 60 123, clip=true]{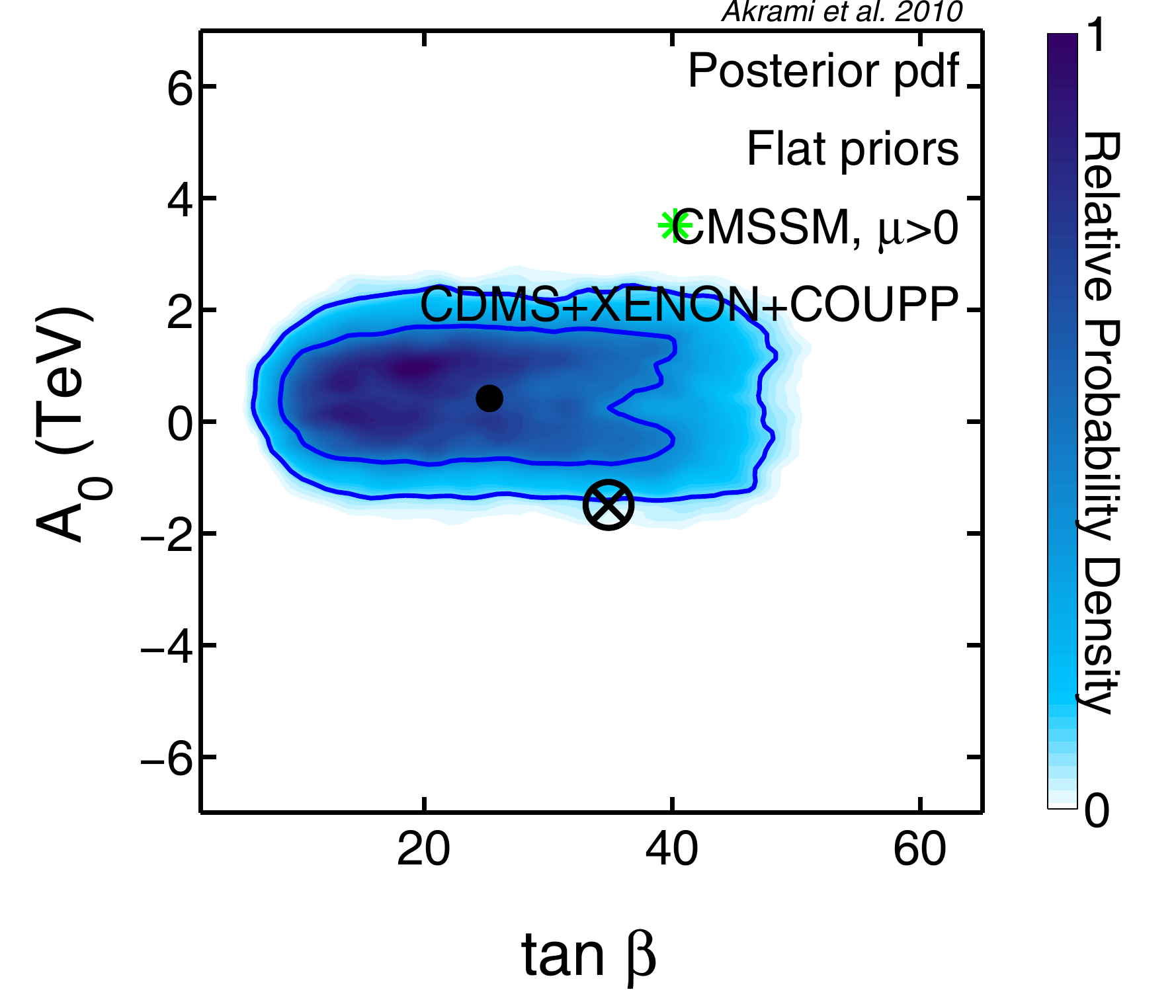}}\\
\end{center}
\caption[aa]{\footnotesize{Upper two rows: Reconstruction of the mass and scattering cross-sections of the neutralino for a CMSSM benchmark when likelihoods from ton-scale versions of three direct detection experiments CDMS, XENON and COUPP are combined and synthetic data are used in the scan. Lower two rows: Constraints on the CMSSM parameters from the combination of the experiments. Left panels: $2$-dimensional profile likelihoods (yellow and red indicate $68.3\%$ and $95.4\%$ confidence regions, respectively). Right panels: $2$-dimensional marginal posteriors (inner and outer contours represent $68.3\%$ and $95.4\%$ confidence levels, respectively). Black dots and crosses show the posterior means and best-fit points, respectively, and benchmark values are marked with green stars. Adapted from \AkramiDD.}}\label{fig:LH}
\end{figure}

We show that when the likelihoods from all three experiments are combined, the primary characteristics of the neutralino, namely its mass and cross-sections, can be determined with high certainty if the neutralino sits at low masses and high cross-sections (first and second rows in \fig{fig:LH}). In particular, it can be seen from our results that when COUPP is added, it can break degeneracies in the parameter space and therefore substantially help us pin down the actual WIMP properties. This can be achieved in cases where a substantial fraction of the signal event rate comes from SD interactions of WIMPs and nuclei. The results also indicate that the uncertainties on the halo parameters may have relatively large effects on the reconstruction of both Bayesian credible and frequentist confidence regions. The uncertainties on the hadronic matrix elements on the other hand do not have significant effects on the credible and confidence regions.

As far as the constraints on the CMSSM parameters are concerned, our analysis indicates that using direct detection data alone does not place strong constraints upon the values of those parameters. We can see this from the third and fourth rows in \fig{fig:LH} where $1\sigma$ and $2\sigma$ contours are given in the $m_0-m_{1/2}$ and $A_0-\tan\beta$ planes for the benchmark WIMP with low mass and high cross-sections: direct detection experiments only determine the gaugino mass parameter $m_{1/2}$ with relatively high precision. This is mainly due to the fact that this parameter has a strong correlation with the neutralino mass that can be measured almost accurately by direct detection experiments.

Perhaps the bottom line of \AkramiDD~is yet another indication that no single type of experiments is able to present conclusive information about the fundamental properties of a dark matter model (\eg supersymmetry in our case). There is always a high degree of degeneracy between different parameters of the model that cannot be broken by only one type of experimental data. In other words, in order to test and characterise \eg weak-scale SUSY as a valid extension of the SM that also provides ``the solution'' to the dark matter problem, a combination of information from complementary experiments such as direct, indirect and collider searches would be required. It is however important to note here that future generations of direct detection experiments are highly promising in this endeavour and constitute one of the main strategies for identifying the nature of dark matter as well as characterising various models beyond the SM that include a dark matter candidate.

In \AkramiGA~\cite{Akrami:2009hp} and \AkramiCOV~\cite{Akrami:2010cz}, we discuss the ability of existing scanning techniques in correctly exploring the parameter spaces of SUSY models in order to provide favoured regions when the models' predictions are compared with observational data. We are in particular interested in how well they find the confidence regions in a frequentist approach.

As we have discussed in section~\ref{sec:algorithms} of this thesis, the most powerful algorithms that are currently employed in SUSY parameter estimation are MCMCs and nested sampling. The structures of these techniques are such that both of them are optimised for Bayesian statistics, \ie they can accurately enough map the posterior PDFs and accordingly the marginal posteriors. In the absence of advanced scanning methods that are specifically designed for frequentist statistics, MCMCs and nested sampling are nowadays also exploited in the latter case, \ie to map profile likelihoods in supersymmetric analyses. It is known that (see section~\ref{sec:statframeworks}) profile likelihoods and marginal posteriors provide very similar results given that high-statistics data are available (\ie when the effects of priors are dominated by likelihoods). This also implicitly means that in these cases, the scanning algorithms that are optimised for one statistical framework can be appropriately used for the other. However, in many realistic cases (at least in SUSY phenomenology), the model at hand exhibits a very complex parameter space and experimental data are not strong enough. As a result, Bayesian credible and frequentist confidence regions are in general very different from each other. As we have remarked explicitly in \Scott, we think that in order ``to gain as complete a picture as possible of the preferred regions in an insufficiently-constrained parameter space like the CMSSM'', both statistical measures should be considered. That is, rather than arguing that one statistical approach is more `correct' than the other, we think that additional insight can be obtained if one considers both for analysing any given parameter space. This is however a useful strategy only if one can accurately map the confidence and credible regions. This is precisely where our points in \AkramiGA~and \AkramiCOV~stand: The methods that work properly for one approach do not necessarily work appropriately for the other.

In \AkramiGA, we propose a new scanning technique based on Genetic Algorithms (GAs) (see section~\ref{sec:algorithms}) that is optimised for frequentist profile likelihood analysis of complex supersymmetric models. We use existing cosmological and collider constraints on SUSY predictions (the same data as in \Scott~except for \emph{Fermi} observations of Segue 1) in a global-fit setup and analyse the CMSSM as a testbed. We implement the GA code \textsf{PIKAIA}~\cite{pikaia} (available from ref.~\cite{PIKAIA:web}) in \textsf{SuperBayeS} and compare our results with those of \MN~(with the standard configuration as is employed \eg in ref.~\cite{Trotta:2008bp}) when the same set of experimental data is used.

\begin{figure}
\begin{center}
\subfigure{\includegraphics[width=0.4\linewidth, trim = 70 0 70 50, clip=true]{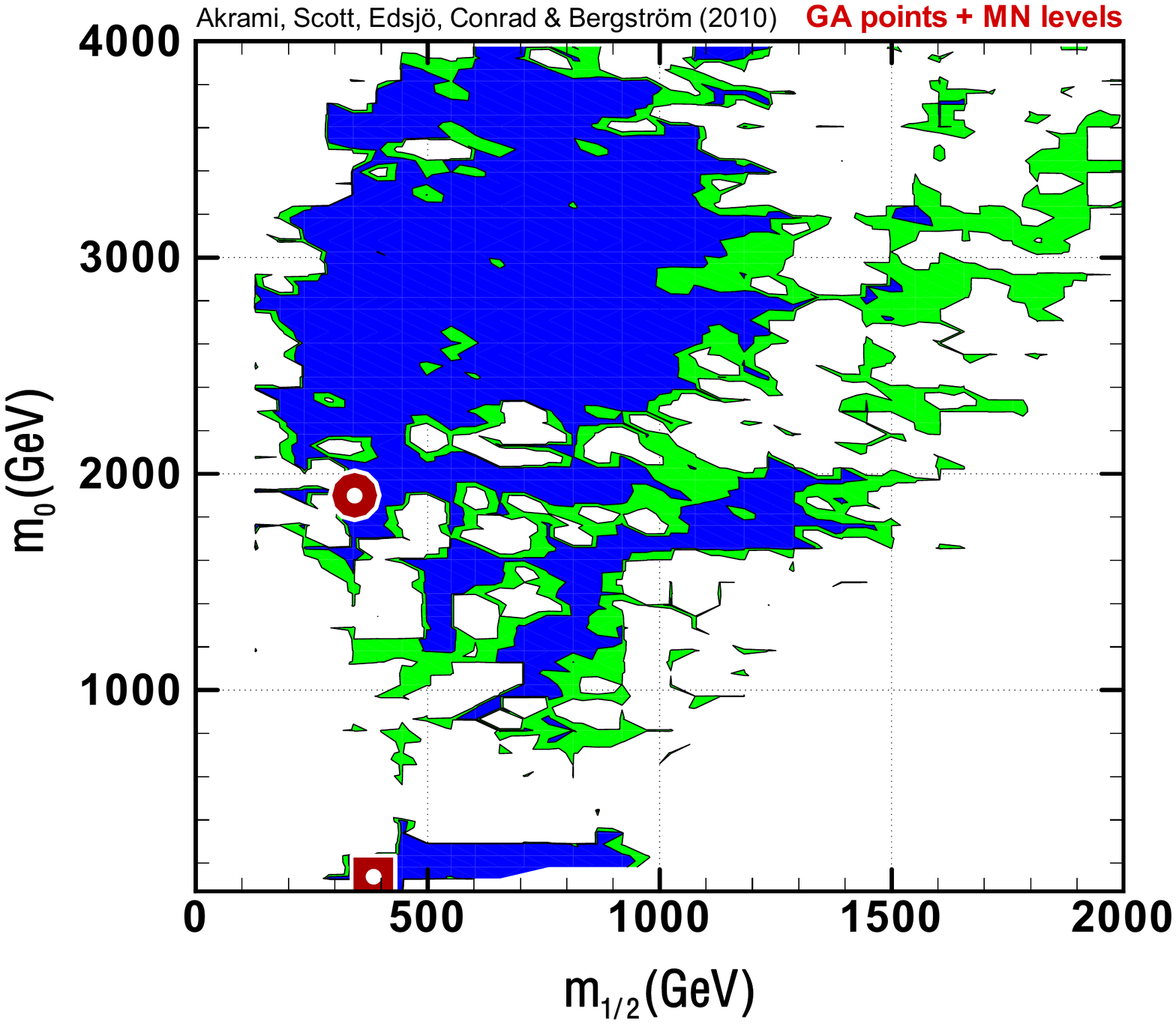}}
\subfigure{\includegraphics[width=0.4\linewidth, trim = 70 0 70 50, clip=true]{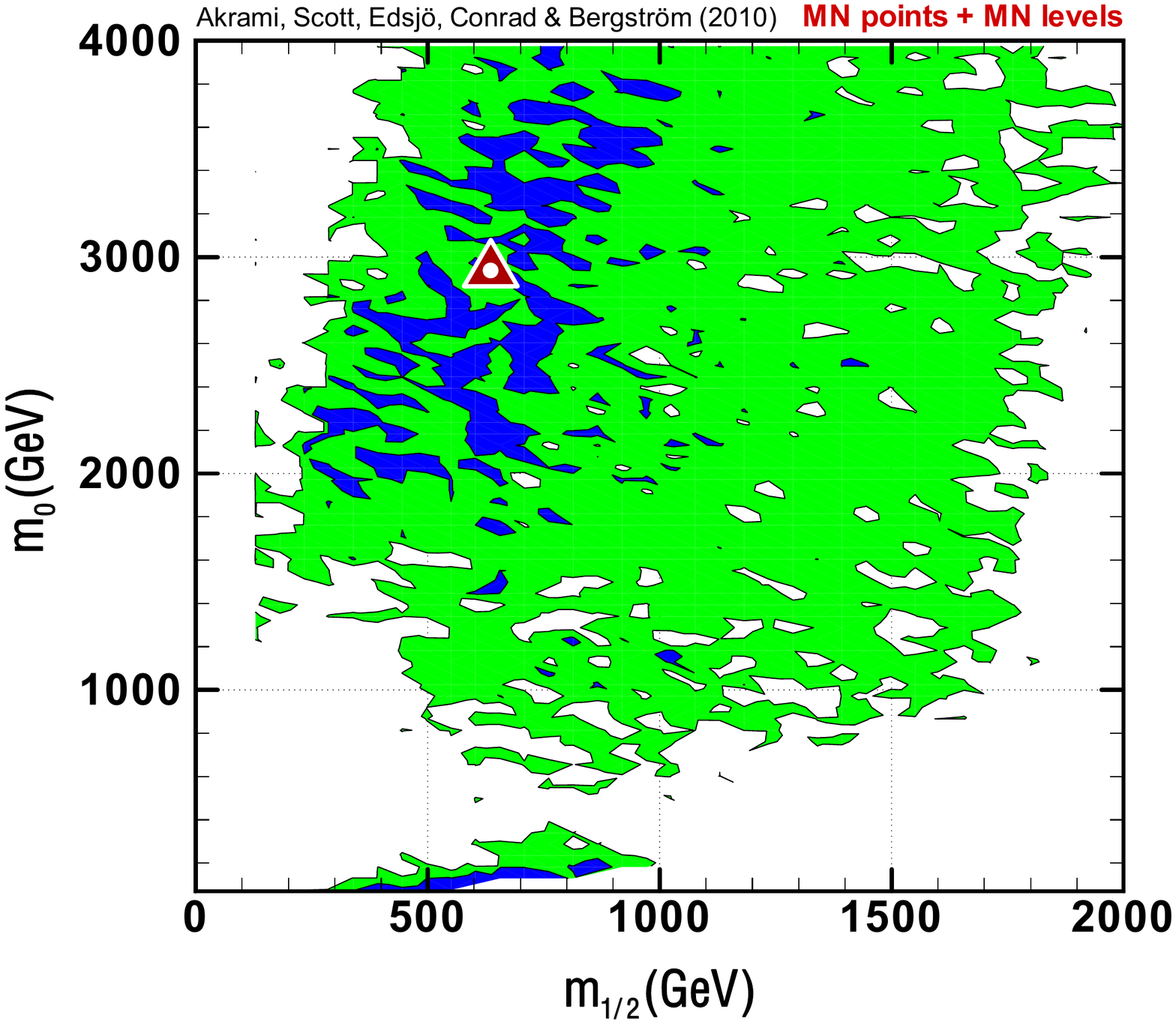}} \\
\subfigure{\includegraphics[width=0.4\linewidth, trim = 70 0 70 50, clip=true]{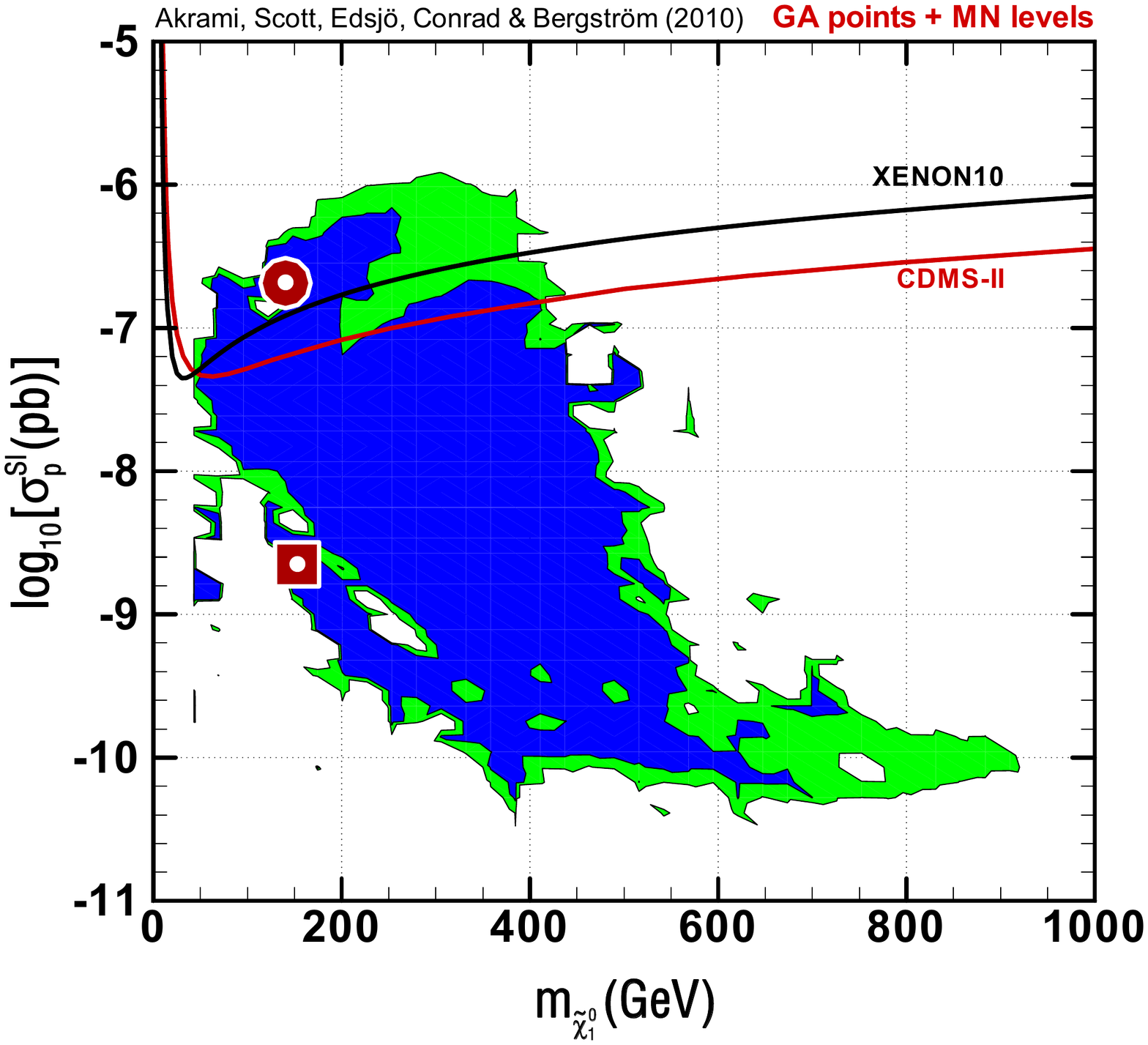}}
\subfigure{\includegraphics[width=0.4\linewidth, trim = 70 0 70 50, clip=true]{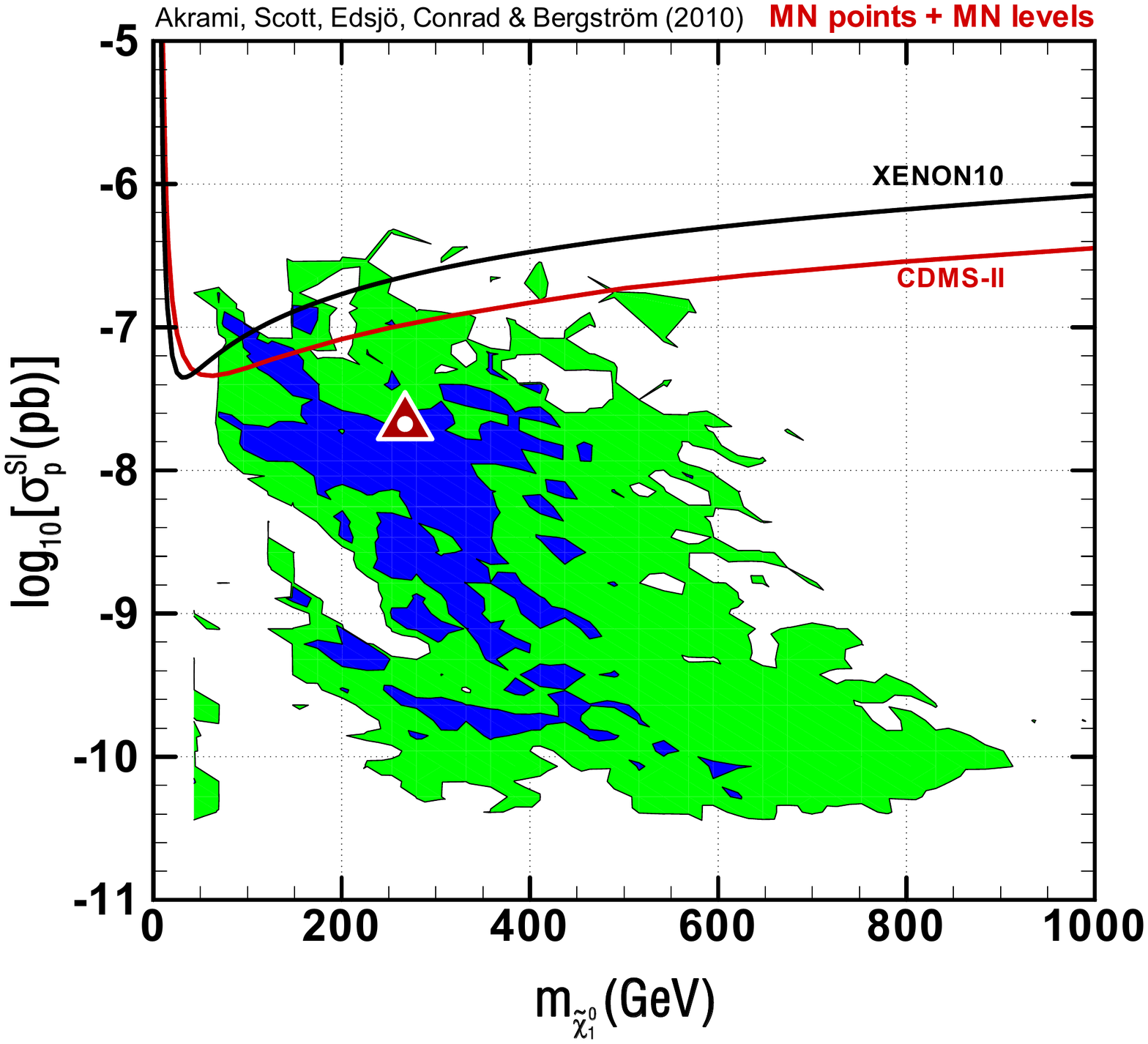}} \\
\subfigure{\includegraphics[width=0.4\linewidth, trim = 70 0 70 50, clip=true]{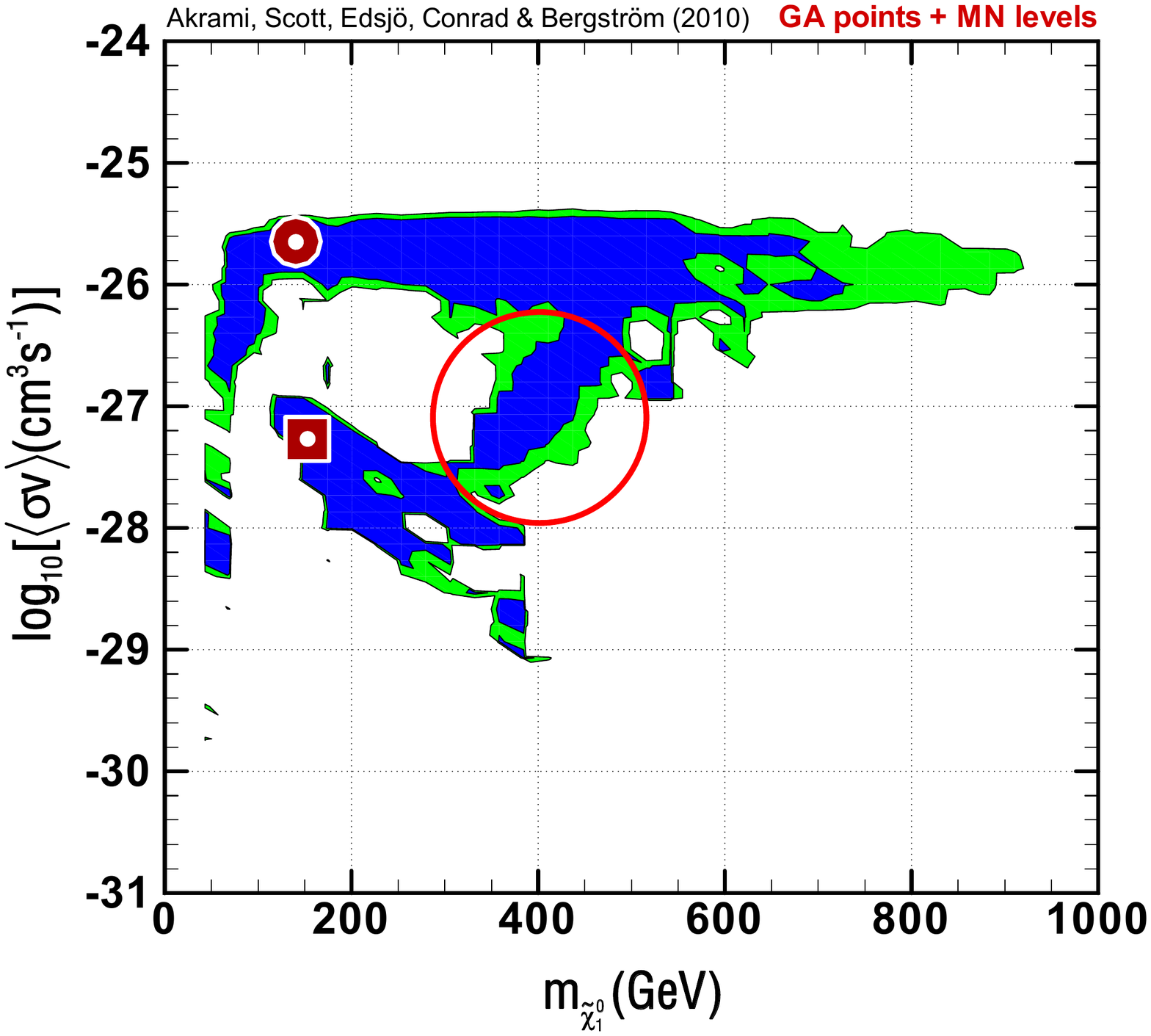}}
\subfigure{\includegraphics[width=0.4\linewidth, trim = 70 0 70 50, clip=true]{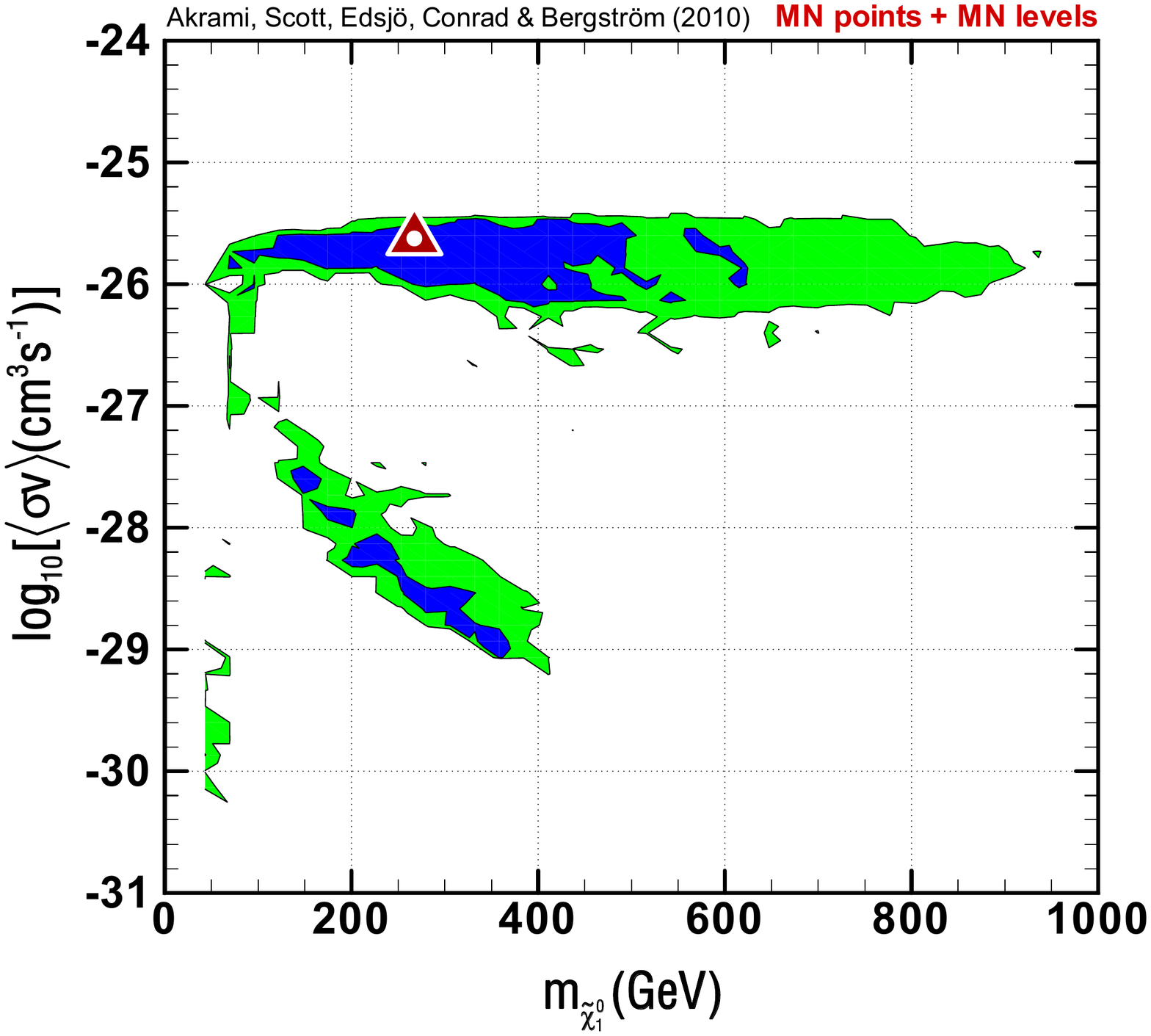}} \\
\caption[aa]{\footnotesize{Best-fit points in the CMSSM parameter space (upper panels), as well as in the $\sigma^{SI}_p-m_{\tilde\chi^0_1}$ (central panels) and $\left\langle \sigma v\right\rangle-m_{\tilde\chi^0_1}$ (lower panels) planes found by Genetic Algorithms (left panels) and \MN~(right panels). The blue and green regions represent iso-likelihood contours that correspond to $1\sigma$ and $2\sigma$ confidence regions in the \MN~scan. The dotted circles and triangles indicate the global best-fit points in the GA and \MN~scans, respectively, and the dotted square shows the best-fit co-annihilation point in the GA scan. The GAs find more high-likelihood points, while \MN~better maps the likelihood around the moderately good fits that it finds. The big circle in the lower left panel indicates a section of the stau co-annihilation region found by GAs and missed in previous scans. Adapted from \AkramiGA.}} \label{fig:GAs}
\end{center}
\end{figure}

Our results are quite surprising: the best-fit point found by GAs has a substantially higher likelihood value compared to the one found by \MN. In addition, many new CMSSM points with high likelihood values show up in our scans. These together dramatically impact the inferred confidence regions for the CMSSM parameters, as well as some other derived quantities such as the neutralino mass and cross-sections (see \fig{fig:GAs}). Our investigation therefore indicates that the conventional Bayesian scanning techniques, including nested sampling (which is arguably the best existing one), may give unsatisfactory results in a frequentist context, except they become appropriately modified or reconfigured. See \eg ref.~\cite{Feroz:2011bj} where a more appropriate \MN~configuration for profile likelihoods is presented and its results are in excellent agreement with the results of \AkramiGA. This new configuration however requires a significantly larger computational effort. We make the point here that even though our application of GAs for the profile likelihood analysis of the CMSSM appears to be quite successful, it is not flawless. The algorithm is primarily designed to find the global maximum of a complex function and is not optimised to accurately `map' the function around the global maximum. This is why the confidence regions found in our results (see \fig{fig:GAs}) look somewhat noisy. Perhaps this displeasing property of the algorithm can be remedied by combining it with other scanning algorithms that are designed to map a function in the vicinity of a given point.

\AkramiGA~also has some physics implications. One of the most interesting ones is the observation that contrary to the findings from some MCMC analyses, our best-fit point lies in the focus point region (\ie at high $m_0$) rather than the stau co-annihilation region. We however think that this discrepancy is likely to come from different physics codes used in calculating different observables rather than from the differences in the employed scanning techniques. In addition, our results uncover the existence of a section of the stau co-annihilation region at large $m_0$ that seems to have been commonly neglected in all previous scans (see \fig{fig:GAs}). The paper also presents some other implications for Higgs and sparticle masses at the LHC, as well as quantities important in direct and indirect searches for dark matter.

We go on to further investigate the ability of scanning algorithms in correctly providing the confidence regions for SUSY models. This time, in \AkramiCOV, we look at a central requirement for properly constructed confidence regions, namely the `statistical coverage'. This means that a region corresponding to a specific confidence level must include the `true' values of the model parameters at the stated confidence level when the experiments are repeated infinitely many times. We study the coverage for the CMSSM when its parameter space is explored with Bayesian scanning techniques, in particular \MN. We perform the analysis for a simple case where only constraints from a direct detection experiment are imposed on the CMSSM parameter space. Two CMSSM benchmarks are chosen and their corresponding parameters are used as hypothetically true parameters. We then generate $100$ sets of synthetic direct detection data in each case and scan over the CMSSM parameters for each set. In order to obtain the degree of coverage, we construct one-dimensional confidence intervals and count how many times (out of $100$) a true parameter falls within the intervals. We assess the coverage when two types of priors are imposed on the parameter space: (1) flat priors on all parameters of the model and (2) logarithmic priors on the scalar and gaugino mass parameters $m_0$ and $m_{1/2}$. For comparison, we also examine the coverage for Bayesian credible intervals although we do not expect proper coverage for these cases.

\begin{figure}[t]
\begin{center}
\subfigure{\includegraphics[scale=0.25, trim = 40 230 130 123, clip=true]{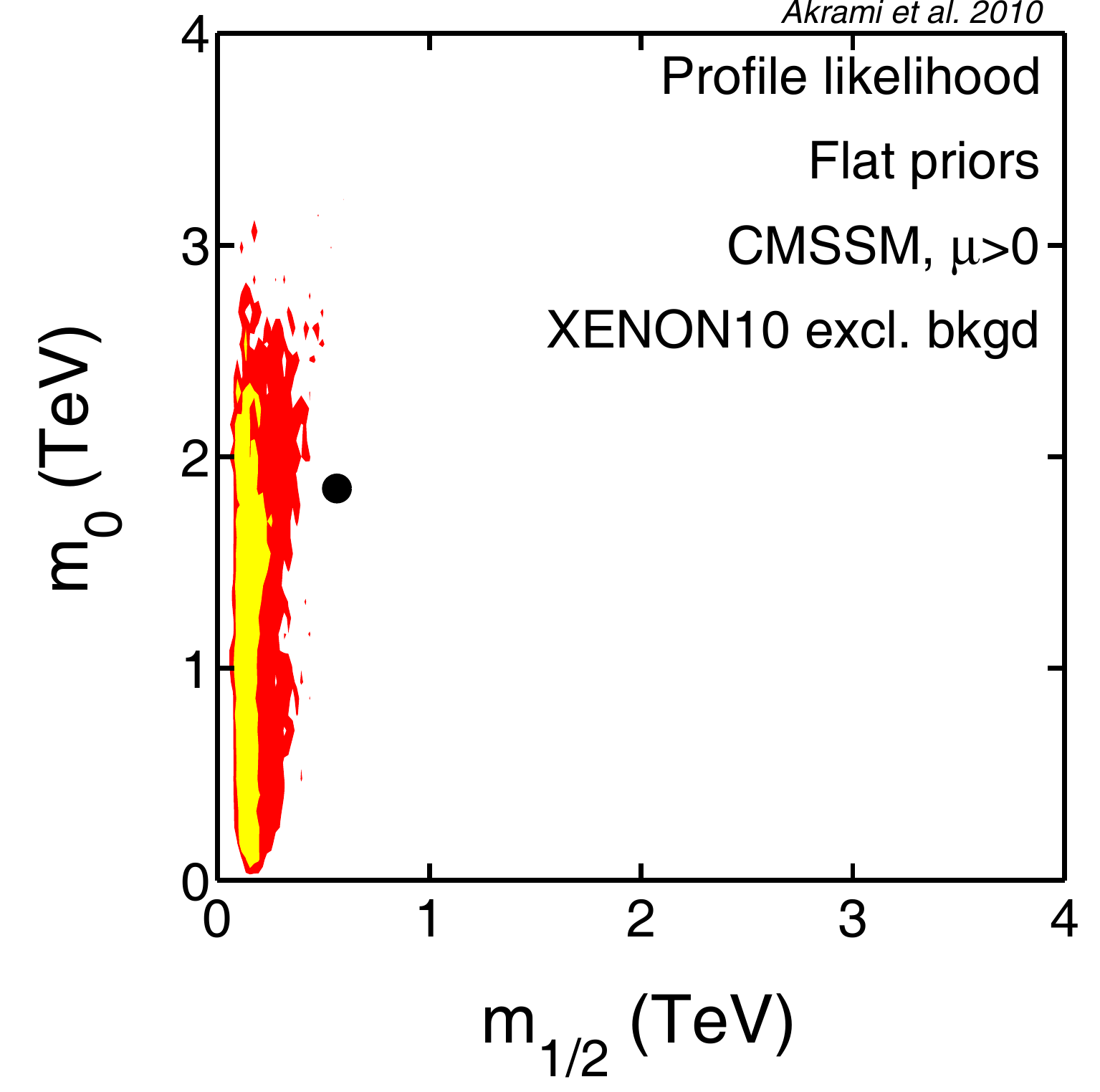}}
\subfigure{\includegraphics[scale=0.25, trim = 40 230 60 123, clip=true]{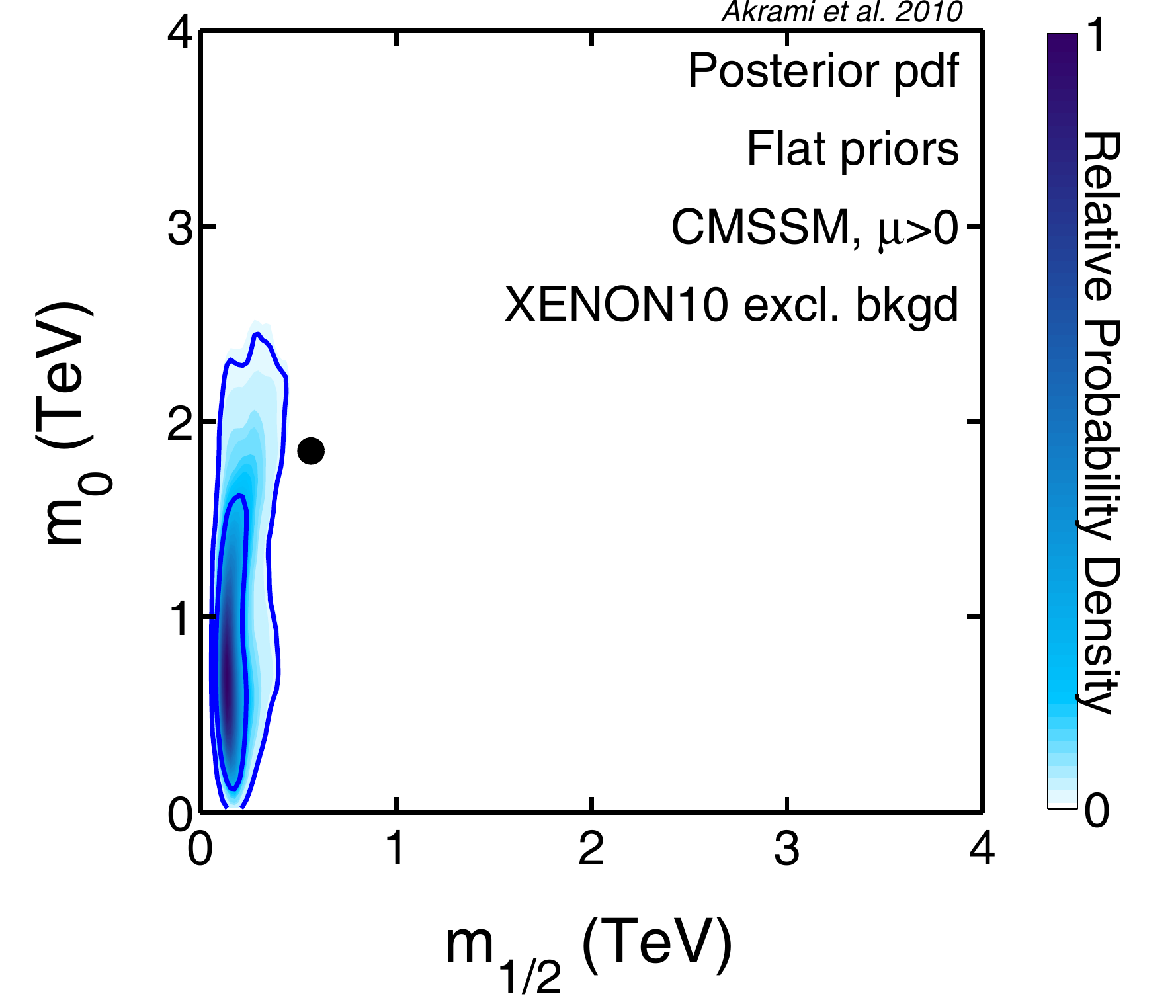}}\\
\subfigure{\includegraphics[scale=0.25, trim = 40 230 130 123, clip=true]{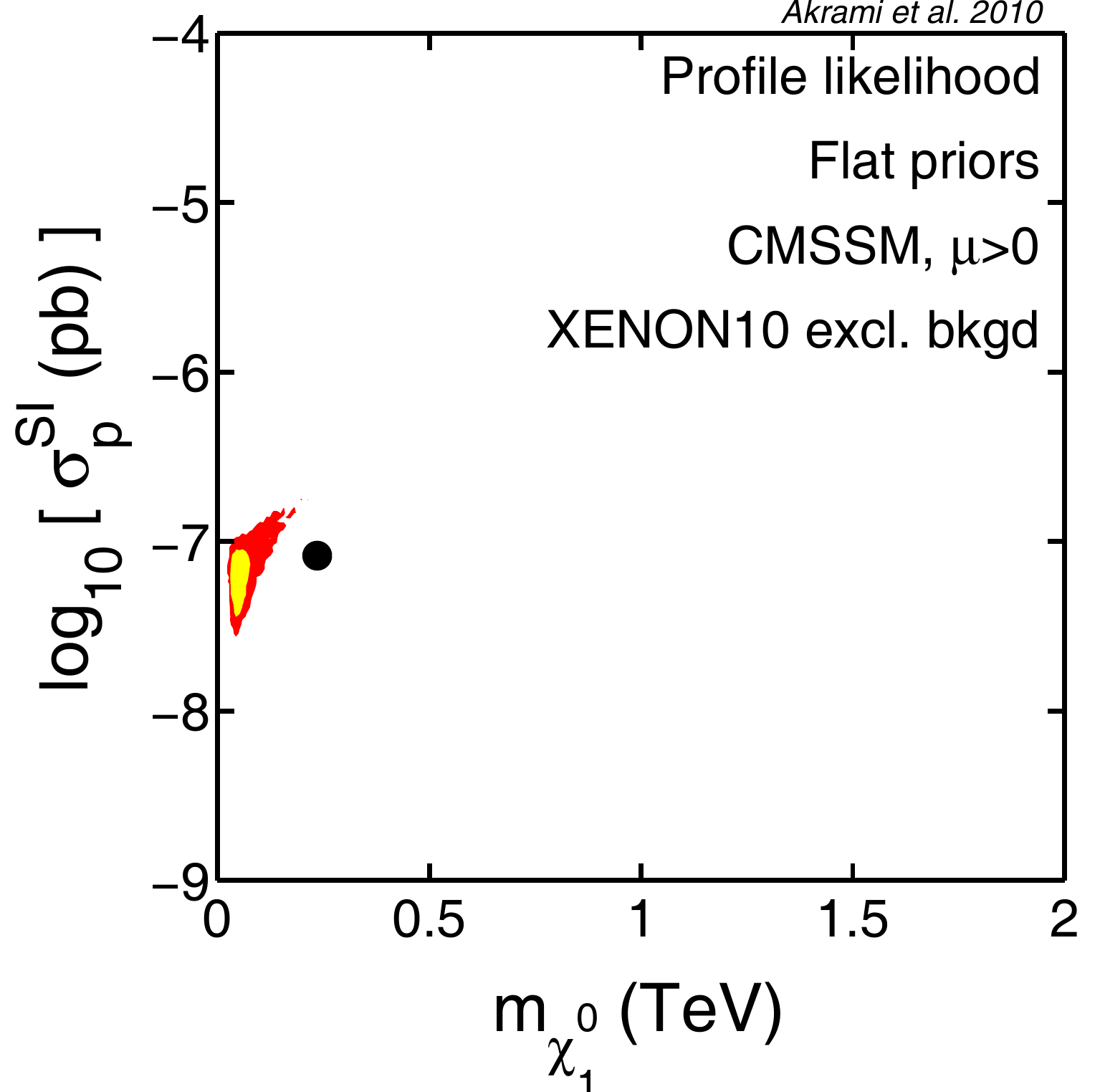}}
\subfigure{\includegraphics[scale=0.25, trim = 40 230 60 123, clip=true]{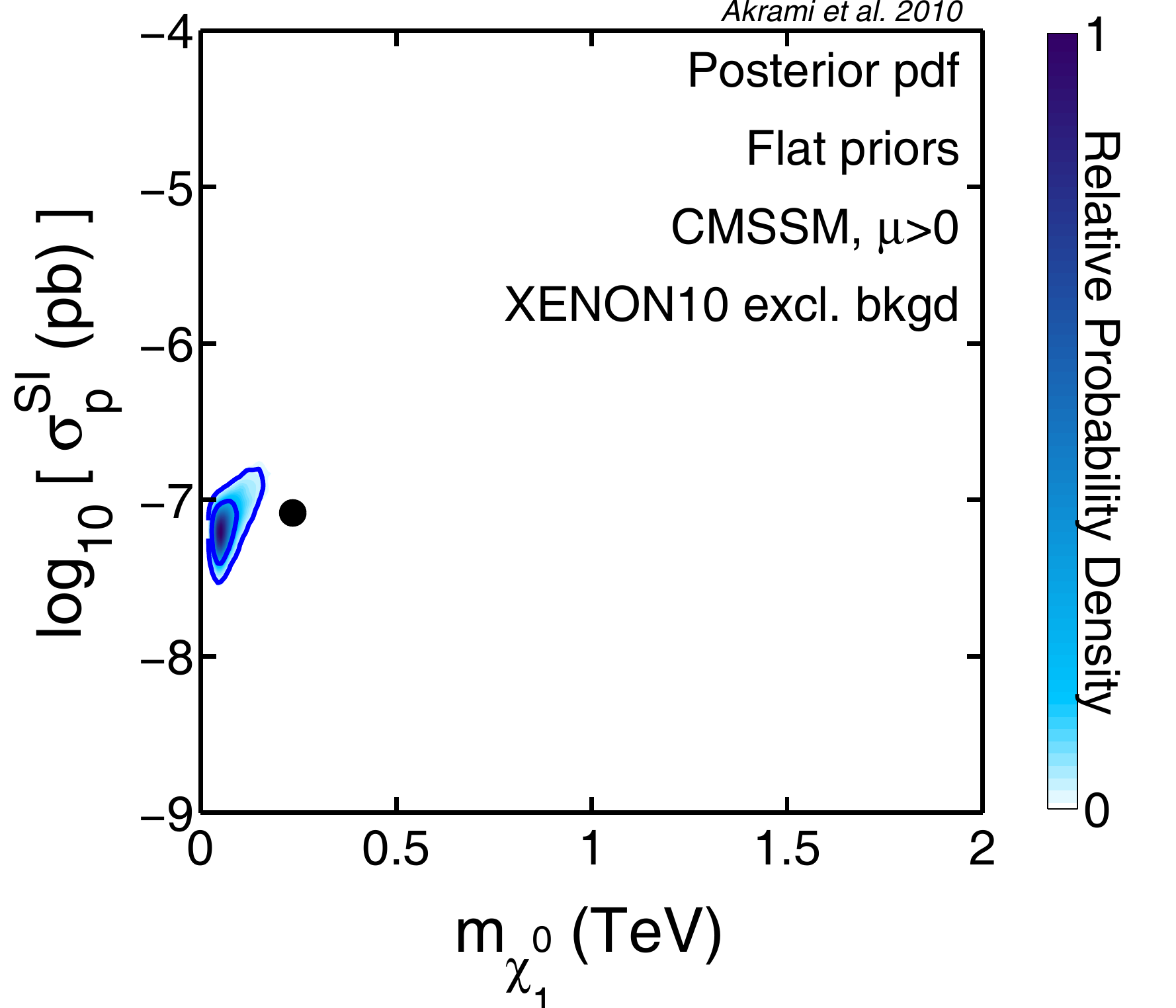}}\\
\caption[aa]{\footnotesize{Two-dimensional profile likelihoods (left) and marginalised posterior PDFs (right) for one typical coverage scan over the CMSSM parameter space with significant under-coverage. The inner and outer contours in each panel represent $68.3\%$ ($1\sigma$) and $95.4\%$ ($2\sigma$) confidence levels, respectively. Plots are shown in $m_0$-$m_{1/2}$ and $\sigma^{SI}_p-m_{\tilde\chi^0_1}$ planes. Black dots indicate the benchmark values which in this case lie outside the $1\sigma$ and $2\sigma$ regions.}}\label{fig:badcoverage}
\end{center}
\end{figure}

Our results indicate both over- and under-coverage that for some cases strongly vary when benchmarks or priors are changed (see \fig{fig:badcoverage} for an example of scans with under-coverage). The possible reasons for the observed poor coverage can be twofold: (1) The profile likelihood is mapped correctly by the scanning algorithm, but its validity as a proper approximation to the full Neyman construction (for details, see \eg \AkramiCOV~or ref.~\cite{Neyman,Feldman:1998}) of the frequentist confidence intervals breaks down for complex parameter spaces such as the CMSSM. In this case more sophisticated methods such as the `confidence belt' construction~\cite{Feldman:1998} could significantly improve the coverage. The problem with these alternative techniques is that they are rather difficult to implement numerically. (2) The profile likelihood is still a good approximation but the scanning algorithm has failed to correctly map it. This can clearly lead to poor coverage because the confidence regions and intervals are not properly constructed. Our analysis in \AkramiCOV~indicates that in our particular case of study, while option 1 may have played some role, it is more likely that a substantial fraction of the over- and (especially) under-coverage comes from the fact that the employed scanning algorithm is not optimised for the profile likelihood analysis (\ie option 2). For another coverage study of the CMSSM using a different type of experimental likelihood, see ref.~\cite{Bridges:2010de}.

\section{Outlook}

The coming decade in particle physics and cosmology will definitely be an exciting period in the history of mankind's great endeavour for understanding Nature at its most fundamental level. There are currently a large number of theories and models as extensions of the Standard Model that equally well describe all known phenomena while give different descriptions of physics at high energies. These theories are all waiting for experimental verification, and it would be of extreme interest to know whether any of them could be confirmed or excluded. Fortunately, many of these new physics frameworks have several predictions that could be tested observationally. With the advent of many new experiments and computational techniques in recent years with unprecedented power and precision, the prospects for verifying and constraining these theories are quite promising.

Most of the new physics models possess rather complex structures with large numbers of free parameter. This means that any phenomenological analysis that aims to properly compare new physics theories with observations should in principle be composed of four main elements: (1) a concrete theoretical \textbf{model} whose predictions for any set of free parameters are calculable, (2) relevant observational \textbf{data} corresponding to the theoretical predictions with the associated uncertainties, (3) the \textbf{statistical framework} and formalism for the analysis, and (4) an appropriate \textbf{scanning algorithm}.

The present thesis centres upon one of the most interesting new physics frameworks, namely weak-scale supersymmetry, and presents a number of powerful strategies and approaches for comparing its predictions with observations. A concrete SUSY model is studied (the CMSSM), various experimental data are employed (from direct and indirect searches for particle dark matter as well as collider constraints), different statistical frameworks are used and advanced scanning techniques are utilised.

Our results (in \Scott~and \AkramiDD) indicate that even though the current data do not give conclusive information about the validity of even the simplest versions of supersymmetric models and the preferred values of their parameters, the situation could dramatically change in the near future. We explicitly investigate the prospects for dark matter direct detection experiments, and our results show great promise for the next generations of these experiments in characterising supersymmetric models. Future indirect detection experiments are also potentially very encouraging especially because there are various targets to look at and different annihilation channels to use. However, perhaps the landmark in these directions would be the LHC results from searches for physics beyond the SM, including SUSY. Analyses similar to ours could be performed using data from the LHC (see \eg refs.~\cite{Roszkowski:2009ye,Bertone:2010rv,Buchmueller:2011aa}). Ideally, all different sets of cosmological, astroparticle (direct and indirect searches) and collider constraints should be eventually combined in a statistically consistent global-fit setup where are sources of uncertainties are taken into account. 

As far as the statistical frameworks and scanning algorithms are concerned, our investigations in \AkramiGA~and \AkramiCOV~imply that there are several subtleties that should be considered when working in any framework and with any algorithm. Supersymmetric models in particular, exhibit quite non-trivial parameter spaces that are constrained differently when different statistical formalisms are employed. This also makes it difficult for a scanning method to properly provide favoured values of the parameters in all statistical frameworks. Methods optimised for one framework usually give unsatisfactory results for the other. Some of the existing scanning techniques are highly powerful, efficient and relatively fast, but their widely-used versions are usually optimised for Bayesian statistics. There have been some successful attempts to reconcile them with frequentist framework, but at the price of (significantly) increasing the computational effort. The latter is indeed a very important point. It should not be forgotten that the full phenomenologically interesting models (such as the MSSM in SUSY extensions) possess very large parameter spaces. Constraining such models requires a sufficiently fast scanning technique that provides the results in reasonable times. We therefore think that the efforts in constructing efficient scanning techniques for both statistical frameworks should still continue.

Finally, we should make the point here that what we have done so far have all been in the context of SUSY `parameter estimation', namely that a SUSY model is given (the CMSSM in our case) and we attempt to fit it to the data so as to find the most favoured values for the free parameters of the model. The statistical frameworks we have used can however be employed equally well in a slightly different context, namely `model selection'. As we broadly discuss in this thesis, there are several models of supersymmetry that are constructed based on different assumptions and motivations. Perhaps a natural way to go, when more constraining data become available, would be to compare different models in a statistically consistent way so as to see which models are completely excluded by experiments or at least which ones are more favoured. Ideally, this should not be restricted to SUSY models and other physics-beyond-the-SM theories should also be analysed. We do not go through a discussion of various techniques and strategies in this direct and only note that such attempts have already been started in the community although most of the current results appear to be inconclusive (see \eg ref.~\cite{AbdusSalam:2009tr}).

\end{fmffile}

\appendix
\include{appendix}

\part{Papers}
\label{papers}

\pagestyle{plain}

\paper{Pat Scott, Jan Conrad, Joakim Edsj\"o, Lars Bergstr\"om, Christian Farnier \& Yashar Akrami}{Direct constraints on minimal supersymmetry from Fermi-LAT observations of the dwarf galaxy Segue 1}{\jcap}{01}{031}{2010}{arXiv:0909.3300}{PaperI}{papI}
\paper{Yashar Akrami, Pat Scott, Joakim Edsj\"o, Jan Conrad \& Lars Bergstr\"om}{A profile likelihood analysis of the constrained MSSM with genetic algorithms}{\jhep}{04}{057}{2010}{arXiv:0910.3950}{PaperII}{papII}
\paper{Yashar Akrami, Christopher Savage, Pat Scott, Jan Conrad \& Joakim Edsj\"o}{How well will ton-scale dark matter direct detection experiments constrain minimal supersymmetry?}{\jcap}{04}{012}{2011}{arXiv:1011.4318}{PaperIII}{papIII}
\paper{Yashar Akrami, Christopher Savage, Pat Scott, Jan Conrad \& Joakim Edsj\"o}{Statistical coverage for supersymmetric parameter estimation: a case study with direct detection of dark matter}{\jcap}{07}{002}{2011}{arXiv:1011.4297}{PaperIV}{papIV}

\end{document}